# Memory Effects in Quantum Processes

Philip Taranto

Dr. Kavan Modi & Dr. Felix A. Pollock

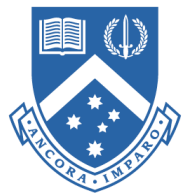

Master of Science (Research)
School of Physics and Astronomy
Monash University

September 10, 2019

Philip Taranto: *Memory Effects in Quantum Processes,* © September 10, 2019.

To my family,

for their unwavering support.

ABSTRACT

Understanding temporal processes and their correlations in time is of paramount importance for the development of near-term technologies that operate under realistic conditions. Capturing the complete multi-time statistics defining a stochastic process lies at the heart of any proper treatment of memory effects. This is well-understood in classical theory, where a hierarchy of joint probability distributions completely characterises the process at hand. However, attempting to generalise this notion to quantum mechanics is problematic: observing realisations of a quantum process necessarily disturbs it, breaking an implicit, and crucial, assumption in the classical setting. This issue can be overcome by separating the experimental interventions from the underlying process, enabling an unambiguous description of the process itself and accounting for all possible multi-time correlations for any choice of interrogating instruments.

In this thesis, using a novel framework for the characterisation of quantum stochastic processes, we first solve the long standing question of unambiguously describing the *memory length* of a quantum processes. This is achieved by constructing a *quantum Markov order* condition that naturally generalises its classical counterpart for the quantification of finite-length memory effects. As measurements are inherently invasive in quantum mechanics, one has no choice but to define Markov order with respect to the interrogating instruments that are used to probe the process at hand: different memory effects are exhibited depending on how one addresses the system, in contrast to the standard classical setting. We then fully characterise the structural constraints imposed on quantum processes with finite Markov order, shedding light on a variety of memory effects that can arise through various examples. Lastly, we introduce an instrument-specific notion of *memory strength* that allows for a meaningful quantification of the temporal correlations between the history and the future of a process for a given choice of experimental intervention.

These findings are directly relevant to both characterising and exploiting memory effects that persist for a finite duration. In particular, immediate applications range from developing efficient compression and recovery schemes for the description of quantum processes with memory to designing coherent control protocols that efficiently perform information-theoretic tasks, amongst a plethora of others.

# PUBLICATIONS

Many of the ideas and figures contained in this thesis have appeared in the following publications: [1–3].

[1] P. Taranto, F. A. Pollock, S. Milz, M. Tomamichel, and K. Modi, Phys. Rev. Lett. **122**, 140401 (2019).

[2] P. Taranto, S. Milz, F. A. Pollock, and K. Modi, Phys. Rev. A **99**, 042108 (2019).

[3] P. Taranto, F. A. Pollock, and K. Modi, arXiv:1907.12583 (2019).

*Memory is not an instrument for surveying the past but its theatre*
— Walter Benjamin.

ACKNOWLEDGEMENTS

First of all, I must thank my supervisors, Kavan Modi and Felix Pollock, for their support and guidance throughout my journey. Both of them have provided a perfectly balanced atmosphere of ongoing encouragement, diligent rigour and healthy scepticism that has undoubtedly helped me develop as a scientist and as a person. Kavan's quick-witted counterexamples to my many outlandish claims saved me from pursuing countless dead-ends and taught me that often *the problem is your friend.* Felix's attention to details and grammatical pedantry immensely improved my abilities in scientific communication and cured my unhealthy obsession with semi-colons.

Secondly, my deepest thanks goes out to my de facto supervisor, Simon Milz. It has been a pleasure to have learnt many fun facts about an undisclosed European nation thanks to you... oh, and a thing or two about physics. It's been a wild ride so far and now we set sail for tomorrow!

I would like to express my gratitude to the School of Physics and Astronomy at Monash University. I couldn't have asked for a more supportive environment in which to study both my undergraduate and masters degrees. In addition, I appreciate the financial support of the Australian Government Research Training Program (RTP) and the J. L. William Scholarships.

I especially thank the entire MonQIS group for their friendship, in particular Simon Milz, Francesco Campaioli, Josh Morris, Top Notoh and the honorary member, Cody Duncan. I am yet again thankful to Simon Milz for his thorough revisions of this thesis, Lee Miles for providing inspiration in times of need, and Ashley Bransgrove, Vanessa La Delfa and John Farrugia for their thoughtful comments on an early draft.

Lastly, thanks to my family and friends for their support throughout this journey.

Melbourne, 13 March, 2019.

# Contents

# List of Figures

# List of Tables

## ACRONYMS

ME      Master equation

P       Positive

CP      Completely-positive

TP      Trace-preserving

CJI     Choi-Jamiołkowski isomorphism

CPTP    Completely-positive and trace-preserving

KET     Kolmogorov extension theorem

IC      Informationally-complete

POVM    Positive operator-valued measure

CMI     Conditional mutual information

GKSL    Gorini, Kossakowski, Sudarshan and Lindblad

## Part I

## A KIND OF GLORY

Sometimes a kind of glory lights up the mind of a man. It happens to nearly everyone. You can feel it growing or preparing like a fuse burning toward dynamite. It is a feeling in the stomach, a delight of the nerves, of the forearms. The skin tastes the air, and every deep-drawn breath is sweet.

— John Steinbeck, *East of Eden*.

# 1

INTRODUCTION

Although physical laws are fundamentally local in time, memory effects are ubiquitous in processes observed in nature [4–6]. We see such effects when we try to predict the weather or the stock market, describe transport processes at the microscopic level or understand the random motion of particles suspended in a fluid, to name but a few examples. Memory arises because, in reality, no system is truly isolated; our inability to capture interactions between the system of interest and its environment leads to dynamics that can exhibit complex temporal correlations. Through these interactions, information about the system's past can be stored in the environment, which carries it forward to dictate the future evolution of the system.

In classical physics, should an experimenter be equipped with sufficient resources to track the evolution of *all* relevant degrees of freedom—including those of the environment—a deterministic, memoryless description of the evolution could, in principle, be derived. In practice, however, resource constraints quickly banish such lofty ambitions to the realm of the idyllic. *A priori*, one does not know the structure of system-environment interactions concerning complex phenomena in precise detail; even if one did, for sufficiently large environments, such a description rapidly defies feasibility with respect to the amount of data required to be stored and manipulated in a reasonable time on a reasonable computer.

From an operational perspective, it is desirable to understand properties of a stochastic process from information that can be inferred from probing the system of interest alone. Intuitively, in this sense a stochastic process refers to the joint probability distribution that expresses the likelihood of a quantity taking certain values: the probability for a certain stock to have price $x_1$ *and* $x_2$ *and* $x_3$ on three consecutive days, for instance. Fluctuations of the stock price on any given day can directly influence the rest of the market (*i. e.,* its environment), which in turn impacts the price of the original stock at a later time.

Such memory effects are encoded as correlations in the joint probability distribution over the relevant timesteps and can manifest themselves as genuinely multi-time correlations. In the simplest non-trivial memoryless scenario, a process can exhibit only two-point correlations: the probability distribution of tomorrow's stock price only depends upon today's price, and not any further back in the history. However, more generally, when memory effects are at play, all multi-time correlations must be considered to unambiguously describe the process, as joint effects between sequences of events can play a significant role in the future evolution.

When attempting to generalise this understanding of stochastic processes to the quantum realm, perhaps unsurprisingly, the intricacies involved come to light. Just like in the classical realm, any realistic phenomenon must be described within the theory of *open systems* to account for the stochasticity that arises due to our subjective ignorance of the degrees of freedom of the environment [7–9]. In this formalism—as is the case for classical processes—dynamical phenomena are described solely in terms of accessible quantities that are derivable from the system of interest alone. The open systems framework has enjoyed tremendous recent success, translating fundamental theories into real-world predictions, and has led to a multitude of tools and techniques for manipulating quantum processes which have fostered many technological advances.[1] However, in stark distinction to the classical setting, here we must also deal with the fact that additional randomness arises at an elementary level through the very nature of *measurement* at the microscopic scale.

In quantum mechanics, measurements are fundamentally invasive, which seemingly leads to incompatibility of measurement statistics observed in time [13, 14]. Indeed, such invasiveness leads directly to a breakdown of the *Kolmogorov extension theorem* [15], which importantly links the operational description of a classical stochastic process in terms of joint statistics measured over a set of times to an underlying continuous-time mechanism [16]. Whenever measurements can directly influence the state of the system, it becomes seemingly impossible to define the process independently of the experimental interventions. This scenario is pertinent to both classical causal modelling [17], where an experimenter intervenes with the system in order to deduce causal relations between events, and quantum theory more generally [15, 18–20]. In any such theory, the hierarchy of joint probability distributions alone does not tell the entire story regarding the underlying process; thus, we are seemingly at a loss in answering *what we really mean when we talk about a quantum stochastic process.*

---

1 See, for instance, early work regarding the development of error-correcting codes for resilient quantum computing and the dynamical decoupling protocol for effective quantum control in Refs. [10, 11] and [12] respectively.

The lack of equivalence between the accessible and the underlying physical descriptions of quantum stochastic processes has irked the open systems and quantum information theory communities for some time, leading to nonequivalent definitions of key concepts pertinent to open quantum evolution; the most relevant to our present interests being that of *memorylessness*. Classically, a memoryless process is one for which the statistics observed at any point in time only depend on the most recent state of the system. Satisfaction of this condition has profound implications that importantly lead to a significantly simplified description of the process; indeed, the study of memoryless processes forms an entire branch of mathematics, and the evolution of many systems is frequently approximated to be memoryless, because of both the relative ease with which they can be analytically and computationally manipulated and the experimental evidence supporting this simplification [4–6, 21, 22].

However, the complications discussed above regarding a multi-time description for general quantum processes make it challenging to define the process independently of the interventions applied by the experimenter. Without such a description, there is no meaningful way to take intermediate measurements into account and check for conditional independence of the future evolution from the history. Thus, until recently, there has been no unique condition to define memorylessness in quantum mechanics. Nonetheless, many traditional approaches of the open systems formalism focus on the time-evolving state of the system of interest, which can provide valid witnesses for the presence of memory effects [23–25]. Whilst of immense practical importance, such descriptions do not serve to fully characterise the process, as they are limited in scope to capturing two-time correlations, specifically those between that of the state prepared at some point in time and the subsequent measurement statistics deduced at any later time. As is also true in the classical setting [26], when memory effects play a non-negligible role, all multi-time correlations must be considered.

Various frameworks circumvent the crucial problem of formalism that arises due to the invasiveness of measurements in quantum mechanics by actively taking measurements and *controllable* manipulations of the system of interest into account, enabling the separation of the underlying, *uncontrollable* process *per se* and the influence enacted by the experimenter. These were initially introduced by Lindblad [27] and Accardi, Frigerio and Lewis [28]. Modern variants of similar formalisms have been applied to study general quantum circuit architectures [29–32], foundational aspects of causal-ity in quantum theory [33–38], quantum causal modelling [18–20, 39], quantum theory in spacetime [40–42], quantum game theory [43, 44], generalised communication protocols [45], non-equilibrium quantum thermodynamics [46, 47], and quantum processes with memory [1, 2, 15, 48–56].

Indeed, this school of thought has led to the development of a generalised Kolmogorov extension theorem that holds in any generalised probabilistic theory—including quantum theory—crucially giving rise to an operational definition of quantum stochastic processes [15, 28]. Most relevant to our present purposes, the breakthrough result of Ref. [50] provides an unambiguous criterion for a quantum process to be memoryless, thereby unifying all previous approaches. With this comprehensive mathematical language that captures all possible multi-time correlations at hand, one can properly describe quantum stochastic processes with memory independently of the experimenter and accurately understand important phenomena in which temporal correlations play a significant role, such as, *e. g.,* the emission spectra of quantum dots [57] and the vibrational modes of interacting fluids [58].

The discussion so far has centred on the existing literature; now we move to focus on the developments of this thesis. While the previously discussed frameworks are perfectly tailored to unambiguously define memory effects in quantum mechanics—and describe processes that display them—a thorough analysis of their structure, length, and strength is still missing. In particular, the concept of Markov order, which is regularly invoked in the study and simulation of classical stochastic processes with finite-length memory effects, has not been generalised to, or analysed in the context of, quantum mechanical processes. Such an investigation is of tremendous practical importance, as, although temporal correlations in complex phenomena are exhibited over various timescales, Markov order provides a natural notion of *memory length* that emerges in the context of statistical modelling, namely the amount of a system's history that directly affects its future dynamics [5].

Classically, the concept of *Markov order*, $\ell$, dictates that the statistics describing a system of interest at a given time only depend upon knowledge of its past $\ell$ observed states. Processes with finite Markov order therefore permit a significant reduction in modelling complexity: one must only estimate the conditional transition probabilities from the most recent set of observations, rather than the exponentially many more parameters for each additional timestep further back in the history [59–61]. Fortunately, many complex processes typically have an effectively finite-length memory, allowing for an efficient description that only considers the relevant portion of history necessary to optimally predict the future [4–6]. In these cases, it is our understanding of the *memory strength* across a given duration of time that justifies the suitability of invoking finite-memory approximations

A thorough understanding and characterisation of memory effects in general stochastic processes is of crucial importance from both a theoretical and practical standpoint. Many theoretical developments regarding quantum systems have shown a superiority over their classical counterparts with respect to the successful enactment of certain information-processing tasks [12, 62–64]. Explicit understanding of memory effects will be of ever increasing importance and it is clear that future quantum technologies will need to embrace memory in order to display these advantages under realistic conditions [65].

The aforementioned modern frameworks pave the way for an unambiguous foray into the study of memory effects in quantum processes and provide the starting point for this thesis. There are, as we see it, a number of aspects regarding memory that must be considered for a holistic comprehension, namely the duration of time over which memory effects persist, the strength of their impact and the complexity of their simulation. The main concern and achievement of the present work regards extending the concept of finite memory length and the quantification of memory strength to quantum stochastic processes, and investigating the subsequent implications. Leveraging this vantage point, these concepts serve to provide a cohesive framework pertaining to the characterisation, simulation and exploitation of memory.

## 1.1 READING GUIDE

The outline of this thesis is as follows. In Part II, we present the story so far: the current understanding of physical processes in the presence of noise and memory effects therein. No new results are presented here; rather, the purpose of this part is to interrogate why a proper characterisation of memory effects in quantum processes has hitherto seemed an insurmountable task, and elucidate how recent theoretical advancements have provided a resolution, thereby suitably equipping us to overcome this concern. We begin in Chapter 2, where we first discuss classical stochastic processes in order to lay the foundations of some core concepts, before turning our attention to the quantum realm. Here, as we shall see, things become slightly more nuanced: although the standard approaches of the open systems paradigm are adequate to describe memoryless processes, they turn out to be fraught with problems in describing those with memory, as explored throughout the culminating discussion of this chapter. Chapter 3 formally presents the mathematical formalism that circumvents the issues at hand by way of the process tensor framework. The first half of this thesis is thus deliberately pedagogical: many of these modern ideas have emerged in the context of a variety of disparate studies, but with a recently revamped understanding of quantum stochastic processes, we are now in a position to synthesise them into a cohesive story, as we attempt to undertake here.

With this unambiguous description of quantum stochastic processes at hand, we are then ready to present the novel results of this thesis, which are contained in Part III. We begin, in Chapter 4, by formalising the notion of memory length: we first generalise the concept of Markov order to open quantum processes, before exploring in detail the structure of processes with finite-length memory in Chapter 5. Many of the results presented here are developed through a number of illustrative examples that aim to build intuition rather than through general mathematical analysis, although the conclusions hold generically. Following this examination, in Chapter 6 we quantify the memory strength of a given process, analysing our proposed measures through application to an exactly solvable dynamics.

The main original contributions in this work are summarised as follows. In Chapter 4, we show that the introduced notion of quantum Markov order can be expressed as a constraint on the process tensor, provided in Eq. (4.4). We prove that only memoryless quantum processes can display finite Markov order with respect to all possible sequences of interrogations in Theorem 4.2. Nonetheless, it is sensible to consider the memory length in quantum processes with respect to specified choices of instruments; indeed, the classical notion of Markov order is one such special case. In Chapter 5, we provide the general description of a process with finite quantum Markov order in Theorem 5.1. Examples 5.1–5.3 highlight the implications for processes with finite memory length with respect to natural choices of manipulations applied to the system, including unitary transformations, measurements followed by independent repreparations, and sharp, projective measurements. In light of the general structure deduced, we subsequently show that such processes do not necessarily have vanishing quantum conditional mutual information through Theorem 5.3. In Chapter 6, we introduce a number of measures for the memory strength of a process with finite quantum Markov order, particularly focusing on various important operational scenarios, such as when an experimenter does nothing to the system or, on the other extreme, actively tries to erase temporal correlations. The chapter culminates with an explicit study of the behaviour of this instrument-specific memory strength for a tuneable two-qubit model.

Lastly, in Part IV, we conclude with a summary of our present work and a brief outlook. Note that in Appendix A we provide a synopsis of the notational conventions, as well as explicitly define a number of common functions and outline a conceptually important colour-coding scheme, that we employ throughout this thesis in order to aid the reader. Before proceeding, we wish to make the following disclaimer: to enhance legibility, we tend to cite references in which conceptual points of interest have been studied at their initial introduction, and refrain from repeatedly citing such references throughout subsequent discussion, except for the cases where direct results are pertinent or the citation is particularly poignant.

# Part II

# NARRATIVE ON THE EDGE

The nature of parties has been imperfectly studied. It is, however, generally understood that a party has a pathology, that it is a kind of individual and that it is likely to be a very perverse individual. And it is also generally understood that a party hardly ever goes the way it is planned or intended.

— John Steinbeck, *Cannery Row*.

# 2

## CLASSICAL AND QUANTUM DYNAMICS WITH NOISE

A NY REALISTIC MODEL OF OBSERVED PHENOMENA must take into account the possibility of randomness. That we are often interested in quantities that are not deterministically predictable in advance, but rather exhibit some random variation, dictates that our descriptions be built from probabilistic models. Stochastic processes are ubiquitous in nature: they arise in a number of ways throughout the physical, biological and social sciences [4–6]. One way is through our *subjective ignorance.* Consider, for instance, a coin toss. Here, the classical laws of mechanics deem the underlying process to be deterministic—if an experimenter were to flip a coin in *exactly* the same fashion with *exactly* the same external conditions every time, they could predict each outcome with certainty.

However, realistically, this is not possible to achieve. Even if the experimenter could precisely apply the same force at the same point on the coin when initiating each flip, in each trial the coin is subject to different external factors that influence its trajectory, *e. g.,* those due to relentless collisions with molecules in the air. Although it is possible, in principle, to track enough variables to account for all of the additional degrees of freedom that the coin interacts with—thereby deterministically modelling the evolution—this is practically unfeasible due to sheer weight of numbers. Although the process is fundamentally deterministic, we are forced to deal with the fact that the outcomes observed look random to us due to our subjective ignorance, which is the root of all classical randomness.

Classical stochastic processes are thus characterised by a joint probability distribution over sequences of events in time. Crucially, from the correlations encoded in this multi-time distribution, all memory effects of the process can be deduced. For instance, a special case that has been extensively studied due to its particularly simple structure are *Markovian* or *memoryless* processes, in which only two-point correlations between events on adjacent timesteps are present. Here, the statistics observed at any point in time are completely determined by the most recently observed state. A more general scenario where memory effects persist for a finite length of time is captured by the notion of *Markov order*, $\ell$, in which the conditional statistics at any point in time only depend upon the most recent $\ell$ observations. In this case, multi-time effects can play a significant role, and the complexity of describing such processes grows exponentially in the length of the memory. Nonetheless, when the memory length is substantially less than the number of timesteps over which the process is defined, higher-order Markov models provide an alluring reduction in the computational resources required for accurate simulation [4–6].

Perhaps unsurprisingly, describing stochastic processes in quantum mechanics is somewhat more challenging. Here, in addition to randomness that arises from subjective ignorance, the outcomes of measurements are *fundamentally* random. Even in the static scenario, *i. e.,* when repeatedly measuring one and the same quantum state, we cannot describe the properties of any quantum system with certainty—the best we can do is to repeatedly perform experiments and make measurements to build up statistics that allow us to infer the quantum state. Just as in the classical case, our subjective ignorance can also lead to randomness, which necessitates treating realistic evolutions with the theory of open quantum systems [8, 9]. This formalism considers a joint system-environment state evolving in accordance with the laws of quantum mechanics, with the environmental degrees of freedom being regarded as inaccessible.

One way to understand correlations in time is via the dynamical map formalism. It bypasses the underlying details of the system-environment dynamics by expressing the effective mapping of the instantaneous quantum state of the system from some point in time to a later one. By preparing input states and performing subsequent measurements on the output states, the dynamical map prescribing the evolution between the two timesteps can be uniquely determined [9, 52, 66]. Importantly, any such description of a stochastic process, which only captures correlations between *pairs* of timesteps—as are frequently used in the experimental study of open quantum systems [23–25]—do not tell the whole story. Indeed, many non-Markovian processes can lead to two-point measurement statistics that could be incorrectly classified as Markovian [26, 50]. Specifically, when there is no memory in the process, two-point characterisations suffice to provide the correct description of the process at hand, but are insufficient otherwise.

As summarised by van Kampen [26]: "non-Markovian processes...cannot be considered merely as corrections to the class of Markov processes but require special treatment". What is meant by this quote is that, in order to properly characterise processes with genuine memory effects, one must take into consideration all of the possible multi-time correlations, as two-point descriptions necessarily lack such information, which may prove vital in the evolution. However, as discussed in Chapter 1, the invasiveness of measurements in quantum theory make it difficult to delineate between the active influence of the experimenter and the underlying process, which proves problematic in developing a multi-time description of quantum processes [15]. Since our aim is to understand memory effects in such processes, the first problem we must address in this thesis is an unambiguous understanding and operationally meaningful description of *quantum stochastic processes* that explicitly captures multi-time correlations.

In this chapter, we formally introduce classical stochastic processes and the formalism of open quantum systems in Sections 2.1 and 2.2 respectively. Stochastic processes are well-understood in classical theory, but not so in the quantum realm: there, for instance, a hierarchy of notions of non-Markovianity abound, which do not agree with each other in general. We present a brief literature review of such witnesses for non-Markovianity in order to highlight important gaps in the previous state of knowledge in Section 2.3. The incompatibility of the myriad of such definitions has been artificially reconciled by many through the belief that there exists no unique condition for Markovianity in the quantum realm. The overarching goal of this chapter is to explain why this is not true: we reiterate how the problems at hand arises from a breakdown of the standard formalism to properly describe quantum processes with memory, rather than a fundamentally irreconcilable problem. This motivation leads us naturally to an operationally meaningful framework for unambiguously describing quantum stochastic processes, which is the focus of Chapter 3.

## 2.1 CLASSICAL STOCHASTIC PROCESSES

We begin this section by presenting the key concepts needed to describe states of classical physical systems, before focusing on how these evolve over time. Most of the notions introduced here are explored within various contexts in a number of excellent textbooks, especially Refs. [4–6, 16, 21, 22].

### 2.1.1 *Probability Spaces*

The first primitive concept we need is that of a *sample space*, $\Omega$, denoting the set of all possible outcomes for an experiment. This set can either be discrete—*e.g.,* the experiment of flipping two coins, each with outcomes heads (H) or tails (T), has the sample space $\Omega = \{\text{HH}, \text{HT}, \text{TH}, \text{TT}\}$—or continuous—*e.g.,* the experiment of randomly selecting a chord on a circle of given radius, which has the sample space $\Omega = \{a, b :$ $a$ and $b$ are points on the circle$\}$.

Any subset of the sample space $\omega \subseteq \Omega$ can correspond to an *event*, which represents something that an experimenter can resolve or might be interested in. In the experiment of tossing two coins, the event that *at least one head occurs* is represented by $\omega = \{\mathtt{HH}, \mathtt{HT}, \mathtt{TH}\}$. The collection of such events describes the overall *event space* of the experiment, $\mathcal{G}$, which constitutes a $\sigma$-algebra.

**Definition 2.1** ($\sigma$-Algebra)**.** Given some set $\Omega$, let $\wp(\Omega)$ represent its power set. A subset $\mathcal{G} \subseteq \wp(\Omega)$ is a $\sigma$-algebra if it satisfies:

1. $\mathcal{G}$ contains the set $\Omega$ itself: $\Omega \in \mathcal{G}$.
2. $\mathcal{G}$ is closed under complement: if $g \in \mathcal{G}$, then $(\Omega \backslash g) \in \mathcal{G}$.
3. $\mathcal{G}$ is closed under countable unions: for $\{g_i\} \in \mathcal{G}$, their union $g = \cup_i g_i \in \mathcal{G}$.

These properties immediately imply that the emptyset is an element of $\mathcal{G}$ and that $\mathcal{G}$ is also closed under countable intersections (via De Morgan's law) [5].

A set $\Omega$ and a $\sigma$-algebra $\mathcal{G}$ together constitute a *measurable space*, $(\Omega, \mathcal{G})$. The specific $\sigma$-algebra pertinent to an experiment is dictated by the events of interest, or, put differently, the questions deemed important by the experimenter or by what can be resolved. Regarding the example of tossing two coins and asking question: *did at least one head occur?*, the particular $\sigma$-algebra chosen is $\mathcal{G} = \{\emptyset, \mathtt{TT}, \{\mathtt{HH}, \mathtt{HT}, \mathtt{TH}\}, \Omega\}$, with the interpretation of individual events respectively being: *the experiment was not performed, no, yes*, and *the experiment was performed*.

The final necessary ingredient is the assignment of probabilities to events through a *probability measure*, $\mu : \mathcal{G} \to \mathbb{R}^+ \cup \{0\}$, which maps each event to non-negative real numbers in the following logical manner.

**Definition 2.2** (Probability measure)**.** A measure $\mu$ defined on a measurable space $(\Omega, \mathcal{G})$, where $\Omega$ is some set and $\mathcal{G} \subseteq \wp(\Omega)$ is a $\sigma$-algebra, is a probability measure if it satisfies:

1. The probability of any event is non-negative and real: $\mu(g) \in \mathbb{R}^+ \cup \{0\} \ \forall \ g \in \mathcal{G}$.
2. Some event occurs with certainty: $\mu(\Omega) = 1$.
3. The probability associated to any union of pairwise disjoint subsets of the event space is the sum of probabilities associated to each subset ($\sigma$-additivity): $\mu(\cup_i g_i) = \sum_i \mu(g_i)$.

These axioms immediately imply the following elementary properties:

1. *Probability of the Emptyset*: $\mu(\emptyset) = 0$.
2. *Monotonicity*: if $g_A \subseteq g_B \in \mathcal{G}$ then $\mu(g_A) \leq \mu(g_B)$.
3. *Boundedness*: $0 \leq \mu(g) \leq 1$ for all $g \in \mathcal{G}$.

Overall, the triple $(\Omega, \mathcal{G}, \mu)$ formally describe a *probability space*, providing the mathematical underpinning of any probabilistic theory [16].

In describing properties of a physical system, an experimenter is often interested in some function of the outcomes of an experiment rather than the outcomes themselves, such as average values. These scenarios are best encapsulated with the notion of a *random variable*, $Y : \Omega \to \Gamma$ (these are usually taken to be real-valued functions, *i. e.*, $\Gamma = \mathbb{R}$, as is assumed from now on), which labels possibly abstract events in a meaningful way, *i. e.*, such that

$$Y^{-1}(y) = g \in \mathcal{G} \quad \forall \, y \in \mathcal{Y}, \tag{2.1}$$

where $\mathcal{Y}$ is a new $\sigma$-algebra generated by a collection of subsets of $\Gamma$. Since $\Omega$ is the sample space of a probability space, the new measurable space $(\Gamma, \mathcal{Y})$ naturally inherits the probability measure $\mu$ through a pushforward relation, which defines a *probability distribution* $\mathbb{P} : \mathcal{Y} \to [0,1]$ via[1]

$$\mathbb{P}(y) = \mu(Y^{-1}(y)). \tag{2.2}$$

The properties of the probability measure ensure that any such probability distribution is non-negative, normalised and additive over its $\sigma$-algebra. In summary, a random variable provides a logical relabelling, allowing us to begin with some abstract probability space $(\Omega, \mathcal{G}, \mu)$ and end up with a probability space $(\Gamma, \mathcal{Y}, \mathbb{P})$ that is potentially more amenable to further computational manipulation. Although random variables introduce versatility into the description of stochastic processes, we will ultimately only be concerned in probabilities of events and so the choice of random variable will not be of particular interest. The probability distribution $\mathbb{P}$, which is directly accessible via statistical measurements, defines what we will call a *state* of a classical system.

---

1 Typically, such a distribution is labelled $\mathbb{P}_Y(y)$; in this thesis, we omit such unnecessary labels wherever possible for the sake of notational economy. Moreover, we are somewhat lax in often allowing the values $x, y, \ldots$ to represent either events or individual elements in the sample space, and using the same symbol $\mathbb{P}$ to denote either a probability measure or a probability density.

2.1.2 *Stochastic Processes*

In describing *stochastic processes*, one can build time into the picture by way of a parameterised random variable. Just as a classical state is described by a probability distribution assigning probabilities to events, a classical stochastic process can be defined as a function allocating probabilities to sequences of events over time. This formulation is as follows. Firstly, upon defining a random variable, $Y$, an infinite collection of random variables automatically arises, namely any quantity $X := f(Y)$ that is some function of $Y$. Selecting a value $Y = y \in \mathcal{Y}$ gives a deterministic value for $X = f(y) =: x \in \mathcal{X}$; specification of an event in this way is referred to as a *realisation* or *sample*. The transformation law relating the probability distributions over two such random variables is

$$\mathbb{P}(x) = \int \mathrm{d}y\, \delta(x - f(y))\, \mathbb{P}(y), \tag{2.3}$$

where $\delta$ denotes the Dirac-delta distribution. One could then consider a function that involves $Y$ and some auxiliary parameter, $t \in \mathbb{R}$, usually taken to denote time: $X_Y(t) := f(Y, t)$. Upon inserting for $Y$ one of its possible values $y$, we obtain a deterministic function of time $X_{Y=y}(t) = f(y, t) =: x(t) \in (\mathcal{X}, \mathbb{R})$, representing a realisation of the process or sample trajectory. The stochastic process itself is, overall, regarded as the probability distribution over trajectories

$$\mathbb{P}(x(t)) = \int \mathrm{d}y\, \delta(x(t) - f(y, t))\, \mathbb{P}(y(t)). \tag{2.4}$$

So far, we have introduced classical stochastic processes on axiomatic grounds. We now consider their relation to what is typically measured in experimental procedures. In practice, it is often sensible to store information about the sample trajectories observed in a discrete manner by specifying a finite number $n \in \mathbb{N}$ of timesteps, denoted $\Lambda_n := \{t_1, \ldots, t_n\}$, at which observations are made in order to build up a statistical description of the process.[2] The choice of these timesteps is, in principle, entirely up to the experimenter, although may be dictated by experimental capabilities. By repeatedly performing experiments, the experimenter yields statistics that describe the probability for the system of interest to take the value $x_1$ at $t_1$, $x_2$ at $t_2$, and so on, until $x_n$ at $t_n$.

---

[2] When such a set $\Lambda_n$ is taken to be an ordered sequence, we will indicate this with the subscript label $n:1$. In some cases, we will find it convenient to consider unordered sets of timesteps of a given cardinality, which we will indicate with the label $\Lambda_n$.

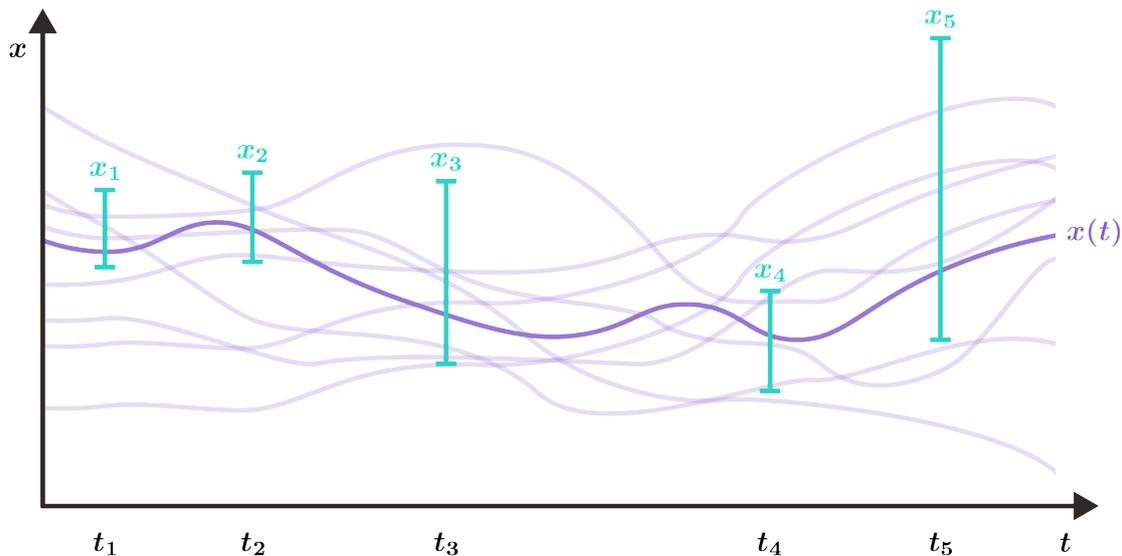

**Figure 2.1:** *Continuous and discrete-time stochastic processes.* In the continuous time picture, a stochastic process is characterised by a probability density $\mathbb{P}(x(t))$ over all possible sample trajectories $x(t)$ (purple). Alternatively, by measuring the system to be some regions $x_1, \ldots, x_5$ at times $t_1, \ldots, t_5$ (green), for instance, one deduces the multi-time statistics of the discrete-time joint probability distribution $\mathbb{P}_{5:1}(x_5, t_5; \ldots; x_1, t_1)$ that characterises the process on the chosen timesteps. Only the bold trajectory can be measured to be in the regions depicted, and hence occurs with unit probability, although considering all possible different spatial measurement settings assigns non-zero probabilities for each trajectory to be realised.

This gives rise to a hierarchy of joint probability distributions written as

$$\mathbb{P}_1(x_1, t_1) \tag{2.5}$$
$$\mathbb{P}_2(x_2, t_2; x_1, t_1)$$
$$\ldots$$
$$\mathbb{P}_{n:1}(x_n, t_n; \ldots; x_1, t_1).$$

For example, $\mathbb{P}_{n:1}(x_n, t_n; \ldots; x_1, t_1)$ might represent the probability to find a randomly moving particle in regions $\{x_1, \ldots, x_n\}$ respectively at times $\{t_1, \ldots, t_n\}$. We call a *joint probability distribution* over $n$ timesteps an $n$-point distribution, since it contains the necessary information to calculate all $n$-point correlations. Insofar as the experimenter is concerned, the hierarchy of finite joint probability distributions above serves to characterise the stochastic process over the chosen timesteps.

It is not clear, *a priori*, that these two notions of stochastic processes are equivalent. This is a subtle point, but worth considering in some detail, as it will become crucial to our understanding of quantum stochastic processes. For the moment, we will distinguish between an underlying continuous stochastic process, as described by Eq. (2.4), and that which is constructed on a finite number of timesteps, characterised by Eq. (2.5). Since the fundamental laws of physics are continuous in time, we are always implicitly assuming the existence of an underlying process that leads to the discrete-time statistics observed. In other words, we assume the existence of a continuous stochastic process that has all the finite ones that the experimenter measures as marginals. Beginning with Eq. (2.4) and specifying a finite set of timesteps, we can derive the hierarchy of joint distributions for all $t_j \in \Lambda_n$ as

$$\mathbb{P}_{j:1}(x_j, t_j; \ldots; x_1, t_1) = \int dy\, \delta(x_j - f(y, t_j)) \ldots \delta(x_1 - f(y, t_1))\, \mathbb{P}(y). \tag{2.6}$$

It is easy to show that Eq. (2.6) implies that the $n$-point distribution $\mathbb{P}_{n:1}$ contains within it the correct descriptor of the process on any subset of times, which is deducible via *marginalisation*. To derive the statistics of the process on any subset $\Lambda_k \subseteq \Lambda_n$, one simply marginalises over the outcomes on the timesteps that are no longer of interest

$$\mathbb{P}_{\Lambda_k}(x_{\Lambda_k}, \Lambda_k) = \sum_{\Lambda_n \setminus \Lambda_k} \mathbb{P}_{\Lambda_n}(x_n, t_n; \ldots; x_1, t_1). \tag{2.7}$$

Here the summation $\sum_{\Lambda_n \setminus \Lambda_k}$ runs over all realisations on the timesteps included in $\Lambda_n$ but not $\Lambda_k$ and $x_{\Lambda_k}$ refers to the subset of possible outcomes corresponding to the times $\Lambda_k$. Thus, if a joint probability distribution arises from an underlying physical process, it necessarily satisfies this so-called *consistency* or *containment* property of Eq. (2.7), which allows one to derive the entire hierarchy of joint distributions in Eq. (2.5).

One of the pioneers of probability theory, Kolmogorov, was concerned with proving equivalence between the continuous and discrete descriptions of stochastic processes by establishing the alternate implication. That is, he endeavoured to understand the conditions a collection of finite joint probability distributions must satisfy in order for an underlying continuous process, with all of the finite ones as marginals, to exist. The *Kolmogorov extension theorem* (KET) says that satisfaction of the consistency condition outlined above implies the existence of such an underlying process, thereby proving equivalence between the descriptions given in Eqs. (2.4) and (2.5) [16]. Thus, practical motivations notwithstanding, the KET bridges the gap between the experimental reality we must face and solid mathematical underpinnings, importantly providing a definition of stochastic processes as the limit of families of finite probability distributions in time.

Intuitively, satisfaction of the consistency conditions means that once a process has been characterised on a set of timesteps, all behaviour on any subset of timesteps can be deduced by marginalising over the outcomes at the excessive times, as per Eq. (2.7) and depicted in Fig. 2.2. Thus, for a stochastic process over $n$ timesteps, the $n$-point joint probability distribution $\mathbb{P}_{n:1}$ completely characterises the process since it contains within it the entire hierarchy of Eq. (2.5). The fact that marginalisation is the correct way to obtain the contained description of a classical stochastic process is, loosely speaking, because there is no difference between having measured the statistics over all timesteps $\Lambda_n$ and then discarding the observations we no longer care about, *i.e.,* summing over those on $\Lambda_n \backslash \Lambda_k$ (as per the r.h.s of Eq. (2.7)), and not having measured outcomes at those irrelevant timesteps anyway (as per the l.h.s).

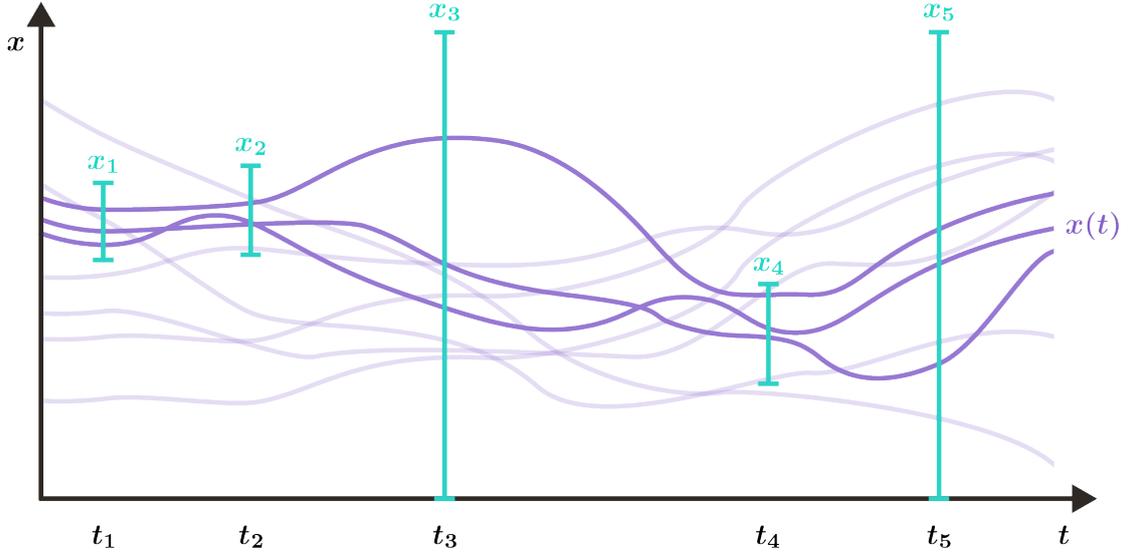

**Figure 2.2:** *Containment property for classical stochastic processes.* Given a joint probability distribution describing a classical stochastic process over some finite set of timesteps, *e.g.,* $\mathbb{P}_{5:1}(x_5, t_5; \ldots; x_1, t_1)$, the correct description of the process on any subset of timesteps is calculated by marginalising over the outcomes at the excessive times. For example, the 3-step process over timesteps $\Lambda_3 = \{t_1, t_2, t_4\}$ is characterised by $\mathbb{P}_{\Lambda_3}(x_{\Lambda_3}, \Lambda_3) = \sum_{x_3 x_5} \mathbb{P}_{5:1}(x_5, t_5; \ldots, x_1, t_1)$. Only the bold paths depicted have non-zero probabilities to be measured in the intervals $x_{\Lambda_3}$ shown above, the weights of each of which are calculated from the original description by summing the five-step joint probabilities over all possible spatial intervals at $t_3, t_5$, shown here as the extended vertical lines labelled $x_3, x_5$.

To see this, consider tossing three fair coins consecutively, described by the joint distribution $\mathbb{P}_{3:1}(x_3, t_3; x_2, t_2; x_1, t_1) = \frac{1}{8}$ for each possible length-3 binary outcome sequence (*e.g.,* HHT). Given this distribution, if we are subsequently interested in describing the process at timesteps $t_1$ and $t_3$ only, we simply marginalise over the outcomes at $t_2$, which gives: $\mathbb{P}_{3,1}(x_3, t_3; x_1, t_1) = \sum_{x_2} \mathbb{P}_{3:1}(x_3, t_3; x_2, t_2; x_1, t_1) = \frac{1}{4}$ for each length-2 binary outcome sequence. This procedure indeed provides the correct description of a process where two coins are tossed at times $t_1$ and $t_3$. Marginalisation works because the state of the coin at time $t_2$ was *some* outcome, and it makes no difference whether we average over all such possibilities or they are not even measured in the first place. The containment property, and indeed Eq. (2.6), implicitly assumes that we can measure realisations of outcomes throughout the process consistently without influencing the state of the system upon each interrogation. This is not fulfilled in more general stochastic frameworks such as classical causal modelling and quantum mechanics [15].

In summary, the connection between the continuous and discrete descriptions of a stochastic process is provided by the KET, which justifies our ability to work in either picture. For the purpose of this thesis, due to our operational perspective, we consider a classical stochastic process to be defined as follows.

**Definition 2.3** (Classical stochastic process)**.** A classical stochastic process over $n$ timesteps is characterised by a joint probability distribution $\mathbb{P}_{n:1}(x_n, t_n; \ldots; x_1, t_1)$ satisfying the containment property of Eq. (2.7) for all subsets of timesteps.

From now on, we drop the explicit labelling of the timesteps as arguments of the distribution and write $\mathbb{P}_{n:1}(x_n, \ldots, x_1)$ to describe the process, with the subscripts indicating the timesteps on which the outcomes are observed. Additionally, although the sample space can be infinite, throughout this thesis we restrict our focus to the finite case.[3]

### 2.1.3 *Modelling Stochastic Processes*

Perhaps the most important reason for understanding stochastic processes from a practical perspective is to model them. Developing models that accurately simulate the statistics observed allows us to predict future behaviour of complex systems, *e. g.,* future stock market fluctuations or evolving population dynamics, amongst other applications [4–6]. While the joint probability distribution $\mathbb{P}_{n:1}(x_n, \ldots, x_1)$ *characterises* the process at hand, containing all possible multi-time correlations between outcomes observed at different times, we are rarely, if ever, privy to such a detailed and resource-exhaustive description [4, 22]. Rather, what is typically feasible is to measure statistics describing the state of the system at each timestep, *i. e.,* the single-point marginals $\mathbb{P}_k(x_k) \; \forall \; t_k \in \Lambda_n$, and perhaps some lower order correlation terms (two- or three-point marginals), which describe how the state at some time is correlated with that at some others.

The complete description of the process contains significantly more information than can be deduced from such lower-order marginals; Indeed, any higher-order marginal such as $\mathbb{P}_{\Lambda_k}(x_{\Lambda_k})$ for any proper subset $\Lambda_k \subset \Lambda_n$ remains insufficient to characterise a generic stochastic process [26]. Consequently, a computationally feasible *model* of a stochastic process does not generally specify the process at hand. Our assumption when modelling stochastic processes is that we can approximate the joint statistics from a manageable set of lower-order distributions. Put simply, the aim of modelling is to accurately reconstruct the joint distribution of the process from smaller ones [4, 5].

---

[3] Almost all of the results presented in this chapter are extendable to the continuous case by replacing sums with integrals.

An important concept pertinent to modelling is that of *conditional probability distributions*. Consider a process characterised over $n$ timesteps by $\mathbb{P}_{\Lambda_n}(x_{\Lambda_n})$ and suppose we observe a specific realisation over a subset of times $\Lambda_k \subseteq \Lambda_n$. Then

$$\mathbb{P}_{\Lambda_n \setminus \Lambda_k}(x_{\Lambda_n \setminus \Lambda_k} | x_{\Lambda_k}) = \frac{\mathbb{P}_{\Lambda_n}(x_{\Lambda_n})}{\mathbb{P}_{\Lambda_k}(x_{\Lambda_k})} \tag{2.8}$$

represents the conditional probability distribution over outcomes on the remaining timesteps $\Lambda_n \setminus \Lambda_k$ given that it took the values $x_{\Lambda_k}$ on timesteps $\Lambda_k$. That we can calculate conditional distributions for stochastic processes in this way implicitly relies on the fact that $\mathbb{P}_{\Lambda_k}(x_{\Lambda_k})$ on the r.h.s is the correct descriptor of the process on timesteps $\Lambda_k$, *i.e.*, that the KET holds.

From Eq. (2.8), we can iteratively decompose any joint distribution as

$$\begin{aligned}
\mathbb{P}_{n:1}(x_n, \ldots, x_1) &= \mathbb{P}_n(x_n | x_{n-1}, \ldots, x_1) \mathbb{P}_{n-1:1}(x_{n-1}, \ldots, x_1) \\
&= \mathbb{P}_n(x_n | x_{n-1}, \ldots, x_1) \mathbb{P}_{n-1}(x_{n-1} | x_{n-2}, \ldots, x_1) \mathbb{P}_{n-2:1}(x_{n-2}, \ldots, x_1) \\
&= \cdots = \mathbb{P}_n(x_n | x_{n-1}, \ldots, x_1) \ldots \mathbb{P}_2(x_2 | x_1) \mathbb{P}_1(x_1).
\end{aligned} \tag{2.9}$$

It is clear from this decomposition that we can build up the joint distribution from its constituent conditional distributions. Although a conditional distribution such as $\mathbb{P}_{\Lambda_n \setminus \Lambda_k}(x_{\Lambda_n \setminus \Lambda_k} | x_{\Lambda_k})$ is technically an $(n-k)$-point joint probability distribution, we will refer to them as $n$-point *correlations*, since we assume the ability to calculate the $k$-point distribution $\mathbb{P}_{\Lambda_k}(x_{\Lambda_k})$ over the conditioning argument, which together provide the $n$-point distribution $\mathbb{P}_{\Lambda_n}(x_{\Lambda_n})$.

Such conditional distributions are often referred to as *transition probabilities*, as, *e.g.*, $\mathbb{P}_2(x_2|x_1)$ represents the probability for the system to change its state from $x_1$ to $x_2$. For a complete description of the process, one requires successively higher-order transition probabilities, *e.g.*, $\mathbb{P}_{3:1}(x_3, x_2, x_1) = \mathbb{P}_3(x_3|x_2, x_1)\mathbb{P}_2(x_2|x_1)\mathbb{P}_1(x_1)$, and so forth. Clearly, a complete description of a stochastic process contains a large amount of information. We can understand the complexity of the process in terms of the number of parameters needed to describe it. Suppose that a system of interest can take $d$ distinct outcomes. For the initial distribution $\mathbb{P}_1(x_1)$, we must specify $d-1$ numbers (*i.e.*, the probability associated to each outcome, with the normalisation condition constraining one value); the joint distribution over two steps $\mathbb{P}_{2:1}(x_2, x_1) = \mathbb{P}_2(x_2|x_1)\mathbb{P}_1(x_1)$ requires $d^2 - 1$ specifications; and so on. In general, modelling an $n$-timestep evolution of a $d$-level classical system requires estimating $\mathcal{O}(d^n)$ transition probabilities. This exponential scaling in the number of timesteps quickly becomes intractable, hence the allure of developing accurate models built upon estimating lower-order transitions. Indeed, modelling means to fix the transition probabilities, which can be meaningfully done under the assumption that the memory is finite in length. We now consider the simplest case of such finite-memory processes.

2.1.4 *Markovian Stochastic Processes*

**Example 2.1** (Perturbed Coin). Consider the toy classical process of a perturbed coin, depicted in Fig. 2.3. Here we have a coin resting on a piece of cardboard, which is being gently shaken at discrete times $t_k \in \Lambda_n$, resulting in a time-independent probability, $p > 1 - p$, for the coin to retain its previous orientation between each shake; with probability $1 - p$, the coin flips from H to T, or vice versa. The probability of the coin being in a particular state at arbitrary timestep $t_k$ depends entirely on its most recent state, *i. e.,* the process is completely characterised by the family of two-point conditional distributions

$$\mathbb{P}_k(\mathtt{H}_k|\mathtt{H}_{k-1}) = \mathbb{P}_k(\mathtt{T}_k|\mathtt{T}_{k-1}) = p \qquad (2.10)$$
$$\mathbb{P}_k(\mathtt{H}_k|\mathtt{T}_{k-1}) = \mathbb{P}_k(\mathtt{T}_k|\mathtt{H}_{k-1}) = 1 - p.$$

Of course, temporal correlations between observations can be exhibited over various timescales; if one begins such a process with the coin facing H up, a few steps later it is more likely than not to be found in the same state. The crucial point is that, *once we know* the state at timestep $t_k$, we may as well discard any observations of previous states since they tell us no additional information.

The dependence of the future statistics on only the most recent outcome dramatically simplifies the complexity of any algorithm predicting its behaviour. This type of process is known as a *Markovian* or *memoryless* process, since the process itself stores no memory of historic outcomes; the only temporal correlations that can arise are mediated through the most recent state of the system. Markovian processes are formally defined as follows.

**Definition 2.4** (Classical Markov process)**.** A classical Markov process is described by a joint probability distribution $\mathbb{P}_{n:1}(x_n,\ldots,x_1)$ whose conditional distribution at each timestep $t_k \in \Lambda_n$ only depends on the statistics of the most recent distribution at time $t_{k-1}$:

$$\mathbb{P}_k(x_k|x_{k-1},\ldots,x_1) = \mathbb{P}_k(x_k|x_{k-1}). \tag{2.11}$$

Note that for a Markovian process, the statistics depend *conditionally* on no more than the most recent timestep. A completely memoryless process is one in which the conditional statistics are independent of even the current state, such as a fair coin flip. Such processes are often said to be *super-Markovian* [5]; in this thesis, we generally drop this distinction and refer to both types of processes as Markovian or memoryless synonymously.

In contrast to the general case (see Eq. (2.9)), Markovian processes can be completely characterised with only an initial con-figuration $\mathbb{P}_1(x_1)$ and the collection of two-point transition probabilities $\mathbb{P}_k(x_k|x_{k-1})$, which specify the present state in terms of the most recent realisation of the system at each timestep. This is because for a Markovian process the overall joint distribution factorises as

$$\mathbb{P}_{n:1}(x_n,\ldots,x_1) = \mathbb{P}_n(x_n|x_{n-1})\ldots\mathbb{P}_2(x_2|x_1)\mathbb{P}_1(x_1). \tag{2.12}$$

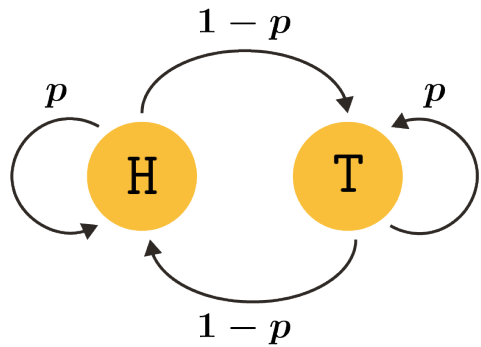

**Figure 2.3:** *Perturbed coin.* The perturbed coin is characterised by the transition probabilities prescribing the likelihood to retain its state or to flip at each shake.

Since each transition map is specified by $d^2 - d$ conditional probabilities and the initial single-point distribution by $d-1$, we require at most $(n-1)(d^2-d)+d-1$ parameters to describe a memoryless process, providing a significant reduction in complexity; this is perhaps the primary reason for the popularity of invoking the Markov assumption when modelling stochastic processes.

Equivalently, considering a discrete-time Markovian process on $\Lambda_n$ and specifying an initial condition at $t_1$, the state at an arbitrary later time $t_k \in \Lambda_n$ can be calculated via

$$\mathbb{P}_k = \mathcal{S}_{k,1}\mathbb{P}_1. \tag{2.13}$$

Here, $\mathcal{S}_{k,1}$ is a $d \times d$ matrix filled with conditional probabilities representing the likelihood of transition from each $x_1$ value to each $x_k$ and $\mathbb{P}_k$ is the $d$-dimensional vector representing the probability distribution at $t_k$. Comparing Eq. (2.13) with the entire description of the process up until the time specified, *i. e.,* Eq. (2.12) written up to $t_k$ and marginalised over the outcomes on all timesteps except $t_k$, it is clear that transition matrix from some $t_1$ to $t_k$ is simply a matrix multiplication of all of the intermediary transition matrices

$$\mathcal{S}_{k,1} = \mathcal{S}_{k,k-1}\mathcal{S}_{k-1,k-2}\ldots\mathcal{S}_{2,1}. \tag{2.14}$$

Clearly, by grouping any sequence of matrix multiplications on the r.h.s of Eq. (2.14) into two stochastic matrices, we have satisfaction of the following *divisibility* property [5]

$$\mathcal{S}_{k,1} = \mathcal{S}_{k,j}\mathcal{S}_{j,1} \quad \forall\, t_k > t_j > t_1. \tag{2.15}$$

That the transition maps for a Markovian process are indeed conditional probability distributions dictates that their matrix representations have non-negative entries whose columns sum to unity, known as *stochastic matrices.* From an axiomatic perspective, stochastic matrices are of significant importance in classical physics, as their properties ensure that for any input probability distribution, the output of the map is always a valid probability distribution, thereby providing the most general unconditional transformation allowable in classical theory between two points in time. For reasons that will become clear, we refer to such processes that can be described by a divisible family of stochastic maps as (*positive*) P-*divisible* processes [67, 68].

As a brief aside, note that for a stationary P-divisible process, one can derive a closed form continuous-time equation of motion for the instantaneous state of the system, as is provided in Appendix B.1 for the sake of completeness. The classical *master equation* (ME) derived provides the most general form of an equation whose solution is guaranteed to provide a divisible single-parameter semi-group of stochastic maps [69].

We have seen already that a Markovian process lends itself to a complete description in terms of a divisible family of such two-point stochastic transition maps. However, it is well-known that P-divisibility is insufficient to classify a process as Markovian [68, 70–73]. The fact that non-Markovian processes can satisfy the P-divisibility criteria is emblematic of a deeper issue, namely that two-point information does not adequately capture the dynamics at hand. In other words, the core reason that the P-divisibility criteria does not imply Markovianity is because the former is based solely on two-point correlations which fail to capture multi-time effects and therefore do not provide full information about a generic process. Consequently, an experimental reconstruction of the two-point stochastic maps does not necessarily correspond to the actual (potentially multi-time) conditional probabilities of the process. In other words, satisfaction of Eq. (2.15) does not imply Eq. (2.11): for a given process, one might be able to construct a stochastic map description satisfying Eq. (2.15) even when the process is non-Markovian.

The crucial point is that in such cases, in contradistinction to Markovian processes, the constituent divided portions of the stochastic maps *cannot* be identified with the actual conditional probabilities of the process; for a non-Markovian P-divisible process, these sequences of stochastic maps are sensitive to initial conditions. The conditional probabilities in a non-Markovian process generally depend on multiple previous outcomes, which is necessarily overlooked by the two-point P-divisibility criteria. We will return to this point with a more detailed discussion in Section 2.3; for now, the key message is that a proper description of non-Markovian processes must take into account all multi-time correlations and cannot simply be an extension to the study of Markovian processes as we know them.

### 2.1.5 *Non-Markovian Stochastic Processes*

From the discussion until now, we have seen that dealing with Markovian processes is much easier than with their non-Markovian counterparts, with the proper description of the latter requiring an exponentially-scaling amount of resources. This begs the question: *to what extent do we need to worry about a proper understanding of non-Markovian processes?* As noted by van Kampen: "non-Markov is the rule; Markov is the exception" [26]. Formally, in the space of all stochastic processes, the Markovian ones form a non-convex set of measure zero (with respect to any meaningful, non-singular measure); in other words, they are isolated special cases. The question of whether or not Markov processes exist in nature—and under which circumstances—is thus of significant interest. For instance, a random walk with the choice of each step being independently and identically distributed is a Markov process, so too is an experiment of picking coloured marbles out of a bag with replacement. However, as soon as non-negligible and realistic effects come into play, such as a direction bias for a random walker or the lack of replacement of marbles, such processes are almost always rendered non-Markovian.

To illustrate, consider the situation pictured in Fig. 2.4. Suppose an experimenter picks marbles from a bag whose initial distribution comprises a fraction $r$ of red, $b$ of blue and $g$ of green coloured marbles, such that $r + b + g = 1$. Suppose that, according to whether the marble chosen at some timestep $t_k$ is red, blue or green, a random walker will respectively step left ($s_k = -1$), stay put ($s_k = 0$) or move right ($s_k = 1$). The position of the walker at $t_n$ can be described by the random variable $X_n := \sum_{t_k}^{t_n} s_k$. If the marble chosen at each timestep is replaced before the next one picked out, then the random walk, $\mathbb{P}_{n:1}(x_n, \ldots, x_1)$, is Markovian, since the probability for the walker to step in each direction at a given timestep remains unchanged by withdrawing any marble from the bag. If, on the other hand, the marbles are not replaced, then the probability for the walker to step in each direction changes with each marble withdrawal, since the relative proportions of the contents are modified. We cannot determine this change from the instantaneous position of the walker alone; instead, we require knowledge of the entire history up until the relevant point in time to determine the current composition of the bag, hence the probabilities for the walker to step in each direction.

Indeed, in many other physically realistic situations of interest, similar finite-sized effects lead to non-negligible memory effects in the process. For example, processes whose microscopic equations of motion are second order differential equations in time require two initial conditions (*i.e.,* position and velocity) to determine the statistics over future positions. Hence, processes considering the position of an initial distribution of particles can only be considered Markovian at sufficiently long timescales that the initial velocity is forgotten [5]. This is a special case of a Langevin-type equation; other continuous time Markovian processes are Wiener processes and Cauchy processes. Their extensive study has shed light on the type of underlying dynamics that give rise to Markovian processes [5, 6]. Loosely speaking, the essential idea is that the system of interest couples weakly to a large environment: the weakness of the coupling ensures that the environment is relatively unperturbed by interactions with the system and the largeness of the environment ensures that from each timestep to the next, the system interacts with a fresh portion of the environment. Modelling and designing mesoscopic processes, where such assumptions are no longer satisfied and finite-sized effects come into significance, clearly requires an understanding of non-Markovian processes.

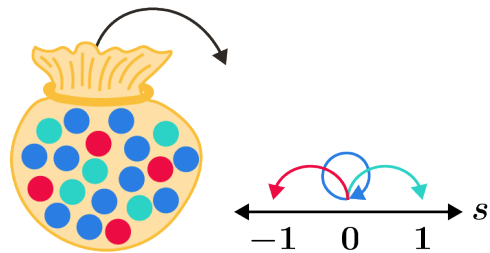

**Figure 2.4:** *Random walk conditioned on marbles drawn.* A bag is filled with a fraction of different coloured marbles: here we have $\frac{1}{5}$ red, $\frac{1}{2}$ blue and $\frac{1}{4}$ green. A marble is drawn at each time and depending on its colour, a walker steps left (red), stays put (blue) or steps right (green). If the marble is replaced after each step, the process is Markovian; otherwise it is not, since the extracted marbles disrupt the relative proportions of the bag.

2.1.6 *Classical Markov Order*

Although generic non-Markovian processes might require knowledge of their entire course of history to predict the future evolution, there is an important class of non-Markovian processes that are nonetheless feasible to model with a reasonable amount of resources: those with *finite memory length*. Formally, the natural way to account for such memory effects that are finite in duration is through the notion of *Markov order*, $\ell$, which dictates that the statistics observed at any given time only depend upon knowledge of the past $\ell$ outcomes. For example, we can generalise the random walk process conditioned on marbles considered above to incorporate longer memory effects by holding out the marbles drawn for a certain number of timesteps before they are replaced. In this case, although the statistics of the next state depend conditionally upon the most recent sequence of outcomes, just like in the Markovian case, the salient point is that once we have knowledge of the most recent sequence of outcomes, we may as well discard observations from the prior history.

By grouping together timestep sequences of length $\ell$, such non-Markovian processes can be tamed into ones that behave like Markovian processes, albeit on a larger state space extending over a period of time. Although processes with finite Markov order are close in spirit to Markovian processes, unlike their truly Markovian counterparts, multi-time correlations can play a crucial role in the future evolution of dynamics, giving rise to a potentially complex memory that is nonetheless limited in duration. To capture such behaviour, rather than begin with a straightforward extension of Markovian processes which do not account for such effects, we are forced to begin with a proper description of non-Markovian stochastic processes, *i.e.,* Eq. (2.5), and study the circumstances in which the process displays finite-length memory.

Intuitively, the concept of Markov order boils down to the following question: *is knowledge of a portion of the history of a process sufficient to predict the statistics of its future evolution?* In other words, given knowledge of the statistics over a sequence of $\ell$ timesteps, $\{t_{k-\ell}, \ldots, t_{k-1}\}$, one can perfectly predict, in principle, the statistics to be expected at timestep $t_k$ for such a process. The conditional distribution for arbitrary $t_k \in \Lambda_n$ of a process with Markov order-$\ell$ is therefore expressed as

$$\mathbb{P}_k(x_k|x_{k-1}, \ldots, x_1) = \mathbb{P}_k(x_k|x_{k-1}, \ldots, x_{k-\ell}). \tag{2.16}$$

In the interest of developing an economical notation for the remainder of this thesis, given an $n$-step stochastic process, we demarcate the timesteps into three intervals: the future $F := \{t_n, \ldots, t_k\}$, the memory $M := \{t_{k-1}, \ldots, t_{k-\ell}\}$, and the earlier history $H := \{t_{k-\ell-1}, \ldots, t_1\}$ (in principle, the history and future can extend to infinitely long times). The realisations observed over these timesteps $x_j$ describing the system of interest are grouped together similarly as $\{x_F, x_M, x_H\}$. The Markov order of the process is defined in terms of the conditional statistics of outcomes as follows.

**Definition 2.5** (Classical Markov order). A classical stochastic process has Markov order-$\ell$ if the conditional probability for any outcomes $x_F$ at or beyond any time $t_k \in \Lambda_n$ depends only on the realisations $x_M$ over the previous $\ell$ timesteps, and not on those $x_H$ of the earlier history

$$\mathbb{P}_F(x_F | x_M, x_H) = \mathbb{P}_F(x_F | x_M). \tag{2.17}$$

As special cases, $\ell = 1$ corresponds to a Markovian process (see Def. 2.4) and $\ell = 0$ a super-Markovian (completely random) process.

The property of Markov order-$\ell$ constrains the underlying joint probability distribution characterising the process, from which the above conditional distributions arise. To highlight the important reduction of complexity in the description of a process with finite Markov order, compare the following decomposition with the general expression of Eq. (2.9)

$$\mathbb{P}_{n:1}(x_n, \ldots, x_1) = \prod_{j=\ell+1}^{n} \mathbb{P}_j(x_j | x_{j-1} \ldots, x_{j-\ell}) \mathbb{P}_{\ell:1}(x_\ell, \ldots, x_1). \tag{2.18}$$

To reiterate, $\ell$ determines the number of timesteps over which one must observe states in order to optimally predict, in principle, the next state, thereby providing a natural and fundamental timescale for *memory length* in stochastic processes. This property is of tremendous practical importance, as processes with finite Markov order can be effectively reduced to Markovian processes upon a suitable grouping of timesteps, allowing for efficient simulation [26].

An alternative but equivalent way of expressing the notion of Markov order is that any statistics that an experimenter might deduce over the history and the future timesteps are *conditionally independent* with respect to an intermediate sequence of realised values

$$\mathbb{P}_{FH}(x_F, x_H | x_M) = \mathbb{P}_F(x_F | x_M) \mathbb{P}_H(x_H | x_M). \tag{2.19}$$

Importantly, it is knowledge of the outcomes on the memory block that renders the future and history conditionally independent. To prelude the discussion to follow in our extension to the quantum realm, an alternative way to think of this is as a sequence of interventions (for instance, measurements) on the system that serves to block any possible historic influence on the future dynamics for each outcome realised.

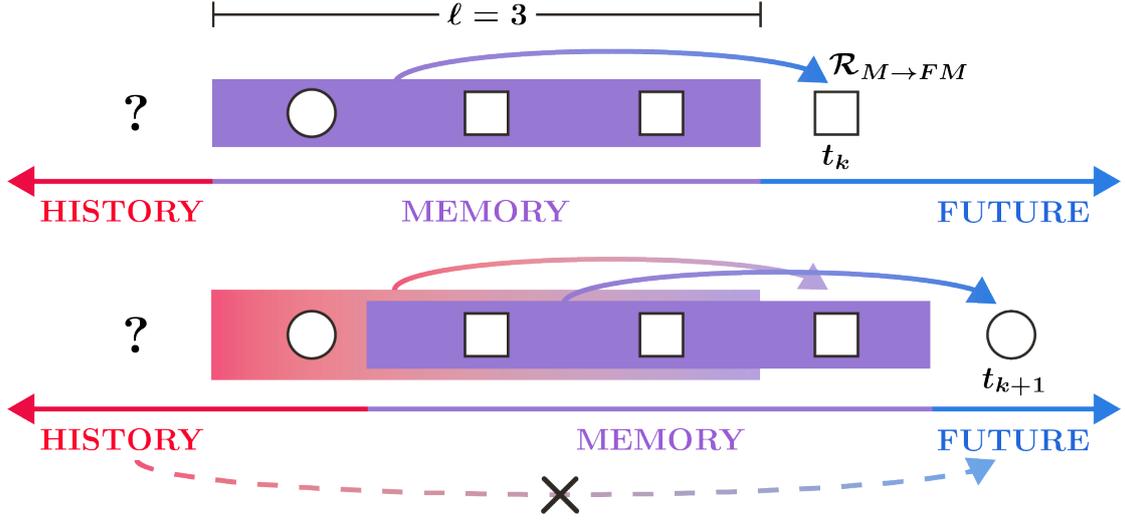

**Figure 2.5:** *Markov order as the natural notion of memory length.* Knowledge of the $\ell = 3$ states in the memory block is sufficient to predict the probabilities of future states by way of the recovery map, $\mathcal{R}_{M \to FM}$. In particular, no information about the prior history is required to determine these probabilities (indicated by the question mark). This property is independent of which timestep is being considered; holding equally well for steps $t_k$ and $t_{k+1}$. At every step, any influence the history (beyond $\ell$ timesteps ago) has on the future must be mediated through the memory blocks. Importantly, states observed in the memory can jointly influence the future statistics. Nonetheless, conditional on the statistics realised in the most recent block, there can be no correlations between history and future, as indicated by the faded, dashed arrow.

Importantly, while Markov order-$\ell$ means that the state of the process at any time only depends conditionally upon the previous $\ell$ states, it does not imply an absolute separation of the timesteps into blocks of memory and irrelevant history. In other words, the probability distribution $\mathbb{P}_{FMH}(x_F, x_M, x_H)$ factorises conditionally, but we do not necessarily have $\mathbb{P}_{FMH}(x_F, x_M, x_H) = \mathbb{P}_F(x_F)\mathbb{P}_M(x_M)\mathbb{P}_H(x_H)$, or $\mathbb{P}_{FMH}(x_F, x_M, x_H) = \mathbb{P}_{FH}(x_F, x_H)\mathbb{P}_M(x_M)$. Instead, the memory blocks corresponding to different timesteps overlap, allowing for the existence of unconditional correlations between timesteps with a separation greater than $\ell$ in general. These are themselves often referred to as memory, however, these temporal correlations are always mediated through overlapping memory blocks as show in Fig. 2.5.

Lastly, a classical stochastic process with Markov order-$\ell$ can be equivalently characterised by the following two statements. Firstly, from an operational perspective, the significance of finite Markov order is best encapsulated through the existence of a so-called *recovery* map $\mathcal{R}_{M \to FM}$, which acts only on $M$ to give the correct future statistics: $\mathbb{P}_{FMH} = \mathcal{R}_{M\to FM}(\mathbb{P}_{MH})$. This map can be directly used to simulate future dynamics, and the complexity of any predictive model is fundamentally upper-bounded by the length of the block $M$ on which it acts (as well as by the number of possible values for each $x_j$). Secondly, an entropic characterisation that is convenient to check in practice states that the classical *conditional mutual information* (CMI) vanishes: $I^{\text{cl}}(F:H|M) := H(\mathbb{P}_{FM}) + H(\mathbb{P}_{MH}) - H(\mathbb{P}_{FMH}) - H(\mathbb{P}_M) = 0$, where $H(\mathbb{P}) := -\sum_x \mathbb{P}(x)\log\mathbb{P}(x)$ is the Shannon entropy. The equivalence between these statements is trivial: satisfaction of Eq. (2.17) implies the distribution factorises as $\mathbb{P}_{FMH}(x_F, x_M, x_H) = \mathbb{P}_F(x_F|x_M)\mathbb{P}_{MH}(x_M, x_H)$; the recovery map $\mathcal{R}_{M\to FM}$ can then be chosen to act as multiplication by the higher-order stochastic transition map $\mathbb{P}_F(x_F|x_M)$. Equivalence to vanishing classical CMI is obvious by writing the CMI as a relative entropy between probability distributions (Kullback-Liebler divergence) as follows $I^{\text{cl}}(F:H|M) = \mathcal{D}^{\text{cl}}(\mathbb{P}_{FH|M}\|\mathbb{P}_{F|M}\mathbb{P}_{H|M})$, where $\mathcal{D}^{\text{cl}}(\mathbb{P}|\mathbb{Q}) := -\sum_x \mathbb{P}(x)\log\frac{\mathbb{P}(x)}{\mathbb{Q}(x)}$. The relative entropy vanishes iff the arguments are identical. Thus, in the classical setting, vanishing CMI is equivalent to finite Markov order.

We now briefly summarise the key points of this section that we will examine in detail when attempting to understand quantum stochastic processes. We have seen that classical stochastic processes can be characterised completely by a joint probability distribution over a discrete set of timesteps satisfying a natural consistency condition, with the existence of an underlying continuous-time physical process guaranteed by the KET. This allows for unambiguous calculation of conditional statistics, which are crucial to defining memoryless processes by way of the Markov condition. This special class of processes are the ones in which two-point information is sufficient for their characterisation. However, such two-point descriptions are destined to overlook multi-time memory effects and any criteria based upon them is not suitable for addressing non-Markovian processes. A particular example of this point is that divisibility does not imply Markovianity. Following this, we considered some properties of a non-trivial but practically important subset of non-Markovian processes which exhibit finite-length memory, namely those with finite Markov order. Our goal in the next section is to consider the extensions of these ideas into the quantum setting, where we will see that a number of subtleties must be addressed for a meaningful description of quantum stochastic processes.

## 2.2 OPEN QUANTUM DYNAMICS

In the classical setting, the finite-length memory approximation underpins the success of the often-invoked order-$\ell$ Markov models, which make use of information from only the past $\ell$ states to predict the next. However, even in the simplest non-trivial case of memoryless dynamics (*i.e.*, $\ell = 1$), the study of stochastic processes is vastly different in the quantum realm than its classical counterpart. This is mainly because, in quantum mechanics, one must necessarily disturb the system in order to observe realisations of the process, breaking an implicit assumption of the classical setting. In quantum mechanics, there is a continuous family of possible non-commuting observables that could be measured, and the choice of measurement at one point in time (or even whether to measure at all) can directly affect the future statistics [15, 52, 66, 74–76]. This is in stark contrast to the consistency conditions satisfied by the joint probability distributions corresponding to classical stochastic processes.

To highlight this issue explicitly, consider the following quantum experiment which is represented schematically in Fig. 2.6 (we follow the example presented in Ref. [15]). Begin with a spin-$\frac{1}{2}$ particle initially prepared in an equal superposition $\frac{1}{\sqrt{2}}(|\downarrow\rangle + |\uparrow\rangle)$ in the $z$-direction. Suppose an experimenter were to set up a sequence of Stern-Gerlach apparata that allows them to measure the spin orientation at successive timesteps in any direction of their choosing. Consider the case where they measure the system at times $\{t_1, t_2, t_3\}$ respectively in the $z$-, $x$-, and $z$-directions, whose outcomes are represented by $\{\downarrow, \uparrow\}$ for $z$-direction measurements and $\{\rightarrow, \leftarrow\}$ for $x$-direction measurements. We assume that the quantum system undergoes trivial dynamics in between measurements. The first measurement in the $z$-direction has equal probability of $\frac{1}{2}$ to yield the result $\downarrow$ or $\uparrow$; in either case, once the outcome is observed, the post-measurement state is either $|\downarrow\rangle$ or $|\uparrow\rangle$ respectively. These states can be expressed as an equal superposition in the $x$-basis: $|\downarrow\rangle = \frac{1}{\sqrt{2}}(|\rightarrow\rangle + |\leftarrow\rangle)$ and $|\uparrow\rangle = \frac{1}{\sqrt{2}}(|\rightarrow\rangle - |\leftarrow\rangle)$. Thus, the spin measurement in the $x$-direction at the second timestep again yields each possible outcome $\rightarrow$ or $\leftarrow$ with equal probability, and the subsequent state is either $|\rightarrow\rangle = \frac{1}{\sqrt{2}}(|\downarrow\rangle + |\uparrow\rangle)$ or $|\leftarrow\rangle = \frac{1}{\sqrt{2}}(|\downarrow\rangle - |\uparrow\rangle)$ accordingly. The statistics of the final measurement are then identical to the first one. Thus, the probability to measure any sequence of outcomes in this case is uniformly distributed over the possibilities, *e.g.*, $\mathbb{P}_{3:1}(\downarrow_3, \leftarrow_2, \downarrow_1) = \mathbb{P}_{3:1}(\downarrow_3, \rightarrow_2, \downarrow_1) = \frac{1}{8}$.

On the other hand, consider an alternative experiment where the experimenter does *not* perform the $x$-direction measurement at the second timestep. Then, suppose for concreteness that the outcome $\downarrow$ was observed at the first timestep, which occurs with probability $\frac{1}{2}$ and leaves the system in the $|\downarrow\rangle$ state. When subsequently measured in the $z$-direction at the third timestep, without any intermediary measurement at $t_2$, the outcome $\downarrow$ is realised with certainty; thus we have $\mathbb{P}_{3,1}(\downarrow_3, \downarrow_1) = \frac{1}{2}$. This is not equal to a marginalisation over the possible outcomes of the second timestep, which gives $\mathbb{P}_{3:1}(\downarrow_3, \leftarrow_2, \downarrow_1) + \mathbb{P}_{3:1}(\downarrow_3, \rightarrow_2, \downarrow_1) = \frac{1}{4}$. The statistics of the process over the first and third timestep *cannot* be deduced by simply marginalising the 3-point distribution $\mathbb{P}_{3:1}$ over the outcomes of the second timestep; the standard KET clearly does not hold on the level of statistics measured in quantum processes.

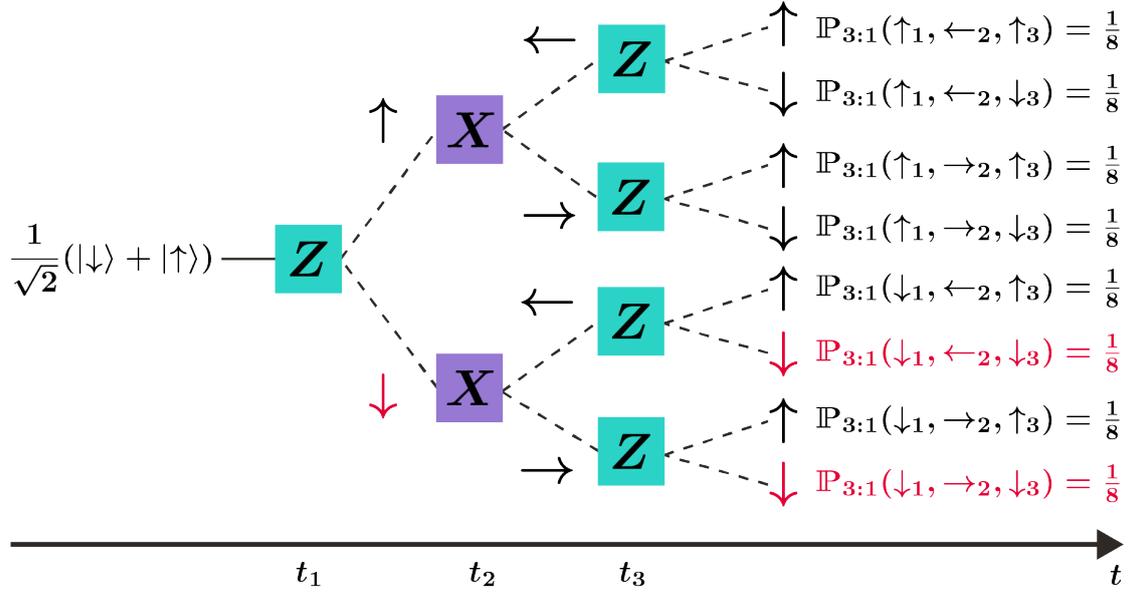

**Figure 2.6:** *Breakdown of KET in a Stern-Gerlach experiment.* An initial state prepared in an even superposition is subject to three sequential measurements, in the $Z, X, Z$ directions. The joint statistics for each possible sequence of outcomes are equal to $\frac{1}{8}$ (shown on the right). However, marginalising over the outcomes observed at the second timestep does not provide the correct probabilities that are predicted by theory in the case where *no* measurement is performed there, highlighting the breakdown of the KET. If the first outcome is $\downarrow$ and no intervention is made at $t_2$, the measurement at $t_3$ yields $\downarrow$ with certainty. This gives $\mathbb{P}_{3,1}(\downarrow_3, \downarrow_1) = \frac{1}{2}$, in contradiction with the marginalised statistics computed as the sum of probabilities displayed in red.

The KET breaks down here on the level of measured statistics because implicitly assumes that there is only one method of probing the degrees of freedom of interest to observe outcomes, and that this probing does not actively change the state of the system. Neither of these assumptions are fulfilled in quantum theory (or, *e. g.*, more general classical stochastic theories with interventions—including causal modelling—which we briefly return our attention to in Section 4.2) [15, 16, 21]. Loosely speaking, it is the fact that for quantum processes, there *is* a difference between averaging over all possible measurement outcomes and not having made a measurement at all, which essentially leads to violation of the KET [15]. Without a consistent method for deriving the contained descriptors of a process from that of its description defined over more timesteps, there is inherent ambiguity in what we mean when we talk about a quantum stochastic process. This leads directly to problems in defining conditional probabilities, and thus an unambiguous classification of memory in quantum processes.

This problem has irked the open systems and quantum information communities for some time, leading to various incompatible descriptions of quantum stochastic processes [77]. Conventional approaches attempt to sidestep this issue by describing properties of the process in terms of the time-evolving density operator of the system of interest, inherently failing to capture memory effects that only appear in multi-time correlations [23–25]; whilst others impose constraints on general system-environment interactions to facilitate a specific mechanism for memory transfer throughout the dynamics [78–80]. In either case, both types of approach fail to yield a comprehensive framework of quantum stochastic processes, and the corresponding definitions of memory are necessary but not sufficient to characterise Markovian processes [50].

As is evident in the classical case, one can only make limited assertions about a stochastic process from, *e. g.,* functions evaluated in terms of correlations between the state of the system at any two timesteps, as is provided by the solutions of generalised master equations. Subsequently, for completeness, we provide a traditional account of the open quantum systems approach to describing quantum stochastic processes. Along the way, we will highlight some critical shortcomings in order to motivate the necessity for the more general and operationally meaningful *process tensor* formalism to properly describe quantum processes [51], which we introduce in Chapter 3. Many of the notions presently introduced can be found in a number of excellent textbooks on the subject, for instance Refs. [7–9, 81–83].

### 2.2.1 *Open Quantum Systems*

The open systems paradigm acknowledges the fact that it is generally impossible to isolate a system of interest from its surroundings, and, as such, we must consider the effects of the external environment on the system [7, 8]. As we are typically unable to track the total evolution of the system and its environment—either by way of experimental limitation or lack of computational resources—the aim is to understand how the system of interest evolves dynamically. While the overall system and environment evolves unitarily according to the standard Schrödinger equation, the evolution on the level of the system alone need not, due to the environmental influence.

We begin with a brief description of closed quantum dynamics which leads to the open setting by eventually restricting our consideration to a subsystem of interest. States of a quantum system, $S$, are described by bounded linear operators on an associated Hilbert space $\rho \in \mathsf{BL}(\mathcal{H}^S)$ satisfying the following three properties

$$\rho = \rho^\dagger, \quad \rho \geq 0 \quad \text{and} \quad \text{tr}\,[\rho] = 1. \tag{2.20}$$

This *density operator* description is nicely tailored for our present purposes as it allows for the preparation of classical mixtures of pure quantum states which are pertinent to any stochastic theory, *e. g.*, a stochastic preparation procedure of a spin-$\frac{1}{2}$ particle where the $|\uparrow\rangle$ state is prepared with probability $p$ and the $|\downarrow\rangle$ state is prepared with probability $1-p$. Suppose we have a device preparing a system in one of a number of pure states $\{|\psi^{(i)}\rangle\}$ with probabilities $\{p_i\}$. The density operator of such an ensemble encodes all of the physically meaningful information about the system and is constructed as follows

$$\rho = \sum_i p_i |\psi^{(i)}\rangle\langle\psi^{(i)}|. \tag{2.21}$$

Pure states correspond to extremal states that cannot be written as a convex sum as above. In general, a density operator decomposition in terms of an ensemble of pure states is non-unique. The properties of Eq. (2.20) provide an intrinsic characterisation that holds independently of the ensemble interpretation, guaranteeing that its spectrum always represents a valid probability distribution. The Hermiticity condition ensures the eigenvalues are real; the positivity condition ensures they are all non-negative, thus corresponding to probabilities; and the trace condition stipulates an overall normalisation of probabilities. Importantly, density operators form a convex set: consider an ensemble of density operators $\{\rho^{(i)}\} \in \mathsf{BL}(\mathcal{H}^S)$ distributed with probabilities $\{p_i\}$. Then

$$\rho = \sum_i p_i \rho^{(i)} \tag{2.22}$$

is a valid representation of a quantum state.

In the density operator picture, the closed unitary evolution described by the Schrödinger equation translates into the *von Neumann* equation

$$\frac{\partial \rho_t}{\partial t} = -i[H_t, \rho_t], \tag{2.23}$$

where $[A, B] := AB - BA$ represents the commutator and we set $\hbar = 1$. Integrating the von Neumann equation admits a formal solution in terms of a unitary map, $\mathcal{U}_{t:0}$, that expresses the time-evolution of the system

$$\rho_t = \mathcal{U}_{t:0}\rho_0 := U_{t:0}\rho_0 U_{t:0}^\dagger, \tag{2.24}$$

where $\rho_0$ is the initial system state and the family of unitary operators are generated by a Hamiltonian $H_{t'}$

$$U_{t:0} = \mathcal{T}\left\{\exp\left(-i\int_0^t \mathrm{d}t'\, H_{t'}\right)\right\}, \tag{2.25}$$

where $\mathcal{T}$ represents the time-ordering operator.

The theory of open quantum systems concerns the evolution of a system, $S$, and some uncontrollable and inaccessible environment, $E$. The environment can describe any degrees of freedom that develop in time with the state of the system, *e. g.,* the system of interest could be a spin-$\frac{1}{2}$ particle and its environment an infinite field of bosonic modes. The joint Hilbert space of the system and environment is the tensor product of each subspace, $\mathcal{H}^{SE} = \mathcal{H}^S \otimes \mathcal{H}^E$, with the joint state space described accordingly as the subset of $\mathsf{BL}(\mathcal{H}^{SE})$ satisfying Eq. (2.20). The tensor product operation thus provides a natural way to represent composite systems: given some quantum states $\rho^S \in \mathsf{BL}(\mathcal{H}^S)$ and $\tau^E \in \mathsf{BL}(\mathcal{H}^E)$, the composition $\rho^S \otimes \tau^E \in \mathsf{BL}(\mathcal{H}^{SE})$ represents the joint system-environment state. On the other hand, there are elements $\rho^{SE} \in \mathsf{BL}(\mathcal{H}^{SE})$ of the joint state space that *cannot* be written as a tensor product of individual constituent quantum states, *e. g.,* convex mixtures $\sum_i p_i \rho^S \otimes \tau^E$ of product states, which are known as *separable* states; or those states that are not of separable form, known as *entangled* states. In either case, given knowledge of the joint state of some composite system, $\rho^{SE}$, we can deduce a description of the subsystems that adequately describes the local properties of each subsystem, *i. e.,* provides the correct statistics for any local measurement, via the partial trace operation

$$\rho^S = \mathrm{tr}_E\left[\rho^{SE}\right] \in \mathsf{BL}(\mathcal{H}^S) \quad \text{and} \quad \rho^E = \mathrm{tr}_S\left[\rho^{SE}\right] \in \mathsf{BL}(\mathcal{H}^E). \tag{2.26}$$

The open dynamics framework considers the entire system-environment to evolve according to Eq. (2.23), and the dynamics of the system alone is deduced by taking the partial trace over the environment

$$\frac{\partial \rho_t^S}{\partial t} = -i\,\mathrm{tr}_E\left[[H^{SE}, \rho_t^{SE}]\right], \tag{2.27}$$

which admits the solution

$$\rho_t^S = \mathrm{tr}_E\left[\mathcal{U}_{t:0}^{SE}\rho_0^{SE}\right] = \mathrm{tr}_E\left[U_{t:0}^{SE}\rho_0^{SE}U_{t:0}^{SE\dagger}\right]. \tag{2.28}$$

In contradistinction to Eq. (2.24), this is not a closed-form equation for the system state, as it depends on the dynamics of the joint state.

In addition to the evolution of quantum states, an important concept in quantum theory is that of *measurement*. A measurement on a quantum system is described by a collection of operators, $\mathcal{J} = \{K^{(x)}\} \in \mathsf{BL}(\mathcal{H}^S)$, such that $\sum_x K^{(x)\dagger} K^{(x)} = \mathbb{1}^S$. In anticipation of later developments, we refer to the entire collection as a *measurement instrument*: the index $x$ refers to the possible measurement outcomes that can be observed given that an experimenter interrogates the system with the instrument $\mathcal{J}$. If the pre-measurement state of the system is $\rho$, then the probability that outcome $x$ occurs is given by

$$\mathbb{P}(x|\mathcal{J}) = \mathrm{tr}\left[K^{(x)\dagger} K^{(x)} \rho\right]. \tag{2.29}$$

Upon recording outcome $x$ when using the instrument $\mathcal{J}$ to interrogate the system, its state undergoes the transformation

$$\rho \overset{x}{\mapsto} K^{(x)} \rho K^{(x)\dagger}. \tag{2.30}$$

As a special case of this general notion of quantum measurement is when the post-measurement state is not of interest, but merely the statistics associated to measurement outcomes are. In this case, it is sufficient to consider only collections of operators defined via $\Pi^{(x)\mathrm{T}} := K^{(x)\dagger} K^{(x)} \in \mathsf{BL}(\mathcal{H}^S)$, where the transpose $(\bullet)^\mathrm{T}$ is added to the standard definition to better align with later notation. By definition, such a collection satisfies $\sum_x \Pi^{(x)\mathrm{T}} = \mathbb{1}$ and contains only Hermitian, positive semidefinite operators, leading to the standard *Born rule*

$$\mathbb{P}(x|\mathcal{J}) = \mathrm{tr}\left[\Pi^{(x)\mathrm{T}} \rho\right]. \tag{2.31}$$

Overall, any such set $\{\Pi^{(x)}\}$ is known as a *positive-operator valued measure* (POVM), with each constituent operator referred to as a POVM-*element*. The linear functional induced on quantum states to yield probabilities via the Born rule, *i. e.*, $\mathcal{E}^{(x)}(\bullet) := \mathrm{tr}\left[\Pi^{(x)\mathrm{T}} \bullet\right]$, is known as an *effect*. A POVM is a special case of a measurement instrument that contains the necessary information to determine the observed statistics of any measurement on a normalised quantum state, but is insufficient to deduce the post-measurement state. Another special case considers projective measurements, where each operator associated to a measurement instrument is a projector that is orthogonal to every other, *i. e.*, $\{\Pi^{(x)}\}$ such that $\Pi^{(x)} = \Pi^{(x)\dagger}$ and $\Pi^{(x)} \Pi^{(x')} = \delta_{xx'} \Pi^{(x)}$.

2.2.2 *Dynamical Maps*

To reiterate an earlier point, what we often desire when studying a physical evolution is simply a description of how to map quantum states from one point in time to another. In the open dynamics framework, where we assume only access to the system, we would like such a map to act only on the space of the system, in analogy to the action of the unitary map for closed dynamics in Eq. (2.24). However, due to potentially dissipative system-environment interactions in the open setting, such a map is no longer restricted to being unitary. We first aim to understand the properties such a map must have in order to represent a physical evolution within quantum theory from an abstract perspective, before providing a connection to the standard open systems framework.

In short, we aim to understand the physically allowable quantum transformations between two points in time, *i.e.*, the set of maps $\mathcal{C} : \mathsf{BL}(\mathcal{H}^S) \to \mathsf{BL}(\mathcal{H}^S)$ that take an arbitrary initial quantum state $\rho \in \mathsf{BL}(\mathcal{H}^S)$ to a valid output quantum state $\sigma \in \mathsf{BL}(\mathcal{H}^S)$, schematically depicted in Fig. 2.7, as follows[4]

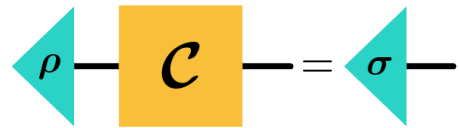

**Figure 2.7:** *Dynamical map.* A dynamical map, $\mathcal{C}$ (yellow), takes input states $\rho$ to output states $\sigma$ (green). Any such transformation in quantum theory must be linear, completely-positive and trace-preserving.

$$\sigma = \mathcal{C}(\rho). \tag{2.32}$$

In the classical setting, the requirement that a map takes arbitrary input probability distributions to valid output distributions serves to constrain their structure to be stochastic maps. Similarly, the analogous demand in quantum theory imposes structural constraints on the allowable maps. Any meaningful transformation must preserve the key properties of the density operator, *i.e.*, it must preserve trace, Hermiticity and positivity, in order to ensure that $\sigma$ indeed represents a valid quantum state. In addition, its action must preserve convex mixtures of states

$$\mathcal{C}\left(\sum_i p_i \rho^{(i)}\right) = \sum_i p_i \mathcal{C}\left(\rho^{(i)}\right) = \sum_i p_i \sigma^{(i)}. \tag{2.33}$$

This is not a requirement that stems from the linearity of quantum mechanics; rather, it follows from the *linearity of mixing* principle that must be satisfied by any statistical theory. The importance of this principle can be highlighted through the following

---

[4] The present restriction of our attention to maps taking states of some input system to those of *the same* system, *i.e.*, living on the same Hilbert space, as its output is by no means necessary and will be relinquished in the forthcoming chapter.

*Gedankenexperiment*: suppose Alice prepares a quantum system in either state $\rho_1$ or $\rho_2$, which she sends to Bob, with the transmission represented by the map $\mathcal{C}$. Bob then performs state tomography [9, 84] to determine which state Alice sent him, with the entire protocol repeated many times. Suppose first that Alice only sends $\rho_1$ on Monday and $\rho_2$ on Tuesday. Bob will then conclude that he receives $\sigma_1 = \mathcal{C}(\rho_1)$ on Monday and $\sigma_2 = \mathcal{C}(\rho_2)$ on Tuesday. Now suppose instead that Alice sends these two states at random, with probabilities $p$ and $1-p$ respectively. Without knowledge of which state was sent in each run, Bob would conclude that he receives the average state $\sigma = \mathcal{C}(\rho)$, where Alice's preparation corresponds to the average state $\rho = p\rho_1 + (1-p)\rho_2$. Consider now the scenario where Alice reveals to Bob which state she sent in which run; surely now Bob concludes that he received the states $\sigma_1$ or $\sigma_2$ whenever Alice sent him $\rho_1$ or $\rho_2$ respectively. On the other hand, averaging over the trials would amount to Bob receiving $\sigma$. Thus, it must be the case that the transformation map acts linearly: $\sigma = p\sigma_1 + (1-p)\sigma_2 = p\mathcal{C}(\rho_1) + (1-p)\mathcal{C}(\rho_2)$.

In analogy to the classical case, trace-preservation of the density operator corresponds straightforwardly to the classical notion that the output of a stochastic map is normalised; likewise, Hermiticity and positivity preservation echo the demand that stochastic maps take probability distributions to probability distributions. However, a distinct departure in the quantum setting arises due to the existence of entangled states within the theory, which enforces us to strengthen the notion of positivity, which means that the output state of a map is always positive semidefinite, *i. e.*, $\sigma = \mathcal{C}(\rho) \geq 0 \ \forall \ \rho$, to the stricter one of *complete-positivity*.

Consider the situation were the initial state of some bipartite system $\rho^{AB} \in \mathsf{BL}(\mathcal{H}^{AB})$ is represented by an entangled density operator

$$\rho^{AB} \neq \sum_i p_i \alpha^{(i)} \otimes \beta^{(i)}, \tag{2.34}$$

where $\{\alpha^{(i)}\} \in \mathsf{BL}(\mathcal{H}^A)$ and $\{\beta^{(i)}\} \in \mathsf{BL}(\mathcal{H}^B)$ are density operators describing the states of the subsystems $A$ and $B$ respectively and $\{p_i\}$ are probabilities. Suppose these subsystems are sufficiently well-separated and undergo separate evolutions; one can easily construct examples for which the post-evolution joint state is not a positive semidefinite operator. For example, if $A$ undergoes a trivial evolution $\mathcal{I}^A(\bullet)$ whilst $B$ is subject to the transposition map, *i. e.*, and $\mathcal{T}^B(\bullet) := (\bullet)^{\mathrm{T}}$. Both of these maps are positive, *i. e.*, $\mathcal{I}^A(\rho^A) \geq 0 \ \forall \ \rho^A \in \mathsf{BL}(\mathcal{H}^A)$ and $\mathcal{T}^B(\rho^B) \geq 0 \ \forall \ \rho^B \in \mathsf{BL}(\mathcal{H}^B)$. However, when applied jointly to an entangled state $\rho^{AB} \in \mathsf{BL}(\mathcal{H}^{AB})$, a negative operator, representing no physical state, can result $\mathcal{I}^A \otimes \mathcal{T}^B(\rho^{AB}) < 0$.

The natural condition required to describe meaningful quantum evolutions of composite systems is that their local implementation leads to a valid joint output state in the presence of any innocuous ancillary system (see Fig. 2.8). Formally, when a map acting on some system, $\mathcal{C}^S$, is extended to act trivially on an arbitrarily-sized ancillary space, $\mathcal{H}^R$, via the identity map $\mathcal{I}_d^R$ acting on $d$-dimensional quantum states, its action on an arbitrary joint state $\rho^{SR} \in \mathsf{BL}(\mathcal{H}^{SR})$ must lead to a positive semidefinite output

$$\sigma^{SR} = \mathcal{C}^S \otimes \mathcal{I}_d^R(\rho^{SR}) \geq 0 \quad \forall\, d \in \mathbb{N}. \tag{2.35}$$

We refer to maps satisfying the above condition (which also implies Hermiticity preservation [36]) as *completely-positive* (CP) maps.[5]

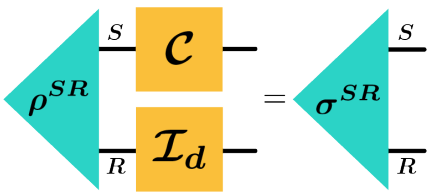

**Figure 2.8:** *Complete-positivity.* To represent a valid evolution in quantum theory, a dynamical map $\mathcal{C}$ must be completely-positive. Consider a joint system, $SR$, with $R$ of dimension $d$, in an arbitrary state $\rho^{SR}$. Complete-positivity of $\mathcal{C}$ means that when it is implemented locally on $S$, whilst $R$ evolves trivially under $\mathcal{I}_d$, the joint output is a valid quantum state $\sigma^{SR} \in \mathsf{BL}(\mathcal{H}^{SR})$, for all $d \in \mathbb{N}$.

Lastly, the trace-preservation requirement, stemming from the fact that the probabilities of any measurement on the output state sum to one, is simply $\mathrm{tr}\,[\sigma] = \mathrm{tr}\,[\mathcal{C}(\rho)] = \mathrm{tr}\,[\rho] \;\forall\; \rho \in \mathsf{BL}(\mathcal{H}^S)$. An alternative way to consider trace-preservation is that the transformation $\mathcal{C}$ taking $\rho$ to $\sigma$ occurs overall with certainty. Indeed, one can envisage situations where a process only occurs conditionally with some probability, such as that on a quantum state upon recording a measurement outcome, in which case the trace-preservation condition must be slightly modified to capture such scenarios (although the dynamics must remain CP). We will focus on such maps in detail within the broader framework introduced in the coming chapter.

To summarise, the most general form of maps describing overall deterministic transformations in quantum theory are those that are both CP and *trace-preserving* (TP). Due to their importance in the study of open dynamics, these maps are referred to synonymously as CPTP maps, *dynamical maps* or *quantum channels* [7, 81]. Since the set of density operators is convex, and CPTP maps act linearly, the space of CPTP maps is also a convex set. So much for the mathematical properties of the allowable maps taking input states to output states in open quantum theory. We now make the connection with an underlying joint system-environment evolution.

---

[5] We note that although maps that are positive but not completely-positive, such as the partial transpose considered above, are useful for witnessing entanglement in quantum states [85, 86], for the purposes of describing quantum evolution in a meaningful way, we require the associated map to be CP.

A major result concerning the axiomatic considerations above and the vantage point of open quantum systems dynamics is Stinespring's dilation theorem [87], which guarantees that any CPTP map on the system can be thought of as arising from some underlying unitary dynamics of the system with some environment. That is, we can represent any CPTP map $\mathcal{C}^S : \mathsf{BL}(\mathcal{H}^S) \to \mathsf{BL}(\mathcal{H}^S)$ in terms of a dilated picture of unitary dynamics in $\mathcal{H}^{SE}$, with a fiducial initial environment state $\tau^E \in \mathsf{BL}(\mathcal{H}^E)$ and the environment being ignored following the evolution, depicted in Fig. 2.9, as

$$\sigma^S = \mathcal{C}^S(\rho^S) = \mathrm{tr}_E\left[\mathcal{U}^{SE}\rho^S \otimes \tau^E\right]. \tag{2.36}$$

The dilation for a quantum channel is non-unique: many configurations of $\{\mathcal{U}^{SE}, \tau^E\}$ can give rise to the same dynamical map $\mathcal{C}^S$. Nonetheless, a dilation can always be found for an environment with dimension $d_E^2$ [87]. It is often quipped that this correspondence allows us to always *go to the Church of the Larger Hilbert Space*, considering any overall deterministic but irreversible (*i. e.,* non-unitary) transformation of the density operator of the system to be the manifestation of our subjective ignorance of part of a reversible (unitary) transformation in the larger joint Hilbert space.

Stinespring's theorem tells us that any CPTP map can always be dilated to a fixed (though non-unique) quantum circuit comprising a joint unitary transformation through which the system interacts with a fiducial environment state, the latter of which is finally discarded. On the other hand, a CP map can always be thought of as arising similarly through a joint unitary interaction with some initial environment state, although in this case, rather than tracing over the degrees of freedom of the environment following the evolution, one hypothetically performs a measurement and postselects on an outcome [29]. Since such a procedure can only occur probabilistically, there can be no such *fixed* underlying dilation model attributed to the map, in contrast with the case for CPTP maps; thus, the property of trace-preservation may be interpreted as one of an *overall deterministic* implementation of the transformation.

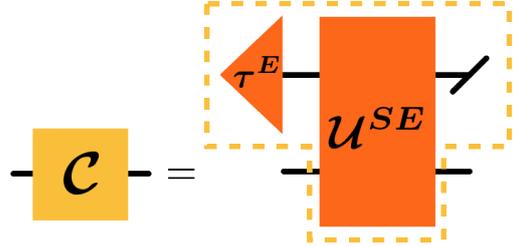

**Figure 2.9:** *Stinespring dilation of a quantum channel.* Any CPTP map $\mathcal{C}$ can be dilated in terms of a unitary interaction $\mathcal{U}^{SE}$ of the system with some fiducial environment state $\tau^E$, which is finally disregarded (indicated by the diagonal slash). A dilation is non-unique: the channel contains all information enclosed in the yellow border on the right, but individual contributions of the environment state and the joint unitary (orange) cannot be delineated.

### 2.2.3 *GKSL Equation*

The above discussion culminates in the most general structure of a quantum evolution taking density operators to density operators. We can bring time back into the picture to see how a continuous-time evolution can naturally lead to CPTP dynamics on the system level. Although the exact evolution of the system is governed by its microscopic interactions with the environment, we can invoke the *approximation* that the evolution of the system only depends on its current state via some linear map, $\mathcal{L}_t$, and write down the following equation of motion for the system density operator [8]

$$\frac{\partial \rho_t^S}{\partial t} = \mathcal{L}_t(\rho_t^S). \tag{2.37}$$

When the generator is time independent, the formal solution to this equation constitutes a single-parameter semi-group, *i.e.*, a family of norm-continuous maps $\mathcal{C}_t^S = \exp(t\mathcal{L})$ satisfying $\mathcal{C}_{t+s}^S = \mathcal{C}_t^S \mathcal{C}_s^S$.[6] This allows us to express the system state at arbitrary time $t$ in terms of a map acting on an arbitrary initial state via

$$\rho_t^S = \mathcal{C}_{t:0}^S(\rho_0^S). \tag{2.38}$$

The family of dynamical maps $\mathcal{C}_{t:0}^S$ naturally inherit the properties of linearity, complete-positivity and trace-preservation. Furthermore, the semi-group property leads to the divisibility of the dynamical maps, *i.e.*, they satisfy $\mathcal{C}_{t:0}^S = \mathcal{C}_{t:s}^S \mathcal{C}_{s:0}^S \ \forall \ t \geq s \geq 0$. Intuitively, this means that we can calculate the system density operator at any time via a successive composition of previous dynamical maps. We will return to a deeper discussion of this concept in the coming section.

A breakthrough result for the field of open quantum systems was presented independently by *Gorini, Kossakowski and Sudarshan* [88] and *Lindblad* [89] (GKSL)[7], who provided the most general form of the generator $\mathcal{L}_t$ in Eq. (2.37) such that the resulting solution forms a divisible CPTP semi-group, yielding the GKSL equation

$$\frac{\partial \rho_t}{\partial t} = -i[H, \rho_t] + \sum_i \gamma_i \left( L_i \rho_t L_i^\dagger - \frac{1}{2} \left\{ L_i L_i^\dagger, \rho_t \right\} \right). \tag{2.39}$$

Here, everything is understood to act on the space of the system only. Note that the anti-commutator is represented by $\{A, B\} := AB + BA$, the $H$ represents a Hamiltonian (Hermitian) contribution to the system evolution, the $\{L_i\}$ are known as Lindblad operators and the $\{\gamma_i\} \geq 0$ are non-negative rates.

---

6 For a time dependent $L_t$, the solution would involve some time-ordered exponential and time integral, and would depend on two time parameters.

7 This equation has also been independently discovered by Franke in 1976 [90], although his name has not made it into the famous acronym yet.

To reiterate, the GKSL equation is the quantum generalisation of the classical ME (see Appendix B.1): the solution of the latter is a P-divisible family of stochastic maps, whereas that of the former is a divisible family of CPTP maps taking the density operator from one point in time to any other later time in a physically acceptable manner, as per Eq. (2.38) [88, 89]. Moreover, one can arrive directly at the form of the GKSL equation when starting from a system-environment model and invoking the Born-Markov and secular/rotating wave approximations [8, 83].

### 2.2.4 *Tomographic Reconstruction of Quantum Channels*

Recall that in the open systems framework, we assume that we cannot probe the state of the environment and do not know the joint unitary transformations. We now consider how one can experimentally reconstruct a description of the dynamical map for the evolution between pairs of timesteps by probing the system alone. By considering a fixed environment state in Eq. (2.36), the linear dynamical map is induced on the state space of the system. This property of linearity importantly permits a procedure to reconstruct quantum channels, known as quantum *process tomography* [9, 84, 91].

To understand process tomography, it is useful to first briefly consider quantum *state tomography*, where an unknown quantum state is inferred through measurement statistics. Given a state $\rho \in \mathsf{BL}(\mathcal{H}^S)$, one begins by choosing a POVM that spans $\mathsf{BL}(\mathcal{H}^S)$. Such a set of operators is called *informationally-complete* (IC) and necessarily contains at least $d^2$ POVM elements. By recording the probability for each measurement outcome, the quantum state can be uniquely reconstructed due to linearity [9, 84].

In direct analogy to state tomography, the protocol for quantum process tomography, depicted graphically in Fig. 2.10, is as follows [91]: **i)** take a basis of input states $\{\hat{\rho}^{(i)}\}_{i=1}^{d^2}$ that span the operator space of the system;[8] **ii)** each of these input states are sent through the process; **iii)** the corresponding output states $\{\hat{\sigma}^{(i)}\}_{i=1}^{d^2}$ are determined via state tomography; and **iv)** the input-output relations deduced can be linearly inverted to uniquely determine the dynamical map.

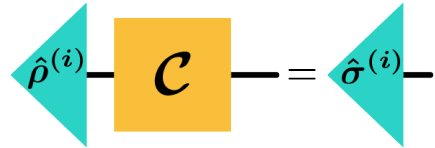

**Figure 2.10:** *Tomographic reconstruction of a quantum channel.* A quantum channel $\mathcal{C}$ can be reconstructed by determining the output states $\{\hat{\sigma}^{(i)}\}$ corresponding to an IC set of inputs $\{\hat{\rho}^{(i)}\}$.

---

8  Here, we use the caret notation to indicate that the object belongs to a fixed basis, which is not necessarily normalised or orthogonal.

Mathematically, we make use of the natural inner product on $\mathsf{BL}(\mathcal{H}^S)$, *i.e.*, for any elements $\mu, \nu \in \mathsf{BL}(\mathcal{H}^S)$, we have the Hilbert-Schmidt inner product $(\mu, \nu) := \mathrm{tr}\left[\mu^\dagger \nu\right]$. We can always construct a basis of $\mathsf{BL}(\mathcal{H}^S)$ with a set of $d^2$ density operators. For any such basis, there exists a dual set of objects, $\{\hat{D}^{(i)}\}_{i=1}^{d^2}$, such that [52, 75]

$$\mathrm{tr}\left[\hat{D}^{(i)\dagger} \hat{\rho}^{(j)}\right] = \delta_{ij} \quad \forall \, i, j. \tag{2.40}$$

We provide an explicit construction of the dual set to an arbitrary basis in Appendix B.2.

By linearity, determination of the output states for a basis of input states uniquely specifies the map. Thus the action of the CPTP map $\mathcal{C}^S$ on an arbitrary input state $\rho \in \mathsf{BL}(\mathcal{H}^S)$ can be linearly extended and expressed as

$$\mathcal{C}^S(\rho) = \sum_{i=1}^{d^2} \hat{\sigma}^{(i)} \, \mathrm{tr}\left[\hat{D}^{(i)\dagger} \rho\right] \tag{2.41}$$

The elements of the dual basis to a basis of density operators are elements of $\mathsf{BL}(\mathcal{H}^S)$, but not necessarily density operators, and the overall construction above is guaranteed to yield a positive semidefinite output.

Lastly, note that in this procedure it is assumed that the initial state of $SE$ is uncorrelated, allowing us to treat the reduced initial state of the environment $\tau^E$ as a fixed constant of the problem. This property means that the dynamics is independent of the initial state of the system, allowing us to vary the inputs to the channel freely, as is a tenet of the process tomography protocol. For instance, GKSL dynamics allows for this clear delineation between the instantaneous state of the system and the subsequent dynamics at all times (see Eq. (2.37)), therefore leading to a CP-divisible description.

## 2.3 A PROBLEM OF FORMALISM

So far, we have explored the open quantum systems formalism and the dynamical map in describing the evolution of a quantum state to one at a later time. The usefulness of the dynamical map picture comes to the fore when considering its ability to be tomographically reconstructed in a finite number of experiments, providing an unambiguous description of the process in terms of accessible quantities. However, the dynamical map description does not accommodate for intermediary interventions on the level of the system. In order to understand memory effects in quantum stochastic processes, we would like to develop a similar operational framework that actively accounts for multi-time correlations. As we shall see throughout this section, a number of subtleties arise when attempting to do so.

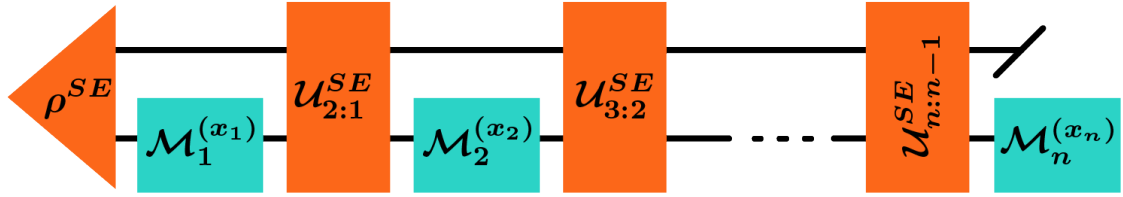

**Figure 2.11:** *Dilation of a quantum process interrogated in time.* A joint system-environment state $\rho^{SE}$ is interrogated in time by measurements $\mathcal{M}_j^{(x_j)}$ interspersed throughout periods of joint unitary evolution $\mathcal{U}_{2:1}^{SE}$.

Consider an experimenter making measurements on some system over a set of timesteps, $\Lambda_n$, described by the measurement operators $\{M_j^{(x_j)}\}$ interspersed throughout periods of joint unitary dynamics, with $t_j \in \Lambda_n$ denoting the timestep of each measurement. The conditionally realised transformation of the system associated to observing the outcome $x_j$ is described by the CP map $\mathcal{M}_j^{(x_j)}(\rho^{SE}) := M_j^{(x_j)} \rho_j^{SE} M_j^{(x_j)\dagger}$, where the measurement operators act on the system alone and identity operators on the environment are implied. The joint probability distribution of the statistics observed given a sequence of measurements applied over time is depicted in Fig. 2.11 and written

$$\mathbb{P}_{n:1}(x_n,\ldots,x_1) = \mathrm{tr}\left[\mathcal{M}_n^{(x_n)} \mathcal{U}_{n:n-1}^{SE} \ldots \mathcal{M}_2^{(x_2)} \mathcal{U}_{2:1}^{SE} \mathcal{M}_1^{(x_1)} \rho^{SE}\right]. \tag{2.42}$$

To reiterate, here, the unitary maps $\mathcal{U}_{j:j-1}^{SE}$ are understood to act on the joint system-environment space, while the CP maps $\mathcal{M}_j^{(x_j)}$ act only on the system, and we have chosen not to include nested parentheses to avoid notational clutter, as we will often do throughout this thesis, with the understanding that all maps act on everything to the right of them.

As exemplified previously through the Stern-Gerlach example at the beginning of Section 2.2, a major problem of this description is that the resulting joint distributions do not satisfy the containment property of the KET. Logically, from a proper description of the dynamics over multiple timesteps, we expect to be able to deduce the correct description of the dynamics on any subset of timesteps. Its breakdown arises due to the necessarily invasive nature of measurements in quantum mechanics [15]: in contrast to classical theory, here, choosing not to interrogate the system is different from averaging over all possible measurement outcomes.

An additional point of concern is that for a system evolving in an open fashion, performing a measurement on the system can condition the state of the environment. This is not a purely quantum mechanical feature, but rather a symptom of stochastic processes with memory. For example, consider applying a rank-1 projective measurement $\mathcal{P}^{(x)}(\bullet) := P^{(x)} \bullet P^{(x)}$ on the subsystem $S$ of a correlated $SE$ state, as shown in Fig. 2.12. Upon recording outcome $x$, the joint state maps to

$$\rho^{SE} \xmapsto{x} P^{(x)} \otimes \tau^{(x)}, \tag{2.43}$$

where $\tau^{(x)} = \mathrm{tr}_S\left[P^{(x)} \rho^{SE}\right] \in \mathsf{BL}(\mathcal{H}^E)$. Although the post-measurement state is uncorrelated for each measurement outcome—it is a tensor product state—the post-measurement environment state is correlated with the outcome observed; as such, the future evolution of the system can depend upon knowledge of previous outcomes.

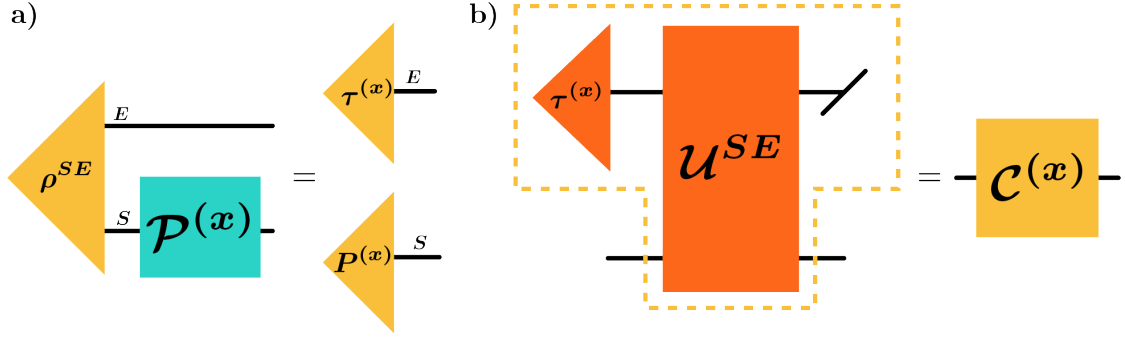

**Figure 2.12:** *Measuring the system conditions the environment and influences future dynamics.* In panel **a)**, the $S$ subsystem of an initially correlated system-environment state $\rho^{SE}$ is subject to a projective measurement, which has the effect of conditioning the environment into some state $\tau^{(x)}$. In panel **b)**, we consider the scenario where some system-environment dynamics $\mathcal{U}^{SE}$ subsequently occurs. As the environment state that the system goes on to interact with is conditioned by the measurement outcome (rather than being some fixed, fiducial state), a *set of many different* dynamical maps $\{\mathcal{C}^{(x)}\}$ that describe the evolution is induced, one for each measurement outcome. Each of these maps has a non-deterministically occurring environment state in its dilation, represented by the orange objects enclosed by the yellow dashed line.

The points raised above have led many to the conclusion that an "intrinsic characterization and quantification of memory effects in the dynamics of open quantum systems... has to be based solely on the properties of the dynamics of the open system's *density matrix*" [24]. In the coming chapter we defy this statement, highlighting that it is a problem of *formalism* that must be overcome to obtain a fully fledged description of quantum stochastic processes, rather than a fundamental issue. To put the cart before the horse, there exist operational frameworks that can account for multi-time correlations, superseding traditional open systems frameworks and subsequently allowing for a proper description of memory effects. We will first examine in detail how criteria based on two-point descriptions of quantum processes fail to accurately represent the underlying process when memory effects are present, motivating the fact that we are forced to go beyond such a paradigm to characterise generic quantum stochastic processes.

2.3.1 *A Hierarchy of Notions of Non-Markovianity*

Due to the breakdown of the KET on the level of observed probabilities in quantum mechanics—and therefore the lack of a well-defined notion of conditional statistics—the concept of Markovianity in the quantum realm has been mainly studied throughout the open systems community in terms of the time-evolution of the density operator, or, more operationally, the dynamical maps describing the evolution between pairs of timesteps. Of course memory effects display signatures that can be gleaned from such two-point considerations, giving rise to a myriad of non-Markovianity *witnesses*. The usefulness of such approaches is not to be underestimated: from a practical perspective, they often provide easy-to-check criteria that purport to verify the presence of memory effects. Prominent examples include those based on: the divisibility of the dynamics [67, 92]; the monotonically non-increasing nature of the distinguishability of quantum states subject to the evolution [93]; the detection of initial correlations [75, 94–99]; the positivity of the dynamical maps [100–102]; changes to quantum correlations or coherence [103, 104]; changes to the Fisher information [105]; and channel capacities and information backflow [106–109]; to name but a few. See Refs. [24, 77] for a thorough overview of these various concepts and their relations.

The main problems with such approaches are that they: **i**) lack a clear operational interpretation; **ii**) do not coincide with Markovianity in the classical limit; and **iii)** do not agree on the characterisation of whether a given process is Markovian or not. Consequently, different introduced "measures" of non-Markovianity disagree on both the degree of non-Markovianity and whether or not memory effects are present at all [50, 68, 107, 110, 111].

For instance, one of the most widely used criteria is based on the notion of CP-*divisibility* [67], which intuitively means that the open dynamics can be broken into a piecewise composition of dynamical maps satisfying[9]

$$\mathcal{C}_{k:i} = \mathcal{C}_{k:j}\mathcal{C}_{j:i} \quad \forall\, t_k > t_j > t_i. \tag{2.44}$$

The property of CP-divisibility implies some of the other aforementioned concepts, *e.g.*, the non-increasing distinguishability criteria is a direct consequence of the contractivity of the trace-distance under CPTP maps, although the converse does not hold [67]. Consequently, there are many processes that the former characterisation would deem to be non-Markovian, but the latter would disagree, *e.g.*, see Refs. [110, 111].

---

9 Note that CP-divisibility obviously implies P-divisibility.

Moreover, considering how an experimenter might test for CP-divisibility in practice illuminates that there are (at least) two inequivalent definitions of divisibility [55]. This is because of the difficulties that arise when attempting to tomographically reconstruct the maps $\mathcal{C}_{k:j}$. By time $t_j$, the system will generally be correlated with its environment, hindering ones ability to vary the input state to the map independently of the environment and therefore breaking a crucial requirement of quantum process tomography. To circumvent this problem, under the assumption that no initial correlations are present at $t_i$ so that the maps $\mathcal{C}_{k:i}$ and $\mathcal{C}_{k:j}$ can be reconstructed and that $\mathcal{C}_{j:i}$ is invertible, one may compute an artificial map $\mathcal{A}_{k:j} = \mathcal{C}_{k:i}\mathcal{C}_{j:i}^{-1}$ and test if it is CPTP.

Notwithstanding the fact that neither of these assumptions might be satisfied, it is unclear operationally what dynamics the inverse map $\mathcal{C}_{j:i}^{-1}$ and therefore the derived map $\mathcal{A}_{k:j}$ actually represents. An operational way around this is to simply discard any possible system-environment correlations present at $t_j$ by, *e. g.,* making a measurement and preparing the output state in a fixed state that is independent of measurement outcomes, which are then averaged over. This allows for the experimental reconstruction of $\mathcal{C}_{k:j}$ which is guaranteed to be CPTP, and one must simply check whether Eq. (2.44) holds. Within the set of processes where the dynamical maps in the former construction are invertible, the latter operational criteria is stricter; in either case, just as in the classical setting, CP-divisibility does not imply Markovianity [28, 55].

Additional examples of a similar flavour can be found in the Supplemental Material of Ref. [50]: in particular, an example where the trace-distance distinguishability criteria would deem the process Markovian, and another in which no system-environment correlations are ever built up in the process, which is also often considered a hallmark of Markovianity [77]; however, both of these examples can display memory effects. In summary, each of the criteria proposed at the beginning of this section are based upon an inadequate description of processes with memory, giving rise to the aforementioned inconsistencies and leading some to the conclusion that there can be no unique condition for Markovianity for quantum stochastic processes. This is not true: the problem we must overcome is one of formalism, as we now discuss.

### 2.3.2 *Limitations of Traditional Approaches*

The reason that the traditional approaches considered above fail is because Markovianity is, at its core, a statement concerning multi-time conditional statistics: in its truest sense, determining whether a process is Markovian requires testing for conditional in-dependence between the current statistics measured from those deduced at all earlier times. This demands an exponentially large set of conditions to be satisfied, even in the classical case, *i.e.*, Eq. (2.11) must hold for all $x_1, \ldots, x_n$. Despite the added complication of the incompatibility of measurement statistics in quantum mechanics described above, the fact that any two-point description of dynamics cannot suffice to characterise Markovianity is more of a logical statement: it simply cannot be used to test all conditions, and therefore is inadequate at describing the complete story. In short, a proper treatment of non-Markovian processes *cannot* be a simple extension of the tools used to describe Markovian ones [26].

The fact that the dynamical map formalism fails to capture multi-time memory effects highlights its major shortcoming when it comes to describing quantum processes with memory. Moreover, the dynamical map description cannot even accurately address all two-point dynamics allowable in quantum theory; for instance, it is well-known that it fails to describe open dynamics in the presence of initial system-environment correlations [66, 74–76, 112–115]. As discussed earlier, the traditional treatment and reconstruction of open quantum system dynamics assumes an initially uncorrelated system-environment state. This hypothesis assigns a peculiar role to the initial time; even if it happens to be true at some time, which could be taken as the initial time leading to subsequent CPTP dynamics to any later time, the system will, at this later time, generally be correlated to its environment and we immediately face the problem of describing the dynamics beyond that later timestep. Thus, the proper description of quantum processes across multiple timesteps inherently concerns the initial correlation problem, which must be overcome as a starting point.

This problem came to the fore due to technological advances that allowed experimentalists to begin tomographically reconstructing quantum logic gates by means of process tomography [116–120]. Although the gates implemented were expected to be non-ideal, they were expected to be CPTP. However, to the surprise of many, this was not necessarily the case: the dynamical maps constructed were not CP. A notable theoretical explanation for the construction of non-CP quantum gates was immediately put forth by Pechukas, who realised that in the presence of initial system-environment correlations, the dynamical map formalism suggests that the subsequent dynamics of the system need not be CP [112]. This means that a density operator describing a quantum state can evolve, in the presence of some ancillary system, into a non-positive operator, whose physical interpretation is unclear. Fundamentally, this notion is at odds with the fact that we always observe positive probabilities in any experiment. Furthermore, relinquish-ing the requirement of CP means sacrificing many physically important principles, such as the Holevo bound [121], data processing inequality [122], and the entropy production inequality [123]. An alternate approach is to give up the requirement of linearity [114], which also proves problematic: complete tomography is no longer possible in general when the dynamics is non-linear [52, 124]; the data processing inequality is also violated [9]; and numerous problematic implications have been shown to arise (see, *e. g.,* Refs. [125–127]). On the other hand, at first glance, it seems as if we are in a double-bind: forked between sacrificing either CP or linearity for a consistent description of the dynamics [115].

Despite their original operational motivations, dynamical map descriptions do not take active interventions into account, besides the limited scenario where the initial system state is uncorrelated from its environment and active preparation procedures can be enacted *without* influencing the subsequent dynamics. When initial system-environment correlations are present, on the other hand, one cannot probe the system without also affecting the environment (see Eq. (2.43)). Since the environment can be conditioned by an operation on the system and then can feed forward to play a role in the subsequent evolution (see Fig. 2.12 **b)**), any tomographically reconstructed description of the dynamical map will depend upon the interventions applied to the system, seemingly implying the lack of a process that exists independent of the experimenter. This is why, for instance, even non-Markovian generalisations of MEs with memory kernels, which are useful for simulating processes with memory [25], are insufficient to characterise them: unless they allow for active interventions on the system, their operational consequences are unclear. As we shall soon see, this blurriness between the the dynamics governed by the process and the transformations applied by a probing experimenter is directly related to the breakdown of the KET in quantum theory.

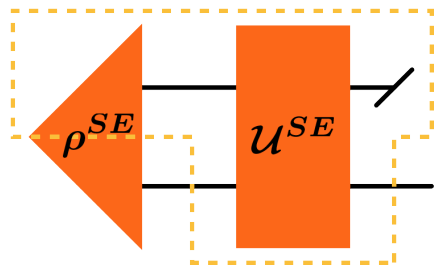

**Figure 2.13:** *Initial correlation problem.* The dynamical map formalism assumes the ability to consider the environment as some fixes state. This is not the case when initial correlations are present, since we cannot separate system states that are input to the subsequent dynamics from the dynamics themselves. This artefact is depicted by the yellow dashed line which (wrongly) attempts to 'cut' the initially correlated state.

To summarise, on the one hand, the lack of consistency conditions on the level of probability distributions seemingly imply that there may not exist a unique, fixed process giving rise to the statistics observed. This renders any notion of conditioning, as is required to characterise Markovianity and Markov order more generally, nonsensical. On the other hand, any framework that does not properly account for active interventions on the system, such as those based on dynamical maps, are necessarily inadequate to describe processes with memory. In order to generalise Markov order to the quantum realm, we require a reasonable picture of multi-time correlations. We stress, finally, that in any such case, these problems arise due to inadequacy of *formalism* rather than fundamental physical truths: a proper description of a stochastic process should ameliorate these aforementioned issues. To correctly describe the statistics observed for sequences of measurement outcomes, thereby capturing multi-time memory effects, we are forced to go beyond such traditional approaches and develop a framework that meaningfully accounts for active interventions, as we consider in the coming chapter.

# 3

## QUANTUM STOCHASTIC PROCESSES

Throughout the previous chapter, we explored in detail the nature of both classical and quantum stochastic processes, highlighting some key difficulties that arise when attempting to characterise memory in the quantum case. Here, we introduce a general formalism that accommodates an unambiguous study of processes with memory by making explicit the role of the *experimenter*.[1] By separating the underlying, uncontrollable system-environment dynamics of the process at hand from the controllable interventions an experimenter might choose to apply on the level of the system, we come to a robust operational framework for describing stochastic processes that actively takes interventions into account, thereby solving the aforementioned issues.

One such framework that accounts for possible interventions across multiple timesteps is that of the *process tensor* [50, 51], which is the fundamental mathematical object of the operational formalism we employ throughout the remainder of this thesis to describe quantum stochastic processes. It provides a multi-linear mapping from sequences of controllable operations applied by an experimenter to the final output density operator of the evolution. Additionally, by way of a generalised spatio-temporal Born rule, the process tensor yields the correct joint statistics for any temporal operation sequence an experimenter might implement [35]. It thereby encapsulates all (multi-time) memory effects in the process, which are crucial for a proper treatment of non-Markovian processes; indeed, within the context of open system dynamics, the process tensor was developed specifically to generalise traditional approaches of the open systems formalism [50, 51], which are limited in scope to two-point correlations.

Before properly introducing this versatile framework, it is worth mentioning that similar formalisms have been developed within various other contexts, initially by Lind-blad [27] and Accardi, Frigerio and Lewis [28], with more modern incarnations going under the guises of: *quantum combs* in the context of generic quantum circuit architectures [29–31], *causal automata* or *non-anticipatory quantum channels* to address memory effects in quantum processes [48, 49], *process matrices* to study the nature of causality in quantum foundations [18, 33, 34], *operator tensors* [40, 41] and *superdensity operators* [42] pertaining to the development of quantum mechanics in spacetime, *quantum strategies* for quantum game theory [43, 44], and *causal boxes* regarding secure communication protocols [45]. While the motivations and subtle details behind these notions vary slightly, the common thread is that they all separate the controllable from the uncontrollable influence on the system; making perspicuous that it is everything that is *out of control* of an experimenter that constitutes the *process* itself.

---

1 Although we speak explicitly of an experimenter, we wish to be clear that they need not actually be there; the notion of an experimenter simply provides an artifice that lends itself nicely to mental imagery and language.

In short, these frameworks describe a quantum stochastic process as a *collection* of joint probability distributions over the outcomes of *any possible sequence of measurements*. For example, the dynamics of a spin-$\frac{1}{2}$ particle can be uniquely described by recording the probability for the spin to be found in alignment with any sequence of independent directions an experimenter might choose to measure (at the timesteps of interest). Once this data has been recorded, generalised Kolmogorov conditions hold, and a generalisation of the KET can be recovered for quantum (and more general) stochastic processes [15, 28], thereby unifying previous approaches to open system dynamics [52]. Crucially, the process tensor framework provides both an unambiguous definition of quantum stochastic processes and a suitable notion of marginalisation in quantum theory. Perhaps most importantly for our present purposes, the formalism permits the development of a set of necessary and sufficient conditions for a quantum process to be classified as Markovian [50, 51]. In this chapter, we introduce the process tensor formalism, its properties, and other tools necessary for the remainder of this thesis.

## 3.1 PROCESS TENSOR FRAMEWORK

The operational perspective to quantum mechanics embraces the philosophy perhaps best stated by Peres [128]: "The simple and obvious truth is that quantum phenomena do not occur in a Hilbert space. They occur in a laboratory. If you visit a real laboratory, you will never find there Hermitian operators. All you can see are emitters (lasers, ion guns, synchrotrons and the like) and detectors. The experimenter controls the emission process and observes detection events." An experimenter has access–in principle–to everything that can be measured on the level of the system, and hence a proper description of the process is one that reproduces the correct measurement statistics. We now take up this perspective and show it resolves the problem of describing multi-time quantum dynamics by way of the process tensor.

### 3.1.1 *Multi-time Quantum Experiments*

We consider the scenario in which an experimenter probes a quantum system that is evolving according to some dynamics that they wish to characterise. Importantly, the experimenter is presumed to have complete instantaneous control over the choice of operations that they implement on the system over a number of timesteps, but no control over the intermediary dynamics; for this reason, we refer to the setting as a *multi-time quantum experiment*. A conceptual schema that describes any such physical experiment depicts it as the composition of the following steps.

**Definition 3.1** (Multi-time quantum experiment). A multi-time quantum experiment proceeds according to the following protocol:

1. The initially unknown state of a system is prepared into a known state (which could be statistical in nature, *i.e.,* not pure).
2. The system is subsequently subject to some physical evolution.
3. An experimenter has access to probe the system.
4. Steps 2 and 3 repeat a number of times, with the system finally being measured.

We can concretely relate this multi-time experiment scheme to a dilated system-environment picture, as depicted in Fig. 3.1. In general, the initially unknown quantum state in step (1) of Def. 3.1 can display system-environment correlations and is described by a density operator $\rho_{1^i}^{SE} \in \mathsf{BL}(\mathcal{H}_{1^i}^{SE})$. We label the state with the subscript $1^i$ to denote that it is the *input* state to the first interrogation procedure applied by the experimenter, namely the initial preparation procedure applied to the system. The initial preparation is only different in name to any of the subsequent probing operations that an experimenter will be allowed to implement in the multi-time setting; the sole reason for the distinction is to emphasise that in the special case where there are no initial system-environment correlations, any system state can be prepared by the experimenter independently of the process, and so the description of the process will begin on the space associated to the *output* of the preparation map, labelled $1^o$. However, this is not generally the case: the possibility of initial system-environment correlations make the role of preparation of significant importance [66, 74, 75]. We will now examine each of the individual elements in this setting in order to motivate the computation of joint probabilities. We begin with the two-time case, where an experimenter applies a preparation procedure (on a potentially initially correlated state) and a subsequent measurement, which is the first generalisation of quantum channels.

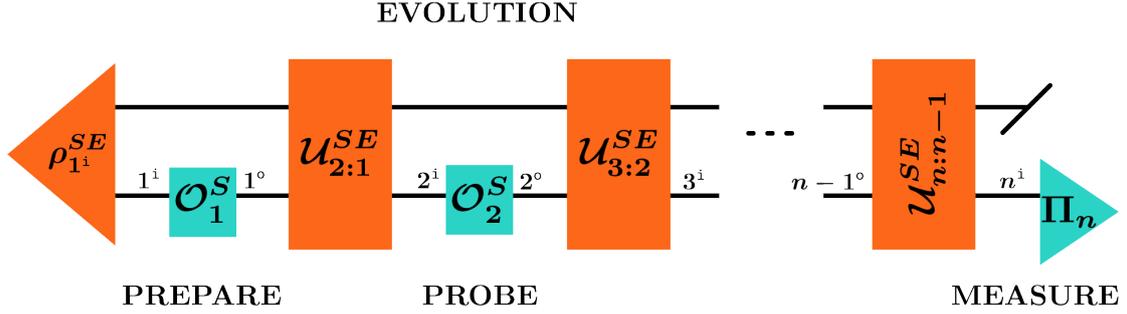

**Figure 3.1:** *Multi-time quantum experiment.* According to the steps of Def. 3.1, first an initially unknown state of a system, potentially correlated with its environment, is prepared according to the preparation procedure $\mathcal{O}_1^S$ in Eq. (3.2). The system and environment subsequently evolve via the joint unitary evolution $\mathcal{U}_{2:1}^{SE}$. Then the experimenter probes the system, applying another CP map $\mathcal{O}_2^S$. The probing repeats a number of times, with the final state being measured with the POVM $\Pi_n$. We colour the operations that an experimenter can control, namely the preparation, the probing instruments and the final measurement, in green. In contrast, the underlying process over which an experimenter has no control, consisting of the initial system-environment state and the subsequent joint unitary evolutions, is coloured in orange. Note that each timestep is associated to an input and an output space, labelled from the perspective of the experimenter.

For instance, the experimenter might apply a projective measurement to yield a known system state with some probability. As we briefly touched on in Subsection 2.2.2, since such conditional transformations between states must occur with at most unit probability, any valid physical preparation procedure on a quantum system $S$ must be a CP and trace-non-increasing map.[2] Specifically, these are transformations $\mathcal{O}_1^S :$ $\mathsf{BL}(\mathcal{H}_{1^\mathrm{i}}^S) \to \mathsf{BL}(\mathcal{H}_{1^\mathrm{o}}^S)$, which takes input states $\rho_{1^\mathrm{i}}^S \in \mathsf{BL}(\mathcal{H}_{1^\mathrm{o}}^S)$ to subnormalised output states $\rho_{1^\mathrm{o}}^S \in \mathsf{BL}(\mathcal{H}_{1^\mathrm{o}}^S)$ via

$$\rho_{1^\mathrm{o}}^S = \mathcal{O}_1^S\left(\rho_{1^\mathrm{i}}^S\right) \quad \text{such that} \quad \mathrm{tr}\left[\rho_{1^\mathrm{o}}^S\right] \leq \mathrm{tr}\left[\rho_{1^\mathrm{i}}^S\right]. \tag{3.1}$$

The trace of the output state encodes the probability of the specific CP transformation being realised (for the given input state). Preparation procedures thus defined can correspond to any physically implementable transformation, including unitary evolutions, CPTP transformations and measurements. After the preparation procedure, the system state is a subnormalised density operator. More generally, in the multi-time scenario, keeping track of success probabilities encoded in the trace of the associated output state will prove helpful for the computation of joint measurement statistics.

---
2 Such maps are referred to as CP maps, with their trace-non-increasing nature implicit.

Any such preparation on the system acts non-trivially on correlations that the system shares with its environment. Given some initially correlated $\rho_{1^\text{i}}^{SE} \in \mathsf{BL}(\mathcal{H}_{1^\text{i}}^{SE})$, a CP preparation procedure $\mathcal{O}_1^S : \mathsf{BL}(\mathcal{H}_{1^\text{i}}^S) \to \mathsf{BL}(\mathcal{H}_{1^\text{o}}^S)$ yields the joint output state

$$\rho_{1^\text{o}}^{SE} = \mathcal{O}_1^S \otimes \mathcal{I}_1^E \left( \rho_{1^\text{i}}^{SE} \right), \tag{3.2}$$

where $\rho_{1^\text{o}}^{SE} \in \mathsf{BL}(\mathcal{H}_{1^\text{o}}^{SE})$ typically exhibits system-environment correlations.

Following this first interrogation by the experimenter, the joint system-environment state is subject to some uncontrollable unitary evolution, as per step (2) of Def. 3.1. The evolution of the joint state from time $t_1$ to $t_2$ is represented by the unitary map $\mathcal{U}_{2:1}^{SE} : \mathsf{BL}(\mathcal{H}_{1^\text{o}}^{SE}) \to \mathsf{BL}(\mathcal{H}_{2^\text{i}}^{SE})$. Note the inevitable awkwardness of the labelling convention that we run into here, which denotes the output of the unitary transformation with an input label: this is due to the fact that input and output labels are written from the perspective of the experimenter, and so, naturally, the output state of the uncontrollable unitary evolution becomes the input to the next operation applied by the experimenter. Following the first two stages of Def. 3.1, the system-environment state is

$$\rho_{2^\text{i}}^{SE} = \mathcal{U}_{2:1}^{SE} \left( \mathcal{O}_1^S \otimes \mathcal{I}_1^E \left( \rho_{1^\text{i}}^{SE} \right) \right). \tag{3.3}$$

As a brief aside, note that it is clear that the act of initial preparation can strongly influence the dynamics of the state of the system, because the specific transformation realised can be correlated with the state of the environment, which in turn influences the subsequent evolution. The state of the system following the first two steps of a quantum experiment described above is related to the CP preparation procedure $\mathcal{O}_1^S : \mathsf{BL}(\mathcal{H}_{1^\text{i}}^S) \to \mathsf{BL}(\mathcal{H}_{1^\text{o}}^S)$ in the dilated picture via

$$\rho_{2^\text{i}}^S = \text{tr}_E \left[ \mathcal{U}_{2:1}^{SE} \left( \mathcal{O}_1^S \otimes \mathcal{I}_1^E (\rho_{1^\text{i}}^{SE}) \right) \right]. \tag{3.4}$$

To clearly see the influence of the initial preparation, first suppose that the experimenter applies a rank-1 projective measurement on the system, where observation of any particular outcome $x$ corresponds to the conditional transformation $\mathcal{P}^{(x)}(\bullet) = P^{(x)} \bullet P^{(x)}$ being realised with probability $p_x = \text{tr}\left[P^{(x)} \rho\right]$. If initial correlations are present, the post-preparation state is[3]

$$\mathcal{P}^{(x)} \otimes \mathcal{I}^E(\rho_{1^\text{i}}^{SE}) = P_{1^\text{o}}^{(x)} \otimes \widetilde{\tau}_{1^\text{o}}^{(x)}, \tag{3.5}$$

---

3 Here we drop the subsystem labels on conditional states and transformations to avoid clunky notation, with the understanding that the measurement acts on the system alone.

where $\widetilde{\tau}_{1^\circ}^{(x)} = \text{tr}_S\left[\mathcal{P}^{(x)} \otimes \mathcal{I}^E(\rho_{1^\text{i}}^{SE})\right] \in \mathsf{BL}(\mathcal{H}_{1^\circ}^E)$ and $P_{1^\circ}^{(x)} \in \mathsf{BL}(\mathcal{H}_{1^\circ}^S)$ describe the post-measurement states. We immediately see that the state of the environment is conditioned by the measurement enacted by the experimenter: the dynamical map description of the subsequent unitary evolution would make it seem as if the process itself is dependent on the preparation, as a distinct quantum channel is induced on the level of the system for each conditional environment state. Consequently, this seems to imply that there is no proper process *per se* that is independent of the experimenter, which is indeed the reason why traditional approaches to open quantum dynamics break down when initial correlations are present.

If, on the other hand, the system is initially uncorrelated from its environment, then any preparation has no influence on the state of the latter. The initially product system-environment hypothesis of the tomographic scheme discussed in Subsection 2.2.4 corresponds to this special case, where states of the system can be prepared without affecting the environment, which retains its unique, fixed state $\tau_{1^\text{i}}^E \in \mathsf{BL}(\mathcal{H}_{1^\text{i}}^E)$ throughout the preparation accordingly

$$\sigma_{1^\circ}^S \otimes \tau_{1^\circ}^E = \mathcal{O}_1^S \otimes \mathcal{I}_1^E\left(\rho_{1^\text{i}}^S \otimes \tau_{1^\text{i}}^E\right). \tag{3.6}$$

Satisfaction of Eq. (3.6) means that the dynamical map describing the subsequent evolution of the experiment is independent of the preparation procedure and uniquely defined, since the environment state can be treated as a constant of the problem. However, the assumption of an initial product state is not satisfied in many realistic circumstances, especially beyond the weak-coupling regime [46, 47, 66, 129].

Returning to the main point of this subsection, at step (3) of the procedure in Def. 3.1 the experimenter has access to the system part of $\rho_{2^\text{i}}^{SE}$ given in Eq. (3.3). Again, they can apply any CP map of their choosing, before the joint system-environment state is subject to another portion of unitary evolution, and so on. For a process where the experimenter has access to the system at $n$ timesteps, the final state of the system after applying a sequence $\mathcal{O}_1^S, \ldots, \mathcal{O}_{n-1}^S$ of CP maps is

$$\rho_{n^\text{i}}^S = \text{tr}_E\left[\mathcal{U}_{n:n-1}^{SE}\mathcal{O}_{n-1}^S \ldots \mathcal{U}_{2:1}^{SE}\mathcal{O}_1^S\left(\rho_{1^\text{i}}^{SE}\right)\right], \tag{3.7}$$

where all maps act upon everything to their right. Since the initial system-environment state, the subsequent joint unitary evolution and the discarding of the environmental degrees of freedom all occur deterministically, the trace of the output state $\rho_{n^\text{i}}^S$ is sub-normalised with respect to the probability of realising the sequence of transformations applied by the experimenter; by tracking the operations applied, we come to a meaning-ful description of conditional quantum states. At the conclusion of the experiment, the final state is measured with a POVM $\Pi_n$.

### 3.1.2 *Process Tensor*

The important shift in perspective that allows for a consistent description of multi-time quantum dynamics is to separate what an experimenter has control over from what they do not. Although they cannot generally know, in advance, the initially correlated system-environment state, nor the subsequent portions of joint unitary evolution, what *can* be controlled is the choice of instrument interrogating the system at each timestep. An operationally meaningful framework for describing quantum evolution therefore necessarily consists of two parts: **i)** the *uncontrollable* underlying process which governs the joint unitary evolution of the system with some inaccessible environment, and **ii)** the interleaved *controllable* changes to the state of the system, effected by the probing operations implemented by an experimenter.

For instance, in describing open quantum dynamics in the presence of initial correlations (*i.e.,* the most general two-time quantum experiment as per Def. 3.1), what is desirable is *not* a dynamical map taking initial system density operators—which are uncontrollable—to final ones; indeed, such maps are not linear in general [112], or cannot be reconstructed [130]. Instead, we seek a *map on a map*, which takes any choice of the preparation map as input and outputs the final density operator of the system. To this end, one can define the *superchannel*, $\mathcal{T}_{2:1}$ [75], to represent everything that is *out of the control of an experimenter*, *i.e.,* everything on the r.h.s of Eq. (3.4) except for $\mathcal{O}_1^S$, as shown in Fig. 3.2

$$\mathcal{T}_{2:1}^S(\bullet) := \mathrm{tr}_E\left[\mathcal{U}_{2:1}^{SE}\left(\bullet \otimes \mathcal{I}_1^E(\rho_{1^\mathrm{i}}^{SE})\right)\right]. \tag{3.8}$$

The superchannel contains information regarding both the initially correlated state $\rho_{1^\mathrm{i}}^{SE}$ and the subsequent joint unitary dynamics $\mathcal{U}_{2:1}^{SE}$; however, from the superchannel alone we cannot delineate the individual contributions of these elements (without access to the environment). On the other hand, the superchannel contains within it, by definition, everything required to determine the output state of a quantum experiment for *any* preparation procedure, making it the object of fundamental operational importance.

The superchannel acts on a CP preparation map and yields the correct output state in analogy to the way the dynamical map acts on an input density operator. It is linear in its argument by construction, which permits its tomographic reconstruction in a similar way to that of a dynamical map (see Subsection 2.2.2), *i. e.,* by linearly extending input-output relations deduced for a basis set to uniquely determine its action on arbitrary inputs [75]. However, in contrast to the dynamical map, the inputs of the superchannel are CP preparation maps rather than density operators. While a set of $d^2$ linearly independent density operators can be chosen to span the space of quantum states, $d^4$ linearly independent CP maps are required to span the space of allowable preparations [52, 124]. Indeed, such an experimental reconstruction of a superchannel has recently been achieved in the laboratory to characterise the evolution of a photonic qubit that is initially correlated with a single-photon environment [131].

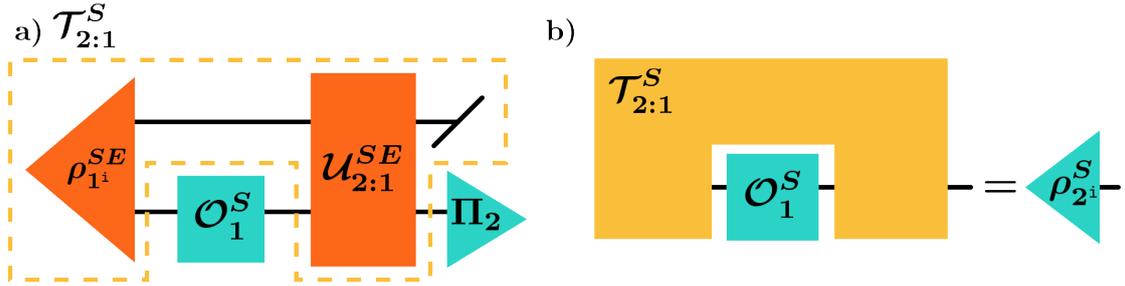

**Figure 3.2:** *Superchannel: resolution to the initial correlation problem.* In panel **a)** we depict a two-step quantum experiment. By separating what is controllable to the experimenter from what is uncontrollable, we can define the superchannel $\mathcal{T}^S_{2:1}$, which is everything enclosed in the dashed, yellow border. In panel **b)**, we depict how this higher-order map acts on any preparation $\mathcal{O}^S_1$ to yield the correct output state $\rho^S_{2^i}$ of the system that is accessible to the experimenter at the second timestep.

Thus, the superchannel is the natural logical extension of the operationally accessible input-output relations that motivated the dynamical maps description, taking controllable inputs to measurable outputs, with the generalisation allowing for the presence of initial correlations. Moreover, the superchannel satisfies natural extensions of the notions of complete-positivity and trace-preservation [75], as we will discuss for the more general process tensor shortly. Indeed, this approach operationally solves the problem of describing quantum dynamics in the presence of initial correlations. Most importantly for our purposes, the superchannel includes all two-point correlations of the process, allowing for a consistent calculation of multi-time statistics for any choice of interrogation by the experimenter.

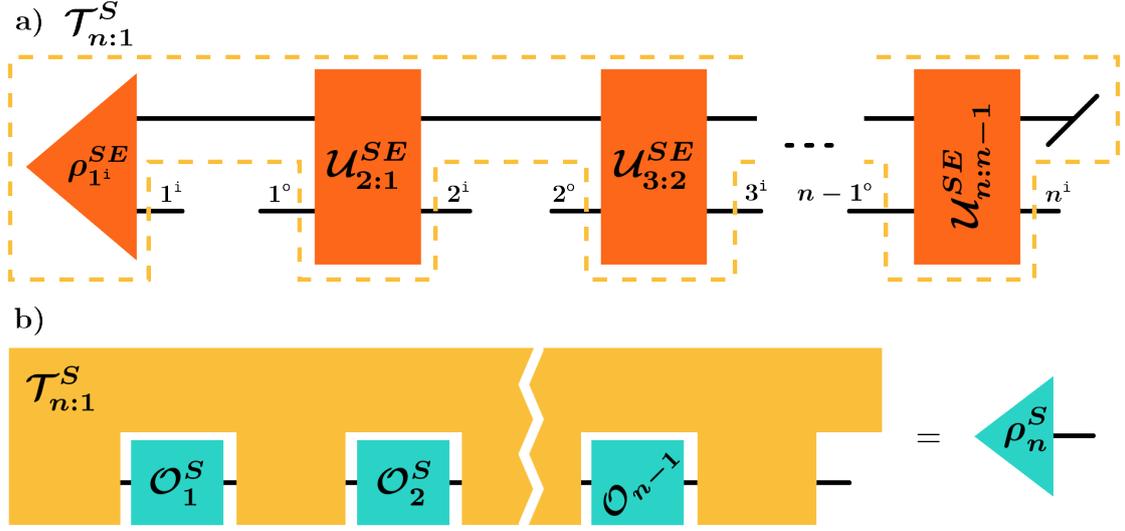

**Figure 3.3:** *Process tensor: an operational description of quantum stochastic processes.* In panel **a)**, we show the abstraction of everything that is uncontrollable to an experimenter, which defines the process tensor, $\mathcal{T}_{n:1}^S$. In panel **b)**, we depict how the process tensor acts on sequences of CP maps applied by an experimenter on the level of the system to the final density operator.

In analogy to the definition of the superchannel for the two-time case, one can abstract all that is uncontrollable in an open process across multiple timesteps as the *process tensor*, $\mathcal{T}_{n:1}$, depicted in Fig. 3.3, as follows[4]

$$\rho_n^S = \mathrm{tr}_E\left[\mathcal{U}_{n:n-1}^{SE}\mathcal{O}_{n-1}^S\ldots\mathcal{U}_{2:1}^{SE}\mathcal{O}_1^S\rho^{SE}\right] =: \mathcal{T}_{n:1}^S(\mathcal{O}_{n-1}^S,\ldots,\mathcal{O}_1^S). \quad (3.9)$$

The process tensor is a multi-linear mapping from the sequences of CP control operations upon which it acts to the quantum states at the final output. Due to this linearity, it follows that the process tensor, like the superchannel and the quantum channels that it generalises, can be experimentally reconstructed in a finite number of experiments by way of an extended tomographic scheme [51, 124]. Since the process tensor acts only upon operations applied on the level of the system, from this point forth we will drop the superscript label $S$ for any maps that act on/states that live on the system (unless potentially ambiguous).

One of the advantages of the process tensor formalism is that it directly relates to an operational picture which clarifies a number of concepts pertinent to open quantum dynamics. For example, recall the breakdown of the consistency conditions of the KET on the level of measured probability distributions in quantum theory. For classical stochastic processes, averaging over measurement outcomes amounts to doing nothing to the system on average, and so the descriptors of the process on any subset of times obtained via marginalisation of deduced statistics are compatible. In quantum theory, a distinction between marginalisation and doing nothing arises because measurements in different bases have different overall effects. As depicted in Fig. 3.4, doing nothing to the system is by no means equivalent to marginalising over the statistics observed for a fixed sequence of measurements.

---

4 Here, on the rightmost side where everything acts on the system alone, we economise the timestep labelling with the understanding that each timestep is associated to an input and output Hilbert space by writing $j = \{j^\mathrm{i}, j^\mathrm{o}\}$ and further compressing ordered sequences of timesteps as $n:1 := \{1,\ldots,n\}$.

It is obvious from the operational formalism developed that the process tensor fulfils the following natural consistency condition [15]: for any two sets of timesteps $\Lambda_k \subset \Lambda_n$, the descriptor of the process over the smaller set of times can be obtained from $\mathcal{T}_{\Lambda_n}$ by letting it act on identity maps $\mathcal{I}_{\Lambda_{n\setminus k}}$ at the times $\Lambda_{n\setminus k}$, as highlighted in Fig. 3.5. Mathematically, the consistency condition reads

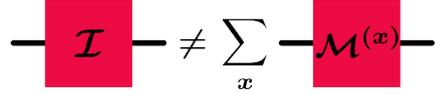

**Figure 3.4:** *Doing nothing vs. averaging over measurements.* In quantum theory there is a difference between doing nothing to the system, $\mathcal{I}$, and averaging over all measurement outcomes, $\{\mathcal{M}^{(x)}\}$.

$$\mathcal{T}_{\Lambda_k}(\bullet) = \mathcal{T}_{\Lambda_n}\left(\mathcal{I}_{\Lambda_{n\setminus k}}, \bullet\right) =: \mathcal{T}_{\Lambda_n}^{|\Lambda_k|}(\bullet) \quad (3.10)$$

where $\bullet$ is a placeholder for operations one could implement at the remaining timesteps in $\Lambda_k$. By separating the controllable influence on the system from the underlying process, the process tensor formalism allows us to recover compatibility for the descriptors of quantum stochastic processes for different sets of times, and with this a generalised version of the KET can be derived [15, 28]. This result serves to define what we mean by a *quantum stochastic process*, paving the way for an unambiguous study of them.

Up until this point, we have stressed the intuitive picture of the process tensor as the object that allows an experimenter to compute all possible multi-time statistics they might deduce via actively probing the system of interest. We now show how these can be calculated directly for any sequence of interrogations. Recall that a measurement is represented by a POVM $\mathcal{J} = \{\Pi^{(x)\mathrm{T}}\}$, where each of the elements corresponds to a possible outcome, and they satisfy the summation condition $\sum_x \Pi^{(x)} = \mathbb{1}$ ensuring that *some* outcome occurs with certainty. In the temporal setting where the post-measurement state of the system is of interest, the natural generalisation of a POVM is an *instrument*, which is a collection of CP maps $\mathcal{J} = \{\mathcal{O}^{(x)}\}$ that overall yields a CPTP map $\sum_x \mathcal{O}^{(x)} = \mathcal{O}^{\mathcal{J}}$ [7, 27]. Intuitively, the requirement that all the CP maps that make up the instrument sum up to a CPTP map means that *some* transformation to the system occurs with certainty. Thus, if an experimenter applies a sequence of CP maps $\mathcal{O}_1^{(x_1)}, \ldots, \mathcal{O}_{n-1}^{(x_{n-1})}$, each of which are elements of an instrument $\mathcal{J}_1, \ldots, \mathcal{J}_{n-1}$, with a measurement instrument $\mathcal{J}_n = \{\Pi_n^{(x_n)\mathrm{T}}\}$ applied to the final state, the joint probability distribution over the outcomes realised can be calculated from the process tensor directly via

$$\mathbb{P}(x_n, \ldots, x_1 | \mathcal{J}_n, \ldots, \mathcal{J}_1) = \mathrm{tr}\left[\Pi_n^{(x_n)\mathrm{T}} \rho_n\right] \quad (3.11)$$
$$= \mathrm{tr}\left[\Pi_n^{(x_n)\mathrm{T}} \mathcal{T}_{n:1}\left(\mathcal{O}_{n-1}^{(x_{n-1})}, \ldots, \mathcal{O}_1^{(x_1)}\right)\right].$$

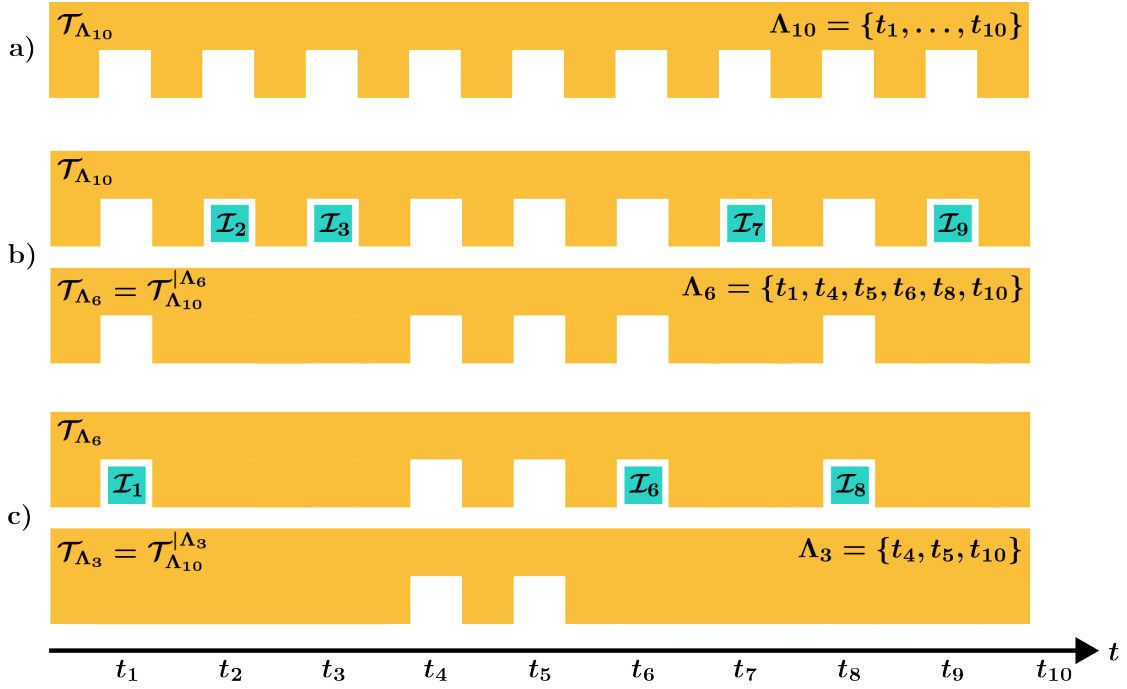

**Figure 3.5:** *Consistency condition for the process tensor.* The process tensor satisfies a natural consistency condition. For concreteness, in panel **a)** we depict a process tensor over ten timesteps. From this, the correct descriptor on any subset of timesteps can be derived by letting the it act on identity maps at the appropriate times. For example, in panel **b)** we show how the correct description over times $\Lambda_6 = \{t_1, t_4, t_5, t_6, t_8, t_{10}\}$ can be obtained in this way from that defined on $\Lambda_{10} = \{t_1, \ldots, t_{10}\}$. Moreover, in panel **c)**, we show the containment of $\mathcal{T}_{\Lambda_3}$ in both descriptors $\mathcal{T}_{\Lambda_6}$ and $\mathcal{T}_{\Lambda_{10}}$, where $\Lambda_3 = \{t_4, t_5, t_{10}\}$. The crucial point is that the unique maximal description contains within it the proper description of the process over any subset of timesteps.

The process tensor contains all joint probability distributions for all possible measurement settings, and is thus the natural generalisation of classical stochastic processes, as well as quantum states (see below).

To summarise the developments so far in this subsection, recall that for classical stochastic processes, it is the hierarchy of compatible joint probability distributions over all timesteps that serves to characterise the underlying process. For quantum processes, each event must be associated to a CP map on the system, and it is the process tensor that characterises the process by mapping any possible multi-time sequence of CP maps to the correct joint statistics via Eq. (3.11). By way of the generalised KET, one can straightforwardly deduce the correct descriptor of the process over any subset of timesteps, thereby alleviating conceptual difficulties regarding the proper characterisation of quantum stochastic processes. Indeed, by accounting succinctly for all possible sequences of interventions on a system of interest, the process tensor encodes all possible multi-time correlations between deducible statistics and therefore, on a sufficiently fine-grained set of timesteps, captures the most general evolutions possible in both quantum and classical physics.

We reiterate the important conceptual departure from traditional approaches to open quantum dynamics: there, descriptions of quantum processes typically involve tracking the state of the system as a function of time, which limits the ability to calculate the outcomes of measurements to at most two timesteps in any given trial of the experiment, inherently failing to capture multi-time memory effects that are critical to understanding processes with memory. Clearly, the system density operator at each timestep can be obtained from the process tensor by simply plugging in identity maps at all of the preceding timesteps, thereby unifying all such two-point descriptions. Besides providing a more intuitive operational picture, the process tensor description also subsumes the standard approaches to quantum processes with memory by way of non-Markovian master equations, which aim to account for the effects of some memory kernel on the evolution of the system. Admittedly, the process tensor is more of a characterisation than a dynamical description; nonetheless, on the timesteps upon which it is defined, it provides a more general description, accounting for all possible multi-time correlations deducible—rather than only those either derived microscopically or deduced phenomenologically—the process tensor goes beyond the realm of applicability of such approaches.

Crucially, we now have a way to meaningfully construct quantum generalisations of statements that are multi-time in nature, such as Markovianity [50, 51] and Markov order [1, 2], allowing for a consistent study of memory in quantum stochastic processes from an operationally sound perspective. Before we do so, we present a brief mathematical interlude to develop a useful representation of the process tensor as a multi-partite quantum state, which will prove fruitful for understanding its properties and proving statements throughout the remainder of this thesis.

## 3.2 Representing linear maps

Up to this point, we have discussed dynamical maps and process tensors on rather abstract grounds as mappings. For explicit statements, it proves helpful to choose a representation that is well-adapted to the respective purpose. There are various different explicit representations of maps describing valid physical evolution in quantum mechanics, each suitable for certain purposes, such as the tomographic representation of Eq. (2.41), the so-called *A*- and *B*-forms of dynamical maps first introduced by Sudarshan *et al.* in Ref. [132], and the Kraus decomposition [81, 133], amongst others; for an overview of the inter-relations between such representations, see, *e.g.,* Ref. [52]. Here, we introduce and use exclusively the *Choi-Jamiołkowski Isomorphism* (CJI) [134, 135].

### 3.2.1 *Choi-Jamiołkowski Isomorphism*

This isomorphism allows us to consider any linear map taking elements of some input vector space to some other output vector space as a single element of the joint input-output vector space. Concretely, consider a linear map $\mathcal{L}$ acting on the bounded linear operators on a Hilbert space $\mathcal{L} : \mathsf{BL}(\mathcal{H}_\mathtt{i}) \to \mathsf{BL}(\mathcal{H}_\mathtt{o})$, where, for generality, we allow for the input and output Hilbert spaces to be distinct. This map can be represented as a bipartite operator $\mathsf{L}_\mathtt{oi} \in \mathsf{BL}(\mathcal{H}_\mathtt{o} \otimes \mathcal{H}_\mathtt{i})$ through its action on half of an unnormalised maximally-entangled state $\Psi := \sum_{ij}^{d_\mathtt{i}} |ii\rangle\langle jj| \in \mathsf{BL}(\mathcal{H}_\mathtt{i} \otimes \mathcal{H}_\mathtt{i})$ as follows

$$\mathsf{L}_\mathtt{oi} := \mathcal{L} \otimes \mathcal{I}(\Psi). \tag{3.12}$$

See Fig. 3.6 for a graphical representation. Note that we consistently use upper-case Roman calligraphic letters to denote maps and their sans-serif variant to denote their corresponding representation in terms of the CJI. We refer to the matrix $\mathsf{L}_\mathtt{oi}$ resulting from Eq. (3.12) as the *Choi operator* associated to $\mathcal{L}$.

The action of the map $\mathcal{L}$ on an arbitrary element of its input space $\eta_\mathtt{i} \in \mathsf{BL}(\mathcal{H}_\mathtt{i})$ can be expressed in terms of its Choi operator $\mathsf{L}_\mathtt{oi}$ via

$$\mathcal{L}(\eta_\mathtt{i}) = \mathrm{tr}_\mathtt{i} \left[ \left( \mathbb{1}_\mathtt{o} \otimes \eta_\mathtt{i}^\mathrm{T} \right) \mathsf{L}_\mathtt{oi} \right], \tag{3.13}$$

where $\mathbb{1}_\mathtt{o} := \sum_i^{d_\mathtt{o}} |i\rangle\langle i| \in \mathsf{BL}(\mathcal{H}_\mathtt{o})$ is the identity operator on the output Hilbert space and we slightly abuse notation by writing $\mathrm{tr}_\mathtt{i}[\bullet] := \mathrm{tr}_{\mathcal{H}_\mathtt{i}}[\bullet]$. The validity of Eq. (3.13) can be shown by direct insertion of Eq. (3.12) as follows

$$\operatorname{tr}_{\mathtt{i}}\left[(\mathbb{1}_{\mathtt{o}}\otimes\eta_{\mathtt{i}})^{\mathsf{T}}\mathsf{L}_{\mathtt{oi}}\right]=\operatorname{tr}_{\mathtt{i}}\left[\left(\mathbb{1}_{\mathtt{o}}\otimes\eta_{\mathtt{i}}^{\mathrm{T}}\right)\mathcal{L}\otimes\mathcal{I}(\Psi)\right] \tag{3.14}$$

$$=\sum_{ij}^{d_{\mathtt{i}}}\operatorname{tr}_{\mathtt{i}}\left[\left(\mathbb{1}_{\mathtt{o}}\otimes\eta_{\mathtt{i}}^{\mathrm{T}}\right)\mathcal{L}(|i\rangle\langle j|)\otimes|i\rangle\langle j|\right]$$

$$=\sum_{ijk}^{d_{\mathtt{i}}}\mathcal{L}(|i\rangle\langle j|)\langle k|\eta_{\mathtt{i}}^{\mathrm{T}}|i\rangle\langle j|k\rangle=\sum_{ij}^{d_{\mathtt{i}}}\mathcal{L}(|i\rangle\langle j|)\langle j|\eta_{\mathtt{i}}^{\mathrm{T}}|i\rangle$$

$$=\sum_{ij}^{d_{\mathtt{i}}}\mathcal{L}(|i\rangle\langle j|)\eta_{\mathtt{i}}^{ij}=\mathcal{L}\left(\sum_{ij}^{d_{\mathtt{i}}}\eta_{\mathtt{i}}^{ij}|i\rangle\langle j|\right)=\mathcal{L}(\eta_{\mathtt{i}}),$$

where the final line of equalities holds by the linearity of $\mathcal{L}$ and the decomposition of an arbitrary element of $\mathsf{BL}(\mathcal{H}_{\mathtt{i}})$ as $\eta_{\mathtt{i}} = \sum_{ij}^{d_{\mathtt{i}}} \eta_{\mathtt{i}}^{ij}|i\rangle\langle j|$.

While the CJI holds for linear maps in general, for the types of evolution that are physically meaningful in quantum theory, Choi operators have particularly nice properties.[5] Consider a CPTP map $\mathcal{C} : \mathsf{BL}(\mathcal{H}_{\mathtt{i}}) \to \mathsf{BL}(\mathcal{H}_{\mathtt{o}})$, where again, for generality, we allow the input and output systems of the map to be distinct. Complete-positivity and trace-preservation for the quantum channel $\mathcal{C}$ translate into the following properties of its Choi operator $\mathsf{C}_{\mathtt{oi}}$

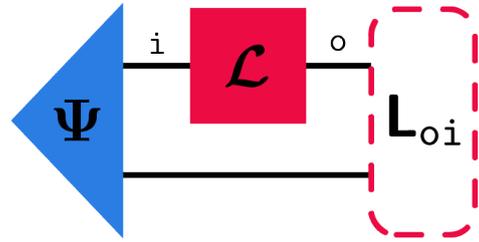

**Figure 3.6:** *CJI of a linear map.* Any linear map $\mathcal{L} : \mathcal{H}_{\mathtt{i}} \to \mathcal{H}_{\mathtt{o}}$ can be represented as a bipartite operator $\mathsf{L}_{\mathtt{oi}} \in \mathsf{BL}(\mathcal{H}_{\mathtt{o}} \otimes \mathcal{H}_{\mathtt{i}})$ through its action on half of a maximally entangled state $\Psi$.

1. **Complete-positivity** : $\mathsf{C}_{\mathtt{oi}} \geq 0$. (3.15)

2. **Trace-preservation** : $\operatorname{tr}_{\mathtt{o}}[\mathsf{C}_{\mathtt{oi}}] = \mathbb{1}_{\mathtt{i}}$.

*Proof.* **1.** Clearly, a CP map $\mathcal{C}$ corresponds to a positive semidefinite Choi operator $\mathsf{C}_{\mathtt{oi}} \geq 0$ by definition. Any Hermitian,[6] positive semidefinite operator admits a singular-value eigendecomposition $\mathsf{C}_{\mathtt{oi}} = \sum_{\alpha=1}^{d_{\mathtt{oi}}} \lambda_\alpha |\alpha\rangle\langle\alpha|$ with each $\lambda_\alpha \geq 0$ and $d_{\mathtt{oi}} := d_{\mathtt{o}}d_{\mathtt{i}}$. The action of $\mathcal{C}$ on an arbitrary state $\rho_{\mathtt{i}} \in \mathsf{BL}(\mathcal{H}_{\mathtt{i}}^S)$ can then be written as

$$\mathcal{C}(\rho_{\mathtt{i}}) = \operatorname{tr}_{\mathtt{i}}\left[\left(\mathbb{1}_{\mathtt{o}} \otimes \rho_{\mathtt{i}}^{\mathrm{T}}\right)\mathsf{C}_{\mathtt{oi}}\right] \tag{3.16}$$

$$= \operatorname{tr}_{\mathtt{i}}\left[\left(\mathbb{1}_{\mathtt{o}} \otimes \rho_{\mathtt{i}}^{\mathrm{T}}\right)\sum_{\alpha=1}^{d_{\mathtt{io}}}\lambda_\alpha|\alpha\rangle\langle\alpha|\right] = \sum_{\alpha=1}^{d_{\mathtt{io}}}\lambda_\alpha\sum_{ij}^{d_{\mathtt{i}}}\langle i|\alpha\rangle\langle j|\rho_{\mathtt{i}}^{\mathrm{T}}|i\rangle\langle\alpha|j\rangle$$

---

[5] Indeed, it turns out that the Choi representation of a quantum dynamical map is equivalent to the $B$-form introduced in Ref. [132] for precisely this reason.

[6] It is straightforward to show that the Hermiticity-preservation property of the dynamical map (which is implied by the CP condition) leads to an Hermitian Choi operator.

$$= \sum_{\alpha=1}^{d_{\text{io}}} \left( \sum_{i}^{d_{\text{i}}} \sqrt{\lambda_\alpha} \langle i | \alpha \rangle \langle i | \right) \rho_{\text{i}} \left( \sum_{j}^{d_{\text{i}}} \sqrt{\lambda_\alpha} | j \rangle \langle \alpha | j \rangle \right) =: \sum_{\alpha=1}^{d_{\text{io}}} K_\alpha \rho_{\text{i}} K_\alpha^\dagger,$$

where we have introduced the orthonormal basis vectors of the input space $\{|i\rangle\}, \{|j\rangle\}$ to perform the partial trace explicitly, made use of the non-negativity of the eigenvalues to write a unique square-root, and each the $K_\alpha$ are $d_{\text{o}} \times d_{\text{i}}$ matrices (since $\langle i | \alpha \rangle \in \mathsf{BL}(\mathcal{H}_{\text{o}})$). The expression of the dynamical map derived above is known as the Kraus or operator-sum representation, with each $K_\alpha$ known as the Kraus operator of the map. A fundamental result given in Ref. [133] states that a map is CP iff it can be written in the Kraus form above, concluding the proof. □

*Proof.* **2.** A TP map $\mathcal{C}$ satisfies $\text{tr}\,[\mathcal{C}(\rho_{\text{i}})] = \text{tr}\,[\rho_{\text{i}}]\ \forall\ \rho_{\text{i}} \in \mathsf{BL}(\mathcal{H}_{\text{i}})$. Writing this out explicitly in terms of the Choi operator $\mathsf{C}_{\text{oi}}$, we have

$$\text{tr}\,[\mathcal{C}(\rho_{\text{i}})] = \text{tr}_{\text{o}} \left[ \text{tr}_{\text{i}} \left[ \left( \mathbb{1}_{\text{o}} \otimes \rho_{\text{i}}^{\text{T}} \right) \mathsf{C}_{\text{oi}} \right] \right] = \text{tr}_{\text{i}} \left[ \rho_{\text{i}}^{\text{T}} \text{tr}_{\text{o}}\,[\mathsf{C}_{\text{oi}}] \right] = \text{tr}\,[\rho_{\text{i}}], \tag{3.17}$$

which holds true for all $\rho_{\text{i}}$ iff $\text{tr}_{\text{o}}\,[\mathsf{C}_{\text{oi}}] = \mathbb{1}_{\text{o}}$. □

The set of physically allowable CPTP maps are therefore equivalent to bipartite operators $\mathsf{C}_{\text{oi}} \in \mathsf{BL}(\mathcal{H}_{\text{o}} \otimes \mathcal{H}_{\text{i}})$ satisfying the conditions outlined in Eq. (3.15). A few remarks are in order.

**i)** It is clear from the trace-preservation property that $\text{tr}\,[\mathsf{C}_{\text{oi}}] = d_{\text{i}}$. Since the Choi operator of a CPTP channel must also be positive semidefinite, it can therefore be regarded as a supernormalised quantum state. That is, any such Choi operator lies in the convex cone of non-negative bounded linear operators of the joint input-output Hilbert space. More precisely, the set of Choi operators of CPTP maps corresponds to the intersection of the cone of positive semidefinite operators $\mathsf{A}_{\text{oi}} \geq 0$ with the hyperplane of those that satisfy $\text{tr}_{\text{o}}\,[\mathsf{A}_{\text{oi}}] = \mathbb{1}_{\text{i}}$.

**ii)** Whilst all CPTP maps can be uniquely identified with a (potentially supernormalised) density operator, not all states in $\mathsf{BL}(\mathcal{H}_{\text{o}} \otimes \mathcal{H}_{\text{i}})$ represent a CPTP evolution, since, although they are positive semidefinite, a generic quantum state does not necessarily satisfy the additional trace-preservation constraint.

**iii)** A general CP map corresponds to a positive semidefinite Choi operator that does not necessarily satisfy the second property in Eq. (3.15), but instead must satisfy $\text{tr}_{\text{o}}\,[\mathsf{C}_{\text{oi}}] \leq \mathbb{1}_{\text{i}}$.

**iv)** Lastly, through the CJI we can think of states and effects as the Choi operators of CP maps with trivial input and output spaces respectively. States can be considered as the Choi operator of a CPTP map $\mathcal{R} : \mathbb{R} \to \mathsf{BL}(\mathcal{H}_{\text{o}})$, and the normalisation of the state $\text{tr}_\texttt{o}[\rho] = 1$ can be understood as a trace-preservation condition. In this sense, all normalised quantum states are TP, which is intuitive since they can be prepared deterministically. Effects $\mathcal{E}^{(x)} : \textsf{BL}(\mathcal{H}_\texttt{o}) \to \mathbb{R}$ map quantum states to real numbers (probabilities) and in this sense positivity of effects is equivalent to complete-positivity, since the output space is trivial.[7] The action of an effect on a quantum state is written in terms of their Choi operators as $\mathcal{E}^{(x)}(\rho) = \text{tr}\left[\rho^\text{T} \Pi^{(x)}\right] = \text{tr}\left[\Pi^{(x)\text{T}} \rho\right]$, which unsurprisingly gives the Born rule (see Eq. (2.31)). In contrast to the case for states, the fact that the output space of effects is trivial implies that there is only one trace-preserving effect, namely the trace map $\mathcal{E}(\rho) = \text{tr}[\rho]$ whose Choi operator is the identity matrix $\mathbb{1}$ (see Fig. 3.7).

As a final brief example, note that we can write the tomographic representation of a quantum channel given in Eq. (2.41) in terms of the dual objects to the set of input spanning states, $\{\hat{\mathsf{D}}_\texttt{i}^{(i)}\}$, and the corresponding output states, $\{\hat{\sigma}_\texttt{o}^{(i)}\}$, as follows

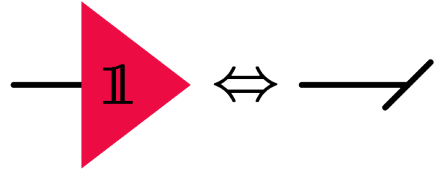

**Figure 3.7:** *The unique trace-preserving effect.* The only trace-preserving effect is the trace map, corresponding to an identity Choi operator $\mathbb{1}$. To abide by traditional notation, we depict this with a slash.

$$\mathsf{C}_\texttt{oi} = \sum_i \hat{\sigma}_\texttt{o}^{(i)} \otimes \hat{\mathsf{D}}_\texttt{i}^{(i)*}, \tag{3.18}$$

where $(\bullet)^*$ denotes complex conjugation. The validity of this construction can be seen directly by insertion

$$\mathcal{C}(\rho_\texttt{i}) = \text{tr}_\texttt{i}\left[\left(\mathbb{1}_\texttt{o} \otimes \rho_\texttt{i}^\text{T}\right) \mathsf{C}_\texttt{oi}\right] = \text{tr}_\texttt{i}\left[\left(\mathbb{1}_\texttt{o} \otimes \rho_\texttt{i}^\text{T}\right) \sum_i \hat{\sigma}_\texttt{o}^{(i)} \otimes \hat{\mathsf{D}}_\texttt{i}^{(i)*}\right] \tag{3.19}$$

$$= \sum_i \hat{\sigma}_\texttt{o}^{(i)} \text{tr}\left[\rho_\texttt{i}^\text{T} \hat{\mathsf{D}}_\texttt{i}^{(i)*}\right] = \sum_i \hat{\sigma}_\texttt{o}^{(i)} \text{tr}\left[\hat{\mathsf{D}}_\texttt{i}^{(i)\dagger} \rho_\texttt{i}\right].$$

### 3.2.2 *Choi Representation of Process Tensor*

We now employ the CJI to explicitly represent process tensors. Recall that its action is to take sequences of CP maps as its input and map them to some final output density operator. For generality, we allow the input and output systems of the CP maps to be distinct, as too the final output of the process tensor. With the CJI, we can represent any CP map $\mathcal{O}_j : \textsf{BL}(\mathcal{H}_{j^\texttt{i}}) \to \textsf{BL}(\mathcal{H}_{j^\texttt{o}})$ as a Choi operator $\mathsf{O}_j \in \textsf{BL}(\mathcal{H}_{j^\texttt{o}} \otimes \mathcal{H}_{j^\texttt{i}})$. The process tensor acts on sequences of these as $\mathcal{T}_{n:1} : \textsf{BL}(\bigotimes_{j=1}^{n-1} \mathcal{H}_{j^\texttt{o}} \otimes \mathcal{H}_{j^\texttt{i}}) \to \textsf{BL}(\mathcal{H}_{n^\texttt{i}})$. As for the case of quantum channels, we can invoke the CJI to represent the process tensor as a positive matrix on the correct space, that satisfies certain trace conditions.

---
[7] Note that the input space of the effect is labelled with $\texttt{o}$, in line with previous notation.

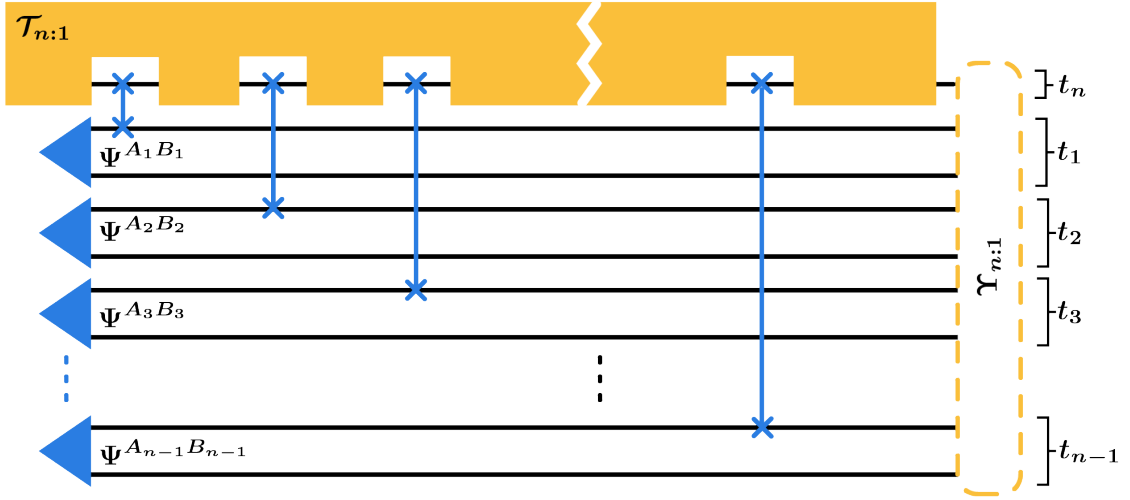

**Figure 3.8:** *CJI for the process tensor.* The multi-linear process tensor map, $\mathcal{T}_{n:1}$, can be represented as a many-body Choi operator, $\Upsilon_{n:1}$ through the CJI. At each timestep $t_j \in \{t_1, \ldots, t_n\}$, half of an (unnormalised) maximally entangled state, $\Psi^{A_j B_j}$, is swapped into the process (blue crosses). The resulting $2n-1$ body quantum state $\Upsilon_{n:1}$ contains equivalent information to the multi-linear map $\mathcal{T}_{n:1}$, with temporal correlations of the process being mapped to spatial ones between subsystems of its Choi operator. This is depicted by the brackets on the right, which signify the degrees of freedom of $\Upsilon_{n:1}$ that correspond to different times of the process $t_j$.

In detail, the many-body Choi operator corresponding to the process tensor is depicted in Fig. 3.8 and can be physically prepared (up to normalisation) as follows. Begin with $2(n-1)$ ancillary systems $\{A_j, B_j\}$ of appropriate dimension $d_{j^\mathrm{i}} := \dim(\mathcal{H}_{j^\mathrm{i}})$ in unnormalised maximally entangled pairs, $\{\Psi^{A_j B_j}\}$, where each $\Psi^{A_j B_j} = \sum_{\alpha\beta}^{d_j} |\alpha\alpha\rangle\langle\beta\beta|$ with $d_j := d_{j^\circ} d_{j^\mathrm{i}}$. At each timestep of the process, half of each pair is swapped with the system state through $\mathcal{G}_j^{SA_j}$. The resultant $d^{2n-1}$ dimensional system-ancillary operator $\Upsilon_{n:1} \in \mathsf{BL}(\mathcal{H}_{n^\mathrm{i}} \bigotimes_{j=1}^{n-1} \mathcal{H}_{j^\circ} \otimes \mathcal{H}_{j^\mathrm{i}})$ encodes equivalent information as the temporal map $\mathcal{T}_{n:1}$, and can be explicitly written as[8]

$$\Upsilon_{n:1} := \mathrm{tr}_E\left[\mathcal{U}_{n:n-1}^{SE}\mathcal{G}_{n-1}^{SA_{n-1}}\ldots\mathcal{U}_{2:1}^{SE}\mathcal{G}_1^{SA_1}\left(\Psi^{A_{n-1}B_{n-1}} \otimes \ldots \otimes \Psi^{A_1 B_1} \otimes \rho_{1^\mathrm{i}}^{SE}\right)\right]. \tag{3.20}$$

It is straightforward to show that the action of the process tensor—in clear analogy to the action of quantum maps in terms of their Choi operators—on a arbitrary sequences of CP maps can be expressed in terms of both the Choi operator of the process tensor itself and the maps on which it acts as follows

---

[8] Due to the importance of the process tensor as the fundamental object in the operational view of open quantum dynamics, we denote its Choi operator by the special symbol $\Upsilon$ instead of a sans serif letter.

$$\mathcal{T}_{n:1}(\mathcal{O}_{n-1},\ldots,\mathcal{O}_1) = \mathrm{tr}_{n-1:1}\left[\left(\mathbb{1}_{n^\mathrm{i}}\bigotimes_{j=1}^{n-1}\mathsf{O}^\mathrm{T}_{j^\mathrm{o}j^\mathrm{i}}\right)\Upsilon_{n:1}\right]. \quad (3.21)$$

More generally still, one could apply a sequence of interrogations correlated across timesteps, *e.g.*, by sending forward the ancilla that was used to implement an earlier operation. The corresponding correlated map $\mathcal{O}_{n-1:1}$ is a similar object to the process tensor (in a way that we will soon make explicit) and represents the most general kind of transformation one could implement over a sequence of timesteps. The action of the process tensor on such a correlated sequence of operations is, again, given by

$$\mathcal{T}_{n:1}(\mathcal{O}_{n-1:1}) = \mathrm{tr}_{n-1:1}\left[\left(\mathbb{1}_{n^\mathrm{i}}\otimes \mathsf{O}^\mathrm{T}_{n-1:1}\right)\Upsilon_{n:1}\right]. \quad (3.22)$$

3.2.3 *Properties of the Process Tensor*

Since the process tensor acts on sequences of CP maps it makes no sense to speak of complete-positivity and trace-preservation in the original sense; however, from its definition, meaningful extensions of these concepts are satisfied [29, 51]. The following properties follow directly from the definition of the process tensor, but can also be motivated on more axiomatic grounds [29–31].

Complete-positivity means that if the process tensor acts on some sequence of CP operations $\mathcal{O}_1^{SA_1},\ldots,\mathcal{O}_{n-1}^{SA_{n-1}}$, where each operation $\mathcal{O}_j^{SA_j}:\mathsf{BL}(\mathcal{H}_{j^\mathrm{i}}^S\otimes\mathcal{H}_{j^\mathrm{i}}^{A_j})\to \mathsf{BL}(\mathcal{H}_{j^\mathrm{o}}^S\otimes\mathcal{H}_{j^\mathrm{o}}^{A_j})$ acts on the system and some ancilla, the resulting transformation $\mathcal{T}_{n:1}^S\otimes\mathcal{I}_{n:1}^A(\mathcal{O}_{n-1}^{SA_{n-1}},\ldots,\mathcal{O}_1^{SA_1}):\mathsf{BL}(\mathcal{H}_{n-1^\mathrm{i}}^{A_{n-1}}\otimes\ldots\otimes\mathcal{H}_{1^\mathrm{i}}^{A_1})\to\mathsf{BL}(\mathcal{H}_{n^\mathrm{i}}^S\otimes\mathcal{H}_{n-1^\mathrm{o}}^{A_{n-1}}\otimes\ldots\otimes\mathcal{H}_{1^\mathrm{o}}^{A_1})$ is a CP map, no matter the size of the ancillary spaces; this is represented graphically in Fig. 3.9 for the superchannel. As a special case, this implies that the output of the process tensor, for any physically allowed transformations the experimenter might apply, is always a valid quantum state (up to normalisation). The trace preservation property of quantum channels on the level of process tensors translates to the statement that for any overall deterministic sequences of operations applied, the output quantum state must have unit trace.

These properties are encoded naturally in the Choi operator of the process tensor as

1. **Complete-positivity** : $\Upsilon_{n:1} \geq 0.$ \hspace{2em} (3.23)

2. **Trace-preservation** : $\mathrm{tr}_{j^\mathrm{i}}[\Upsilon_{j:1}] = \mathbb{1}_{j-1^\mathrm{o}}\otimes\Upsilon_{j-1:1}\quad \forall\, t_1 < t_j \leq t_n.$

Importantly, each $\Upsilon_{j-1:1}$ in the second condition is a proper process tensor describing the process on the timesteps $t_1,\ldots,t_{j-1}$ preceding each $t_j$. Thus the second property encapsulates an entire hierarchy of trace conditions, which implies that the trace of a process tensor is equal to the product of the dimension of the system on all of its output Hilbert spaces: $d^\circ_{n:1} := d_{1^\circ} \times \ldots \times d_{n^\circ}$. Clearly, the process tensor constructed via Eq. (3.20) satisfies both properties; conversely, every Choi operator satisfying them corresponds to a fixed open dynamics and thereby represents a valid process [51].

The CJI for the process tensor maps *temporal* correlations into *spatial* ones; thus, although almost all of the results to follow are presented in terms of the Choi operators of processes, these statements fundamentally address temporal properties of processes, such as correlations between observables measured over time on some evolving quantum system. As already encountered for the case of CPTP maps, all processes can be represented in this way as supernormalised many-body quantum states, but not all such quantum states represent valid processes [38]. The set of possible temporal correlations are restricted, compared to their spatial counterparts, because the process tensor must satisfy the above hierarchy of trace-conditions in order that it can be dilated to a *fixed* system-environment model [29–31, 51].

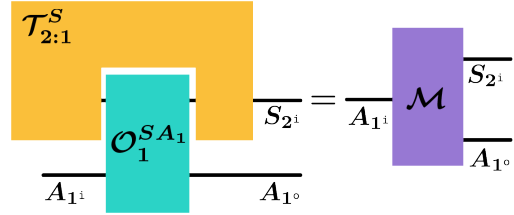

**Figure 3.9:** *Complete-positivity for the superchannel.* Complete-positivity for the superchannel $\mathcal{T}^S_{2:1}$ means that when acting on part of a CP map $\mathcal{O}^{SA_1}$, the resulting map $\mathcal{M} : \mathsf{BL}(\mathcal{H}^{A_1}_{1^i}) \to \mathsf{BL}(\mathcal{H}^S_{2^\circ} \otimes \mathcal{H}^{A_1}_{1^\circ})$ is a CP map, independent of the size of the ancilla.

The trace-preservation property on the process tensor can equivalently be viewed as a statement about causality. It is straightforward to show that if the hierarchy of trace conditions in Eq. (3.23) is satisfied, any later choice of instruments cannot influence earlier measurement statistics, and vice versa. This condition can be relaxed without leading to paradoxical situations [31, 33], but all process tensors naturally satisfy causality. We will only sometimes encounter non-ordered objects later on in this thesis; these are obtained from post-selection, which does not enforce causal order [31, 36, 40, 136].

From the vantage point provided by the Choi representation of processes and their properties, we now briefly reconsider from a more formal perspective a number of key concepts that have previously been loosely discussed. This will allow us to concretely calculate the joint probability distributions that are encoded in the process tensor with respect to any interrogating instrument sequence by way of a generalised spatio-temporal Born rule defined on the respective Choi operators.

## 3.3 SPATIO-TEMPORAL BORN RULE

Quantum theory is, at its core, about measurement statistics observed through experiment. Here we will take a closer look at POVM and instruments to make the connection between quantum stochastic processes and classical stochastic processes perspicuous.

It is well-known that quantum mechanics cannot be adequately described within standard probability theory due to inherently non-classical features, such as non-commutativity and contextuality. As discussed, a measurement instrument (*i.e.*, a POVM) is a set of such operators $\mathcal{J} = \{\Pi^{(x)\mathrm{T}}\}$ that sum to the identity $\sum_x \Pi^{(x)\mathrm{T}} = \mathbb{1}$. Here, each operator corresponds to an outcome, with the summation condition ensuring that *some* outcome occurs with certainty. With this, the probability for a measurement outcome to be realised for a given quantum state $\rho$ are computed via the Born rule

$$\mathbb{P}(x|\mathcal{J}) = \mathrm{tr}\left[\Pi^{(x)\mathrm{T}}\rho\right]. \tag{3.24}$$

In the above equation, $\rho$ is some *fixed, overall deterministic* object (*i.e.*, unit-trace quantum state) that contains all of the statistical information about the system of interest. This information can be deduced by means of the measurement instrument $\mathcal{J}$, where each of the possible outcomes corresponds to a (positive) POVM element, as depicted in Fig. 3.10. Here, the choice of measurement corresponds to what is controllable to an experimenter, and the state upon which it is implemented constitutes that which is uncontrollable. In the language of a generalised probability theory, intuitively, the measurement instrument plays the role of providing a $\sigma$-algebra on the event space, with each constituent POVM element corresponding to an event.[9]

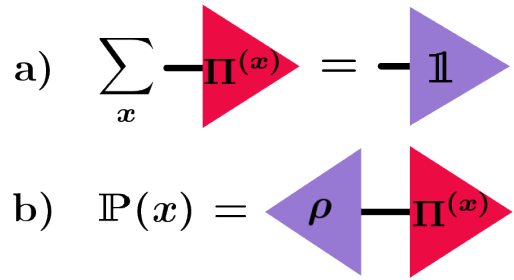

**Figure 3.10:** *Graphical representation of a POVM.* A POVM is a collection of positive operators $\mathcal{J} = \{\Pi^{(x)\mathrm{T}}\}$ that sum to the identity operator $\sum_x \Pi^{(x)\mathrm{T}} = \mathbb{1}$, shown in panel **a)**. That the overall implementation is TP allows us to interpret the effect of each POVM element on a state, *i.e.*, the outputs of the Born rule, as probabilities, as per panel **b)**.

However, the Born rule, in its original form, does not properly assign joint probabilities to consecutive events [35]; in order to study temporal processes, one must track the transformations of the system over time upon observation of outcomes, which cannot be accounted for by POVM. As briefly discussed previously, in the temporal setting individual conditionally-realised events are elevated to CP transformations, and a collection of these that ensures *some* transformation occurs overall constitutes an instrument, which plays the analogous role to a POVM, as illustrated in Fig. 3.11 and formally defined as follows.

---

9 Indeed, this is how the POVM got its name.

**Definition 3.2** (Instrument [7, 27])**.** An *instrument*, $\mathcal{J}$, is a collection of physical transformations that overall correspond to a deterministic transformation. Concretely, an instrument is represented by a set of CP maps, $\{\mathsf{O}^{(x)}\}$, that sum to a CPTP map, *i.e.*, $\mathsf{O}^{\mathcal{J}} := \sum_x \mathsf{O}^{(x)}$ satisfies both conditions in Eq. (3.15).

Intuitively, the CP operation $\mathsf{O}^{(x)}$ describes how the state of the system is changed upon measuring outcome $x$, given that the instrument $\mathcal{J}$ was used to interrogate the system. The concept of an instrument captures the most general overall deterministic transformation allowable within quantum theory that an experimenter could invoke at some point in time, including instruments with only a single deterministic 'outcome' corresponding to, *e.g.*, a unitary transformation. It is clear that the summation constraint imposed on POVMs is a special case of an instrument with a trivial output space, since the only trace-preserving effect is the identity operator. In contrast to the case of POVMs, however, the summation constraint on an instrument is non-unique; different instruments can correspond to different (unconditional) CPTP transformations.

We have already seen that the most general two-time quantum experiment is described by a superchannel $\mathsf{S}_{2^\mathrm{i}1^\mathrm{o}1^\mathrm{i}} \in \mathsf{BL}(\mathcal{H}_{2^\mathrm{i}} \otimes \mathcal{H}_{1^\mathrm{o}} \otimes \mathcal{H}_{1^\mathrm{i}})$. Consider then an experimenter applying an instrument $\mathcal{J}_1 := \{\mathsf{O}^{(x_1)}_{1^\mathrm{o}1^\mathrm{i}}\}$ followed by a POVM on the final output $\mathcal{J}_2 := \{\Pi^{(x_2)}_{2^\mathrm{i}}\}$ (since the process ends at $t_2$, the post-measurement state is irrelevant and a POVM suffices to calculate all statistics). For any specific realisation of the preparation procedure applied, the final output state is subnormalised with respect to the probability of the said preparation occurring and is given by

$$\rho^{(x_1)}_{2^\mathrm{i}} = \mathrm{tr}_1\left[\left(\mathbb{1}_{2^\mathrm{i}} \otimes \mathsf{O}^{(x_1)\mathrm{T}}_{1^\mathrm{o}1^\mathrm{i}}\right)\mathsf{S}_{2^\mathrm{i}1^\mathrm{o}1^\mathrm{i}}\right]. \tag{3.25}$$

The probability for implementation of a particular CP map at $t_1$ *and* the realisation of a measurement outcome at $t_2$ to occur is computed by applying the POVM on the output state via the standard Born rule, giving the joint statistics (see Eq. (3.11))

$$\mathbb{P}_{2:1}(x_2, x_1|\mathcal{J}_2, \mathcal{J}_1) = \mathrm{tr}\left[\Pi^{(x_2)\mathrm{T}}_{2^\mathrm{i}}\rho^{(x_1)}_{2^\mathrm{i}}\right] = \mathrm{tr}_{2^\mathrm{i}}\left[\Pi^{(x_2)\mathrm{T}}_{2^\mathrm{i}}\mathrm{tr}_1\left[\left(\mathbb{1}_{2^\mathrm{i}} \otimes \mathsf{O}^{(x_1)\mathrm{T}}_{1^\mathrm{o}1^\mathrm{i}}\right)\mathsf{S}_{2^\mathrm{i}1^\mathrm{o}1^\mathrm{i}}\right]\right]$$
$$= \mathrm{tr}\left[\left(\Pi^{(x_2)\mathrm{T}}_{2^\mathrm{i}} \otimes \mathsf{O}^{(x_1)\mathrm{T}}_{1^\mathrm{o}1^\mathrm{i}}\right)\mathsf{S}_{2^\mathrm{i}1^\mathrm{o}1^\mathrm{i}}\right]. \tag{3.26}$$

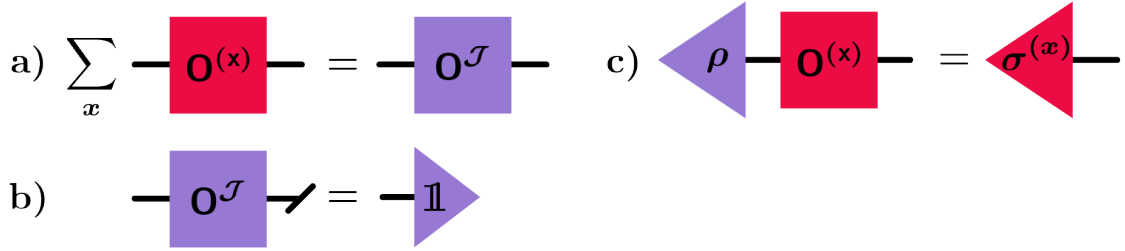

**Figure 3.11:** *Graphical representation of an instrument.* An instrument generalises the notion of a POVM. An instrument is a collection of CP maps $\mathcal{J} = \{\mathsf{O}^{(x)}\}$ that sum to a CPTP map, shown in panel **a)**. The trace-preservation constraint on CPTP maps entails that tracing over the output space of the map yields an identity operator effect, shown in panel **b)**. Although in each run of the experiment a specific transformation is only conditionally realised $\mathcal{O}^{(x)}(\rho) = \sigma^{(x)}$, as shown in panel **c)**, the summation constraint ensures that overall the instrument transforms the state into a normalised one $\sigma = \sum_x \sigma^{(x)}$.

It is instructive to compare the structure of Eq. (3.26) with that of the standard Born rule in Eq. (3.24). In the two-time scenario, it is the superchannel that now plays the role of the density operator, containing the information about all the probabilities for all the ways in which a two-time process can be interrogated. In this sense, Eq. (3.25) constitutes a spatio-temporal generalisation of the Born rule, at least for uncorrelated probing operations.

In order to describe more general scenarios, we must first extend the definition of instruments to *testers*, after which we will provide a generalised Born rule that allows calculation of all multi-time probabilities. In the setting envisaged, an experimenter could apply a sequence of instruments that are correlated across timesteps, *e.g.,* by sending forward the ancilla that was used to implement an earlier operation. Since the experimenter does something with overall certainty, the (potentially correlated) CP operation sequence they implement corresponds to a realisation or trajectory of a process which, when summed over all possible outcomes, yields an overall deterministic transformation. In the previous subsection, we saw that any such multi-time process must have the structure of a process tensor, *i.e.,* satisfy both constraints of Eq. (3.23). We therefore introduce the following definition that generalises the notion of an instrument to the multi-time setting where operations are permitted to be temporally correlated, depicted in Fig. 3.12.

**Definition 3.3** (Tester [31] / Instrument sequence). An *n-tester* or *instrument sequence*, $\mathcal{J}_{n:1}$, is a collection of physical transformations, which may be temporally correlated across $n$ timesteps, that overall correspond to a deterministic transformation. Concretely, a tester is represented by a set of multi-time CP maps defined on $n$ timesteps, *i. e.*, positive Choi operators $\{\mathsf{O}_{n:1}^{(x_{n:1})}\}$, that sum to a process tensor. That is, $\mathsf{O}_{n:1}^{\mathcal{J}_{n:1}} := \sum_{x_{n:1}} \mathsf{O}_{n:1}^{(x_{n:1})}$ satisfies both conditions in Eq. (3.23).

The individual tester elements need not satisfy the trace condition in Eq. (3.23) and represent the most general probing apparata one could implement over a sequence of timesteps. To summarise, in decreasing levels of generality, a tester element is to a tester what a CP map is to an instrument what a POVM element is to a POVM. As an example, the (uncorrelated) instrument sequence considered in Eq. (3.26) is a 2-tester, $\mathcal{J}_{2:1} := \{\Pi_2^{(x_2)} \otimes \mathsf{O}_1^{(x_1)}\}$, since $\sum_{x_2 x_1} \Pi_2^{(x_2)} \otimes \mathsf{O}_1^{(x_1)} = \Pi_2^{\mathcal{J}_2} \otimes \mathsf{O}_1^{\mathcal{J}_1} = \mathbb{1}_2 \otimes \mathsf{O}_1^{\mathcal{J}_1}$, with $\mathsf{O}_1^{\mathcal{J}_1}$ CPTP by assumption; thus, the overall transformation implemented by this 2-tester is a two-step process comprising a CPTP transformation followed by a measurement. Here, the fact that the measurement instrument implemented does not depend on the realisation of the preparation—*i. e.*, that the respective probing instruments are independent—is reflected in the tensor product structure of each tester element.

Although the mathematical structure of the process itself and the tester used to probe it are identical, we emphasise the distinction between the underlying, *uncontrollable* process and the *controllable* tester. Lastly, to keep the language consistent with that which is most commonly employed throughout the community, we will often be lax and simply use the term 'instrument' to dub both instruments and testers.

In a straightforward generalisation of Eq. (3.26) to the case where a process, $\Upsilon_{n:1}$, is interrogated over $n$ timesteps via a tester $\mathcal{J}_{n:1} = \{\mathsf{O}_{n:1}^{(x_{n:1})}\}$, one has the following spatio-temporal Born rule [35]

$$\mathbb{P}_{n:1}(x_{n:1}|\mathcal{J}_{n:1}) = \mathrm{tr}\left[\mathsf{O}_{n:1}^{(x_{n:1})\mathrm{T}} \Upsilon_{n:1}\right]. \tag{3.27}$$

The above equation is nothing more than than a restatement of Eq. (3.11) in terms of solely the Choi operators of the process tensor and tester elements applied. This is perhaps the most pivotal expression in this thesis, concretely relating the description of a given process to the statistics observed with respect to any meaningful probing schema. Although we have arrived at it from an open systems perspective, Eq. (3.27) can be derived on axiomatic grounds via a generalisation of Gleason's theorem applied to the process matrix [35].

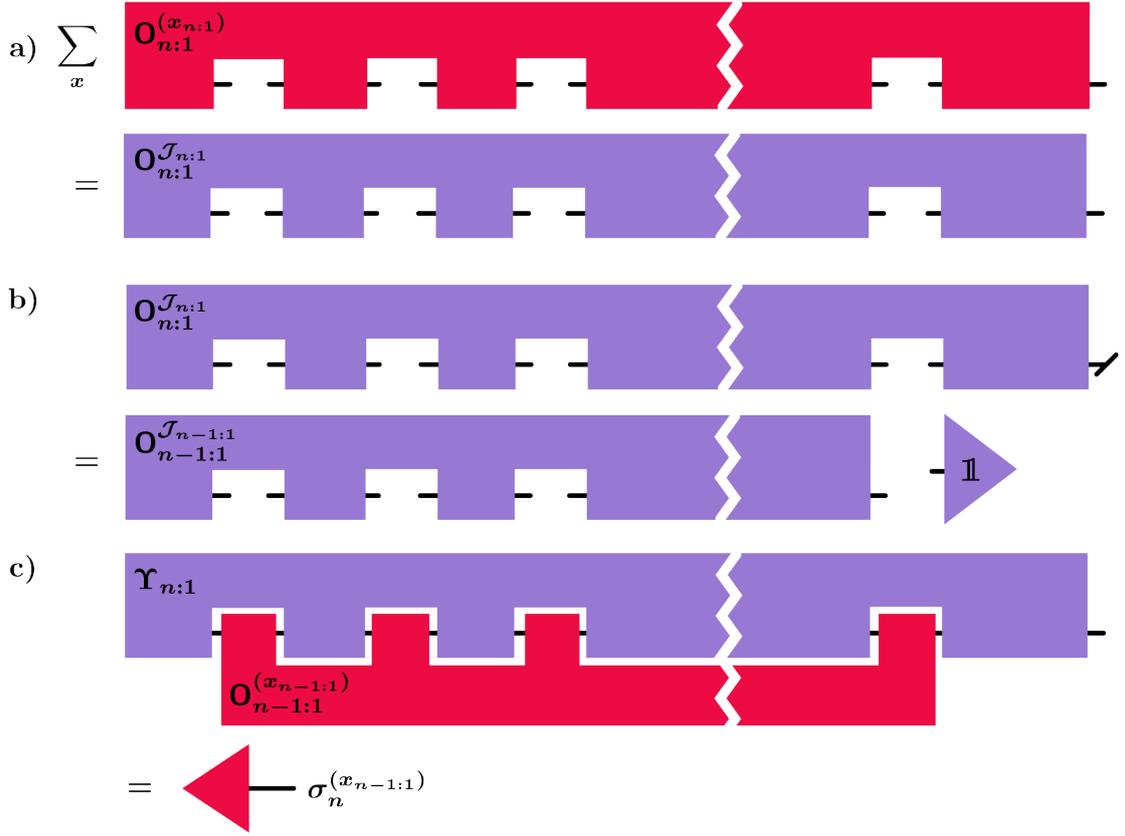

**Figure 3.12:** *Graphical representation of a tester.* A tester generalises the notion of an instrument to the case where the transformations may be correlated in time. An *n*-tester is a collection of positive Choi operators $\mathcal{J}_{n:1} = \{\mathsf{O}_{n:1}^{(x)}\}$ that sum to a multi-time CPTP transformation, *i.e.*, a process tensor, shown in panel **a)**. The summation condition ensures some transformation happens with certainty and entails a hierarchy of constraints on the structure of the process tensor, illustrated in panel **b)**. Given a process tensor $\Upsilon_{n:1}$, upon applying any sequence of operations $\mathsf{O}_{n-1:1}^{(x)}$ that constitute an $(n-1)$-tester, some probabilistically-realised output state results $\sigma_n^{(x_{n-1:1})}$. Overall, a unit-trace state $\sigma_n = \sum_{x_{n-1:1}} \sigma_n^{(x_{n-1:1})}$ is the final output.

In clear analogy to Eq. (3.24), here tester elements correspond to a realisation of the process, *i.e.*, a sequence of potentially correlated events across the specified timesteps, and the tester itself plays an analogous role to a choice of $\sigma$-algebra over the space of possible trajectories. The spatio-temporal Born rule therefore provides the mapping to probabilities for interrogations applied to some fixed process tensor. Thus, the process tensor generalises the notion of the density operator to the temporal setting inasmuch as it contains sufficient information to deduce all possible joint probabilities corresponding to the realisation of any valid instrument sequence applied.

As mentioned previously, the system density operator at each timestep can be obtained from the process tensor, and hence the latter contains all existing descriptors of open quantum processes as special cases. However, importantly, the process tensor additionally contains all of the multi-time information relevant to describing temporal correlations exhibited in any possible joint probability distribution that an experimenter might deduce. In this sense, quantum stochastic processes are described exactly like classical ones, as mappings from sequences of outcomes to probabilities, but with the crucial difference, that in quantum mechanics the concept of an instrument has to be inserted 'in the middle' to account for the possible invasiveness of measurements.

## 3.4 MARKOVIAN QUANTUM PROCESSES

With our ability to calculate joint statistics for quantum stochastic processes by way of the process tensor formalism and the generalised spatio-temporal Born rule, we now have an operational way to characterise Markovianity in quantum processes. Markovianity is, at its core, a multi-time statement regarding the conditional independence between the statistics observed at any given point in time and those of its entire history, given knowledge of its most recent state. We begin this section by considering how to meaningfully condition in quantum theory.

Since probabilities in quantum theory can only be calculated with respect to choices of probing instruments, a seemingly natural generalisation of Def. 2.4 is to demand that the following holds true for all times $t_k$ on which a quantum stochastic process is defined

$$\mathbb{P}_k(x_k|\mathcal{J}_k; x_{k-1}, \mathcal{J}_{k-1}; \ldots; x_1, \mathcal{J}_1) = \mathbb{P}_k(x_k|\mathcal{J}_k; x_{k-1}, \mathcal{J}_{k-1}). \tag{3.28}$$

Here $\mathbb{P}_k(x_k|\mathcal{J}_k; x_{k-1}, \mathcal{J}_{k-1}; \ldots; x_1, \mathcal{J}_1)$ denotes the probability to measure $x_k$ *given* that the outcomes $x_{k-1}, \ldots, x_1$ were previously observed, with the instruments $\mathcal{J}_k, \ldots, \mathcal{J}_1$ used to probe the dynamics. Note that on the right hand side, the probability conditioned on only the most recent outcome can only be meaningfully calculated with respect to the overall instruments applied throughout the prior history, $\mathcal{J}_{k-2}, \ldots, \mathcal{J}_1$. Nonetheless, requiring Eq. (3.28) to hold for arbitrary sequences of historic instruments seems to provide a sensible notion of Markovianity, since it guarantees conditional independence of observed statistics. However, as we will now see, a subtlety arises when attempting to condition on the knowledge of previous measurement outcomes in quantum theory.

Given a quantum stochastic process, $\Upsilon_{n:1}$, we seek to calculate the conditional statistics above. Suppose the probing instruments chosen at each timestep are $\mathcal{J}_j = \{\mathsf{O}_j^{(x_j)}\}$. We must first calculate the joint statistics up to some specified time $t_k \in \{t_1, \ldots, t_n\}$. By causality, the choice of future instruments $\mathcal{J}_j$ for $t_j > t_k$ cannot influence the statistics measured up until $t_k$, and it follows that the joint statistics can be computed from the reduced process tensor $\Upsilon_{k:1} = \text{tr}_{n:k^\circ}[\Upsilon_{n:1}]$ as follows

$$\mathbb{P}_{k:1}(x_k, \ldots, x_1 | \mathcal{J}_k, \ldots, \mathcal{J}_1) = \text{tr}\left[\left(\Pi_k^{(x_k)} \otimes \mathsf{O}_{k-1}^{(x_{k-1})} \otimes \ldots \otimes \mathsf{O}_1^{(x_1)}\right)^\text{T} \Upsilon_{k:1}\right], \quad (3.29)$$

where we specify the POVM at $t_k$ as $\mathcal{J}_k = \{\Pi_k^{(x_k)}\}$.

In order to condition on all of the previously observed statistics and calculate the l.h.s of Eq. (3.28), one simply wishes to divide the probabilities in Eq. (3.29) by those of the previous realisations

$$\mathbb{P}_{k-1:1}(x_{k-1}, \ldots, x_1 | \mathcal{J}_{k-1}, \ldots, \mathcal{J}_1) = \text{tr}\left[\left(\mathsf{O}_{k-1}^{(x_{k-1})} \otimes \ldots \otimes \mathsf{O}_1^{(x_1)}\right)^\text{T} \Upsilon_{k-1:1}\right]. \quad (3.30)$$

However, here we immediately run into the problem that $\mathsf{O}_{k-1}^{(x_{k-1})}$ is an operator on $\mathsf{BL}(\mathcal{H}_{k-1^\circ} \otimes \mathcal{H}_{k-1^\text{i}})$, whereas the reduced process tensor $\Upsilon_{k-1:1}$ only meaningfully provides information up to $t_{k-1^\text{i}}$, since $\text{tr}_{k^\text{i}}[\Upsilon_{k:1}] = \mathbb{1}_{k-1^\circ} \otimes \Upsilon_{k-1:1}$ (see Eq. (3.23)). Continuing from Eq. (3.30), we therefore have

$$\mathbb{P}_{k-1:1}(x_{k-1}, \ldots, x_1 | \mathcal{J}_{k-1}, \ldots, \mathcal{J}_1) \quad (3.31)$$
$$= \text{tr}\left[\text{tr}_{k-1^\circ}\left[\mathsf{O}_{k-1}^{(x_{k-1})\text{T}}\right] \otimes \left(\mathsf{O}_{k-2}^{(x_{k-2})} \otimes \ldots \otimes \mathsf{O}_1^{(x_1)}\right)^\text{T} \Upsilon_{k-1:1}\right].$$

The operator $\text{tr}_{k-1^\circ}\left[\mathsf{O}_{k-1}^{(x_{k-1})\text{T}}\right]$ necessarily does not capture any information about the output state of the interrogation at $t_{k-1}$, which could crucially impact the measurement outcomes observed at $t_k$. For instance, one can envisage situations where the experimenter performs a measurement $\{\Pi_{k-1^\text{i}}^{(r_{k-1})}\}$ and then prepares one of a set of (subnormalised) states with some outcome-dependant probability $\{\rho_{k-1^\circ}^{(s_{k-1}, r_{k-1})}\}$ to feed forward, implementing the instrument $\mathcal{J}_{k-1} = \{\Pi_{k-1^\text{i}}^{(r_{k-1})} \otimes \rho_{k-1^\circ}^{(s_{k-1}, r_{k-1})}\}$. Here, Eq. (3.29) would be sensitive to the choices of prepared states, whereas Eq. (3.31) would not. Thus, to divide the former by the latter would certainly not provide a meaningful notion of conditioning.

Indeed, conditioning necessarily breaks the information flow between past and future, while a generic operation need not. To ameliorate this issue we require the concept of a *causal break*, which is a particular type of instrument that clearly separates information about the inputs and outputs of its transformations. Intuitively, it corresponds to the scenario where an experimenter makes a measurement and then—in contrast to the example above—prepares a fresh, *independent* known state. A causal break instrument $\mathcal{J}_j = \{\mathsf{B}_j^{(x_j)}\}$ comprises elements (see Fig. 3.13)

$$\mathsf{B}_j^{(x_j)} := \rho_{j^\circ}^{(s_j)} \otimes \Pi_{j^\text{i}}^{(r_j)}, \quad (3.32)$$

where the labels $x_j$ of the overall CP maps in the causal break instrument are split into those labelling the measurement outcome, $r_j$, and the subsequent independent repreparation, $s_j$. Importantly, the Choi operators on the input and output spaces are completely uncorrelated; more generally, any operation whose output is independent of its input in the sense above constitutes a causal break.

Returning our consideration to the calculation of conditional statistics, we see that demanding the most recent instrument applied at time $t_{k-1}$ to be a causal break immediately resolves the problem, allowing us to meaningfully condition on prior knowledge obtained through probing a quantum stochastic process. Now, Eq. (3.31) reads

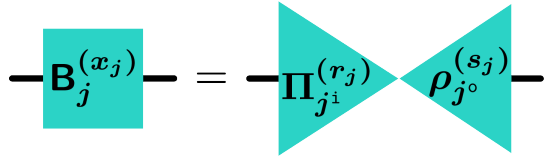

**Figure 3.13:** *Causal break.* A causal break at time $t_j$ is a POVM $\{\Pi_{j^i}^{(r_j)}\}$ followed by an independent repreparation into one of a known set of states $\{\rho_{j^\circ}^{(s_j)}\}$. This breaks any information/memory flow on the level of the system.

$$\mathbb{P}_{k-1:1}(s_{k-1}, r_{k-1}, \ldots, x_1 | \mathcal{J}_{k-1}, \ldots, \mathcal{J}_1) \tag{3.33}$$
$$= \mathbb{P}(s_{k-1} | \mathcal{J}_{k-1}) \, \mathrm{tr}\left[\left(\Pi_{k-1}^{(r_{k-1})} \otimes \ldots \otimes \mathsf{O}_1^{(x_1)}\right)^{\mathrm{T}} \Upsilon_{k-1:1}\right]$$
$$= \mathrm{tr}\left[\left(\rho_{k-1^\circ}^{(s_{k-1})} \otimes \Pi_{k-1}^{(r_{k-1})} \otimes \ldots \otimes \mathsf{O}_1^{(x_1)}\right)^{\mathrm{T}} \Upsilon_{k-1:1}\right],$$

where $\mathbb{P}(s_{k-1} | \mathcal{J}_{k-1}) = \mathrm{tr}\left[\rho_{k-1^\circ}^{(s_{k-1})}\right]$ is simply the probability that the experimenter chooses to prepare the particular state $\rho_{k-1^\circ}^{(s_{k-1})}$, which is independent of the process. Similarly, Eq. (3.29) reads

$$\mathbb{P}_{k:1}(x_k, s_{k-1}, r_{k-1}, \ldots, x_1 | \mathcal{J}_k, \ldots, \mathcal{J}_1) \tag{3.34}$$
$$= \mathrm{tr}\left[\left(\Pi_k^{(x_k)} \otimes \rho_{k-1^\circ}^{(s_{k-1})} \otimes \Pi_{k-1^i}^{(r_{k-1})} \otimes \ldots \otimes \mathsf{O}_1^{(x_1)}\right)^{\mathrm{T}} \Upsilon_{k:1}\right],$$

By breaking the flow of information on the level of the system, a causal break allows us to compare like with like for each run of the experiment and therefore meaningfully condition in the quantum setting. Dividing Eq. (3.34) by Eq. (3.33) gives the probability to observe outcome $x_k$ given knowledge of all previous outcomes for the instruments chosen, *i.e.*, the conditional probability $\mathbb{P}_k(x_k | \mathcal{J}_k; s_{k-1}, r_{k-1}, \mathcal{J}_{k-1}, \ldots, x_1, \mathcal{J}_1)$.

Intuitively, conceptual introduction of the causal break is necessary to demarcate those memory effects that arise from the process, *i.e.*, are the manifestation of environmental influence, rather than as a result of the applied control operations feeding-forward information about the system's past. As the fresh preparation is independent of the previous measurement outcome, a causal break ensures that no temporal correlations are transmitted through the system itself, breaking the causal link on the level of the system between the past $t_i \leq t_{k-1^\text{i}}$ and the future $t_j \geq t_{k-1^\text{o}}$. By the very nature of the causal break, the system itself cannot possibly transmit information beyond time $t_{k-1}$ concerning the measurement outcome associated to $\Pi_{k-1}^{(r_{k-1})}$ *or* the operations implemented throughout its earlier history. In light of this, it is sensible to slightly revise Eq. (3.28) to demand that a Markovian quantum process should display statistics that are conditionally independent of all historic outcomes *and* their instruments, including the measurement outcome $r_{k-1}$ of the causal break realised at $t_{k-1^\text{i}}$, given knowledge of the probabilities with which the output states of the causal break are prepared.

**Definition 3.4** (Quantum Markov condition [50])**.** Consider a causal break $\mathcal{J}_{k-1}$ at timestep $t_{k-1}$ in which the measurement outcome is labelled by $r_{k-1}$ and the prepared state by $s_{k-1}$. A quantum stochastic process is Markovian when the statistics observed with respect to an arbitrary measurement instrument $\mathcal{J}_k$ at $t_k$ are conditionally independent of all historic outcomes $r_{k-1}, x_{k-2}, \ldots, x_1$ for any possible historic instruments $\mathcal{J}_{k-2}, \ldots, \mathcal{J}_1$, given knowledge of only the state prepared during the causal break:[10]

$$\mathbb{P}_k(x_k|\mathcal{J}_k; s_{k-1}, r_{k-1}, \mathcal{J}_{k-1}; \ldots; x_1, \mathcal{J}_1) = \mathbb{P}_k(x_k|\mathcal{J}_k; s_{k-1}, \mathcal{J}_{k-1}). \tag{3.35}$$

The definition above lends itself to an operational criterion for classifying Markovian quantum processes: a quantum process is non-Markovian iff there exists at least two different historic testers, $\mathcal{J}_{k-1^\text{i}:1} = \{\mathsf{O}_{k-1^\text{i}:1}^{(x_{k-1^\text{i}:1})}\}$ and $\mathcal{J}'_{k-1^\text{i}:1} = \{\mathsf{O}_{k-1^\text{i}:1}'^{(x'_{k-1^\text{i}:1})}\}$ such that, for some choice of preparation in the causal break, the density operator of the system at $t_k$ is different, implying the violation of Eq. (3.35). This provides a valid and unambiguous method to witness memory effects. Conversely, fixing the preparation in the causal break and finding the subsequent density operator to be constant for all linearly independent historic tester elements implies that the process is Markovian.

Although we have made extensive use of the notion of causal breaks to provide an operational picture, the question of whether or not a process is Markovian does not depend on it. Indeed, Def. 3.4 directly leads to an unambiguous constraint on the structure of the process tensor itself, which is depicted in Fig. 3.14.

---

10 Interestingly, the same definition was introduced in Ref. [18], at roughly the same time as ours, using the process matrix formalism in the context of quantum causal modelling.

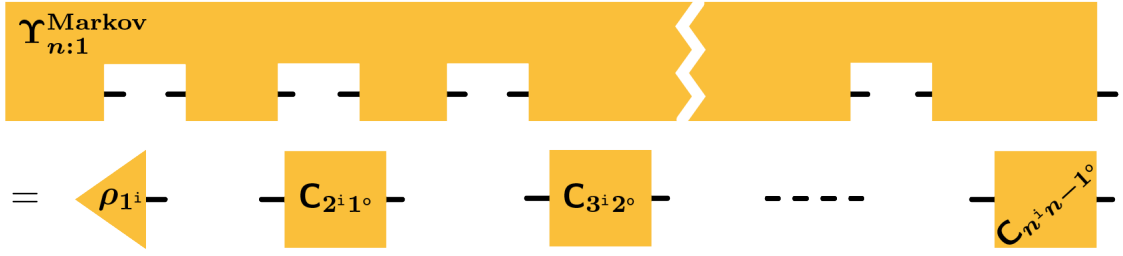

**Figure 3.14:** *Markovian process tensor.* The Choi operator of a Markovian process tensor is a tensor product of CPTP maps, representing uncorrelated evolution between timesteps.

**Theorem 3.5** (Markovian quantum process). *A process tensor represents a Markovian process iff it has the following tensor product structure:*

$$\Upsilon_{n:1}^{\mathrm{Markov}} = \mathsf{C}_{n^{\mathrm{i}}n-1^{\mathrm{o}}} \otimes \ldots \otimes \mathsf{C}_{2^{\mathrm{i}}1^{\mathrm{o}}} \otimes \rho_{1^{\mathrm{i}}}, \qquad (3.36)$$

*where each* $\mathsf{C}_{j^{\mathrm{i}}j-1^{\mathrm{o}}} \in \mathsf{BL}(\mathcal{H}_{j^{\mathrm{i}}} \otimes \mathcal{H}_{j-1^{\mathrm{o}}})$ *is the Choi operator of a CPTP map and* $\rho_{1^{\mathrm{i}}} \in \mathsf{BL}(\mathcal{H}_{1^{\mathrm{i}}})$ *is the initial average system state.*

*Proof.* The l.h.s of Eq. (3.35) is calculated as

$$\frac{\mathrm{tr}_{k:1}\left[\left(\Pi_k^{(x_k)} \otimes \rho_{k-1^{\mathrm{o}}}^{(s_{k-1})} \otimes \Pi_{k-1^{\mathrm{i}}}^{(r_{k-1})} \otimes \ldots \otimes \mathsf{O}_1^{(x_1)}\right)^{\mathrm{T}} \Upsilon_{k:1}\right]}{\mathrm{tr}_{k-1:1}\left[\left(\rho_{k-1^{\mathrm{o}}}^{(s_{k-1})} \otimes \Pi_{k-1^{\mathrm{i}}}^{(r_{k-1})} \otimes \ldots \otimes \mathsf{O}_1^{(x_1)}\right)^{\mathrm{T}} \Upsilon_{k-1:1}\right]} \qquad (3.37)$$

$$= \frac{\mathrm{tr}_{k:1}\left[\left(\Pi_k^{(x_k)} \otimes \rho_{k-1^{\mathrm{o}}}^{(s_{k-1})} \otimes \Pi_{k-1^{\mathrm{i}}}^{(r_{k-1})} \otimes \ldots \otimes \mathsf{O}_1^{(x_1)}\right)^{\mathrm{T}} \Upsilon_{k:1}\right]}{\mathbb{P}(s_{k-1}|\mathcal{J}_{k-1})\,\mathrm{tr}_{k-1:1}\left[\left(\Pi_{k-1^{\mathrm{i}}}^{(r_{k-1})} \otimes \ldots \otimes \mathsf{O}_1^{(x_1)}\right)^{\mathrm{T}} \Upsilon_{k-1:1}\right]}$$

$$= \frac{\mathrm{tr}_{k:1}\left[\left(\Pi_k^{(x_k)} \otimes \rho_{k-1^{\mathrm{o}}}^{(s_{k-1})} \otimes \Pi_{k-1^{\mathrm{i}}}^{(r_{k-1})} \otimes \ldots \otimes \mathsf{O}_1^{(x_1)}\right)^{\mathrm{T}} \Upsilon_{k:1}\right]}{\mathbb{P}(s_{k-1}|\mathcal{J}_{k-1})\,\mathbb{P}_{k-1:1}(r_{k-1},\ldots,x_1|\mathcal{J}_{k-1},\ldots,\mathcal{J}_1)}.$$

The r.h.s cannot be calculated readily by the spatio-temporal Born rule without first considering some historic instruments $\mathcal{J}_{k-2},\ldots,\mathcal{J}_1$. At all of these timesteps, we use their associated CPTP maps to calculate the probabilities; these are arbitrary and will eventually be varied freely to prove the result. Note also that summing over all of the most recent measurement outcomes, $r_{k-1}$, leads to the identity operator on the input space, $\mathbb{1}_{k-1^{\mathrm{i}}}$. We thus have

$$\frac{\mathrm{tr}_{k:1}\left[\left(\Pi_k^{(x_k)} \otimes \rho_{k-1^{\mathrm{o}}}^{(s_{k-1})} \otimes \mathbb{1}_{k-1^{\mathrm{i}}} \otimes \ldots \otimes \mathsf{O}_1^{\mathcal{J}_1}\right)^{\mathrm{T}} \Upsilon_{k:1}\right]}{\mathrm{tr}_{k-1:1}\left[\left(\rho_{k-1^{\mathrm{o}}}^{(s_{k-1})} \otimes \mathbb{1}_{k-1^{\mathrm{i}}} \otimes \ldots \otimes \mathsf{O}_1^{\mathcal{J}_1}\right)^{\mathrm{T}} \Upsilon_{k-1:1}\right]} \qquad (3.38)$$

$$= \frac{\mathrm{tr}_{k:1}\left[\left(\Pi_k^{(x_k)} \otimes \rho_{k-1^{\mathrm{o}}}^{(s_{k-1})} \otimes \mathbb{1}_{k-1^{\mathrm{i}}} \otimes \ldots \otimes \mathsf{O}_1^{\mathcal{J}_1}\right)^{\mathrm{T}} \Upsilon_{k:1}\right]}{\mathbb{P}(s_{k-1}|\mathcal{J}_{k-1})}.$$

Comparing Eqs. (3.37) and (3.38) shows that for a Markovian process we have

$$\operatorname{tr}_{k:1}\left[\left(\Pi_k^{(x_k)} \otimes \rho_{k-1^\circ}^{(s_{k-1})} \otimes \mathbb{1}_{k-1^\mathrm{i}} \otimes \ldots \otimes \mathsf{O}_1^{\mathcal{J}_1}\right)^\mathrm{T} \Upsilon_{k:1}\right] \qquad (3.39)$$

$$= \frac{\operatorname{tr}_{k:1}\left[\left(\Pi_k^{(x_k)} \otimes \rho_{k-1^\circ}^{(s_{k-1})} \otimes \Pi_{k-1^\mathrm{i}}^{(r_{k-1})} \otimes \ldots \otimes \mathsf{O}_1^{(x_1)}\right)^\mathrm{T} \Upsilon_{k:1}\right]}{\mathbb{P}_{k-1:1}(r_{k-1},\ldots,x_1|\mathcal{J}_{k-1},\ldots,\mathcal{J}_1)}.$$

This is clearly satisfied by the Markov process tensor structure given in Eq. (3.36), as can be seen by direct insertion. In the coming chapters, we will use similar arguments to derive structural properties of processes with finite Markov order.

In the converse direction, recall that this condition must hold for arbitrary choices of historical instruments. For any specific $x_k$ and $s_{k-1}$, we can consider varying the outcomes for a fixed sequence of historical instruments, leading to the r.h.s changing but not the l.h.s; on the other hand, we can consider a fixed sequence of CP maps and vary the overall instruments, leading to the l.h.s changing but not the r.h.s. The only remaining way that Eq. (3.39) can be satisfied is if the process tensor itself splits as a tensor product $\mathsf{C}_{k^\mathrm{i}:k-1^\circ} \otimes \Upsilon_{k-1^\mathrm{i}:1}$. That $\mathsf{C}_{k^\mathrm{i}:k-1^\circ}$ is a CPTP map follows directly from the properties of the process tensor. Lastly, repeating the argument for all times $t_k$ on which the process is described yields Eq. (3.36). $\square$

Through Theorem 3.5 we finally have an intrinsic characterisation of Markovian quantum processes. Intuitively, the Choi operator of a Markovian process only displays correlations between adjacent times, meaning that the only temporal correlations in the process are those between preparations and the subsequent measurements at the next timestep. This is analogous to the fact that Markovian process have non-trivial transition probabilities between adjacent timesteps only. Because the process tensor description includes the system density operator at all timesteps, the strict condition of Def. 3.4 presents both a unification and generalisation of previous definitions of Markovianity throughout the literature. For instance, this result confirms the well-known fact that Markovian processes are divisible; however, the converse direction fails to hold, as divisibility is not a strict enough criteria to force the process tensor into the required tensor product structure [55].

Moreover, the process tensor formalism can be used to explicitly calculate any of the measures of non-Markovianity introduced in the literature; several examples of memory effects that are not detected by conventional approaches but are within our operational framework are presented in the Supplemental Material of Ref. [50]. Indeed, from the structure of Markovian processes in Eq. (3.36), it is straightforward to define a measure that quantifies the amount of non-Markovianity, evaluated in terms of the distance between the Choi operator of the process at hand and the nearest Markovian Choi operator with respect to any suitable distance metric, $\mathbb{D}$ [50], as

$$\mathcal{N}(\Upsilon_{n:1}) := \min_{\Upsilon_{n:1}^{\text{Markov}}} \mathbb{D}\left(\Upsilon_{n:1} \| \Upsilon_{n:1}^{\text{Markov}}\right). \tag{3.40}$$

For instance, the measure chosen could be in terms of the relative entropy, as we will use in Subsection 6.1.2, or the Schatten 1-norm (trace distance), as used in Ref. [46] to study the distinguishability between Markovian and non-Markovian processes.

To summarise the story so far, we have here shown how the process tensor formalism naturally leads to a necessary and sufficient condition for a quantum process to be considered Markovian. This condition is stricter than those based on traditional approaches since the process tensor accounts for potential multi-time memory effects. By shifting to an operational description of stochastic processes, we have overcome the problem of formalism that led many to believe that there could be no unique criteria of Markovianity in open quantum dynamics. Indeed, the constraint on the structure of Markovian processes provides an intrinsic characterisation of a memoryless process.

We finally have a clear picture of what it means for a quantum process to have memory. As in the classical case, the description of generic processes grows rapidly in terms of complexity with the length of the memory; fortunately for us, many processes found in nature display an effectively finite-length memory. The remainder of this thesis presents novel work pertaining to characterising, quantifying and exploiting by way of simulation such finite memory processes.

# Part III

## UNSPOKEN WORDS

Not the weekend dance where you two-step to music you've heard before and always know... but the Daily Dance with the wilder step, to a tune as soundless as the eelgrass tune, to an echo of a song, or a song still unechoed.

— Ken Kesey, *Sometimes a Great Notion.*

# 4

## MEMORY LENGTH

W<small>E HAVE SEEN THAT FOR</small> M<small>ARKOVIAN PROCESSES</small>, once the state is known, the future is independent from the past. A central obstacle in predicting the future of a dynamical system is to understand how much of the system's past acts as a relevant influence, which crucially determines the resources required for simulation. Although non-Markovianity is the rule rather than the exception when it comes to stochastic processes [26], their characterisation is resource intensive, both in the classical and quantum setting. Fortunately for us, in reality, even the most complex processes have a finite effective range—that is, a cause can only noticeably affect the future for a certain length of time—providing a natural notion of memory length, formally captured by the concept of Markov order.

A process with finite Markov order allows for an efficient description compared to the general non-Markovian case, as it only requires a portion of the history to optimally predict the future. Indeed, the substantial reduction in modelling complexity that ensues underpins the often-invoked higher-order Markov models to simulate non-Markovian stochastic processes [59–61]. The motivation for understanding processes with finite Markov order is thus two-fold: on the one hand, they exhibit genuine memory effects; on the other, these effects are constrained in time, rendering their description tractable [26].

Akin to the joint probability distribution describing a generic classical stochastic process, the description of a general quantum stochastic process has an exponentially increasing complexity with respect to the length of history that must be retained, with the added complication that all possible sequences of external interventions need to be accounted for. This naturally begs the question that will be the central focus of the present chapter: *are there quantum processes with finite-length memory, and hence a significantly reduced complexity?* We have already seen that the process tensor formalism naturally lends itself to a proper classification of Markovianity in the quantum regime; the first line of pursuit here is a similar generalisation of Markov order.

We will see that a reasonable and logically sound definition of Markov order for quantum processes can be formulated in terms of a constraint on the structure of the process tensor. However, in quantum mechanics, we have no choice here but to propose an *instrument-specific* notion of Markov order. Intuitively, the prospective definition implies that with respect to a given instrument sequence specified across $\ell$ timesteps of the memory, any deducible statistics across the history and future timesteps (*i.e.*, for arbitrary choices of history and future probing instruments) are guaranteed to be conditionally independent for each realisation of the memory tester in question. The logical consequence of this characterisation is that—perhaps unsurprisingly—quantum processes with memory exhibit different memory effects when probed in different ways. The aim of the present chapter is to precisely formulate and justify this notion of quantum Markov order which formally captures such behaviour.

Following the introduction of this instrument-specific notion of quantum Markov order, we provide two crucial pieces of supporting evidence of its plausibility. Firstly, that the definition reduces to the classical one in an appropriate fashion. Indeed, classical Markov order is implicitly defined in an instrument-specific way—with respect to the statistics deduced via sequences of *sharp* measurements. As soon as we allow for *fuzzy* measurements that coarse-grain information—or, more broadly, any active experimental interventions—the length of the memory depends on the instruments used to interrogate the process at hand, just like in the quantum case.

Within quantum theory, however, there exist a much richer array of instruments that an experimenter might apply to the system of interest than are available in the classical world, including generalised measurements, unitary transformations, and temporally correlated sequences implemented via a genuinely quantum control (*i.e.,* the most general testers). The second key result of this chapter shows that demanding a quantum process to have finite Markov order with respect to all possible instruments immediately trivialises the theory into only admitting Markovian processes. This result, in turn, leads to the realisation that any quantum process with memory has infinite Markov order with respect to generic instrument sequences and therefore requires the complete description. This fact notwithstanding, for many practical purposes, such a comprehensive understanding of the memory is often either unachievable—due to experimental or computational limitations—or otherwise unnecessary with regard to the specificity of the experimenter's concerns.

Indeed, the landscape of memory effects in quantum processes is vast and ripe for exploration, as will become apparent in Chapter 5 where we analyse the structure of quantum processes with finite Markov order with respect to familiar classes of history-blocking instruments. Finally, note that many of the concepts introduced, results derived and discussion that entails in the coming chapter are presented in Refs. [1, 2].

## 4.1 QUANTUM MARKOV ORDER

Before we begin the main discussion of this chapter, it is worthwhile to briefly consider some alternative suggestions that arise naturally when considering an extension of Markov order into quantum mechanics. This aside should both illuminate some crucial details that must be addressed in the temporal setting and distinguish our approach from a number of related concepts studied throughout the community.

Recall that, as discussed in Subsection 2.1.6, there are three equivalent characterisations for classical Markov order:

1. The joint probability distribution across the future and history, given events in the memory, factorises: $\mathbb{P}_{FH}(x_F, x_H | x_M) = \mathbb{P}_F(x_F | x_M) \mathbb{P}_H(x_H | x_M)$.
2. There exists a stochastic map $\mathcal{R}_{M \to FM}$ acting only on the memory block that can 'recover' the future statistics correctly: $\mathbb{P}_{FMH} = \mathcal{R}_{M \to FM}(\mathbb{P}_{MH})$.
3. The CMI between the history and the future given the memory vanishes: $I^{\text{cl}}(F : H | M) = 0$.

For some time, the recovery map has featured in the quantum information literature: here, quantum Markov chains have been defined as tripartite quantum *states* $\rho^{ABC} \in \mathsf{BL}(\mathcal{H}^A \otimes \mathcal{H}^B \otimes \mathcal{H}^C)$ satisfying a quantum generalisation of recoverability [137–143]. Equivalently, quantum Markov chains have been defined as those states with vanishing quantum CMI, which is identical to those that saturate the strong subadditivity inequality [137, 144]. In the case of the former, the natural extension posits the existence of a CPTP recovery map $\mathcal{R}_{B \to BC} : \mathsf{BL}(\mathcal{H}^B) \to \mathsf{BL}(\mathcal{H}^B \otimes \mathcal{H}^C)$ such that $\rho^{ABC} = \mathcal{R}_{B \to BC}(\rho^{AB})$. Interestingly, unlike in the classical setting, the relation between quantum recoverability and the vanishing of the quantum CMI is not at all obvious; the proof of their equivalence is in fact a highly celebrated result [137, 138, 145]. Additionally, in Ref. [138] the authors introduced an important structural characterisation of the set of states with vanishing quantum CMI, which are therefore recoverable. The main theorem highlighting the existence of a projective measurement on the $B$ subsystem such that, for each outcome of the measurement, the $AC$ subsystem is steered into an uncorrelated state, providing a link by way of analogy to the quantum counterpart of the first concept listed above.

However, at first sight, it is unclear how such characterisations relate to *temporal processes*, where an experimenter has access to an evolving quantum system across multiple timesteps. As previously discussed, a number of concerns arise in the temporal setting because of the necessity to track the state of the system throughout the process in a meaningful way. Due to the breakdown of the KET for quantum processes on the level of joint probability distributions, there has heretofore been no sensible way to develop a generalisation of Eq. (2.19)—arguably the most fundamental definition of Markov order—to the temporal realm. The process tensor formalism, by way of encoding all possible joint distributions for all possible sequences of operations in time and permitting calculation of conditional statistics, provides a straightforward and unambiguous way to define Markov order for quantum processes. We will return to the relation between quantum Markov order and vanishing quantum CMI in the coming chapter.

### 4.1.1 *Instrument-specific Quantum Markov Order*

A natural approach to defining Markov order for quantum processes is to require that for every possible way of probing the process, the corresponding joint probability distribution satisfies the classical Markov condition. In other words, just as was the case for Markovian quantum processes, we fix the choice of instrument on the timesteps associated to the memory block and check for conditional independence between the history and the future. Thus, a sensible demand of a quantum process with finite-length memory is that *any* future statistics deducible—*i. e.,* no matter which future instruments are chosen—are conditionally independent of *any* historical instruments applied and their measurement outcomes, given knowledge of a length-$\ell$ instrument sequence applied to the memory block. Grouping together the timesteps as in Def. 2.5, we therefore define quantum Markov order as follows.

**Definition 4.1** (Quantum Markov order [1])**.** A quantum stochastic process has Markov order $\ell = |M|$ with respect to an instrument $\mathcal{J}_M$ when, for all possible instruments $\mathcal{J}_H$ and $\mathcal{J}_F$ on the history and future, the following is satisfied:

$$\mathbb{P}_F(x_F|\mathcal{J}_F; x_M, \mathcal{J}_M; x_H, \mathcal{J}_H) = \mathbb{P}_F(x_F|\mathcal{J}_F; x_M, \mathcal{J}_M), \tag{4.1}$$

where $\mathbb{P}_F(x_F|\mathcal{J}_F; x_M, \mathcal{J}_M; x_H, \mathcal{J}_H)$ denotes the probability to measure $x_F$ given that outcomes $x_M$ and $x_H$ were previously observed, with the instruments $\mathcal{J}_F, \mathcal{J}_M$ and $\mathcal{J}_H$ used to probe the system dynamics.

The above definition is intuitive: we can imagine an experimenter who implements the instruments $\mathcal{J}_F, \mathcal{J}_M$ and $\mathcal{J}_H$. They would consider the process to display finite memory if, for any choice of $\mathcal{J}_F$ and $\mathcal{J}_H$, the statistics observed on $F$ and $H$ are conditionally independent with respect to a *fixed* instrument $\mathcal{J}_M$. For any instruments the experimenter might use to probe the future evolution of the system, the full statistics is then completely determined by the instruments and outcomes of the most recent $\ell$ timesteps. Equivalently, given knowledge of the outcomes of the instrument across the past $\ell$ timesteps, the process governing the future is uncorrelated with that governing the history, guaranteeing that any possible statistics one might deduce on the history and future timesteps are independent in each run of the experiment. Again, it is crucial to note here that the notion of independence here is a conditional one.

Quantum Markov order thus defined is *instrument-specific* in the sense that it fixes the instrument sequence applied across $M$ in order to sensibly calculate conditional statistics, stipulating that meaningful statements regarding memory length must be presented with the caveat regarding the instrument of choice. This instrument-specific definition boils down to the classical one if all instruments are fixed, sharp and classical, as we will discuss in Section 4.2. As we will soon see, demanding a quantum process to have finite Markov order for all possible instruments admissible in quantum theory is too restrictive, immediately trivialising the theory. Such a demand is equivalent to asking the process to have finite Markov order with respect to an arbitrary length-$\ell$ sequence of causal breaks, since they form a basis for the space of valid instrument sequences on the memory block. As this constraint only permits Markovian processes, we are forced to consider the more general instrument-specific scenario in order to meaningfully describe processes with memory, as we do here. Nonetheless, a process can have finite quantum Markov order with respect to an entire family of instruments.

In fact, we have already seen an example of this: Markovian quantum processes have Markov order 1 for all instruments consisting of only causal breaks. In contrast, when memory plays a non-negligible role in the evolution, operations performed on the system generally impact the environment significantly, which, in turn, inevitably influences the future dynamics. Thus, Markov order $\ell$ does not only refer to processes with conditional independence across a length $\ell$ sequence of causal breaks, since these may serve to 'open a pathway' for the history to influence the future via environmental conditioning. Indeed, this is a special case of the instrument-specific definition of quantum Markov order, as we will see in Chapter 5, where we will study a number of processes with finite-length memory with respect to natural families of instruments, including those comprising unitary operations and extended sequences of causal breaks.

### 4.1.2 *Finite Memory Constraint on the Process Tensor*

We saw in Section 3.4 that Markovian quantum processes have a simple tensor product structure, and it is hence an interesting question to explore the implications of Def. 4.1 on the structure of the process tensor. To this end, consider a stochastic process described by the process tensor $\Upsilon_{FMH}$ and denote the probing instrument sequences by $\mathcal{J}_X = \{\mathsf{O}_X^{(x_X)}\}$. Then, the joint probability distribution over statistics observed is calculated via the generalised Born rule as

$$\mathbb{P}_{FMH}(x_F, x_M, x_H | \mathcal{J}_F, \mathcal{J}_M, \mathcal{J}_H) \qquad (4.2)$$
$$= \mathrm{tr}\left[\left(\mathsf{O}_F^{(x_F)} \otimes \mathsf{O}_M^{(x_M)} \otimes \mathsf{O}_H^{(x_H)}\right)^{\mathrm{T}} \Upsilon_{FMH}\right].$$

From this expression, we can calculate well-defined conditional probabilities

$$\mathbb{P}_{FMH}(x_F | \mathcal{J}_F; x_M, \mathcal{J}_M; x_H, \mathcal{J}_H) \qquad (4.3)$$
$$= \frac{\mathrm{tr}\left[\left(\mathsf{O}_F^{(x_F)} \otimes \mathsf{O}_M^{(x_M)} \otimes \mathsf{O}_H^{(x_H)}\right)^{\mathrm{T}} \Upsilon_{FMH}\right]}{\mathrm{tr}\left[\left(\mathsf{O}_M^{(x_M)} \otimes \mathsf{O}_H^{(x_H)}\right)^{\mathrm{T}} \Upsilon_{MH}\right]},$$

where $\Upsilon_{MH} := \frac{1}{d_{F^\circ}} \mathrm{tr}_F[\Upsilon_{FMH}]$ with $d_{F^\circ} := \prod_{j=k}^n d_{j^\circ}$ denoting the total dimension of the output spaces associated to $F$.

In Appendix C.1, we show that Def. 4.1 implies the following product structure on the process tensor, represented graphically in Fig. 4.1 [1]

$$\widetilde{\Upsilon}_{FH}^{(x_M)} := \mathrm{tr}_M\left[\mathsf{O}_M^{(x_M)\mathrm{T}} \Upsilon_{FMH}\right] = \Upsilon_F^{(x_M)} \otimes \widetilde{\Upsilon}_H^{(x_M)} \quad \forall\, \mathsf{O}_M^{(x_M)} \in \mathcal{J}_M. \qquad (4.4)$$

Here, the conditional future process, $\Upsilon_F^{(x_M)}$, is described by a proper process tensor whereas the unnormalised description of the historic process, $\widetilde{\Upsilon}_H^{(x_M)}$, is garnished with a tilde to denote that it is an element of a tester, *i.e.,* when summed over all outcomes of the memory instrument, it yields a proper process tensor. We adhere to this notation throughout and will return to discuss this point shortly. Crucially, finite Markov order does not force the process tensor into an overall tensor product structure, but only conditionally; in analogy to Eq. (2.19), the conditional history and future processes are independent for each realisation of the instrument applied.

A few further comments are in order. Firstly, if Eq. (4.4) is satisfied, we say that the process has Markov order-$\ell$ *with respect to the history-blocking instrument sequence,* $\mathcal{J}_M$. The fact that the process is rendered conditionally independent for each realisation of the instrument—which is, overall, a *deterministic* implementation—means that an experimenter is *guaranteed* to block the effect of the history on the future upon its application (given that they know the outcome). More generally, there may exist individual operation sequences, *i.e.,* tester elements, that block the history. However, in contrast to instruments, since these could only be implemented with some probability, such operations act to *probabilistically* render the future and history conditionally independent. In this thesis, we focus only on overall *deterministic* history-blocking sequences, in which every constituent tester element in an instrument sequence acts to block the effect of history, guaranteeing conditional independence for each run of the experiment.

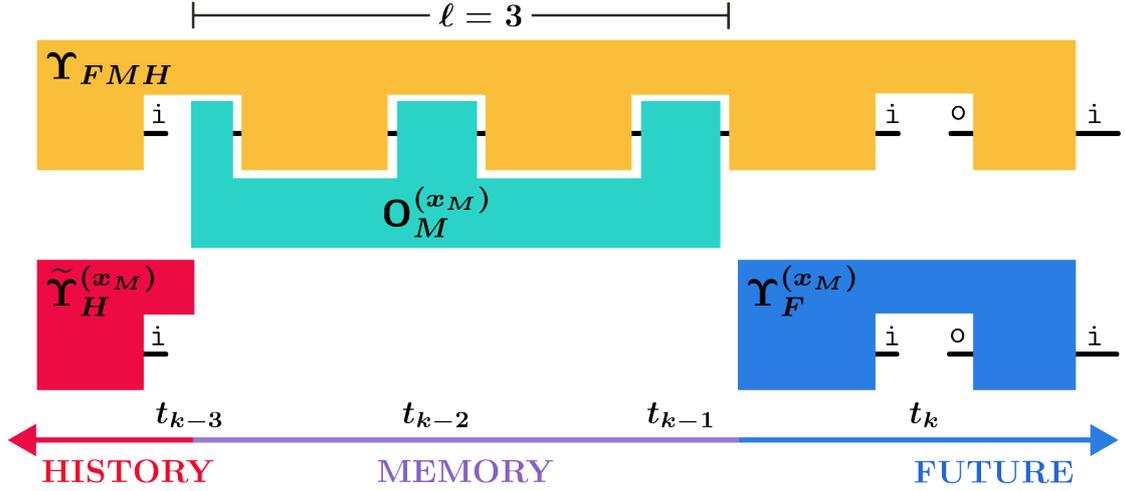

**Figure 4.1:** *Instrument-specific quantum Markov order.* An instrument sequence $\mathcal{J}_M$, comprising (temporally correlated) CP operations $\{\mathsf{O}_M^{(x_M)}\}$ (green) across a sequence of timesteps of length $\ell = 3$, is applied to a process $\Upsilon_{FMH}$. The process is said to have Markov order $\ell$ with respect to this instrument sequence if, for each possible realisation of the instrument, labelled by $x_M$, the history (red, $\widetilde{\Upsilon}_H^{(x_M)}$) and future (blue, $\Upsilon_F^{(x_M)}$) parts of the process are rendered conditionally independent. Note that here, the memory length of 3 timesteps corresponds to 5 Hilbert spaces on which the process tensor is defined (from $t_{k-3}\circ$ to $t_{k-1}\circ$).

Secondly, satisfaction of Eq. (4.4) guarantees the conditional independence of *any* possible statistics one could obtain on the future and history given knowledge of the history-blocking instrument sequence. We can see clearly that the mutual information between the conditional future and history processes for any realisation of $\mathcal{J}_M$ vanishes, since they are of product form

$$I(F:H)_{x_M} := S\left(\Upsilon_F^{(x_M)}\right) + S\left(\widetilde{\Upsilon}_H^{(x_M)}\right) - S\left(\widetilde{\Upsilon}_{FH}^{(x_M)}\right) = 0, \tag{4.5}$$

where $I(F:H)_{x_M}$ denotes the mutual information between the history and future processes given that the sequence corresponding to outcome $x_M$ was realised, and $S(\bullet)$ is the von Neumann entropy.[1] The mutual information upper-bounds all possible correlations between arbitrary observables on $F$ and $H$, and thus its vanishing implies the temporal regions of the future and history are uncorrelated given knowledge of each outcome $x_M$ [146].

Thirdly, note that the conditional future process is a proper process tensor by construction, whilst the conditional history process represents an element of a tester. The former point arises from the fact that for each realisation of the instrument on the memory block, *some* fixed future evolution occurs. Mathematically, this can be seen by considering that the hierarchy of trace conditions in Eq. (3.23) holds for all $x_M$ in Eq. (4.4), as each successive partial trace applied from the future backwards ends up acting only upon $\Upsilon_{FMH}$. On the other hand, the fact that $\Upsilon_{FMH}$ is a proper process tensor no longer ensures the hierarchy of trace conditions hold for each conditional positive object $\widetilde{\Upsilon}_H^{(x_M)}$ prior to the memory, as the multiplication with $\mathsf{O}_M^{(x_M)\mathrm{T}}$ prevents us from continuing to trace back through the hierarchy; each realisation of the instrument in question amounts to a post-selection [1, 29, 36]. Nonetheless, since the overall implementation of the memory instrument is deterministic, when all possible outcomes are summed over, the history is described by a proper process tensor. In other words, on average, the history is described by a positive semidefinite Choi operator satisfying Eq. (3.23); however, the individual tester elements corresponding to each conditional outcome need not obey the causality conditions. In the special cases where they do, the probability of realising the associated sequence of outcomes of the history-blocking instrument can be extracted from the conditional history process, *i.e.,* we can write $\widetilde{\Upsilon}_{FH}^{(x_M)} = \mathbb{P}(x_M|\mathcal{J}_M)\Upsilon_F^{(x_M)} \otimes \Upsilon_H^{(x_M)}$, with $\Upsilon_H^{(x_M)}$ a proper process tensor for each $x_M$, as we do at some points throughout this thesis.

Lastly, it is important to distinguish which input and output spaces constitute a memory block of length $\ell$, as there is evidently some innocuous source of potential confusion in the way that we have discussed memory length so far. Any such block may begin and end on either the input or output Hilbert spaces associated with timesteps $t_{k-\ell}$ and $t_{k-1}$ respectively (see Fig. 4.1: here, the memory block illustrated begins on the output space at $t_{k-\ell}$ and ends at the output space of $t_{k-1}$). To mitigate any possible confusion, one could describe the memory length in terms of the number of Hilbert spaces comprising the memory block. However, intuitively, an experimenter is surely more concerned with the question of *how long ago something happened* rather than *how many Hilbert spaces back something happened*. As such, and to ease notation, we refrain from labelling each of these cases distinctly and simply consider the memory length to be the number of timesteps across which the history-blocking sequence spans. We will provide visual representations of each example considered throughout this thesis to elucidate how the memory block is defined in each case; for instance, in Fig. 4.1, the memory length is 3 timesteps rather than 5 Hilbert spaces.

---

1 Since entropies are only meaningfully-defined for normalised objects, any entropic quantity is calculated using the normalised process tensor, *i.e.,* $\Upsilon/\mathrm{tr}\left[\Upsilon\right]$.

As we have previously alluded to, a quantum process can have finite Markov order with respect to a family of instruments. Demanding that Eq. (4.1) holds with respect to all possible instrument sequences immediately trivialises the theory.

**Theorem 4.2** (Only Markovian processes have finite Markov order for all instruments [1]). *The only quantum processes with finite Markov order with respect to all possible instruments are Markovian.*

The proof follows along the lines of that of Theorem 3.5 and is given in Appendix C.2. We make use of the fact that the set of tester elements forms a vector space to show that the only processes satisfying Eq. (4.4) for all possible instruments have trivial Markov order, *i.e.*, either $\ell = 1$ or 0. Specifically, we show that if a process has finite Markov order with respect to a complete basis of CP maps on $M$, it cannot have the same Markov order with respect to any non-trivial linear combination of them.

Returning to the subtlety in defining memory length in quantum mechanics (*i.e.*, with respect to Hilbert spaces or with respect to timesteps), note that when a 1-step memory block is specified, there are two cases to be considered: the first being where the memory begins and ends on the most recent output Hilbert space (*i.e.*, $M = \{t_{k-1^\circ}\}$), and the second where it extends over both Hilbert spaces associated to the previous timestep (*i.e.*, $M = \{t_{k-1^\circ}, t_{k-1^i}\}$). Demanding the finite Markov order condition to hold for all instruments in the former case is equivalent to the Markovianity condition of Def. 3.4, and since all instruments on only a single output Hilbert space correspond to a unit-trace quantum state, it leads directly to the structure of Eq. (3.36).

On the other hand, requiring finite Markov order to hold for all possible instruments in the latter scenario is an even stricter demand than Markovianity, which only dictates that the condition hold for the family of causal break instruments (see Eq. (3.32)). If the future process is to be rendered independent from the history for all instruments spanning *both* Hilbert spaces of the previous timestep, the process is forced into the following super-Markovian structure, where the future evolution is independent of even the most recent state preparation (like a fair coin toss)

$$\Upsilon_{n:1}^{\text{S-Markov}} = \rho_{n^\text{i}} \otimes \mathbb{1}_{n-1^\circ} \otimes \rho_{n-1^\text{i}} \otimes \ldots \otimes \mathbb{1}_{1^\circ} \otimes \rho_{1^\text{i}}. \tag{4.6}$$

In either case, demanding finite Markov order with respect to all instruments on a memory block ending at some timestep $t_{k-1}$ imposes a tensor product structure on the process tensor (either between $t_{k-1^\text{i}}$ and $t_{k-1^\circ}$ or between $t_{k-1^\circ}$ and $t_{k^\text{i}}$) such that the future and history processes are independent of the measurement outcome; requiring this to hold at all timesteps enforces a Markovian or super-Markovian structure, thereby trivialising the theory by only admitting memoryless processes. This result immediately implies the following property for quantum stochastic processes.

*Remark.* Any non-Markovian quantum process has infinite Markov order with respect to a generic instrument sequence. Furthermore, applying random instruments can almost always witness memory effects.

This is because, if the process uses any memory at all, there always exists some instrument such that Eq. (4.4) does not hold (it only holds for all instruments if the process is Markovian). Applying random choices of operations on a finite-dimensional Hilbert space will eventually span the space and an experimenter can then find the suitable operation to witness memory effects.

It is clear that demanding Def. 4.1 to hold for all instruments is a very strong condition, as it requires the statistics observed by an experimenter to satisfy the Markov order-$\ell$ property *no matter* how they measure realisations of the process. Theorem 4.2 shows that this requirement is too restrictive in the quantum case. Thus, we have no choice but to characterise Markov order for quantum processes in an instrument-specific way. In light of this analysis, we see that classical Markov order is also defined in an inherently instrument-specific way, as it assumes the ability for an experimenter to sharply measure realisations of events. In many practical cases of interest, however, this assumption is either not satisfied due to experimental limitations that lead to noisy measurements [147–153], or insufficient to capture the scenario at hand, such as in causal modelling [17, 18]. In any framework that allows for active interventions, a vast array of memory effects can be captured by probing the system with different instruments, as we now consider.

## 4.2 RELATION TO CLASSICAL MARKOV ORDER

Importantly, quantum Markov order reduces to the classical Markov order statement in the correct limit: when the stochastic process at hand is entirely characterised by a joint probability distribution and the experimenter probes it appropriately via classical means.

Classical stochastic processes in the traditional sense, where interventions are not allowed, assume the existence of only one probing instrument, namely that of a measurement comprising only rank-1 projections onto one of a complete set of orthogonal (classical) states at each timestep: $\mathcal{J}^{\text{cl}} := \{\Pi_{\text{i}}^{(x)} \otimes \Pi_{\text{o}}^{(x)}\}$, where $\Pi_{\text{i}}^{(x)} = \Pi_{\text{o}}^{(x)} := |x\rangle\langle x|$ satisfy $\text{tr}\left[\Pi^{(x)} \Pi^{(y)}\right] = \delta_{xy} \,\forall\, x, y$. Clearly, as this is the only interrogating instrument allowed here, $\mathbb{P}_{FMH}(x_F, x_M, x_H | \mathcal{J}_F^{\text{cl}}, \mathcal{J}_M^{\text{cl}}, \mathcal{J}_H^{\text{cl}}) = \mathbb{P}_{FMH}(x_F, x_M, x_H)$ by definition. Quantum Markov order therefore automatically reduces to the classical statement, since the latter stipulates conditional independence of statistics measured with the instrument above.

This can be readily seen by considering that the complete description of any stochastic process arising from classical physics, *i.e.,* its joint probability distribution, can be encoded in the diagonal of a process tensor written with respect to the local product basis that the classical measurements act in. Thus, the process tensor of a classical stochastic process has the structure $\Upsilon_{FMH}^{\text{cl}} = \sum_y \mathbb{P}_{FMH}(y_F, y_M, y_H) \, \Pi_{F^{\text{i}}}^{(y_F)} \otimes \Pi_{M^{\text{i}}}^{(y_M)} \otimes \Pi_{H^{\text{i}}}^{(y_H)}$. Importantly, these block projectors are tensor products of local projectors. Calculating the probabilities for the classical measurement above gives the correct statistics.

### 4.2.1 *Classical Stochastic Processes with Interventions*

As noted by van Kampen, "a physical process... may or may not be Markovian, depending on the variables used to describe it" [26]; the same is true for the Markov order. The existence of perceived memory effects implicitly depends on our experimental abilities, both in quantum mechanics—where it is generally acknowledged—as well as in classical physics—where it is often forgotten. Indeed, the standard framework for studying classical stochastic processes assumes the ability to measure observations of the system sharply and it breaks down when one allows for fuzzy measurements or interventions more generally [15, 17, 18].

For instance, consider a scenario in which an experimenter is able to sharply measure realisations of a process, deducing the 'true' statistics of its description that satisfy the Markov order-$\ell$ condition, *i.e.*, for all timesteps $t_k \in \{t_{\ell+1}, \ldots, t_n\}$, the following holds

$$\mathbb{P}_k(x_k|x_{k-1}, \ldots, x_1) = \mathbb{P}_k(x_k|x_{k-1} \ldots, x_{k-\ell}). \tag{4.7}$$

Suppose now that, instead of measuring the observations $x$, the experimenter is limited in resolution to only measuring some values $y$ that coarse-grain over subsets of the possible $x$ values. The conditional statistics of the observed outcomes $y$ can be explicitly written in terms of the fine-grained variable $x$ as

$$\begin{aligned}\mathbb{P}_k(y_k|y_{k-1}, \ldots, y_1) &= \frac{\mathbb{P}_{k:1}(y_k, \ldots, y_1)}{\mathbb{P}_{k-1:1}(y_{k-1}, \ldots, y_1)} \\ &= \frac{\sum \mathbb{P}_{k:1}(x_k, \ldots, x_1)}{\sum \mathbb{P}_{k-1:1}(x_{k-1}, \ldots, x_1)} \neq \mathbb{P}_k(y_k|y_{k-1} \ldots, y_{k-\ell}),\end{aligned} \tag{4.8}$$

where the summation runs over the $x$ values that are lumped together for each $y$.

Even if the process displays finite Markov order with respect to the sharply observed events, it does not necessarily do so for their coarse-grained variants. The fact that coarse-graining can increase the memory length observed by an experimenter arises from the well-known property that the space of Markovian processes is not convex [5]. Interestingly, we can also have the opposite occur. A process can exhibit finite Markov order with respect to a fuzzy measurement sequence, but, given access to the system at a finer resolution, the experimenter would attribute a longer memory length to the process. Such a process has finite-length memory on average. To highlight this concretely, explicit examples of both situations are provided in Appendix C.3.

Indeed, when one allows for noisy interventions in the classical case, the product structure of Eq. (2.19) is no longer satisfied for each observation, even when the underlying process is Markovian, in the sense that additional demands must be satisfied to render the future independent from the past with noisy measurements [152].[2] This point raises significant concerns for the practical reconstruction of complex classical dynamics, since an experimenter cannot always be certain that their measurements are distinguishing outcomes to a sufficient level of granularity [155, 156]. Various ways of dealing with such measurement noise in realistic experimental scenarios have been proposed and analysed throughout the literature [147–153].

Formally, a measurement that can pick out a particular physical state with certainty is called *sharp*; one that is not so is called *fuzzy*. In what follows, we slightly generalise the definitions provided in Ref. [157] in order to account for the input and output Hilbert spaces associated to each timestep. In the classical setting, a sharp measurement corresponds to a set of rank-1 projectors that are pairwise orthogonal: $\mathcal{J}^{\text{Sharp}} = \{|y\rangle\langle y|_\text{i} \otimes |x\rangle\langle x|_\text{o}\}$ such that $\langle yx|zw\rangle_\text{io} = \delta_{yz}\delta_{xw} \ \forall \ x, y, w, z$ and $\sum_{yx} |y\rangle\langle y|_\text{i} \otimes |x\rangle\langle x|_\text{o}$ is a stochastic map. Intuitively, this means that if the state is measured to correspond to the value $y$ and the state of the system that is actually sent forward into the process corresponds to $x$, both of these are distinguishable with certainty. Fuzzy classical measurements $\mathcal{J}^{\text{Fuzzy}}$ correspond to instruments made up of higher-rank projectors, as is the case in the example above, where the aggregation of some of the granular $x$ values lead to higher-rank projectors describing the measured $y$ values, with non-zero overlap between the state measured and that sent forward. Similarly, sharp quantum measurements can be defined in terms of a set of pairwise orthogonal rank-1 projectors such that $\sum_{yx} |y\rangle\langle y|_\text{i} \otimes |x\rangle\langle x|_\text{o}$ is CPTP. However, fuzziness arises in quantum theory in two possible ways: firstly, through subjective ignorance that is made manifest in the same way as classical fuzziness, *i. e.,* through higher-rank projectors with non-zero overlap; and secondly through the non-commutativity of measurement operators, which means that even sets of rank-1 POVM elements can give rise to fuzziness, since they are not necessarily projective.

---

2 Classical processes for which the property of Markovianity remains invariant with respect to aggregations of events have been studied formally under the guise of *lumpable* Markov chains and the conditions for satisfaction of lumpability clearly laid out in terms of the stochastic maps of the process [154].

More generally than the case where some outcomes are coarse-grained over, memory effects must be understood with respect to probing instruments whenever experimental interventions that directly influence the state of the system are allowed. Such invasive operations lie at the core of the theory of classical causal modelling [17] (which contains standard classical stochastic processes as a special case [15]). Here, at each timestep an experimenter is permitted to implement transformations that map any probability distributions in the state space to any other—just like in the quantum case, the experimenter could perform non-TP maps. Due to the possibility of different choices of instruments, here too the standard Markov condition must be generalised to the causal Markov condition [17]. This, in turn, is a special case of quantum Markovianity [18].

In any operational scenario where active interventions are considered, unambiguous definitions of memory effects must inherently be made with respect to the instruments used to probe the process at hand. To this end, even in classical physics, we should say that if a classical process is considered to have Markov order-$\ell$, it is *with respect to sharp observations of the process.* The generalisation of Markov order provided in Def. 4.1 reduces to unambiguously characterise memory length in any scenario where classical processes with fuzzy measurements and/or experimental interventions are allowed. In an operational sense, it is more intuitive to think of finite memory length of a process as presupposing the ability for an experimenter to apply a sequence of instruments that serve to block the influence of history on the future statistics. Moreover, the framework of quantum stochastic processes contains within it all generalised classical stochastic processes, including those pertaining to scenarios where arbitrary interventions are allowed such as causal modelling [15, 18]. Thus, statements made in terms of the process tensor reduce naturally to their classical counterparts in the appropriate sense.

However, even in the most general classical setting of causal modelling, this instrument-dependence of Markov order is liftable, in the sense that it can be removed by changing perspective. By incorporating the experimenter *and* their choice of intervention into the description of the process, the standard definitions of Markov order apply on a higher level [16]. In other words, the above concern can be 'explained away' on grounds of principle. On the other hand, in the study of quantum stochastic processes, even sharp quantum measurements *look* fuzzy when they act on a state in general; the measurement-dependence issue is fundamentally unavoidable and must be acknowledged accordingly through an instrument-specific notion of Markov order as per Def. 4.1. We now further explore some of its consequences in terms of a dilated dynamics, following an example presented in Ref. [2], in order to build some intuition.

## 4.3 MEMORY LENGTH OF A GENERALISED COLLISION MODEL

Within the framework of open quantum dynamics, collision models have been introduced to great pedagogical effect to provide a concrete underlying mechanism describing memoryless processes [158–162]. In such models, a system interacts successively with an environment comprising independent ancillary subsystems through successive unitary 'collisions' with each ancilla. Because each ancilla is only interacted with once, there is no way for the environment to act as a mechanism for memory transport by influencing future dynamics. Since no physical model need be prescribed to the framework, it provides a flexible toy model that is applicable to studying a wide range of phenomena.

This setting can be generalised to allow for non-trivial memory effects: the most common approaches include beginning with an initially correlated environment [163, 164], allowing for ancilla-ancilla interactions [165–169], allowing for repeated system-ancilla collisions [170, 171], or some type of hybrid approach [79, 80, 172, 173]. Each of these scenarios can be motivated through realistic physical origins that demand some reasonable assumptions [174]; in any case, the environment acts as a memory by storing information about previous system states to govern future evolution (see Fig. 4.2 for illustration). Here, we focus on a special case of such dynamics with repeated system-ancilla interactions, which has application in studying phenomena with substantial time-delays between repeated interactions, *e. g.,* developing feedback-assisted control protocols [170, 171].

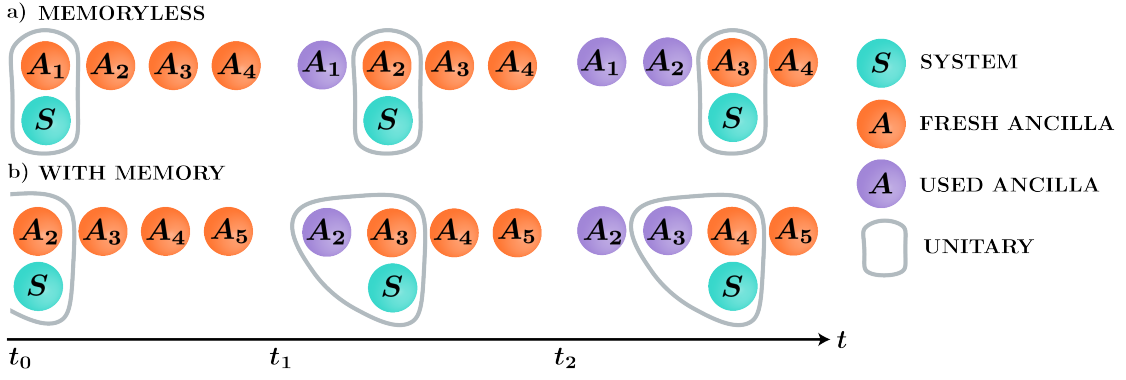

**Figure 4.2:** *Generalised collision model with memory.* The top panel **a)** depicts a standard memoryless collision model. The system $S$ (green) interacts unitarily at each timestep once with each of a number of uncorrelated, fresh ancillary states $A_j$ that constitute the environment (orange); the collision is represented by the grey boundary. Following the dynamics from $t_0$ to $t_1$, the $A_1$ ancilla has been used and so stores information about the initial state of the system, indicated by the purple colour (see $A_1$ after $t_1$). However, each successive portion of evolution proceeds through an interaction with a fresh ancilla that has not yet interacted with the system. Thus, any memory of the system's history cannot influence the future evolution, leading to Markovian dynamics. The bottom panel **b)** shows a generalised collision model, where the system is allowed to interact with multiple ancillas during each period of evolution. Here, between $t_0$ and $t_1$, the system interacts with both $A_1$ (not shown) and $A_2$, meaning that these ancillas can store information about the initial system state. The next portion of dynamics following $t_1$ involves $A_2$ again; thus, the future dynamics depend on the history. In this way, the ancillas serve to propagate memory effects through the process.

Consider specifically the following $n$-step process, depicted as a dilated quantum circuit in Fig. 4.3. A system, $S$, interacts with some inaccessible environment, $E$, which comprises $n + \ell - 1$ initially uncorrelated ancillary systems $\tau_0^E := \bigotimes_{x=1}^{n+\ell-1} \tau^{A_x}$. For simplicity, here we assume no initial system-environment correlations, and so we have a timestep $t_0$ at which an arbitrary system state can be prepared. The overall joint system-environment dynamics between timesteps $t_{j-1}$ and $t_j$ is represented by the map: $\rho_j^{SE} = \widetilde{\mathcal{U}}_{j:j-1}(\rho_{j-1}^{SE})$. In this particular example, the joint evolution is broken up into an ordered sequence of pairwise collisions between the system and ancillary states of the environment as follows

$$\widetilde{\mathcal{U}}_{j:j-1} := \mathcal{U}_{j:j-1}^{SA_j} \ldots \mathcal{U}_{j:j-1}^{SA_{j+\ell-1}},  \tag{4.9}$$

where the superscripts label the subsystems involved in the interaction. Following the dynamics between timesteps $t_{j-\ell}$ and $t_j$, the specific ancilla $A_j$ will have interacted with the system $\ell$ times; it is then discarded and never again involved in the system's evolution. Note that in this model, we have not allowed for any initial system-environment correlations or ancilla-ancilla collisions; the type of evolution proposed here describes a time-translationally invariant microscopic model for processes with memory, which propagates through the $\ell$ ancillas that feed-forward to act like a linear memory tape. By design, we can see how memory effects arise: each ancillary system $A_x$ can store information about the system, acquired during its first interaction through $\mathcal{U}_{x-\ell+1:x-\ell}^{SA_x}$, and use it to influence the future dynamics up until its final interaction with the system mediated through $\mathcal{U}_{x:x-1}^{SA_x}$.

Suppose then that an experimenter wishes to characterise the memory length of such a process. To do so, they must measure realisations of the state of the system at each timestep, immediately facing the problem that any such measurement both conditions the state of the environment and directly affects the state of the system which leads to different future dynamics dependent on both the measurement outcomes observed and the way in which they were measured. As we have discussed, the appropriate question relevant to understanding the memory length of the process is: *how can the experimenter block the effect of the history on the future dynamics over a finite number of timesteps?*

The representation of the process in Fig. 4.3 is particularly illuminating. We can see the possible ways in which information originating from the history, *i.e.*, about the initial system state, can perpetuate forward in time along connected paths (traced in red). For the particular collision model described above, an obvious history-blocking strategy involves discarding the system state emitted by the process and re-preparing one of a known set of states to feed into the process over a sequence of $\ell$ timesteps. It is clear that upon applying such a sequence of *trash-and-prepare* instruments, any possible path connecting the history to the future across $\ell$ timesteps is broken, thereby guaranteeing that the future evolution of the system is independent of anything that happened to it prior to the trash-and-prepare sequence.

In Appendix C.4, we prove that this trash-and-prepare protocol indeed blocks any possible influence that the history can have on the future evolution. Specifically, we show that for this particular process, at arbitrary time $t_k$, all future states of the sys-tem after application of any length-$\ell$ sequence of trash-and-prepare instruments can be uniquely described as a function of only the $\ell$ most recently prepared states, for any prior history. That is, the process has Markov order $\ell$ with respect to the entire family of trash-and-prepare sequences, implying that any possible statistics an experimenter might observe in the history and future are conditionally independent given this particular experimental control. Explicitly expressing the length-$\ell$ trash-and-prepare sequence in terms of operations on the system as

$$\mathcal{D}_{k-1:k-\ell}(\rho_{k-1}^S, \ldots, \rho_{k-\ell}^S) := \sigma_{k-1}^S \mathrm{tr}\left[\rho_{k-1}^S\right] \ldots \sigma_{k-\ell}^S \mathrm{tr}\left[\rho_{k-\ell}^S\right]. \tag{4.10}$$

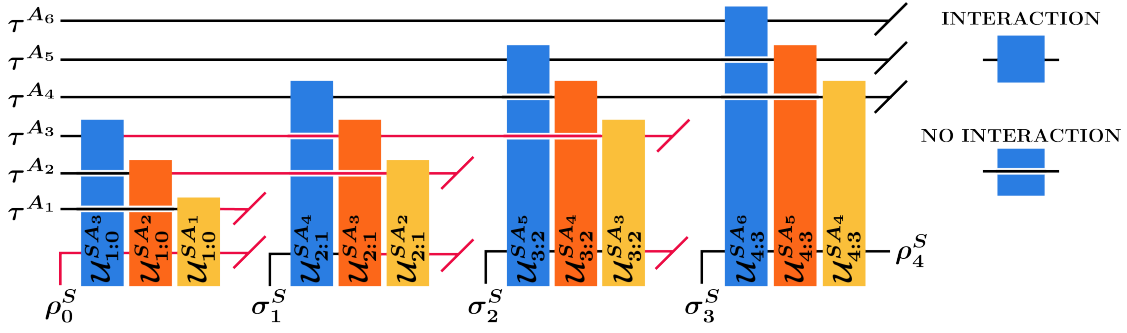

**Figure 4.3:** *Finite memory with respect to trash-and-prepare protocol.* The system-environment dynamics for the generalised collision model described, interspersed with the trash-and-prepare protocol applied to the system. Any possible influence stemming from the history persists to impact the future for at most $\ell = 3$ timesteps before being trashed. For instance, the red paths depicted signify the degrees of freedom that can be affected by the initial preparation, whereas the black ones cannot be. The final state is a function of only the most recent $\ell$ preparations, $\{\sigma_1^S, \sigma_2^S, \sigma_3^S\}$, and entirely independent of the initial system state, $\rho_0^S$. Any other instrument sequence on the system, *e.g.*, a measure-and-prepare rather than a trash-and-prepare instrument, would 'open up' a pathway for the initial state $\rho_0^S$ to influence the future state $\rho_4^S$.

In a slight abuse of notation, we can write

$$I(\{t_n, \ldots, t_k\} : \{t_{k-\ell-1}, \ldots, t_1\})_{\mathcal{D}_{k-1:k-\ell}} = 0. \tag{4.11}$$

By this we mean that the mutual information between any possible statistics recorded on the future and history timesteps, which quantifies any possible correlation between them, vanishes for all length-$\ell$ trash-and-prepare sequences $\mathcal{D}_{k-1:k-\ell}$.

Moreover, the experimenter could discard the states emitted by the process and feed in probabilistically-prepared states, *i.e.*, choose $\sigma_j^S$ at random from the ensemble $\{\sigma_j^{(x_j)}\}$ with corresponding probabilities $\{p^{(x_j)}\}$ and the effect of history would still be blocked deterministically overall. In other words, the process also has finite memory with respect to the trash-and-probabilistically-prepare family of instruments $\mathcal{J}_{k-1:k-\ell} = \{\mathcal{D}_{k-1:k-\ell}^{(x_{k-1}:x_{k-\ell})}\}$ with

$$\mathcal{D}_{k-1:k-\ell}^{(x_{k-1}:x_{k-\ell})}(\rho_{k-1}^S, \ldots, \rho_{k-\ell}^S) := \quad (4.12)$$
$$p^{(x_{k-1})} \ldots p^{(x_{k-\ell})} \sigma_{k-1}^{(x_{k-1})} \text{tr}\left[\rho_{k-1}^S\right] \ldots \sigma_{k-\ell}^{(x_{k-\ell})} \text{tr}\left[\rho_{k-\ell}^S\right].$$

More generally still, the experimenter could even choose to feed in subsystems of an $\ell$-partite entangled state at each timestep sampled from some ensemble $\{\sigma_{k-1:k-\ell}^{(x_{k-1:k-\ell})}\}$ with probabilities $\{p^{(x_{k-1:k-\ell})}\}$, overall implementing an multi-time (quantum) correlated instrument sequence of the form $\mathcal{J}_{k-1:k-\ell} = \{\mathcal{E}_{k-1:k-\ell}^{(x_{k-1:k-\ell})}\}$ with

$$\mathcal{E}_{k-1:k-\ell}^{(x_{k-1:k-\ell})} := p^{(x_{k-1:k-\ell})} \sigma_{k-1:k-\ell}^{(x_{k-1:k-\ell})} \text{tr}\left[\rho_{k-1}^S\right] \ldots \text{tr}\left[\rho_{k-\ell}^S\right]. \quad (4.13)$$

In either case, for any realisation of the instrument sequences defined as the collection of CP operations in Eqs. (4.12) and (4.13), the future dynamics is conditionally independent of the history and we have satisfaction of Eq. (4.5) for all families of instruments in question

$$I(\{t_n, \ldots, t_k\} : \{t_{k-\ell-1}, \ldots, t_1\})_{x_{k-1:k-\ell}} = 0. \quad (4.14)$$

It is clear that the generalised collision model considered implies finite Markov order with respect to any of the aforementioned length-$\ell$ generalised trash-and-prepare sequences, but what can we say in the converse direction? That is, *does every process that is of finite Markov order with respect to trash-and-prepare instruments have a dilation as the one depicted in Fig. 4.3*? It turns out that having finite-length memory with respect to the trash-and-prepare protocol is a necessary but insufficient condition to deduce this generalised collision model dilation.

As a counterexample, consider two timesteps of dynamics in which two ancillary states of the environment are initially entangled, represented by the density operator $\tau^{A_1 A_2}$, and in product with the initial system state $\rho_0^S$. The system first interacts with $A_1$ via $\mathcal{U}_{1:0}^{SA_1}$, before $A_1$ is discarded, and then with $A_2$ via $\mathcal{U}_{2:1}^{SA_2}$, with a trash-and-prepare instrument $\sigma_1^S \text{tr}_S$ applied to the system in between. It is clear that the initial state $\rho_0^S$ can have no influence on the future evolution, since the final system state can be written uniquely as a map acting only on the preparation fed into the process: $\rho_2^S = \text{tr}_{A_2}\left[\mathcal{U}_{2:1}^{SA_2} \sigma_1^S \otimes \tilde{\tau}^{A_2}\right]$, where $\tilde{\tau}^{A_2} := \text{tr}_{SA_1}\left[\mathcal{U}_{1:0}^{SA_1} \rho_0^S \otimes \tau^{A_1 A_2}\right] = \text{tr}_{A_1}\left[\tau^{A_1 A_2}\right]$ represents the reduced state of $A_2$ which, importantly, shows no memory of $\rho_0^S$. Therefore, the dynamics has finite Markov order $\ell = 1$ with respect to the trash-and-prepare protocol, but evidently does not have a dilation of the form depicted in Fig. 4.3; namely, because the ancillas begin in an entangled state. In general, even if $A_1$ and $A_2$ interact with some unitary $\mathcal{V}_{1:0}^{A_2 A_1}$ *after* $\mathcal{U}_{1:0}^{SA_1}$, there is no dilation with initially correlated ancillas that can capture temporal correlations that might arise if some portion of the later dynamics is conditioned on the state of $A_2$ *before* that interaction.

To summarise, in this section we have introduced a specific type of generalised collision model which, by construction, perpetuates information about the history via a particularly simple mechanism. This allows us to study explicitly how memory effects arise from the perspective of the underlying dynamics and build an intuitive understanding of the necessity for instrument-specific quantum Markov order. The salient points to note are as follows.

**i)** The trash-and-prepare protocol does not block every type of memory. For arbitrary system-environment dynamics, following a length-$\ell$ trash-and-prepare sequence, $\rho_k^S$ (and the future process more generally) will, generically, depend on both the known preparations $\{\sigma_{k-1}^S, \ldots, \sigma_{k-\ell}^S\}$ *and* the previous historic states $\{\rho_{k-\ell-1}^S, \ldots, \rho_0^S\}$. Thus, if an experimenter were to measure statistics on the future and history, they would be correlated, leading to a breakdown of Eq. (4.11) and, hence, an appreciable memory effect. In the coming chapter, we provide various examples of processes that exhibit finite Markov order with respect to other sequences of instruments, but not this one.

**ii)** Even for the special case of dynamics described above, application of a different sequence of instruments than the trash-and-prepare protocol would not lead to future dynamics that are independent of the history. For example, suppose that the experimenter were to perform a measurement at an intermediary timestep during a length-$\ell$ trash-and-prepare protocol. Here, the measurement would condition the state of the environment on its outcome, and hence the influence of the history could permeate through the memory block, leading to dependence of the final output on previous dynamics. Lastly, in Appendix C.5, we further explore some of the various other types of memory that can be built into collision models.

From the considerations outlined above, it is clear that knowing the history-blocking sequence for a given process gives us information about the process at hand, but not necessarily all of it. Although we have made no assumptions on the action of the unitaries, the dynamics examined here is a special case of generic quantum evolution and the trash-and-prepare protocol is just one of many possible sequences of instruments an experimenter might apply.

## 4.4 CHAPTER SUMMARY

In this chapter we have used the process tensor framework to formally provide an extension of the notion of Markov order to the quantum realm that reduces to the classical condition in the appropriate case. The intuition behind quantum Markov order remains unchanged—as in the classical case, the question boils down to: *can the future statistical evolution of the system be deduced completely, in principle, from the outcomes of the most recent $\ell$ instruments applied?* When any future evolution of the system is independent of any previous history following the application of some instrument sequence, the process exhibits conditional independence between the future and history. Importantly, whilst the instrument on the memory block must be specified to meaningfully define quantum Markov order, the historic and future instruments remain arbitrary: for each realisation of the instrument sequence in question applied to the memory, any possible statistics deducible on the history and future timesteps are guaranteed to be conditionally independent.

In quantum theory, we have no choice but to allow for such an active description of processes, dictating the necessity for an instrument-specific definition of memory length. We saw that demanding conditional independence between the history and the future for all possible instruments on the memory is too strong a restriction: no non-Markovian quantum process can display finite Markov order with respect to all possible interventions. Put differently, quantum stochastic processes exhibit different memory effects when probed with different instruments. Interestingly, such instrument-specific memory effects have been observed previously (see, *e. g.,* Ref. [175]); our characterisation formally captures and explains such behaviour.

This is also the case for classical stochastic processes where an experimenter can actively intervene with the system. Here, however, the issue is liftable in the sense that, in theory, we can always assume the ability to measure sharply. This is not the case in quantum mechanics: even sharp quantum measurements appear noisy as they do not generally reveal the full state of the system, and thus we have no choice but to account for the probing instruments employed. As luck would have it, by treading the path we are forced to take to understand memory effects in quantum processes, we develop a deeper comprehension of their classical counterparts.

To build intuition regarding quantum Markov order, we concluded the chapter by studying the memory length of a generalised collision model with repeated system-ancilla interactions and showed how the dynamics displays finite Markov order with respect to a seemingly natural information-trashing instrument sequence, which serves to average out the effects that system-level operations have on the environment. The deeper exploration of memory effects in similar models in Appendix C.5 further motivates the necessity of instrument-specific Markov order for quantum processes and a better understanding of the microscopic mechanism for memory propagation.

The example analysed here provides but a special case of a quantum process with finite-length memory. There are a rich arsenal of instruments that an experimenter could, in principle, choose to apply in an attempt to block the effect of history on the future; indeed, quantum theory permits a rich landscape of memory effects, with many properties that distinguish it from the classical setting. Extending this line of investigation, we are now interested in what the satisfaction of Eq. (4.4) for a particular instrument sequence implies for the structure of the underlying process tensor. In the following chapter, we will explore the structure of quantum processes with finite Markov order with respect to certain choices of instruments, shedding light on distinguishing features and the relation between memory length, vanishing quantum CMI and the recoverability of the process. In particular, we ask which kinds of processes can have finite-length memory, and what can be inferred about the underlying process through knowledge of the history-blocking sequence.

# 5

# PROCESSES WITH FINITE MEMORY LENGTH

IN THIS CHAPTER WE EXPLORE THE MATHEMATICAL STRUCTURE of processes with finite quantum Markov order. As was the case for the generalised collision model explored in Section 4.3, knowledge of the history-blocking sequence does not suffice to pin down the process at hand: in general, there can exist potential memory effects that are 'hidden' to the instruments in question but could be uncovered with a different probing scheme. Presently, we will examine the degree to which the structure of a process is constrained by having the property of finite Markov order.

We first detail the most general description of a process satisfying Eq. (4.4) for a particular instrument sequence. Although certain structural constraints are imposed, knowing that a particular instrument blocks the history does not necessarily tell us much about the effect of other instruments. For each key finding that arises, we present an associated example processes to build intuition regarding memory length in quantum processes. These examples are constructed in such a way as to highlight some key peculiarities of quantum Markov order, and their essence applies to processes more broadly.

With this structural understanding at hand, we will go on to analyse the connection between processes with finite Markov order and those with vanishing quantum CMI between the history and the future given the memory. From the structure of processes with vanishing quantum CMI, it follows that there exists an instrument sequence made up of orthogonal projectors that serve to render the history and future conditionally independent. Perhaps surprisingly, the converse does not hold, even in the case where the history-blocking sequence comprises only sharp, orthogonal projectors. This situation cannot happen for classical stochastic processes where, as we have discussed, finite Markov order and vanishing CMI are equivalent. We go on to lay out further restrictions imposed on the history-blocking instruments that ensures the quantum CMI vanishes.

The connection between the constraints placed on processes with finite Markov order and our desire to describe and reconstruct those with *approximately* finite memory efficiently naturally brings us to the main concern of Chapter 6: *how can we meaningfully quantify the strength of memory and what are the subsequent implications for compression and recoverability?* Most of the results of this chapter stem from Ref. [2].

## 5.1 STRUCTURE OF QUANTUM PROCESSES WITH FINITE MARKOV ORDER

Our structural analysis is based on the fact that the process tensor is multi-linear in its arguments. Any $\ell$-step operation sequence it acts upon can be considered as an element of a vector space $\mathbf{W}' := \mathsf{BL}(\mathcal{H}_{k-1} \otimes \ldots \otimes \mathcal{H}_{k-\ell})$ of dimension $d_{W'} := \dim(\mathbf{W}') = d^{4\ell}$, where $d$ is the dimension of the quantum system of interest[1]. As already mentioned, the only constraint on a set of operations that constitute a valid instrument sequence is that they sum to a proper comb, *i.e.*, they are positive operators whose sum yields an operator with the same causal ordering as the process tensor that acts on them, which is enough to guarantee their physicality [31]. In general, the CP elements constituting an instrument sequence need not span the entire space $\mathbf{W}'$, even though they are linearly independent. An instrument sequence that does span $\mathbf{W}$ is called *informationally-complete* (IC), and any such instrument must contain a minimum number of $d_{W'}$ linearly independent elements. On the other hand, an instrument sequence that does not entirely span $\mathbf{W}'$ is referred to as *informationally-incomplete*.

Informational-completeness and history-blocking are two distinct properties of an instrument sequence. In particular, an informationally-incomplete instrument sequence can block the history, *e.g.*, the trash-and-prepare sequence in Section 4.3. Informational-completeness pertains to whether or not an experimenter can completely characterise the process on the corresponding timesteps through knowledge of its action on each element. In the same way that an IC set of measurements must be performed to completely determine a quantum state, a process tensor can be uniquely tomographically reconstructed through knowledge of its action on an IC set of operations [51]. This property is of importance in this section, which identifies structure in the process tensor given that an experimenter knows that a certain instrument sequence blocks the history.

We focus first on the most general case, where one has satisfaction of Eq. (4.4) for an arbitrary instrument sequence. Suppose we have an informationally-*incomplete* history-blocking sequence $\mathcal{J}_M = \{\mathsf{O}_M^{(x)}\}_{x=1}^c$ where $c < d_{\mathbf{W}'}$. We can *complete* such an instrument sequence to span the entire space $\mathbf{W}'$ by appending an additional collection of linearly independent operators, *i.e.*, construct the IC set $\mathcal{A}'_M = \mathcal{J}_M \cup \underline{\mathcal{A}}_M := \{\{\mathsf{O}_M^{(x)}\}_{x=1}^c, \{\underline{\mathsf{O}}_M^{(y)}\}_{y=c+1}^{d_{\mathbf{W}'}}\} = \{\mathsf{O}_M^{\prime(z)}\}_{z=1}^{d_{\mathbf{W}'}}$, where the underline signifies the objects that *are not* part of the original history-blocking instrument (we adhere to this notation throughout this chapter to be explicit). Note that the appended operators $\underline{\mathcal{A}}_M$ are not necessarily CP, nor do they necessarily form an instrument sequence; thus nor is the case for the overall construction $\mathcal{A}'_M$. All that is required is that $\underline{\mathcal{A}}_M$ is chosen as a linearly independent set spanning $\mathbf{W}^\perp$, so that $\mathcal{A}'_M$ forms a basis for the entire space $\mathbf{W}'$ on which the process tensor is defined. Since the entire collection $\mathcal{A}'_M$ forms a linearly independent set (by construction), there exists an associated dual set of objects $\{\Delta_M^{\prime(w)}\}$ such that $\operatorname{tr}\left[\mathsf{O}_M^{\prime(z)} \Delta_M^{\prime(w)\dagger}\right] = \delta_{zw}\ \forall\ z,w$ (see Appendix B.2). In terms of this (generally non-orthonormal) basis, we can (completely) represent any process tensor as

$$\Upsilon_{FMH} = \sum_z^{d_{\mathbf{W}'}} \widetilde{\Upsilon}_{FH}^{\prime(z)} \otimes \Delta_M^{\prime(z)*}. \tag{5.1}$$

---

1 Here, for simplicity, we consider a system of fixed dimension across all timesteps; the extension to the more general case where the dimension of the system varies at each timestep is straightforward. Note further that, in the case where the number of Hilbert spaces in the memory block is odd, $d_{W'} = d^{4\ell-2}$. Although we are not explicit regarding either of these points, their consideration only impacts the dimensionality of the underlying vector space and has no relevance to the results to be presented.

Since the instrument sequence $\mathcal{J}_M$ acts to render the history and future independent for each outcome by hypothesis, we can further decompose the process tensor. We first partition the total dual set $\{\Delta_M^{\prime(z)}\}$ into the elements dual to those operations within the history-blocking sequence, $\{\Delta_M^{(x)}\}_{x=1}^c$, and the rest, $\{\underline{\Delta}_M^{(y)}\}_{y=c+1}^{d_{\mathbf{W}'}}$, such that $\operatorname{tr}\left[\mathsf{O}_M^{(a)} \Delta_M^{(b)\dagger}\right] = \operatorname{tr}\left[\underline{\mathsf{O}}_M^{(a)} \underline{\Delta}_M^{(b)\dagger}\right] = \delta_{ab}\ \forall\ a,b$ and $\operatorname{tr}\left[\mathsf{O}_M^{(a)} \underline{\Delta}_M^{(b)\dagger}\right] = \operatorname{tr}\left[\underline{\mathsf{O}}_M^{(a)} \Delta_M^{(b)\dagger}\right] = 0$. Now, the first $c$ terms in the sum in Eq. (5.1) are $\sum_x \Upsilon_F^{(x)} \otimes \Delta_M^{(x)*} \otimes \widetilde{\Upsilon}_H^{(x)}$. By direct insertion, it is clear that this portion of the process tensor indeed satisfies Eq. (4.4). The remaining terms, which are inaccessible to the history-blocking instrument sequence $\mathcal{J}_M$, can be written as: $\sum_y \widetilde{\underline{\Upsilon}}_{FH}^{(y)} \otimes \underline{\Delta}_M^{(y)*}$. These terms encapsulate future-history correlations that an experimenter might observe upon application of an alternative instrument. This leads to the following theorem, which outlines the most general structure a process with finite quantum Markov order must have.

**Theorem 5.1.** *Processes with finite quantum Markov order with respect to the instrument sequence $\mathcal{J}_M = \{\mathsf{O}_M^{(x)}\}$ are of the form:*

$$\Upsilon_{FMH} = \sum_{x=1}^c \Upsilon_F^{(x)} \otimes \Delta_M^{(x)*} \otimes \widetilde{\Upsilon}_H^{(x)} + \sum_{y=c+1}^{d_{\mathbf{W}'}} \widetilde{\underline{\Upsilon}}_{FH}^{(y)} \otimes \underline{\Delta}_M^{(y)*}, \tag{5.2}$$

*where $c = |\mathcal{J}_M|$ is the number of constituent operations of the history-blocking instrument sequence, $\{\Delta_M^{(x)}\}$ form the dual set to $\{\mathsf{O}_M^{(x)}\}$, satisfying $\operatorname{tr}\left[\mathsf{O}_M^{(x)} \Delta_M^{(y)\dagger}\right] = \delta_{xy}\ \forall\ x,y$, and $\{\underline{\Delta}_M^{(y)}\}$ satisfy $\operatorname{tr}\left[\mathsf{O}_M^{(x)} \underline{\Delta}_M^{(y)\dagger}\right] = 0\ \forall\ x,y$.*

Importantly, each term in the first summation has $F$ and $H$ in tensor product, ensuring Eq. (4.4) is satisfied with certainty for each realisation of the instrument sequence in question. Such a decomposition must hold true for every timestep $t_k$ at which a length-$\ell$ memory block ends (although the terms in it can change for different blocks). A quantum process with infinite Markov order with respect to every instrument cannot be written as per Eq. (5.2) with non-trivial terms in the first summation. The structure outlined makes it clear that, for an informationally-incomplete history-blocking sequence $\mathcal{J}_M$, the experimenter can only make meaningful statements about the memory length of the process with respect to the choice of instrument; the $\underline{\Upsilon}_{FH}^{(y)}$ in the second term represents the portion of the process that can only be revealed through other probing schemes.

The generalised collision model explored in Section 4.3 is an example of such a process, since the trash-and-prepare protocol that blocks the effect of history constitutes an informationally-incomplete instrument sequence. This instrument sequence is, by its very nature, incoherent: an experimenter simply discards whatever states are output by the process and feeds in some of their own choosing. In contrast to this, one might expect that applying sequences of coherent, *i. e.,* unitary, operations to a process would always perpetuate memory effects from the history to the future by way of transmission through the level of the system alone. We now provide an explicit counterexample to this claim, *i. e.,* a process whose history is only blocked upon application of a sequence of coherent operations.

### 5.1.1 *Unitary History-Blocking Instrument Sequences*

**Example 5.1** (Finite Markov order for a sequence of unitaries)**.** Consider the process depicted in Fig. 5.1. It is constructed such that there is exactly one length-$\ell$ sequence of unitary operations that guarantees the history is blocked, such that the Markov order of the process is equal to $\ell$. Between each timestep $t_{j-1}$ and $t_j$, the process prepares an ancillary subsystem, $A$, in a maximally entangled state with $S$, $\psi^{AS} = \frac{1}{d}\sum_{xy}|xx\rangle\langle yy|$, which are together in tensor product with the rest of the environment $E$.

The joint $EAS$ state undergoes dynamics according to some unitary map, $\mathcal{U}_j$, before an operation can be applied to the system $S$ by the experimenter. Following this operation, the process applies the inverse $\mathcal{V}_j^\dagger$ of some other unitary map $\mathcal{V}_j$ on the system alone, where $\mathcal{V}_j^\dagger(\bullet) := V_j^\dagger \bullet V_j$. The joint $EAS$ state then evolves according to the inverse unitary map $\mathcal{U}_j^\dagger$. Lastly, $AS$ is subject to the following *memory cutting protocol*: a Bell basis measurement is implemented within the process, with another ancillary subsystem, $C$, counting the measurement outcomes corresponding to $\psi^{AS}$. When $C$ reaches $\ell$, then the environment at that timestep is discarded and a fresh one prepared to govern the future dynamics, and the counter is reset. If the correct measurement outcome is not observed, the environment is left untouched and the counter is also reset.

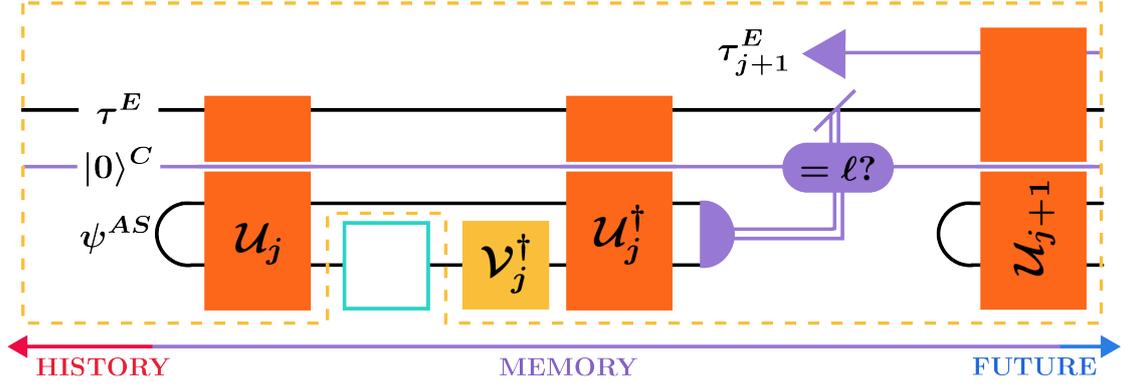

**Figure 5.1:** *Finite memory for a unitary instrument sequence.* A dilation of a single timestep for a process whose historic influence on the future is blocked only by the sequence of unitary operations on the system $\{\mathcal{V}_{k-\ell}, \ldots \mathcal{V}_{k-1}\}$. Everything inside the yellow, dashed boundary, including the unitary operation $\mathcal{V}_j^\dagger$, constitutes the inaccessible process; the experimenter only has the choice of operation applicable in the green box. The cutting protocol described in the main text is depicted here in purple: the ancillary counter, $C$, registers the number of successive successful Bell basis measurements on the $SA$ system, which is re-prepared as a maximally entangled pair, $\psi^{AS}$, at each timestep $t_j$. When the counter system state reaches $\ell$, the current environment state is discarded and a fresh one, $\tau_{j+1}^E$, is prepared to govern the future evolution. If the counter has not reached $\ell$, the environment is left to mediate correlations from the history to the future.

It is evident that only upon application of the entire uncorrelated unitary sequence $\{\mathcal{V}_{k-\ell}, \ldots, \mathcal{V}_{k-1}\}$ are the temporal correlations *guaranteed* to be broken and the history and future processes rendered conditionally independent. If, on the other hand, this correct unitary sequence is not applied, the environment is allowed to mediate correlations between system states of the history and future, breaking the quantum Markov order condition. For any other sequence of operations implemented, although there is a non-zero probability for the counter to reach $\ell$, this is not certain to happen; hence, overall, the influence of the history on the future is not blocked. In other words, unless the total unitary sequence is implemented by the experimenter, correlations between the history and the future can be deduced. This process is of the form of Eq. (5.2) with respect to the informationally-*incomplete* sequence of single-element unitary instruments, with the first sum containing a single term and the remainder of the process description encapsulated in the second term

$$\Upsilon_{FMH} = \frac{1}{d^\ell}\Upsilon'_F \otimes \mathsf{V}'_{k-1} \otimes \ldots \otimes \mathsf{V}'_{k-\ell} \otimes \Upsilon_H + \sum_y \widetilde{\Upsilon}^{(y)}_{FH} \otimes \underline{\Delta}^{(y)}_M, \qquad (5.3)$$

where the $\mathsf{V}'_j/d$ are duals to the Choi states of the unitary maps $\mathcal{V}^\dagger_j$, and the conditional process tensor $\Upsilon'_F$ is the fresh future process initiated by successful implementation of the cutting protocol, *i.e.*, the process that ensues in line with the freshly prepared environment state.

The process tensor in Eq. (5.3) is evidently an expression of Theorem 5.1; however, some remarks are in order. Firstly, note that even in the special case $\ell = 1$, the process is non-Markovian, since it does not have the product structure outlined in Eq. (3.36) (and the coherent unitary operation at timestep $t_{k-1}$ required to block the effect of history on the future operates on $\mathsf{BL}(\mathcal{H}_{k-1^\circ} \otimes \mathcal{H}_{k-1^\mathrm{i}}))$. Loosely speaking, the history-blocking unitary operation does not serve to 'cut' the system line, but rather keep it 'glued together' so that temporal correlations can be transmitted through the system, in contrast to the effect of a causal break. It is these transmitted correlations that are then used to cut the information flow from the history to the future. Secondly, no sequence of unitary operations can be IC; by definition, an informationally-incomplete sequence cannot be used to extract full information about a process. Although we know that any future dynamics will be independent of the history with respect to this sequence, we cannot predict what the next state will be as a function of the history-blocking sequence.

### 5.1.2 *Informationally-Complete History-Blocking Sequences*

Interestingly, in Example 5.1, the influence of the history on the future is blocked only by a sequence of coherent operations. This is somewhat counter-intuitive, as one might expect unitary transformations to perpetuate memory effects. Clearly, the general structural constraint of Theorem 5.1 is rather flexible, since knowledge of such an incomplete history-blocking instrument sequence does not determine the structure of the process at hand. In many cases of interest, an experimenter makes use of an IC set of operations to probe the dynamics, *e.g.*, when attempting to tomographically reconstruct a generic process. In this case, since an IC instrument sequence does in fact span the entire space of operations (by definition), there can be nowhere for potential memory effects correlating the history and future to hide. The memory block can then be completely decomposed onto an IC set of duals, uniquely specifying the entire process for each sequence of outcomes realised on the memory block. In this case, finding the future process to be conditionally independent of the history constrains the structure of the process tensor in a stricter manner than Eq. (5.2); we immediately have the following corollary.

**Corollary 5.2.** *A process with finite Markov order with respect to an informationally-complete instrument sequence must have the following structure:*

$$\Upsilon_{FMH} = \sum_x \Upsilon_F^{(x)} \otimes \Delta_M^{(x)*} \otimes \widetilde{\Upsilon}_H^{(x)}. \tag{5.4}$$

*Note that this structure by no means implies that the process tensor is of a tensor product form, or that the history is necessarily blocked by any instrument sequence other than the IC one in question.*

With this corollary at hand, it is enlightening to re-examine Theorem 4.2, which states that the only processes with finite quantum Markov order with respect to all instrument sequences are memoryless. Its proof begins by demanding Eq. (4.4) to hold for all possible instruments. As such, we can consider an IC instrument sequence, in which case the process tensor must be of the form given by Eq. (5.4). Then, using the fact that one can construct arbitrary operation sequences spanning the space of operations on $M$, we can vary $\Delta_M^{(x)}$ freely. Demanding the structure of Eq. (5.4) to remain intact for arbitrary outcomes forces a tensor product between $M$ and $F$ or $H$ (or both), meaning the process tensor is restricted to a single term in Eq. (5.4), *i.e.*, it is of product form. Requiring this to hold for any timestep leads to a memoryless process (either the Markovian or super-Markovian product structure of Eqs. (3.36) and (4.6) respectively).

An operationally motivated choice for an IC instrument sequence consists of applying a causal break at each timestep: recall that each operation here consists of an IC POVM followed by an independent preparation of one of an IC set of states to feed forward at each timestep (see Eq. (3.32)). The following example is constructed in such a way that the process is non-Markovian; however, it exhibits finite Markov order with respect to such an IC instrument sequence of causal breaks.

**Example 5.2** (Finite Markov order with respect to an informationally-complete instrument sequence (causal breaks))**.** Consider the process depicted in Fig. 5.2, where, for simplicity, we present the case $\ell = 2$ for a 3-step process, with the extension to longer length memory immediate. Initially, the following tripartite state is prepared

$$\rho_{Y2^i 1^i} = \sum_y p_y \rho_Y^{(y)} \otimes \Delta_{2^i}^{(y)*} \otimes \rho_{1^i}^{(y)}, \tag{5.5}$$

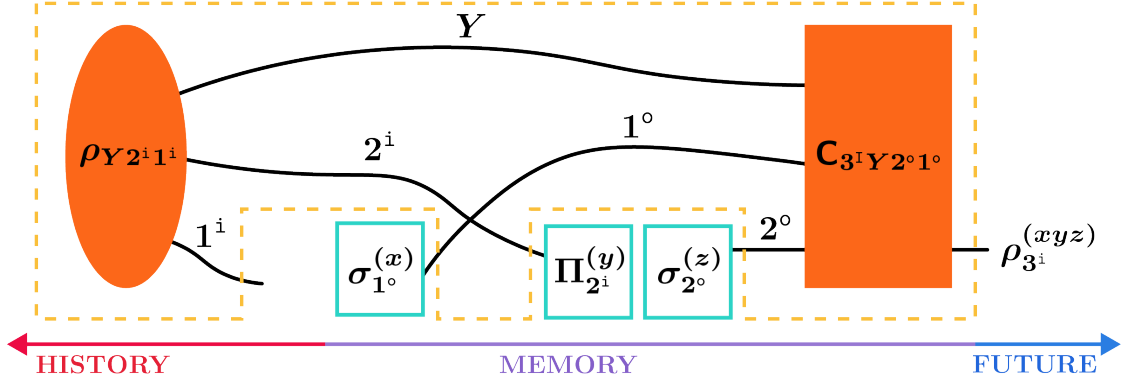

**Figure 5.2:** *Finite memory for an informationally-complete sequence.* Initially, a tripartite state $\rho_{Y2^{\mathrm{i}}1^{\mathrm{i}}}$ is constructed as per Eq. (5.5), with subsystems $1^{\mathrm{i}}, 2^{\mathrm{i}}$ of it fed out at consecutive timesteps as described in the text. The states fed back into the process on spaces $1^{\mathrm{o}}, 2^{\mathrm{o}}$ are fed forward as inputs to the CPTP map $\mathsf{C}_{3^{\mathrm{i}}Y2^{\mathrm{o}}1^{\mathrm{o}}}$ defined in Eq. (5.6). Upon applying any combination of the correct IC causal break sequence $\{\sigma^{(x)}_{1^{\mathrm{o}}}, \Pi^{(y)}_{2^{\mathrm{i}}}, \sigma^{(z)}_{2^{\mathrm{o}}}\}$, one of $d^6$ final output states $\rho^{(xyz)}_{3^{\mathrm{i}}}$ are output by the process in the future, each of which is conditionally independent of the historic $\rho^{(y)}_{1^{\mathrm{i}}}$. If any other operations are applied, correlations can arise between the history and future.

with $\{\Delta^{(y)}_{2^{\mathrm{i}}}\}$ forming the dual set to some IC POVM $\{\Pi^{(y)}_{2^{\mathrm{i}}}\}$ and $Y$ labelling an ancillary Hilbert space of the environment that is never accessible to the experimenter. The $1^{\mathrm{i}}$ part of the initial state is fed out of the process at the first timestep, at which point the experimenter can implement any operation of their choice; similarly, the $2^{\mathrm{i}}$ part is fed out at the second timestep. The output states of the experimenter's operations at timesteps $1^{\mathrm{o}}$ and $2^{\mathrm{o}}$ are mediated forward by the process, along with the $Y$ part of $\rho_{Y2^{\mathrm{i}}1^{\mathrm{i}}}$, as inputs to a CPTP map $\mathcal{C} : \mathsf{BL}(\mathcal{H}_Y \otimes \mathcal{H}_{2^{\mathrm{o}}} \otimes \mathcal{H}_{1^{\mathrm{o}}}) \to \mathsf{BL}(\mathcal{H}_{3^{\mathrm{i}}})$, whose Choi operator is

$$\mathsf{C}_{3^{\mathrm{i}}Y2^{\mathrm{o}}1^{\mathrm{o}}} := \sum_{xyz} \rho^{(xyz)}_{3^{\mathrm{i}}} \otimes \mathsf{D}^{(y)*}_Y \otimes \mathsf{D}^{(z)*}_{2^{\mathrm{o}}} \otimes \mathsf{D}^{(x)*}_{1^{\mathrm{o}}}, \tag{5.6}$$

where $\{\mathsf{D}^{(y)}_Y\}$ are the dual set to $\{\rho^{(y)}_Y\}$, and $\{\mathsf{D}^{(z)}_{2^{\mathrm{o}}}\}, \{\mathsf{D}^{(x)}_{1^{\mathrm{o}}}\}$ respectively form the dual set to some IC set of preparations $\{\sigma^{(z)}_{2^{\mathrm{o}}}\}, \{\sigma^{(x)}_{1^{\mathrm{o}}}\}$. This map acts to take each one of the $\{\sigma^{(x)}_{1^{\mathrm{o}}}, \rho^{(y)}_Y, \sigma^{(z)}_{2^{\mathrm{o}}}\}$ combination of its inputs to one of $d^6$ unique states $\rho^{(xyz)}_{3^{\mathrm{i}}}$, which are the final outputs of the process.

Stipulating the construction of $\rho_{Y2^{\mathrm{i}}1^{\mathrm{i}}}$ in Eq. (5.5) to be a positive semidefinite operator overall, and the map $\mathsf{C}_{3^{\mathrm{i}}Y2^{\mathrm{o}}1^{\mathrm{o}}}$ defined in Eq. (5.6) to represent a valid evolution, requires sufficient mixedness of each $\rho^{(y)}_{1^{\mathrm{i}}}$ and $\rho^{(xyz)}_{3^{\mathrm{i}}}$; additionally, choosing preparations $\rho^{(y)}_Y, \sigma^{(z)}_{2^{\mathrm{o}}}$ and $\sigma^{(x)}_{1^{\mathrm{o}}}$ such that $\sum_{xyz} \mathsf{D}^{(y)}_Y \otimes \mathsf{D}^{(z)}_{2^{\mathrm{o}}} \otimes \mathsf{D}^{(x)}_{1^{\mathrm{o}}} = \mathbb{1}_{Y2^{\mathrm{o}}1^{\mathrm{o}}}$ ensures $\mathsf{C}_{3^{\mathrm{i}}Y2^{\mathrm{o}}1^{\mathrm{o}}}$ satisfies the necessary trace conditions of Eq. (3.15). Importantly, all of these conditions outlined above can be achieved simultaneously. It then follows that there exists an underlying unitary dilation of the map $\mathsf{C}_{3^i Y 2^o 1^o}$. The overall process tensor for the process described above is explicitly given by

$$\Upsilon_{3^i:1^i} = \sum_{xyz} p_y \rho_{3^i}^{(xyz)} \otimes \mathsf{D}_{2^o}^{(z)*} \otimes \Delta_{2^i}^{(y)*} \otimes \mathsf{D}_{1^o}^{(x)*} \otimes \rho_{1^i}^{(y)}. \qquad (5.7)$$

Intuitively, the IC instrument sequence $\mathcal{J}_M = \{\sigma_{1^o}^{(x)}, \Pi_{2^i}^{(y)}, \sigma_{2^o}^{(z)}\}$ blocks any influence from the history to the future, as the measurement performed at $2^i$ leaves the initial state $\rho_{Y 2^i 1^i}$ in a product between $Y$ and $1^i$ for each outcome, such that the final output state is then independent of any operation that could be performed at $t_{1^i}$. Indeed, for any realisation of the instrument sequence, the conditional future and history processes are in the product form of Eq. (4.4)

$$\mathrm{tr}_{2^o 2^i 1^o}\left[\left(\sigma_{2^o}^{(z)} \otimes \Pi_{2^i}^{(y)} \otimes \sigma_{1^o}^{(x)}\right)^{\mathrm{T}} \Upsilon_{3^i:1^i}\right] = p_y \rho_{3^i}^{(xyz)} \otimes \rho_{1^i}^{(y)}. \qquad (5.8)$$

In this sense, the map $\mathsf{C}_{3^i Y 2^o 1^o}$ has no bearing on whether the effect of history is blocked or not: an experimenter could coarse-grain over any of the preparations while applying the correct measurement, *e.g.*, feed in $p\sigma_{1^i}^{(x)} + (1-p)\sigma_{1^i}^{(x')}$, yielding a future state $p\rho_{3^i}^{(xyz)} + (1-p)\rho_{3^i}^{(x'yz)}$ that remains conditionally independent of the history $\rho_{1^i}^{(y)}$ given any measurement outcome $y$ of $\{\Pi_{2^i}^{(y)}\}$ at $t_{2^i}$. Of course, simpler processes can lead to an independent history and future with respect to the outcomes of an IC POVM (see Example D.1 given in Appendix D.1). However, here we construct a more general process with $\mathsf{C}_{3^i Y 2^o 1^o}$ defined as per Eq. (5.6) in order to yield $d^6$ *distinct* future states $\rho_{3^i}^{(xyz)}$ for each possible realisation of the causal break sequence, each one of which is conditionally independent of the history.

Just as in the generalised collision model of Section 4.3, in principle an experimenter can predict the next state of the system as a function of measurements and preparations in the causal break sequence. Furthermore, since the history-blocking sequence is IC, they could perform a process tomography to completely characterise the process as per Eq. (5.7). If, on the other hand, the experimenter were to apply a different instrument on the memory block, then correlations between the future and history would in general arise (but, as already mentioned, they could vary the preparations and not see any influence from the history).

The two examples provided in this section highlight significant properties of memory in quantum processes. Example 5.1 explicitly shows that there exist processes where specific sequences of unitary operations can break all possible temporal correlations between future and history, while Example 5.2 highlights that the operations of a history-blocking instrument sequence can comprise an IC (in general non-orthogonal) set of independent measurements and preparations. So far, through Theorem 5.1 and Corollary 5.2, we have developed the structural constraints that a process tensor must satisfy in order to exhibit finite quantum Markov order for a given instrument sequence. However, this characterisation is difficult to check in practice, due to the non-uniqueness of possible decompositions for a process tensor. It is therefore natural to seek a function of such finite Markov order processes that vanishes iff there are no correlations between the history and future remaining once a memory block of length $\ell$ is specified. For classical stochastic processes (without interventions), it is straightforward to show that the CMI of the underlying joint probability distribution has the desired property. In contrast, in both of the above examples (and also in the generalised collision model of Section 4.3), the quantum CMI evaluated on the Choi operator of the process tensor between the history and future with respect to the memory is non-vanishing. For the sake of comprehensiveness, we explicitly construct the process tensor for a simple example and calculate the quantum CMI to be non-zero in Appendix D.1. This observation is insightful for a number of reasons which we address in the coming section, where we explore in detail the necessary conditions on the history-blocking instrument sequences for processes with vanishing quantum CMI, of which classical processes with finite Markov order are a special case.

## 5.2 QUANTUM MARKOV ORDER AND CONDITIONAL MUTUAL INFORMATION

In light of the observation above, we have the following theorem.

**Theorem 5.3.** *Vanishing quantum CMI guarantees the process has finite quantum Markov order; the converse is not true.*

The structure of processes with vanishing quantum CMI can be deduced from that of quantum states with vanishing quantum CMI, with the additional causality constraint imposed to ensure a valid process. The CMI of a quantum process is defined by $I(F : H|M) := S(\Upsilon_{FM}) + S(\Upsilon_{MH}) - S(\Upsilon_{FMH}) - S(\Upsilon_M)$, and it vanishes iff there exists a block orthogonal decomposition of the composite $M$ Hilbert space as $\mathcal{H}_M = \bigoplus_m \mathcal{H}_{M^L}^{(m)} \otimes \mathcal{H}_{M^R}^{(m)}$, such that [138]

$$\Upsilon_{FMH}^{\text{CMI}=0} = \bigoplus_m p_m \widetilde{\Upsilon}_{FM^L}^{(m)} \otimes \widetilde{\Upsilon}_{M^R H}^{(m)}. \tag{5.9}$$

Here, the decomposition of $\mathcal{H}_M$ does not necessarily respect the temporal ordering of the underlying process; specifically, the Hilbert spaces $\{\mathcal{H}_{M^L}^{(m)}\}$ do not need to describe events that occur strictly before or after those described in $\{\mathcal{H}_{M^R}^{(m)}\}$.

The proof of Theorem 5.3 is given in Appendix D.2; the basic strategy is to explicitly construct a history-blocking instrument sequence for processes of the form in Eq. (5.9), and show that this structure is a special case of Eq. (5.2), meaning that vanishing quantum CMI implies finite Markov order. The history-blocking sequence we construct is, in fact, made up of the set of orthogonal projectors (which form a self-dual set) onto each of the $m$ subspaces in the decomposition above.

An immediate reason for why the converse does not have to hold is that there is no reason why the memory blocking operations should be sharp, *i. e.,* that the projectors are rank-1 and pairwise orthogonal. In the case where the experimenter finds conditional independence with respect to a sequence of higher-rank fuzzy projectors, the future-history correlations hidden within each $m$ subspace need not obey the constraint implied by Eq. (5.9), and hence the process can have non-vanishing quantum CMI, as shown explicitly in Example D.2 of Appendix D.3. As we have already seen in Section 4.2, similar behaviour arises in an operational interpretation of classical stochastic processes: if the experimenter cannot measure realisations of the process sharply then the statistics observed do not necessarily have vanishing classical CMI, even if the true underlying process is one of finite Markov order (see the examples in Appendix C.3).

This immediately begs the question: *do processes with finite Markov order with respect to an instrument sequence comprising only rank-1 orthogonal projectors necessarily have vanishing quantum CMI?* In addition to the proof of Theorem 5.3, we show in Appendix D.2 that—perhaps surprisingly—this is not the case. Intuitively, this is because such projectors that make up an instrument can live on an extended input-output Hilbert space at each timestep; in quantum theory, sharp measurements of composite systems can be fuzzy locally if they are made in an entangled basis. Thus, even if the instrument sequence comprises only rank-1, orthogonal projectors, the structural condition implied on the process tensor is still not strong enough to force a block-diagonal representation as necessary for the quantum CMI to vanish. Here we provide an explicit example that evidences this point.

**Example 5.3** (Process with non-vanishing quantum CMI but finite Markov order for a sequence of rank-1, orthogonal projectors)**.** Consider the process depicted in Fig. 5.3. The 4-dimensional ancilla qudit is initially in a coherent superposition $|\tau\rangle^A = \alpha|0\rangle + \beta|1\rangle + \gamma|2\rangle + \delta|3\rangle$ with $|\alpha|^2 + |\beta|^2 + |\gamma|^2 + |\delta|^2 = 1$ constituting the environment. Controlled on the state of this qudit, the process implements one of the four Pauli maps

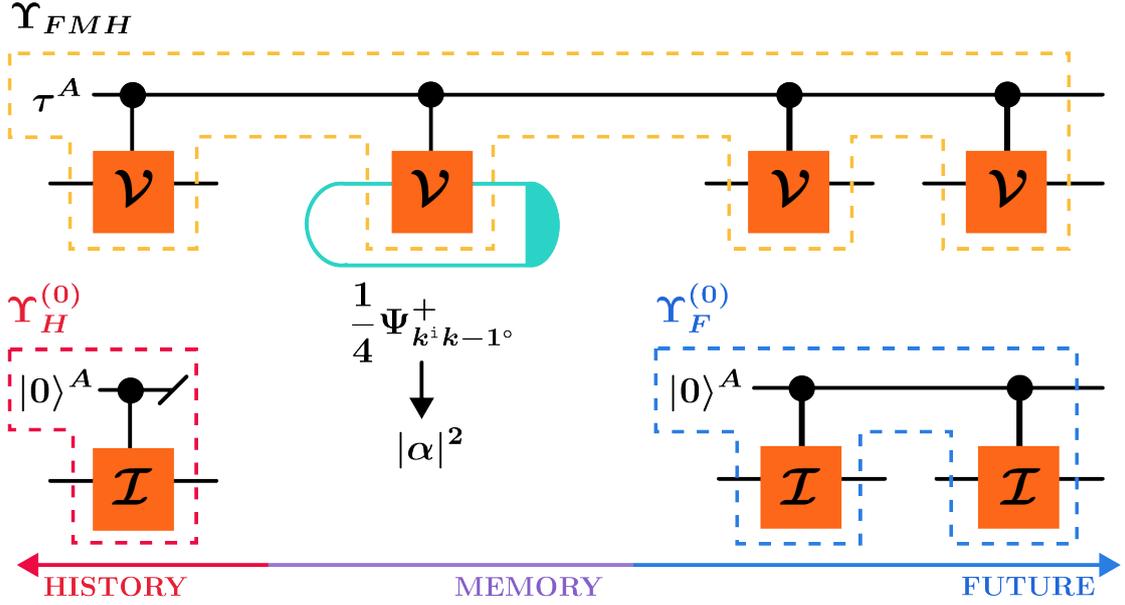

**Figure 5.3:** *Process with non-vanishing quantum CMI.* The environment is a 4-dimensional ancilla. Its initial state is a coherent superposition of the basis states $\{|0\rangle, \ldots, |3\rangle\}$. The system-environment evolution is a control unitary, which implements one of the the four Pauli rotations $\mathcal{V} := \{\mathcal{I}, \mathcal{X}, \mathcal{Y}, \mathcal{Z}\}$ on the system depending on the state of the ancilla (see top panel). The history-blocking instrument sequence consists of feeding in one half of a Bell pair and, at the next step, measuring the system and the other half in the Bell basis at the next timestep. For each outcome of this instrument, one can infer which of the four Pauli rotations was applied, and the history and future processes are conditionally independent. For illustrative purposes, the bottom panel depicts the conditional processes that arise from successful implementation of the operation $\frac{1}{4}\Psi^+$, which occurs with probability $|\alpha|^2$.

(including the identity map), $\mathcal{V} := \{\mathcal{I}, \mathcal{X}, \mathcal{Y}, \mathcal{Z}\}$, on a single qubit system. The Choi operators of these maps are the projectors of the four (unnormalised) Bell pairs

$$|\Psi^\pm\rangle := |00\rangle \pm |11\rangle \quad \text{and} \quad |\Phi^\pm\rangle := |01\rangle \pm |10\rangle. \tag{5.10}$$

Suppose that the process continues for $n$ timesteps and, at the end of the process, the ancilla is fed out with the system in order to retain the quantum features of the process. For simplicity, we also assume that there are no initial system-environment correlations, allowing us to define the process as one beginning on an output wire at $t_1$. The corresponding process tensor is $\Upsilon_{n:1} = |\Upsilon\rangle\langle\Upsilon| \in \mathsf{BL}(\mathcal{H}^A_{n^\mathrm{i}} \otimes \mathcal{H}^S_{n^\mathrm{i}} \otimes \mathcal{H}^S_{n-1^\circ} \otimes \ldots \otimes \mathcal{H}^S_{1^\circ})$, where

$$\begin{aligned}|\Upsilon\rangle := &\alpha|0\rangle^A_{n^\mathrm{i}} \otimes |\Psi^+_{n^\mathrm{i} n-1^\circ} \ldots \Psi^+_{2^\mathrm{i} 1^\circ}\rangle + \beta|1\rangle^A_{n^\mathrm{i}} \otimes |\Phi^+_{n^\mathrm{i} n-1^\circ} \ldots \Phi^+_{2^\mathrm{i} 1^\circ}\rangle \\ &+ \gamma|2\rangle^A_{n^\mathrm{i}} \otimes |\Phi^-_{n^\mathrm{i} n-1^\circ} \ldots \Phi^-_{2^\mathrm{i} 1^\circ}\rangle + \delta|3\rangle^A_{n^\mathrm{i}} \otimes |\Psi^-_{n^\mathrm{i} n-1^\circ} \ldots \Psi^-_{2^\mathrm{i} 1^\circ}\rangle.\end{aligned} \tag{5.11}$$

Note that this is not a Markovian process (it is not of the product form of Eq. (3.36)), nor is it a classical probabilistic mixture of such processes; rather, the process tensor is a pure state representing a coherent superposition of implementing sequences of the four Pauli maps, corresponding to a genuinely quantum memory.

Consider the instrument sequence where, at some timestep $t_{k-1}$, an experimenter inputs half of one of the Bell pairs, feeds the other half forward to the next timestep $t_k$, and then makes a Bell basis measurement on the fed-forward ancilla and the system state output by the process (see Fig. 5.3). This instrument is made up of the Choi states $\mathcal{J}_{k^i k-1^o} = \{\mathsf{O}^{(x)}_{k^i k-1^o}\} := \frac{1}{4}\{\Psi^+_{k^i k-1^o}, \Phi^+_{k^i k-1^o}, \Phi^-_{k^i k-1^o}, \Psi^-_{k^i k-1^o}\}$. Since all cross terms in $\Upsilon_{n:1}$ are orthogonal to any of these, for each outcome observed upon their application, the experimenter observes one of the following four conditional processes

$$\widetilde{\Upsilon}^{(0)}_{FH} = |\alpha|^2 \Psi^+_F \otimes \Psi^+_H, \quad \widetilde{\Upsilon}^{(1)}_{FH} = |\beta|^2 \Phi^+_F \otimes \Phi^+_H \quad (5.12)$$
$$\widetilde{\Upsilon}^{(2)}_{FH} = |\gamma|^2 \Phi^-_F \otimes \Phi^-_H, \quad \widetilde{\Upsilon}^{(3)}_{FH} = |\delta|^2 \Psi^-_F \otimes \Psi^-_H,$$

where $\Psi^+_F := |0\rangle^A_{n^i} \otimes \Psi^+_{n^i n-1^o} \otimes \ldots \otimes \Psi^+_{k+1^i k^o}$, $\Psi^+_H := \Psi^+_{k-1^i k-2^o} \otimes \ldots \otimes \Psi^+_{2^i 1^o}$, and the superscript label corresponds to each possible realisation *e.g.,* the label (0) corresponds to the experimenter feeding in half of the state $\Psi^+/2$ and successfully measuring it, which occurs with probability $|\alpha|^2$, and similarly for the other quantities defined. The conditional tester elements can thus each be normalised to a proper process tensor by simply dividing by the appropriate probability *e.g.,* $\Upsilon^{(0)}_{FH} = \widetilde{\Upsilon}^{(0)}_{FH}/|\alpha|^2$.

Intuitively, once an outcome of the instrument described is observed, the experimenter can deduce which of the four independent control operations were applied to the system and hence the state of the ancilla, which collapses onto one of its computational basis states and does not change further throughout the process. This means that the history and future processes are known with certainty and are thus conditionally independent with respect to knowledge of the instrument outcome. In contrast, suppose that the experimenter were to perform an incoherent operation, such as feeding in the maximally mixed state before averaging over all measurement outcomes at the subsequent timestep. In this case, the conditional future-history process is now a probabilistic mixture of the four control operations being applied *i.e.,* $\Upsilon_{FH} = \sum_x \widetilde{\Upsilon}^{(x)}_{FH}$, with $\{\widetilde{\Upsilon}^{(x)}_{FH}\}$ defined in Eq. (5.12). Such a mixture of Markovian processes is non-Markovian due to the correlations between the future and history: indeed, in this scenario an experimenter could condition the future dynamics by performing certain operations in the history.

A simple calculation shows that the quantum CMI between the history and future given the memory for the process tensor in Eq. (5.11) does not vanish; rather, it is equal to the Shannon entropy of the distribution $\mathbb{P}(x) = \{|\alpha|^2, |\beta|^2, |\gamma|^2, |\delta|^2\}$. Lastly, note that, had we chosen to discard the ancilla, rather than feed it out at the final timestep, the corresponding process tensor is a probabilistic mixture of sequences of the four Pauli maps applied, *i. e.,* the projector of Eq. (5.11) *without any cross terms*. In this case, the process tensor is of the form in Eq. (5.9) and the quantum CMI vanishes.

In summary, here we have an example of a process which has finite Markov order with respect to an instrument sequence comprising only rank-1, orthogonal projectors, but nonetheless has non-vanishing quantum CMI. The example presented here represents a genuinely quantum mechanical memory effect with no classical analog. The intuition behind the distinction is that in the classical setting, $d$ orthogonal projectors are IC for a $d$-level system, whereas this fails to hold true in the quantum realm. An additional restriction on the "off-diagonal" terms of the process tensor must be adhered to in order to ensure the quantum CMI vanishes, which is formulated in Appendix D.2.

## 5.3 CHAPTER SUMMARY

In this chapter, we have outlined some of the key features of memory length in stochastic processes, many of which are peculiar to quantum mechanics. We began by tackling the general problem: *given a sequence of operations that acts to erase the effect of history on the future of a process, what can we say about its overall structure?* In Section 5.1 we detailed the generic constraint on process tensors with finite quantum Markov order, providing the most general structure deducible to an experimenter who knows the history-blocking instrument sequence in question. The first special case of this structure was exhibited in Example 5.1, where we studied a process whose history is blocked by a sequence of unitary operations. Although such unitary sequences can serve to block the effect of history, they provide minimal information to the experimenter about the process at hand and thereby represent the extremal case of informationally-incomplete history-blocking instrument sequences. On the other extreme are IC instruments which stipulate the complete description of the process. We then considered processes with finite-memory with respect to such IC instrument sequences, in particular an IC POVM followed by an independent repreparation of a state from an IC set in Example 5.2.

Following this, we analysed the connection between quantum Markov order and the vanishing of the quantum CMI. As mentioned previously, until the recent introduction of the process tensor, there was no meaningful way to develop a sensible notion of Markov order in the temporal setting, since the statistics observed in time depend upon how an experimenter probes the process and are thus inherently instrument-dependent. Despite this concern, many efforts throughout the literature are concentrated on Markov chains, defined as tripartite quantum states that are recoverable or (equivalently) have vanishing quantum CMI. On the other hand, the general theory of quantum Markov order for processes introduced here is captured by the conditional independence statement of Eq. (4.4); this instrument-dependent statement is in stark contrast with the aforementioned definitions on quantum states, which make no mention of the instrument sequence of choice. Therefore, it is not immediately clear how such characterisations concretely relate to temporal processes with finite quantum Markov order.

Nonetheless, the CJI lets us consider temporal processes in terms of their corresponding Choi operator, allowing us to concretely examine the link between the two inequivalent notions. In Section 5.2, we proved that in the quantum realm, processes with finite Markov order with respect to a sequence of instruments need not necessarily have vanishing quantum CMI. Although a similar departure can arise in the study of classical stochastic processes where fuzzy measurements are permitted, in quantum mechanics there can exist processes with finite Markov order with respect to a sequence of sharp, orthogonal projectors that has non-vanishing quantum CMI—in direct contradistinction to the classical setting. An explicit construction is provided in Example 5.3 and the additional constraints on the process tensor required to guarantee vanishing quantum CMI are detailed in Appendix D.2.

The results uncovered in this chapter raise some interesting avenues for future exploration. For instance, many realistic physical scenarios are often modelled by specific forms of interactions, *e. g.*, nearest-neighbour interaction spin chains evolving in a time-translationally invariant manner. In such a scenario, whilst a generic sequence of instruments such as the trash-and-prepare protocol will typically not act to block the historic influence, in practice it may be the case that such a sequence almost always *approximately* blocks the influence of history. A natural extension to this work would involve a deeper exploration of memory effects in specific physical models with respect to the instrument-specific quantum Markov order formalism.

A first step to understanding processes with approximately finite-length memory is to quantify memory strength, that is, the amount of temporal correlations remaining between the history and the future processes after an instrument sequence of choice is implemented on the memory block. In the coming chapter, we will introduce an instrument-specific notion of memory strength and explore its flexibility in characterising memory effects in an exactly solvable non-Markovian model.

# 6

MEMORY STRENGTH

So far, we have looked at the structural properties of processes with finite Markov order. Now, we will quantify the *strength* of said memory effects, if they exist; that is, the degree to which they influence the observed statistics. This can be quantified by the temporal correlations remaining between the history and the future processes for a specified instrument on the memory block. In other words, we wish to understand the deviation from future-history independence with respect to the instrument sequence in use.

The ultimate goal concerning the quantification of memory strength is to understand the circumstances under which one can effectively describe processes in an efficient manner. Indeed, we are unlikely to find processes with strictly finite Markov order in nature, since, with respect to generic instruments, non-Markovian processes typically exhibit infinite memory length. Understanding how strong the memory effects across a given duration in time are has significant implication for the simulation of processes with *approximately* finite memory length.

For instance, numerical techniques for open dynamics often invoke finite memory approximations, where rapidly vanishing temporal correlations are truncated [25, 176–178]. This is tantamount to treating the process as having finite Markov order with respect to the identity instrument (*i. e.,* doing nothing to the process), although memory approximations involving other choices of instruments can also be made [179]. Another related result concerns quantum *states* with small quantum CMI. Recent bounds on the fidelity of recovery show that these tripartite quantum states allow for the existence of a CPTP map that approximately recovers the total state from only partial information, *i. e.,* by acting upon the conditioning subsystem alone [140]. Although this result does not directly pertain to processes, the CJI permits similar considerations in the temporal setting. However, as we have seen in the previous chapter, the quantum CMI is a poor quantifier for the memory strength, as it does not necessarily vanish for processes with finite Markov order.

As we have laid out so far in this thesis, genuine memory effects in quantum processes must be specified with respect to the instrument sequence used to probe them. It is thus natural that a proper quantification of strength must also be instrument-specific. To this end, in this chapter, we will first develop a number of viable candidates for quantifying memory strength in quantum processes. We will then examine these quantifiers and the resulting measures for memory strength for an exactly solvable system-environment model, introduced in Ref. [180]. This model is amenable to the analysis of memory effects for a large host of physical situations by tuning parameters in the corresponding system-environment Hamiltonian, and therefore allows us to explicitly study many of the main concepts introduced throughout this thesis. We first highlight that multi-time memory effects, which remain uncaptured by two-time witnesses of non-Markovianity, are present throughout the parameter space, before exploring its memory structure with respect to natural families of instrument sequences. Here we see that over a fixed length of the memory block, the process indeed displays different memory strength for different probing instruments.

An important special case is where the instrument sequence is chosen to be the identity transformation on the system, as this pertains to the 'natural' memory of the process at hand, in the sense that it ties in most nicely with existing discussions of memory. For instance, standard numerical open systems techniques can be appropriately implemented when the memory effects in such cases die off rapidly [25]. We go on to study this situation, exploring the length of time over which the memory in the process naturally decays. This chapter contains much of the discussion of Ref. [3].

## 6.1 QUANTIFYING MEMORY STRENGTH

We begin by introducing an instrument-specific notion of memory strength. We build this up by first introducing quantifiers of the memory effects for each specific outcome of the instrument sequence in question, before motivating a number of suitable aggregations that compress these numbers into a single quantifier for the overall memory strength for the instrument.

### 6.1.1 *Instrument-specific Memory Strength*

We have already encountered a quantity that serves as a suitable starting point to the quantification of memory: the mutual information between the conditional history and future processes for each realisation of an instrument on the memory block (see Eq. (4.5)). We take these values as the foundation of the definition of memory strength.

**Definition 6.1** (Outcome-specific memory strength). The memory effects of a process across $\ell = |M|$ timesteps are characterised by the mutual information between the conditional history and future processes for *each outcome* of the instrument:

$$\mathscr{S}_{\mathcal{J}_M}(x_M) := I(F:H)_{x_M} = S(\Upsilon_F^{(x_M)}) + S(\widetilde{\Upsilon}_H^{(x_M)}) - S(\widetilde{\Upsilon}_{FH}^{(x_M)}), \tag{6.1}$$

where $\mathcal{J}_M = \{\mathsf{O}_M^{(x_M)}\}$ and $\widetilde{\Upsilon}_{FH}^{(x_M)} = \mathrm{tr}_M\left[\mathsf{O}_M^{(x_M)} \Upsilon_{FMH}\right]$.

Note that this is not yet a complete definition of memory strength for the full instrument, but a quantifier for each specific outcome. Nonetheless, the quantity vanishes for all $x_M$ iff the process has finite memory with respect to the instrument sequence $\mathcal{J}_M$.

One possible way to aggregate these quantifiers would be to take the average with respect to the probabilities of each realisation $x_M$; however, this is not necessarily valid, as the corresponding probabilities in general depend on the historic probing instruments. Nonetheless, an average with respect to the *uniform distribution* serves to quantify the overall memory effect one might expect for the given instrument, when taking the uniform distribution as the unbiased prior for the occurrence probability of the respective outcomes. In this way, we define the *average* memory strength for $\mathcal{J}_M$ as follows.

**Definition 6.2** (Average instrument-specific memory strength). The average memory strength for the instrument $\mathcal{J}_M$ is computed by taking the average of the outcome-specific mutual informations between the history and future (as per Def. 6.1) with respect to the uniform distribution:

$$\mathscr{S}_{\mathcal{J}_M}^{\mathrm{avg}} := \frac{1}{|\mathcal{J}_M|} \sum_{x_M} \mathscr{S}_{\mathcal{J}_M}(x_M). \tag{6.2}$$

Alternatively, taking the *maximum* over all outcomes of the instrument sequence provides a quantifier for the extremal scenario that an experimenter might see in an individual run of the experiment.

**Definition 6.3** (Maximum instrument-specific memory strength). The maximum memory strength for the instrument $\mathcal{J}_M$ is the maximum of the outcome-specific mutual informations between the history and future (as per Def. 6.1) over all possible outcomes:

$$\mathscr{S}_{\mathcal{J}_M}^{\max} := \max_{x_M} \mathscr{S}_{\mathcal{J}_M}(x_M). \tag{6.3}$$

Both Defs. 6.2 and 6.3 are no longer *outcome-specific*, but rather *instrument-specific* in the sense that they pertain only to quantifying the overall memory effect across a specified instrument. They both vanish iff the process has quantum Markov order $\ell$ with respect to $\mathcal{J}_M$. In contrast, taking, say, the minimum value as a memory strength quantifier and finding it to be zero does not necessarily imply the process to have finite Markov order, since the individual tester element corresponding to the minimum only provides a probabilistic history-blocking, whereas the remaining CP maps belonging to the instrument need not.

The instrument-specific memory strength definitions allow us to deduce the temporal correlations of the process over any time interval with respect to any set of instruments. For a fixed $\ell$, minimising any suitable function of $\mathscr{S}_{\mathcal{J}_M}(x_M)$—such as the average or maximum over outcomes—over all instruments provides a quantification of the *intrinsic* memory strength. That is, the amount of temporal correlations across $\ell$ timesteps that *cannot* be erased by any interrogation sequence, which vanishes iff the process is of Markov order $\ell$. The instrument corresponding to the minimum value, $\operatorname{argmin}_{\mathcal{J}_M}(\mathscr{S}_{\mathcal{J}_M})$, can be interpreted as providing the optimal history-blocking sequence across the given length. On the other hand, $\operatorname{argmax}_{\mathcal{J}_M}(\mathscr{S}_{\mathcal{J}_M})$ relates to the optimal strategy for transmitting information across the memory block; the maximum value provides a novel quantification of the *process capacity*, which is an alternate way of quantifying the channel capacity in the presence of memory [48, 49].

Performing such optimisation is a difficult task in general.[1] Moreover, implementing the optimal sequences in either case may not be possible in practical laboratory setups, as it will generally require the ability to control temporally-correlated quantum operations. However, there are at least two specific instrument sequences that are of immediate relevance, namely the identity instrument and the completely-noisy instrument. The former leaves the system state unchanged and thus relates to the "natural" transmission of information on the system level across the memory block, whereas the latter considers the case where the experimenter tries to actively erase this information by discarding the state emitted by the process and repreparing white noise.

Consequently, we now focus on specifying two particularly important instruments and define the memory strength with respect to them. Note that both the identity map and the completely-noisy instrument have only a single outcome, and thus no aggregation over outcomes is required and we simply have an instrument-specific memory strength denoted $\mathscr{S}_{\mathcal{J}_M}$. We define the *natural* memory strength across a length $\ell$ as follows.[2]

---

1 Since $\mathscr{S}_{\mathcal{J}_M}^{\max}$ simply maximises over all CP maps, and $\mathscr{S}_{\mathcal{J}_M}^{\mathrm{avg}}$ optimises over all CPTP maps—both of which are convex sets—the optimisation should be phraseable as a semidefinite program. However, it is unclear if this is also the case for more general functions of $\mathscr{S}_{\mathcal{J}_M}(x_M)$.

**Definition 6.4** (Natural memory strength). The natural memory strength across a duration of length $\ell = |M|$ is the mutual information between the history and the future upon application of a sequence of identity maps on the memory block:

$$\mathcal{N}_\ell := \mathscr{S}_{\mathcal{I}_M} = I(F:H)_{\Upsilon_{FH}^{\mathcal{I}_M}}, \qquad (6.4)$$

where $\Upsilon_{FH}^{\mathcal{I}_M} := \mathrm{tr}_M \left[ \Psi_M^+ \Upsilon_{FMH} \right]$, with $\Psi_M^+ = \bigotimes_{j=k-\ell}^{k-1} \Psi_{j^\circ j^\mathrm{i}}^+$ denoting the Choi operator of the sequence of identity maps.

We also define the *noise-resistant* length-$\ell$ memory.

**Definition 6.5** (Noise-resistant memory strength). The noise-resistant memory strength across a duration of length $\ell$ is the mutual information between the history and the future upon application of a sequence of maps that discard the state emitted by the process and feed in the maximally-mixed state:

$$\mathcal{R}_\ell := \mathscr{S}_{\mathbb{1}_M} = I(F:H)_{\Upsilon_{FH}}, \qquad (6.5)$$

where the mutual information is simply calculated on the future-history marginal of the process tensor, *i.e.*, $\Upsilon_{FH}^{\mathbb{1}_M} = \Upsilon_{FH} = \frac{1}{d_M^\circ} \mathrm{tr}_M \left[ \Upsilon_{FMH} \right]$, with the normalisation accounting for the preparation of maximally-mixed states.

This instrument sequence corresponds to applying any POVM without recording the outcomes followed by preparing the maximally-mixed state. The Choi operators corresponding to any such erasure sequence amount to the identity matrix on all of the input and output Hilbert spaces associated to each timestep in the memory block. In turn, this implies that $\mathcal{R}_\ell$ quantifies the amount of memory that survives the erasure on average.[3]

The notions introduced in this section are depicted in Fig. 6.1. We can either consider a fixed length $\ell$ and examine how the memory strength behaves over that duration for a variety of instruments, or, on the other hand, consider a fixed instrument and examine the memory as the number of (uncorrelated) sequential applications of said instrument grows with respect to $\ell$. We now explicitly construct a process tensor describing an exactly solvable non-Markovian system-environment model in order to study the behaviour of both the average and maximum memory strength in Eqs. (6.2) and (6.3) respectively for a causal break sequence, and both the natural and noise-resistant memory strength defined in Eqs. (6.4) and (6.5) respectively.

---

[2] Since there is only a single outcome for the instruments considered in the following definitions, we depart from the notation introduced in Def. 6.1. Instead, we resort to the notation prevalent in the literature.

[3] For processes with vanishing quantum CMI, the marginal future-history marginal is separable [138], implying that any faithful entanglement measure $\mathscr{E}(F:H)_{\Upsilon_{FH}}$ (see Ref. [86]) vanishes and any correlation between the history and future upon implementing a length-$\ell$ erasure sequence is classical.

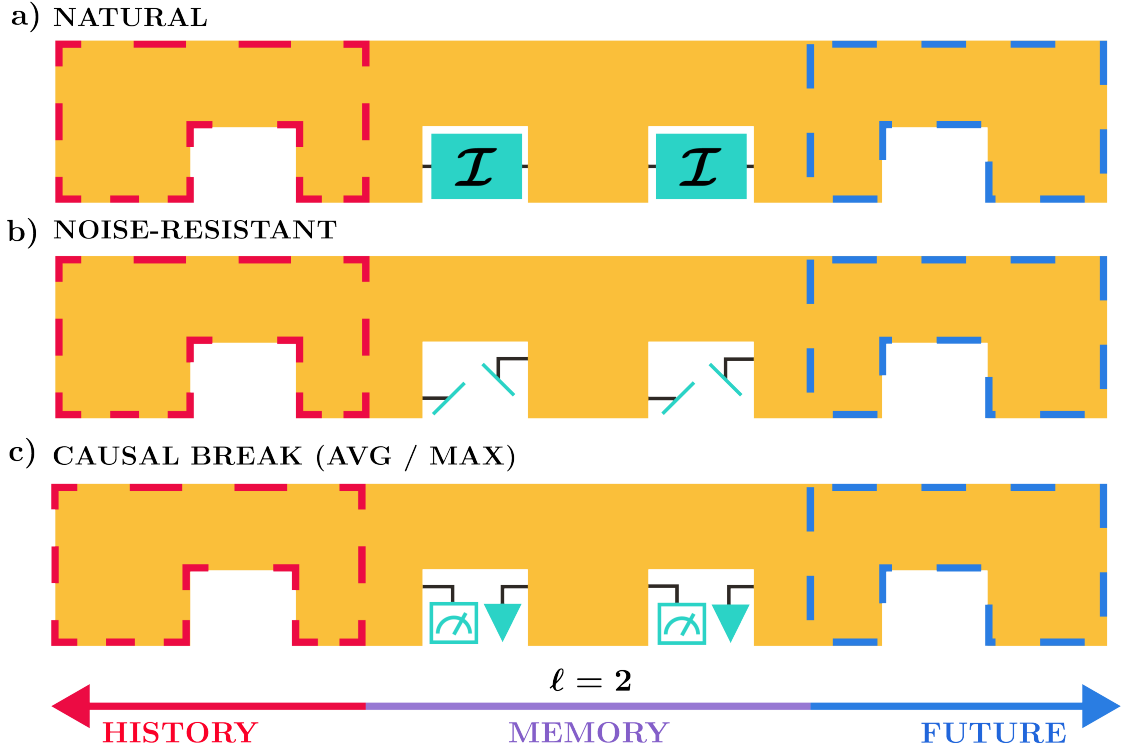

**Figure 6.1:** *Instrument-specific memory strength.* We depict the instrument-specific memory strength across a length $\ell = 2$. In panel **a)** is the *natural* memory strength, which is calculated with respect to the identity instrument. In panel **b)** is the *noise-resistant* memory-strength, in which an experimenter discards the system states and feeds in white noise. In both cases, the instrument has only one outcome, and the memory strength is simply the mutual information between the history (red) and future (blue) conditional processes. Panel **c)** shows the memory strength with respect to a causal break, where one must choose a suitable aggregate over the mutual information for each outcome, such as averaging or maximising over outcomes.

### 6.1.2  *Memory Length for an Exactly Solvable Model*

Here, we focus on a model involving a qubit coupled to another qubit that is interacting with an additional bath (which is regarded as external to the system-environment), as introduced in Ref. [180]. Its low dimensionality makes it easy to solve analytically and therefore permits the investigation of memory effects for different model parameters.

The system-environment evolution proceeds according to the GKSL master equation

$$\frac{\partial \rho_t^{SE}}{\partial t} = -i\xi[\sigma_x^S \otimes \sigma_x^E, \rho_t^{SE}] + \kappa \mathcal{L}[\sigma_-^E](\rho_t^{SE}), \tag{6.6}$$

where the dissipator acts on the environment alone and is defined as: $\mathcal{L}[\sigma_-^E](\rho_t^{SE}) := \sigma_-^E \rho_t^{SE} \sigma_+^E - \frac{1}{2}\{\sigma_+^E \sigma_-^E, \rho_t^{SE}\}$, with $\sigma_\pm^E := \sigma_x^E \pm i\sigma_y^E$. The dynamics describes a qubit system interacting with a qubit environment with $X$–$X$ coupling strength $\xi$ and a cooling process on the environment due to its interactions with the external bath at rate $\kappa$.

The authors of Ref. [180] examined the non-Markovianity of the system dynamics using the breakdown of CP-divisibility, and the increase of the trace-distance distinguishability between arbitrary input states (introduced in Refs. [67] and [93] respectively) as measures of the existence of temporal correlations. Due to the simplicity of the model, the analytic form of the equation of motion on the level of the system alone can be derived and is written [180]

$$\frac{\partial \rho_t^S}{\partial t} = -\frac{\dot{c}_t}{2c_t} \mathcal{L}[\sigma_x^S](\rho_t^S), \tag{6.7}$$

where

$$c_t = \begin{cases} \exp\left(-\frac{\kappa t}{4}\right)\left(\frac{\kappa \sinh \frac{t}{4}\sqrt{\kappa^2 - 64\xi^2}}{\sqrt{\kappa^2 - 64\xi^2}} + \cosh \frac{t}{4}\sqrt{\kappa^2 - 64\xi^2}\right) & \text{for } \kappa^2 > 64\xi^2 \\ \exp\left(-\frac{\kappa t}{4}\right)\left(\frac{\kappa \sin \frac{t}{4}\sqrt{64\xi^2 - \kappa^2}}{\sqrt{64\xi^2 - \kappa^2}} + \cos \frac{t}{4}\sqrt{64\xi^2 - \kappa^2}\right) & \text{for } \kappa^2 < 64\xi^2 \\ \exp\left(-\frac{\kappa t}{4}\right)\left(1 + \frac{1}{4}\kappa t\right) & \text{for } \kappa^2 = 64\xi^2. \end{cases} \tag{6.8}$$

A necessary and sufficient criteria for the dynamics to be CP-divisible is that the coefficients of the dissipation terms in the above master equation for the system, *i.e.*, $-\frac{\dot{c}_t}{2c_t}$, are non-negative for all times [8]. Explicit calculation shows that for $\kappa^2 \geq 64\xi^2$, $-\frac{\dot{c}_t}{2c_t}$ is always non-negative, whereas for $\kappa^2 < 64\xi^2$, $-\frac{\dot{c}_t}{2c_t}$ is negative whenever $\cot \frac{1}{4}t\sqrt{64\xi^2 - \kappa^2} < -\frac{\kappa}{\sqrt{64\xi^2 - \kappa^2}}$. We therefore see an abrupt transition between CP-divisible and non-CP-divisible dynamics across the line $\kappa^2 = 64\xi^2$, as shown in Fig. 6.2.

In the CP-divisible regime, the trace-distance between any two states subject to the evolution is always non-increasing [93]. This fact allows for the quantification of (two-time) non-Markovianity by integrating any increases in the trace distance over all time, which is shown in Ref. [180] to yield the analytic result

$$\mathcal{N}^{\text{Two-time}} = \frac{1}{\exp\left(\frac{\kappa \pi}{\sqrt{64\xi^2 - \kappa^2}}\right) - 1}, \tag{6.9}$$

for $\kappa^2 < 64\xi^2$ and zero otherwise.

However, as discussed, CP-divisibility does not imply Markovianity; as such, Fig. 6.2 does not provide a comprehensive picture of the many prevalent memory effects for different choices of parameters $\kappa$ and $\xi$. Here, we explicitly calculate the process tensor for the dynamics and show that it is non-Markovian for the entire parameter regime, before exploring the behaviour of the instrument-specific memory strength quantifiers introduced in the previous subsection.

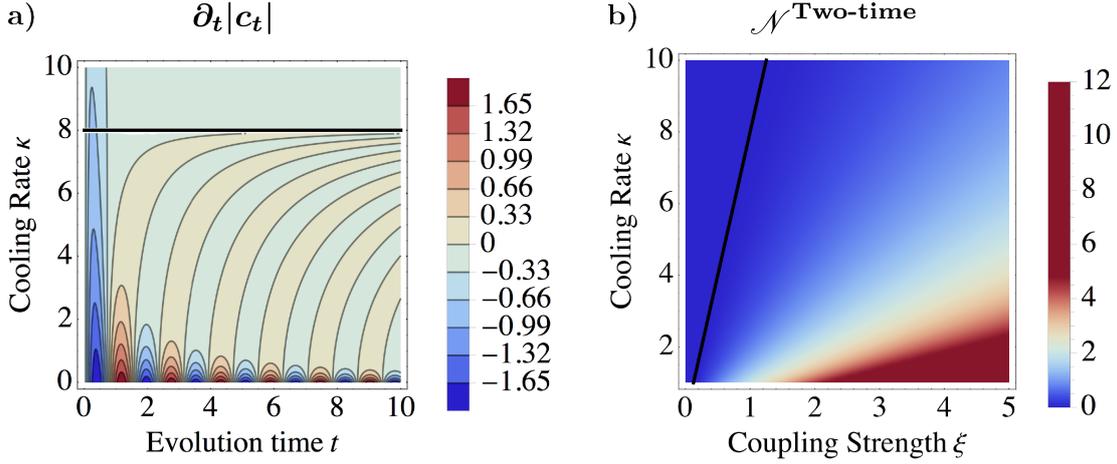

**Figure 6.2:** *Abrupt transition between CP-divisible and non-divisible dynamics.* In panel **a)**, we plot $\partial_t |c_t|$ with $\xi = 1$. As $\text{sgn}\left(\frac{\dot{c}_t}{c_t}\right) = \text{sgn}\left(\partial_t |c_t|\right)$, this implies the dynamics is CP-divisible for $\kappa \geq 8$, but not for $\kappa < 8$. In particular, there is an abrupt transition along the line $\kappa = 8$. In panel **b)**, we plot the two-time non-Markovianity $\mathcal{N}^{\text{Two-time}}$ as per Eq. (6.9). This is plotted in the parameter space $\xi \in [0, 5]$ and $\kappa \in [1, 10]$ to allow for comparison with Fig. 6.3. Note that this measure of non-Markovianity blows up exponentially for small $\kappa$, which is why we have excluded $\kappa \in [0, 1)$ from the plot, and that it vanishes for everything above the black line $\kappa = 8\xi$.

We consider a parameter grid $\xi \in [0, 5]$ and $\kappa \in [0, 10]$ with increments of 0.1 in each direction and construct the $n = 6$ step process tensor, $\Upsilon_{6:1}(\xi, \kappa)$. Here, for simplicity, we assume an initially uncorrelated system-environment state, such that the process tensor begins on an output space. We also choose uniform spacing between timesteps of $dt = 0.3$, which corresponds to the natural timescale over which the trace distance between arbitrary initial system states increases for most values in the parameter space [180]. This means that the final time of the process tensor is $T = 1.5$, which corresponds to where the CP-divisibility criteria would witness non-Markovianity for a range of parameters.

At each point, we can calculate the non-Markovianity in the process by considering the distance to the nearest Markovian process, as per Eq. (3.40). Here, we choose the (pseudo-)distance to be the relative entropy, $\mathcal{D}(\Upsilon_{6:1}(\xi, \kappa) \| \Upsilon_{6:1}^{\text{Markov}})$, in which case the minimum occurs for the Markovian process that is built up from the marginals of the original process tensor [181], *i.e.*, using the relative entropy circumvents the normally necessary minimisation. The corresponding results are depicted in Fig. 6.3, which indicates that the process is non-Markovian for all parameters in the chosen range. In particular, there is no abrupt transition between regimes. Although the non-Markovianity is small above the line $\kappa = 8\xi$—at which the dynamics transitions from CP-divisible to non-divisible—it is non-zero, indicating a weak but detectable memory. The two-time witness of non-Markovianity in Eq. (6.9) is insensitive to such effects, which leads to the abrupt transition between regimes; by capturing all multi-time correlations, the non-Markovianity calculated via the process tensor shows this transition to be artificial. This result begs the question: *how long does the memory persist?* We now move to study the behaviour of some of the memory quantifiers proposed in the previous subsection.

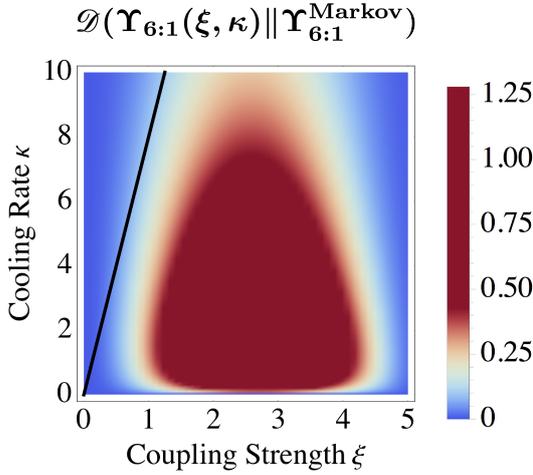

**Figure 6.3:** *Heatmap of non-Markovianity.* Non-Markovianity of $\Upsilon_{6:1}(\xi, \kappa)$. Although the non-Markovianity is small above the line $\kappa = 8\xi$, it is non-zero. Moreover, there is no abrupt transition between regimes, as all memory effects are accounted for.

To this end, we consider three fixed process tensors: one in the almost Markovian and CP-divisible regime, one in the intermediary regime, and one in the strongly non-Markovian regime, respectively defined as follows

$$\Upsilon^{\text{CP}} := \Upsilon_{6:1}(1, 10) \quad (6.10)$$
$$\Upsilon^{\text{Int}} := \Upsilon_{6:1}(1, 8)$$
$$\Upsilon^{\text{SNM}} := \Upsilon_{6:1}(1, 1).$$

We first consider, for fixed $\ell \in \{1, \ldots, 4\}$ ranging from $t_2$ to $t_5$, the natural and noise-resistant memory strength defined in Eqs. (6.4) and (6.5) respectively. We also consider the memory strength for length-$\ell$ sequences of causal breaks, by first constructing the collection of outcome-specific values in Eq. (6.1) and subsequently calculating the average and maximum memory strength in Eqs. (6.2) and (6.3) for the instrument. The causal break is chosen to be a symmetric single-qubit IC POVM (defined in Example D.1) followed by the independent repreparation (with uniform probability) into one of an IC set of states $\{|0\rangle\langle 0|, |1\rangle\langle 1|, |+\rangle^x\langle +|^x, |+\rangle^y\langle +|^y\}$, where $|+\rangle^{x/y}$ is the $+1$ eigenstate of $\sigma_x/\sigma_y$.

The results are summarised in Fig. 6.4. Interestingly, all three of these processes have vanishing memory strength with respect to the completely noisy instrument, meaning that an experimenter can erase the temporal correlations in the process by acting at a sin-gle timestep. The other quantities do not vanish and display stark differences. All three processes display the strongest memory strength with respect to the identity, indicating that a non-negligible amount of memory is transmitted on the level of the system alone. Moreover, it is the process tensor in the strongly non-Markovian regime that exhibits the strongest natural memory strength, with that in the intermediate regime displaying the weakest. The effects of active interventions come to light when we consider the causal break instrument. Here, the CP-divisible and intermediate process tensors display almost vanishing memory strength very quickly with respect to either aggregation, with only the strongly non-Markovian one exhibiting significant temporal correlations. This is not surprising, as a causal break acts to block the flow of information on the level of the system, which should effectively wipe-out any temporal correlations in a CP-divisible process. In all cases, the memory strength decays across longer memory blocks, as is intuitively expected.

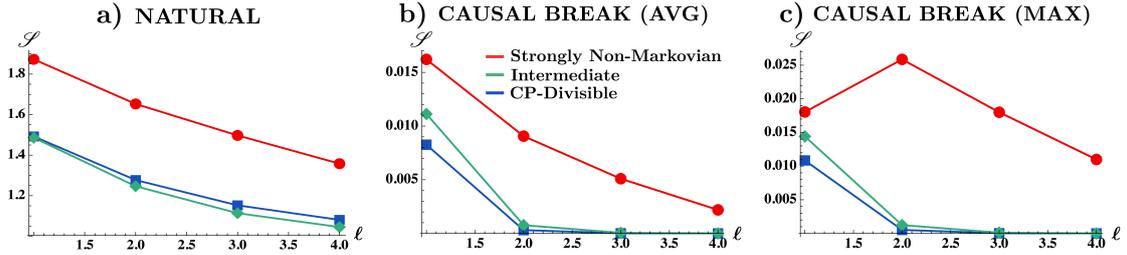

**Figure 6.4:** *Instrument-specific memory strength.* Here we plot the natural memory strength (panel **a)**), and both the average and maximum memory strength with respect to a causal break sequence (panels **b)** and **c)** respectively) as a function of $\ell$. Note the changing vertical scales and the legend in the centre. The strongly non-Markovian process (red) has the strongest memory strength with respect to sequences of the identity instrument, followed by that in the CP-divisible regime (blue) and lastly that in the intermediate regime (green). By comparing panel **a)** with panels **b)** and **c)**, we see the effect of active interventions: for the CP-divisible and intermediate processes, most of the memory strength arises by way of the identity instrument transmitting information on the level of the system; when active probing interventions such as those of a causal break stop this flow of information, their memory strength quickly becomes negligible. On the other hand, for the strongly non-Markovian processes, significant multi-time memory effects are exhibited for all instruments shown. Lastly note that the memory strength is not monotonic with respect to the maximum aggregation, as particularly strong memory effects can occur with small probability, as exhibited here for the strongly non-Markovian process at $\ell = 2$.

## 6.2 chapter summary

In this chapter we have proposed a number of instrument-specific definitions of memory strength. By accounting for all multi-time memory effects that are potentially present in a process, these definitions serve to unambiguously characterise how strong the temporal correlations between the history and the future are for any chosen sequence of instruments applied by an experimenter. Indeed, such instrument-specific notions of memory strength are directly relevant for experimental applications and computational simulation techniques: they provide an operational approach that permits making memory cutoff approximations of choice, rather than relying on the natural timescales of decay imposed by the system-environment dynamics, as are prevalent throughout numerical techniques [25]. Indeed, techniques of a similar flavour are being developed through extensions of the transfer tensor formalism [179, 182–185], bridging the connection between characterisation and efficient simulation of quantum processes with memory.

In studying the exactly solvable model, we showed a prime example of how this might be applied in practice, by examining how the behaviour of memory strength varies for different instruments and across different timescales. By tuning the parameters of the model appropriately, one could simulate dynamics that is amenable to short-time memory approximations with respect to sequences of causal breaks, for instance, by *e. g.,* constructing the CP-divisible process of Eq. (6.10). Indeed, the decay of memory effects over longer sequences of active interventions is related to various operational protocols such as dynamical decoupling [12], erasure of information or transmission of information [48, 49]. The results developed here for the simple two-qubit model are already interesting in their own right, and are suggestive of the possible insights that may be uncovered by considering the memory strength and related approximation techniques for appropriate choices of instruments in a variety of situations.

## Part IV

## ENVOI

You can trip off to places so wild and so wiggy that you don't know where you are until you get back. And sometimes not even know you tripped off at all because you never get back to know that you've left...

— Ken Kesey, *Sometimes a Great Notion.*

# 7

summary

Non-negligible memory effects play a crucial role in phenomena studied throughout a vast range of sciences—physical, biological, chemical, neurological, economical, ...; the list goes on. Many such effects are highly complex and their treatment lies beyond the cutting edge of current knowledge and experimental reach; on the other hand, when adequately understood, memory can be manipulated in order to develop state-of-the-art technologies. Thus, a proper theoretical description of processes with memory is of immense practical importance, marking *a kind of glory* that is deserving of pursuit.

To develop such a comprehension requires, quite naturally, a thorough understanding of the limitations of the prevailing descriptions. This was the focal point of Chapter 2: we first studied classical stochastic processes, which are well-understood and can easily be defined in an unambiguous way. This served as a point of reference against which open quantum evolution stands in juxtaposition when multi-time correlations are of importance. Up until recently, the murkiness surrounding quantum stochastic processes has led to ambiguity in defining many key concepts, culminating in—amongst other confusions—a myriad of incompatible definitions of memoryless quantum processes.

This comparison also made clear that the critical problem is one of formalism, rather than anything fundamental: two-time descriptions of stochastic processes, as they are used ubiquitously in the literature, are inadequate to describe general processes with memory. This is a statement of logic that transcends the physical theory to which it applies, holding equally true in both the classical and quantum realms. In Chapter 3 we saw that in order to capture multi-time correlations in a quantum process, due to the invasive nature of measurement at the nanoscale, we must clearly delineate between the role of the experimenter and that of the underlying physical evolution. This line of thinking naturally led us to a clear definition of quantum stochastic processes by way of the process tensor formalism.

summary

The process tensor ameliorates the aforementioned obstacles, at once providing a linear, completely-positive and trace-preserving (in the correct sense) description that completely captures any causally-ordered evolution on a discrete set of timesteps admissible by quantum theory. Its definition stems from the dilated joint unitary evolution, which offers an intuitive connection to the physical mechanism that drives the process. By abstracting the uncontrollable process at hand from the sequence of controllable interrogations an experimenter might choose to apply, this operational picture captures all possible multi-time statistics that can be deduced with respect to any valid probing schema conceivable. Thereby, it both unifies and generalises previous definitions of memorylessness for quantum processes.

Thus, Part II summarises the story so far, leading us to a *narrative on the edge*: with the correct tools for describing general stochastic processes with memory, we stand at the precipice of a univocal study of memory effects in quantum processes, which has hitherto proved elusive.

The previously *unspoken words* of this thesis follow in Part III. Since the general description of stochastic processes grows, with respect to the length of time considered, to rapidly defy reasonably available computational resources—both in classical and quantum theory—our step towards understanding processes with memory concentrates on examining those for which the memory length is finite in duration. Fortunately, such processes can be efficiently modelled by only taking into account information regarding the most recent $\ell$ timesteps, rather than the entirety of history, when making predictions. Classically, this scenario is captured by the concept of Markov order, which provides a characteristic timescale for the memory length of any stochastic process.

In Chaper 4 we extended Markov order to quantum mechanics. While the idea remains unchanged from the standard intuition, the corresponding phenomenology of quantum processes with finite memory is significantly richer and more complex than its classical counterpart. As in the classical case, we asked if the future evolution of the system can be deduced, in principle, entirely from the most recent sequence of measurement outcomes. However, for quantum stochastic processes, the Markov order—and therefore the memory length of the process—is fundamentally dependent on the instruments used to probe the process.

We formulated the conditions that capture this concise and comprehensive definition of memory length for quantum processes, which is fully reducible to its classical counterpart in the appropriate limit, in terms of a constraint on the process tensor. From this, we saw that no quantum process with memory can have finite Markov order with respect to all possible instruments.

In addition, our analysis highlighted that this dependence of Markov order on the interrogating instruments persists even in classical physics, as soon as active interventions on the system are permitted; in quantum theory, this issue is fundamental and must be acknowledged. In short, quantum processes with memory exhibit distinct memory effects when probed with different instruments. Our work provides the first formal classification of such behaviour.

In Chapter 5, we analysed the structure of quantum stochastic processes that display finite memory effects. We detailed the structural constraints that must be satisfied for the underlying process to have finite Markov order with respect to a given instrument sequence, including some specific natural and experimentally-relevant classes of probing instruments, such as unitary operations and an informationally-complete set of measurements and re-preparations. The connection between finite memory-length and the underlying system-environment dynamics was elucidated through a series of pedagogical examples, which served to outline a broad taxonomy of the memory effects possible in non-Markovian quantum processes. Lastly, we explored the relationship between quantum Markov order and vanishing quantum conditional mutual information, showing explicitly that although the latter implies the former, processes with finite quantum Markov order need not have vanishing quantum CMI.

In Chapter 6, we proposed instrument-specific measures of the memory strength for a quantum process. By explicitly analysing the behaviour of the memory strength for a simple model dynamics with respect to a range of instruments over varying timescales, we uncovered numerous interesting insights. For instance, the approaches developed here can be seen as an operational way to make finite-memory approximations to numerically simulate complex dynamics; or, from the perspective of information-processing, many of the notions discussed are related to finding the optimal sequence of operations to perform a given task. The intriguing results for the toy-model studied already allude to tantalising possibilities that might be harnessed by using such techniques to understand the strength and complexity of memory in quantum processes.

In conclusion, the results of this thesis solve the long-standing problem of extending the definition and quantification of memory effects from the classical to the quantum realm. We expect our approach to fundamentally shift the way memory in quantum processes is described by the community for a number of reasons, including—but not limited to—the following. For one, instrument-dependence of effects measured in quan-tum processes is close to experimental reality, as the boundary of what is experimentally accessible shifts continuously as technology evolves. Moreover, while not all real-world processes will display finite Markov order, many processes are likely do so approximately, and the developed insights will be of fundamental importance for their experimental reconstruction and simulation.

## 7.1 OUTLOOK

We now move to discussing some of the broader implications of our work. On the foundational side, it is clear that, upon the set of timesteps on which it is defined, the process tensor provides the most generic description of causally-ordered processes allowable within quantum theory. Examining properties of its structure, as we have in this thesis, provides fundamental insight into understanding the space of quantum processes and temporal correlations. Indeed, similar frameworks that do away with the axiom of causality, such as those based on the process matrix, are actively being developed to study the most general spatio-temporal correlations allowable [33, 34, 38], shedding light on the distinguishing features of classical, quantum and post-quantum theories.

On the practical side, the process tensor contains all the information one could ever hope to learn about a process. This, unfortunately, can make it computationally daunting to approach. In light of this, its usefulness lies in our ability to develop compression and extraction methods to approximate complex physical evolutions with overlapping process tensors of finite length for efficient simulation of long-term dynamics. Indeed, this is the flavour of many methods proposed throughout the literature, such as the transfer tensor approach [179, 182–185]. A deeper understanding of optimal compression and recovery schemes for processes with approximately finite memory length will have significant consequences for efficient quantum simulation.

Moreover, understanding memory effects has immediate relevance to developing near-term quantum technologies, particularly concerning the construction of error-correcting codes to combat correlated noise [65, 186–190] and the design of feedback protocols for coherent control [170, 171]. The far-reaching implications of these possible avenues of exploration highlight the substantial relevance of our novel approach for researchers interested in open quantum systems, quantum control, stochastic modelling, complexity science and quantum foundations, amongst other fields.

# Part V

# APPENDIX

The reverberation often exceeds through silence the sound that sets it off; the reaction occasionally outdoes by way of repose the event that stimulated it; and the past not uncommonly takes a while to happen, and some long time to figure out.

— Ken Kesey, *Sometimes a Great Notion*.

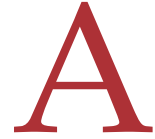

# NOTATION SUMMARY

The mathematical underpinnings of quantum theory involves linear operators on a Hilbert space. We restrict ourselves to finite-dimensional Hilbert spaces; as such, linear operators can be represented as matrices. The notational conventions employed throughout this thesis are summarised in Table A.1 below. Whenever it is unambiguous, to avoid notational clutter, we drop timestep or subsystem labels. For the same reason, we avoid brackets wherever possible, with maps acting on everything to the right of them by convention. An Hermitian operator $X$ is called positive (written $X \geq 0$) whenever its spectrum contains only non-negative values. We use $\log := \log_2$ and set $\hbar = 1$.

Lastly, note the colour-coding schema we employ throughout the figures in this thesis. We consistently use green to denote preparations, transformations and measurements that are controllable by an experimenter; yellow to denote those that are uncontrollable; and orange to represent components of the non-unique dilation of an uncontrollable process. When we wish to describe maps from an abstract mathematical perspective, we colour the components with red, blue or purple to minimise any possible confusion.

| GENERAL | |
|---|---|
| $\mathbb{C}, \mathbb{R}, \mathbb{N}$ | complex, real and natural numbers |
| $t_j, t_k, \ldots$ | lowercase letters used as subscripts represent timesteps |
| $k:j$ | an ordered sequence of timesteps $\{t_j, \ldots, t_k\}$ for $k > j \in \mathbb{N}$ |
| $\Lambda_n$ | a set of timesteps of cardinality $n \in \mathbb{N}$ (not necessarily an ordered sequence, *e.g.*, $\Lambda_3$ could denote $\{t_1, t_4, t_5\}$) |
| $\mathbb{P}_{k:j}, \mathbb{P}_{\Lambda_n}$ | a joint probability distribution defined on the timesteps indicated by the subscript |
| $\mathtt{i}, \mathtt{o}$ | input and output spaces associated to each timestep |
| $A, B, C, \ldots$ | systems are labelled with capital letters |
| $\mathcal{H}^A$ | Hilbert space associated to the system $A$ |
| $\dim(\mathcal{H}^A) =: d_A$ | dimension of the Hilbert space $\mathcal{H}^A$ |
| $\langle \bullet \vert, \vert \bullet \rangle$ | bra and ket |
| $\mathrm{tr}, \mathrm{tr}_A$ | trace and partial trace over $\mathcal{H}^A$ |
| $A^*, A^\mathrm{T}, A^\dagger$ | conjugate, transpose and conjugate transpose of $A$ |
| $\mathrm{Spec}(A)$ | spectrum of $A$ |
| $[A, B]$ | commutator between $A$ and $B$, *i.e.*, $[A, B] := AB - BA$ |
| $\{A, B\}$ | anti-commutator between $A$ and $B$, *i.e.*, $\{A, B\} := AB + BA$ |
| $A \otimes B$ | tensor product of $A$ and $B$ |
| $A \oplus B$ | direct sum of $A$ and $B$ |

OPERATORS

| | |
|---|---|
| $\mathsf{BL}(\mathcal{H}^A)$ | set of bounded linear operators on $\mathcal{H}^A$ |
| $\mathcal{A}, \mathcal{B}, \mathcal{C} \ldots$ | maps are denoted by calligraphic capital letters and act on everything to the right of them |
| $\mathsf{A}, \mathsf{B}, \mathsf{C} \ldots$ | the Choi operators of maps are denoted by their sans-serif counterparts |
| $\mathcal{I}^A, \mathbb{1}^A$ | identity map and identity operator on $\mathcal{H}^A$ respectively |

ENTROPIES

| | |
|---|---|
| $H^\mathsf{cl}(\mathbb{P})$ | Shannon entropy of the probability distribution $\mathbb{P}$, *i.e.*, $H^\mathsf{cl}(\mathbb{P}) := -\sum_{x \in \mathscr{X}} \mathbb{P}(x) \log \mathbb{P}(x)$ |
| $I^\mathsf{cl}(A : B)$ | classical mutual information between $A$ and $B$, *i.e.*, $I^\mathsf{cl}(A : B) := H(\mathbb{P}_A) + H(\mathbb{P}_B) - H(\mathbb{P}_{AB})$ |
| $I^\mathsf{cl}(A : C \vert B)$ | classical conditional mutual information between $A$ and $C$ given $B$, *i.e.*, $I^\mathsf{cl}(A : C \vert B) := H^\mathsf{cl}(\mathbb{P}_{AB}) + H^\mathsf{cl}(\mathbb{P}_{BC}) - H^\mathsf{cl}(\mathbb{P}_{ABC}) - H^\mathsf{cl}(\mathbb{P}_B)$ |
| $\mathscr{D}^{\mathsf{cl}}(\mathbb{P}\|\mathbb{Q})$ | relative entropy (Kullback-Liebler divergence) between two distributions $\mathbb{P}$ and $\mathbb{Q}$, *i.e.*, $\mathscr{D}^{\mathsf{cl}}(\mathbb{P}\|\mathbb{Q}) := -\sum_{x \in \mathscr{X}} \mathbb{P}(x) \log \frac{\mathbb{P}(x)}{\mathbb{Q}(x)}$ |
| $S(\rho)$ | von Neumann entropy of the density operator $\rho$, *i.e.*, $S(\rho) := -\operatorname{tr}[\rho \log \rho] = -\sum_{\lambda \in \operatorname{Spec}(\rho)} \lambda \log \lambda$ |
| $I(A:B)$ | quantum mutual information between $A$ and $B$, *i.e.*, $I(A:B) := S(\rho^A) + S(\rho^B) - S(\rho^{AB})$ |
| $I(A:C|B)$ | quantum conditional mutual information between $A$ and $C$ given $B$, *i.e.*, $I(A:C|B) := S(\rho^{AB}) + S(\rho^{BC}) - S(\rho^{ABC}) - S(\rho^B)$ |
| $\mathscr{D}(\rho\|\sigma)$ | quantum relative entropy between $\rho$ and $\sigma$, *i.e.*, $\mathscr{D}(\rho\|\sigma) := \operatorname{tr}[\rho \log \rho - \rho \log \sigma]$ |

OTHER / EXCEPTIONS

| | |
|---|---|
| $H, M, F$ | reserved to denote the collections of timesteps grouped as the history $\{t_1, \ldots, t_{k-\ell-1}\}$, memory $\{t_{k-\ell}, \ldots, t_{k-1}\}$ and the future $\{t_k, \ldots, t_n\}$ at arbitrary timestep $t_k$ for a memory of length $\ell$ |
| $U, V$ | reserved to denote the unitary matrices associated to unitary maps, *i.e.*, $\mathcal{U}(\bullet) := U(\bullet)U^\dagger$ |
| $\mathbf{W}, \mathbf{W}^\perp$ | boldface capital letter represents a vector space, with the $\perp$ superscript denoting its orthogonal complement |
| $\hat{\rho}$ | a caret is used to denote elements of a fixed basis of a vector space |
| $\mathcal{J}$ | an instrument is a collection of completely-positive maps that sums to a completely-positive and trace-preserving map (these can be higher order objects defined across multiple timesteps, referred to as instrument sequences or testers) |
| $\Upsilon$ | the Choi operator of the process tensor whose mapping is represented by $\mathcal{T}$ |
| $\widetilde{\Upsilon}^{(x)}$ | a tester element of the process tensor $\Upsilon$, *i.e.*, $\sum_x \widetilde{\Upsilon}^{(x)} = \Upsilon$, with $\Upsilon$ a proper process tensor |
| $\underline{\Upsilon}$ | the projection of $\Upsilon \in \mathbf{W} \oplus \mathbf{W}^\perp$ on the $\mathbf{W}^\perp$ subspace |
| $\psi^+, \psi^-, \phi^+, \phi^-$ | normalised maximally entangled two-party Bell states |
| $\mathbf{\Psi}^+, \mathbf{\Psi}^-, \mathbf{\Phi}^+, \mathbf{\Phi}^-$ | unnormalised maximally entangled two-party Bell operators, which are respectively the Choi operators corresponding to the channels that implement the four Pauli rotations $\mathcal{I}, \mathcal{X}, \mathcal{Y}, \mathcal{Z}$ |
| $X, x, \mathscr{X}; Y, y, \mathscr{Y}$ | reserved to denote random variables (capital Roman) and realisations of said random variable (lowercase), which take values from some set (capital script) |
| $\mathtt{H}, \mathtt{T}$ | heads and tails, symbolic of the possible outcomes of any binary experiment |
| $\delta$ | the Dirac-delta distribution |

**Table A.1:** *Notational conventions.* A summary of the notational conventions employed.

# B

## CLASSICAL AND QUANTUM DYNAMICS WITH NOISE

### B.1 CLASSICAL MASTER EQUATION

Given a Markovian process, we can consider the first three timesteps and immediately marginalise over the intermediate variable of the joint distribution to derive

$$\mathbb{P}_{3,1}(x_3, x_1) = \sum_{x_2} \mathbb{P}_{3:1}(x_3, x_2, x_1) \tag{B.1}$$

$$= \sum_{x_2} \mathbb{P}_3(x_3|x_2)\mathbb{P}_2(x_2|x_1)\mathbb{P}_1(x_1).$$

Dividing both sides by $\mathbb{P}_1(x_1)$ yields the *Chapman-Kolmogorov* equation [5]

$$\mathbb{P}_3(x_3|x_1) = \sum_{x_2} \mathbb{P}_3(x_3|x_2)\mathbb{P}_2(x_2|x_1). \tag{B.2}$$

This equation expresses that for a Markovian process that begins with value $x_1$ at $t_1$ and reaches $x_3$ at $t_3$, it must do so in a manner that is specified given knowledge of the value $x_2$ at the intermediate timestep $t_2$.

We can recast this property into an integro-differential equation that the transition probabilities for any Markovian process must satisfy, known as the ME. To do so, for the moment we consider time as a continuous parameter. We further assume that the process is homogeneous in time, meaning that the conditional distributions connecting adjacent timesteps do not explicitly depend on the times themselves, rather, only their difference. This allows us to define

$$\mathbb{P}_2(x_2|x_1) =: S_{\tau'}(x'|x'') \quad \text{and} \quad \mathbb{P}_3(x_3|x_2) =: S_{\tau}(x|x'), \tag{B.3}$$

where $\tau' := t_2 - t_1$, $\tau := t_3 - t_2$, and we have dropped the (now redundant) subscript labels on the outcomes, with the understanding that in each transition probability, the conditioning argument always represents a value measured prior to the remaining argument by the time difference denoted, *i. e.,* $S_\tau(x|x')$ refers to the probability that $x$ will be observed given that $x'$ was the outcome $\tau$ units of time ago. The Chapman-Kolmogorov equation is then expressed as

$$S_{\tau+\tau'}(x|x'') = \sum_{x'} S_\tau(x|x') S_{\tau'}(x'|x''). \tag{B.4}$$

Considering the interval $\tau$ to be small and expanding $S_\tau(x|x')$ about $\tau = 0$ gives

$$S_\tau(x|x') = (1 - a(x)\tau)\delta(x, x') + \tau W(x|x') + \mathcal{O}(\tau^2), \tag{B.5}$$

where $(1 - a(x)\tau)$ represents the probability that the state does not transition from $x'$ to $x$ during the short time interval $\tau$, $W(x|x') := \frac{\partial S_\tau(x|x')}{\partial \tau}|_{\tau=0} \geq 0$ is the instantaneous rate for this transition to occur, and $\mathcal{O}(\tau^2)$ represents higher-order terms that eventually vanish upon taking the limit $\tau \to 0$. By the normalisation constraint on $S_\tau(x|x')$, it follows that

$$a(x) = \sum_{x'} W(x'|x). \tag{B.6}$$

Substituting Eqs. (B.5) and (B.6) into Eq. (B.4) and subsequently taking the limit $\tau \to 0$ yields the classical ME

$$\frac{\partial S_t(x|x'')}{\partial t} = \sum_{x'} \left\{ W(x|x') S_t(x'|x'') - W(x'|x) S_t(x|x'') \right\}. \tag{B.7}$$

## B.2 TOMOGRAPHY OF A DYNAMICAL MAP

The tomographic representation of a dynamical map relies on the concept of duals. Considering the tomographic protocol outlined in Subsection 2.2.2, we begin with a basis set of states $\{\hat{\rho}^{(i)}\}$. Although these need not be orthonormal, they are linearly independent, and hence there exists a dual set of objects $\{\hat{D}^{(i)}\}$ such that

$$\text{tr}\left[\hat{D}^{(i)\dagger} \hat{\rho}^{(j)}\right] = \delta_{ij} \quad \forall \, i, j. \tag{B.8}$$

These can be constructed explicitly as follows [52, 75]. Begin by writing $\hat{\rho}^{(i)} = \sum_j h_{ij} \Gamma^{(j)}$, where $h_{ij} \in \mathbb{C}$ and $\{\Gamma^{(j)}\}$ are a Hermitian, self-dual linearly independent set of operators satisfying $\text{tr}\left[\Gamma^{(i)} \Gamma^{(j)}\right] = 2\delta_{ij}$; for instance, these could be chosen as the generalised Pauli basis. Since $\{\hat{\rho}^{(i)}\}$ are linearly independent, the columns of the matrix $H = \sum_{ij} h_{ij}|i\rangle\langle j|$ are linearly independent vectors, which means that $H$ is invertible. Let $F^\dagger = H^{-1}$ so that $HF^\dagger = \mathbb{1}$, implying that the columns of $F^*$ are orthonormal to those of $H$. Finally, define $\hat{D}^{(i)} = \frac{1}{2}\sum_j f_{ij} \Gamma^{(j)}$, where $f_{ij}$ are the entries of $F$.

For example, in the case of a qubit system, we can use

$$\hat{\rho}^{(1)} = \frac{1}{2}\begin{bmatrix} 1 & 1 \\ 1 & 1 \end{bmatrix}, \qquad \hat{\rho}^{(2)} = \frac{1}{2}\begin{bmatrix} 1 & -i \\ i & 1 \end{bmatrix}, \qquad \text{(B.9)}$$

$$\hat{\rho}^{(3)} = \begin{bmatrix} 1 & 0 \\ 0 & 0 \end{bmatrix}, \qquad \hat{\rho}^{(4)} = \frac{1}{2}\begin{bmatrix} 1 & -1 \\ -1 & 1 \end{bmatrix}.$$

Although they are not orthonormal, these matrices are linearly independent and span the space of qubits. The dual set to the basis defined in Eq. (B.9) is given by

$$\hat{D}^{(1)} = \frac{1}{2}\begin{bmatrix} 0 & 1+i \\ 1-i & 2 \end{bmatrix}, \qquad \hat{D}^{(2)} = \begin{bmatrix} 0 & -i \\ i & 0 \end{bmatrix}, \qquad \text{(B.10)}$$

$$\hat{D}^{(3)} = \begin{bmatrix} 1 & 0 \\ 0 & -1 \end{bmatrix}, \qquad \hat{D}^{(4)} = \frac{1}{2}\begin{bmatrix} 0 & -1+i \\ -1-i & 2 \end{bmatrix}.$$

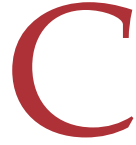

# MEMORY LENGTH

## C.1 QUANTUM MARKOV ORDER CONSTRAINT ON PROCESS TENSOR

Here we prove that Def. 4.1 is equivalent to Eq. (4.4). Explicitly writing out the conditioning over outcomes stipulated in Eq. (4.1) gives

$$\frac{\mathbb{P}_{FMH}(x_F, x_M, x_H | \mathcal{J}_F, \mathcal{J}_M, \mathcal{J}_H)}{\sum_{x_F} \mathbb{P}_{FMH}(x_F, x_M, x_H | \mathcal{J}_F, \mathcal{J}_M, \mathcal{J}_H)} \qquad (\text{C.1})$$
$$= \frac{\sum_{x_H} \mathbb{P}_{FMH}(x_F, x_M, x_H | \mathcal{J}_F, \mathcal{J}_M, \mathcal{J}_H)}{\sum_{x_F x_H} \mathbb{P}_{FMH}(x_F, x_M, x_H | \mathcal{J}_F, \mathcal{J}_M, \mathcal{J}_H)}.$$

On the l.h.s, we can immediately make use of the causal structure of the process tensor to simplify the denominator: since the choice of instruments in the future cannot overall influence the statistics on the history and memory blocks, we have, for any $\mathcal{J}_F$,

$$\sum_{x_F} \mathbb{P}_{FMH}(x_F, x_M, x_H | \mathcal{J}_F, \mathcal{J}_M, \mathcal{J}_H) = \mathrm{tr}\left[\left(\mathsf{O}_F \otimes \mathsf{O}_M^{(x_M)} \otimes \mathsf{O}_H^{(x_H)}\right)^{\mathrm{T}} \Upsilon_{FMH}\right]$$
$$= \mathrm{tr}\left[\left(\mathsf{O}_M^{(x_M)} \otimes \mathsf{O}_H^{(x_H)}\right)^{\mathrm{T}} \Upsilon_{MH}\right], \qquad (\text{C.2})$$

where $\Upsilon_{MH} := \frac{1}{d_{F^\circ}} \mathrm{tr}_F[\Upsilon_{FMH}]$ with $d_{F^\circ} := \prod_{j=k}^{n} d_{j^\circ}$ denoting the joint dimension of the output spaces associated to $F$ and we employ the previously introduced notation for the CPTP map corresponding to an instrument $\mathsf{O}_X := \sum_{x_X} \mathsf{O}^{(x_X)}$ (or, more precisely, the overall deterministic comb corresponding to the tester implemented on the future). Note that we have dropped the explicit labelling of the instrument that the CPTP map $\mathsf{O}_X$ is associated with for compactness; of course, different instruments (in general) correspond to different overall CPTP maps. The fact that $\Upsilon_{MH}$ is a proper, *i. e.,* causally-ordered, process tensor can be seen by simply following the hierarchy of trace conditions assumed to be satisfied by $\Upsilon_{FMH}$ from the future backwards (see Eq. (3.23)). We cannot use a similar trick on the numerator of the r.h.s, since, in general, the statistics observed over the memory and future timesteps depend upon the choice of instrument implemented across the history; although the denominator can be simplified in a likewise manner. Then, expressing Eq. (C.1) in terms of the process tensor and making use of Eq. (C.2), we have

$$\frac{\text{tr}\left[\left(\mathsf{O}_F^{(x_F)} \otimes \mathsf{O}_M^{(x_M)} \otimes \mathsf{O}_H^{(x_H)}\right)^{\text{T}} \Upsilon_{FMH}\right]}{\text{tr}\left[\left(\mathsf{O}_M^{(x_M)} \otimes \mathsf{O}_H^{(x_H)}\right)^{\text{T}} \Upsilon_{MH}\right]} \tag{C.3}$$

$$= \frac{\text{tr}\left[\left(\mathsf{O}_F^{(x_F)} \otimes \mathsf{O}_M^{(x_M)} \otimes \mathsf{O}_H\right)^{\text{T}} \Upsilon_{FMH}\right]}{\text{tr}\left[\left(\mathsf{O}_M^{(x_M)} \otimes \mathsf{O}_H\right)^{\text{T}} \Upsilon_{MH}\right]}.$$

The tensor product structure of Eq. (4.4) is clearly a sufficient condition for Eq. (4.1). Note that the l.h.s of Eq. (C.3) represents $\mathbb{P}(x_F|\mathcal{J}_F; x_M, \mathcal{J}_M; x_H, \mathcal{J}_H)$; considering performing the trace over $M$ first yields

$$\mathbb{P}(x_F|\mathcal{J}_F; x_M, \mathcal{J}_M; x_H, \mathcal{J}_H) \tag{C.4}$$

$$= \frac{\text{tr}_{FH}\left[\left(\mathsf{O}_F^{(x_F)} \otimes \mathsf{O}_H^{(x_H)}\right)^{\text{T}} \text{tr}_M\left[\mathsf{O}_M^{(x_M)\text{T}} \Upsilon_{FMH}\right]\right]}{\text{tr}_H\left[\mathsf{O}_H^{(x_H)\text{T}} \text{tr}_M\left[\mathsf{O}_M^{(x_M)\text{T}} \Upsilon_{MH}\right]\right]}$$

$$= \frac{\text{tr}_{FH}\left[\left(\mathsf{O}_F^{(x_F)} \otimes \mathsf{O}_H^{(x_H)}\right)^{\text{T}} \Upsilon_F^{(x_M)} \otimes \widetilde{\Upsilon}_H^{(x_M)}\right]}{\text{tr}_H\left[\mathsf{O}_H^{(x_H)\text{T}} \widetilde{\Upsilon}_H^{(x_M)}\right]}$$

$$= \text{tr}_F\left[\mathsf{O}_F^{(x_F)\text{T}} \Upsilon_F^{(x_M)}\right] = \mathbb{P}(x_F|\mathcal{J}_F; x_M, \mathcal{J}_M).$$

We now consider the converse direction. Eq. (C.3) must hold for all instruments $\mathcal{J}_H$, and since we can vary the CP maps while keeping the overall CPTP map of the instrument fixed, this implies that we must have

$$\frac{\text{tr}\left[\left(\mathsf{O}_F^{(x_F)} \otimes \mathsf{O}_M^{(x_M)} \otimes \mathsf{O}_H^{(x_H)}\right)^{\text{T}} \Upsilon_{FMH}\right]}{\text{tr}\left[\left(\mathsf{O}_M^{(x_M)} \otimes \mathsf{O}_H^{(x_H)}\right)^{\text{T}} \Upsilon_{MH}\right]} \tag{C.5}$$

$$= \frac{\text{tr}\left[\left(\mathsf{O}_F^{(x_F)} \otimes \mathsf{O}_M^{(x_M)} \otimes \mathsf{O}_H'^{(x_H')}\right)^{\text{T}} \Upsilon_{FMH}\right]}{\text{tr}\left[\left(\mathsf{O}_M^{(x_M)} \otimes \mathsf{O}_H'^{(x_H')}\right)^{\text{T}} \Upsilon_{MH}\right]},$$

for all CP maps $\mathsf{O}_H^{(x_H)}, \mathsf{O}_H'^{(x_H')}$.

We can simplify the numerator on both sides by defining the conditional future process tensor

$$\Upsilon_F^{(x_M, x_H)} := \frac{\text{tr}_{MH}\left[\left(\mathbb{1}_F \otimes \mathsf{O}_M^{(x_M)} \otimes \mathsf{O}_H^{(x_H)}\right)^{\text{T}} \Upsilon_{FMH}\right]}{\mathbb{P}(x_M, x_H|\mathcal{J}_M, \mathcal{J}_H)}, \tag{C.6}$$

where

$$\mathbb{P}(x_M, x_H | \mathcal{J}_M, \mathcal{J}_H) = \text{tr}\left[\left(\mathsf{O}_M^{(x_M)} \otimes \mathsf{O}_H^{(x_H)}\right)^{\text{T}} \Upsilon_{MH}\right]. \tag{C.7}$$

This leads to

$$\text{tr}\left[\mathsf{O}_F^{(x_F)\text{T}} \Upsilon_F^{(x_M, x_H)}\right] = \text{tr}\left[\mathsf{O}_F^{(x_F)\text{T}} \Upsilon_F^{(x_M, x'_H)}\right]. \tag{C.8}$$

Since this must hold true for all future instruments $\mathcal{J}_F$ and the trace corresponds to an inner product on the space on which the Choi operators are defined, we finally have

$$\Upsilon_F^{(x_M, x_H)} = \Upsilon_F^{(x_M, x'_H)} \quad \forall\, x_H, x'_H, \tag{C.9}$$

which implies that the conditional future process tensor defined in Eq. (C.6) is independent of the historic outcomes, *i.e.*, $\Upsilon_F^{(x_M, x_H)} = \Upsilon_F^{(x_M, x'_H)} = \Upsilon_F^{(x_M)}$.

Returning to Eq. (C.6) and considering first taking the partial trace over $M$, we have

$$\frac{\text{tr}_H\left[\left(\mathbb{1}_F \otimes \mathsf{O}_H^{(x_H)}\right)^{\text{T}} \text{tr}_M\left[\mathsf{O}_M^{(x_M)\text{T}} \Upsilon_{FMH}\right]\right]}{\text{tr}_H\left[\mathsf{O}_H^{(x_H)\text{T}} \text{tr}_M\left[\mathsf{O}_M^{(x_M)\text{T}} \Upsilon_{MH}\right]\right]}. \tag{C.10}$$

It is clear from the expression above that in order for it to be independent of the outcome $x_H$, it must be the case that $\text{tr}_M\left[\mathsf{O}_M^{(x_M)\text{T}} \Upsilon_{FMH}\right]$ splits into a tensor product, *i.e.*, $\text{tr}_M\left[\mathsf{O}_M^{(x_M)\text{T}} \Upsilon_{FMH}\right] = \Upsilon_F^{(x_M)} \otimes \widetilde{\Upsilon}_H^{(x_M)}$, where

$$\Upsilon_F^{(x_M)} := \frac{\text{tr}_{MH}\left[\mathsf{O}_M^{(x_M)\text{T}} \Upsilon_{FMH}\right]}{\text{tr}\left[\mathsf{O}_M^{(x_M)\text{T}} \Upsilon_{MH}\right]}, \tag{C.11}$$

and

$$\widetilde{\Upsilon}_H^{(x_M)} := \text{tr}_{FM}\left[\mathsf{O}_M^{(x_M)\text{T}} \Upsilon_{FMH}\right]. \tag{C.12}$$

Note that, by construction, the conditional future process defined in Eq. (C.11) is a proper process tensor for each $x_M$, *i.e.*, a positive semi-definite operator satisfying the hierarchy of trace conditions of Eq. (3.23). Indeed, it is easy to see that the denominator normalises the expression such that for each $x_M$ we have $\text{tr}\left[\Upsilon_F^{(x_M)}\right] = d_{F^\circ}$ (which is necessary for satisfaction of Eq. (3.23)). Conversely satisfaction of Eq. (3.23) for the process tensor $\Upsilon_{FMH}$ means that $\Upsilon_{MH}$ is a well-defined, proper process tensor. On the other hand, such a normalisation for the historic part of the process cannot be acheived, since an object such as $\text{tr}_H[\Upsilon_{FMH}]$ does not represent a well-defined process tensor, as it implicitly dictates a specific choice of instrument on the history (namely, the sequence of information trashing identity operators) that influence the future statistics. This is in contrast to the definition of the conditional future process, where causality ensures that any such implicit choice of instrument on the future is irrelevant to the statistics measured on the history, which permits for the appropriate normalisation to ensure that each $\Upsilon_F^{(x_M)}$ is a proper process tensor. Thus, $\widetilde{\Upsilon}_H^{(x_M)}$ is not necessarily a proper process tensor. Nonetheless, from this object we can calculate the probability to measure certain outcomes on the memory block given a choice of instrument on the memory and the history

$$\mathbb{P}(x_M|\mathcal{J}_M, \mathcal{J}_H) = \text{tr}\left[\mathsf{O}_H^\text{T} \widetilde{\Upsilon}_H^{(x_M)}\right]. \tag{C.13}$$

By the normalisation of total probability, we have that $1 = \sum_{x_M} \mathbb{P}(x_M|\mathcal{J}_M, \mathcal{J}_H) = \sum_{x_M} \text{tr}\left[\mathsf{O}_H^\text{T} \widetilde{\Upsilon}_H^{(x_M)}\right]$, which must hold for all CPTP $\mathsf{O}_H$. This implies that when summed over outcomes, the conditional history process forms a proper process tensor, *i.e.*, $\sum_{x_M} \widetilde{\Upsilon}_H^{(x_M)} = \Upsilon_H^{(x_M)}$ satisfies Eq. (3.23). In other words, each $\widetilde{\Upsilon}_H^{(x_M)}$ is a post-selected tester element of the process tensor describing the history, which is precisely what is indicated by the tilde.

## C.2 DEMANDING FINITE QUANTUM MARKOV ORDER FOR ALL INSTRUMENTS IMPLIES MARKOVIANITY

Here we prove Theorem 4.2. We begin with the following Lemma:

**Lemma C.1.** *The only operators $\Upsilon_{FMH}$ that satisfy Eq. (4.4) for all possible instruments $\mathcal{J}_M$ are those where the M subsystem is in tensor product with F or H (or both).*

Choose a linearly independent, IC set of operators $\mathcal{J}_M = \{\mathsf{O}_M^{(x)}\}$ as the instrument on $M$. Any linearly independent set has an associated dual set of operators $\{\Delta_M^{(y)}\}$ such that $\text{tr}\left[\mathsf{O}_M^{(x)} \Delta_M^{(y)\dagger}\right] = \delta_{xy} \ \forall \ x, y$. Thus, we can write any tripartite state satisfying Eq. (4.4) for each measurement outcome as follows

$$\Upsilon_{FMH} = \sum_x \Upsilon_F^{(x)} \otimes \Delta_M^{(x)*} \otimes \widetilde{\Upsilon}_H^{(x)}. \tag{C.14}$$

Now, consider a different instrument comprising a set of projectors defined via a linear expansion of the original set $\mathcal{J}'_M = \{\mathsf{Q}_M^{(y)} := \sum_x q_{xy} \mathsf{O}_M^{(x)}\}$, with $\{q_{xy}\}$ some non-trivial coefficients. The conditional process upon realisation of any outcome of this instrument is

$$\widetilde{\Upsilon}_{FH}^{(y)} = \text{tr}_M \left[ \mathsf{Q}_M^{(y)\text{T}} \Upsilon_{FMH} \right] \tag{C.15}$$
$$= \sum_x \Upsilon_F^{(x)} \otimes \widetilde{\Upsilon}_H^{(x)} \, \text{tr}\left[ \mathsf{Q}_M^{(y)\text{T}} \Delta_M^{(x)*} \right] = \sum_x q_{xy} \Upsilon_F^{(x)} \otimes \widetilde{\Upsilon}_H^{(x)}.$$

This gives a conditional product state iff either $q_{xy} = \delta_{xy} \,\forall\, x, y$, which is false by construction; or, either $\Upsilon_F^{(x)}$ or $\widetilde{\Upsilon}_H^{(x)}$ (or both) are independent of $x$. Since the original choice of linearly independent CP maps was arbitrary the proof holds for arbitrary instruments on $M$. The only remaining way to satisfy Eq. (4.4) is if either the $F$ or $H$ (or both) parts of the process tensor are in tensor product with the part on $M$.

The proof of Theorem 4.2 is immediate from Lemma C.1, once we consider the fact that the Markov order condition must hold for any block $M$ of length $\ell$. Recall that any such memory block can begin and end on either input or output Hilbert spaces separated by $\ell$ timesteps; for concreteness, here we consider the scenario where $M$ begins and ends on an output Hilbert space, with the proofs for the other cases following the same logic. Consider first the the block $M$ to begin at timestep $t_{k-\ell}$ and end at timestep $t_{k-1}$. Without loss of generality, suppose that, by Lemma C.1, the process tensor factorises into the product $\Upsilon_{n:1} = \Upsilon_{n:k-\ell^\circ} \otimes \widetilde{\Upsilon}_{k-\ell^\text{i}:1}$. Had we chosen the block $M$ to begin one timestep later, the same condition leads to the product $\Upsilon_{n:1} = \Upsilon_{n:k-\ell+1^\circ} \otimes \widetilde{\Upsilon}_{k-\ell+1^\text{i}:1}$. The only way for a single process to satisfy both of these conditions is if there is a CPTP channel $\mathsf{C}_{k-\ell+1^\text{i}:k-\ell^\circ}$ taking whatever an experimenter feeds into the process at timestep $t_{k-\ell}$ to the subsequent output from the process at the next timestep $t_{k-\ell+1}$: $\Upsilon_{n:1} = \Upsilon_{n:k-\ell+1^\circ} \otimes \mathsf{C}_{k-\ell+1^\text{i}:k-\ell^\circ} \otimes \widetilde{\Upsilon}_{k-\ell^\text{i}:1}$. Repeating this argument for all timesteps of the process immediately leads to the Markovian (product) process tensor structure of Eq. (3.36).

## C.3 CLASSICAL MARKOV ORDER WITH FUZZY MEASUREMENTS

The fact that such coarse-graining can increase the memory length observed by an experimenter arises from the well-known property that the space of Markovian processes is not convex, as exhibited in the following example.

**Example C.1** (Fuzzy measurements can increase classical Markov order). Consider the classical process depicted in the left panel of Fig. C.1. At each timestep $t_k$, the system of interest is described by one of three distinct states $x_k \in \{a_k, b_k, c_k\}$. Between each step of dynamics, the time-invariant transition probabilities are given by $\mathbb{P}_k(b_k|a_{k-1}) = \mathbb{P}_k(c_k|b_{k-1}) = 1$, $\mathbb{P}_k(a_k|c_{k-1}) = p$, $\mathbb{P}_k(b_k|c_{k-1}) = 1-p$ (with $p \in (0,1)$) and all

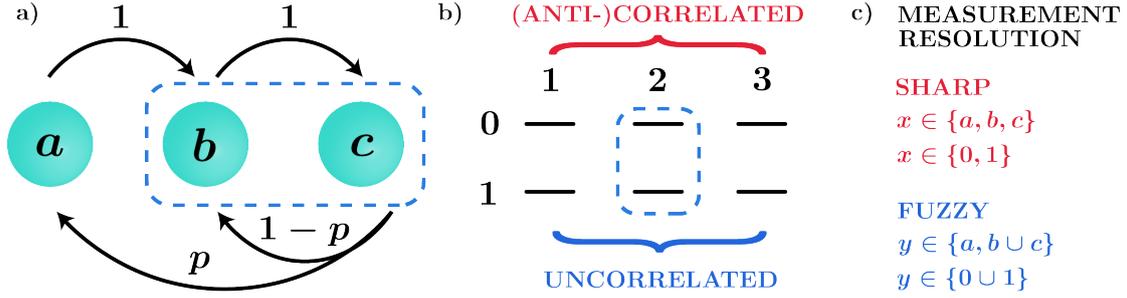

**Figure C.1:** *Instrument-dependence of classical Markov order with fuzzy measurements.* Here we depict two classical processes to highlight the instrument-dependence of Markov order when sharp measurements are not assumed. In panel **a)**, the process of Example C.1 is shown, defined by the transition probabilities depicted at each timestep. Here, if one is able to record observations sharply, *i.e.*, measure the values $x \in \{a, b, c\}$, the process is Markovian; however, if one cannot measure at that resolution and, *e.g.*, the measurement apparatus only records fuzzy statistics of $y \in \{a, b \cup c\}$, as depicted by the blue dashed box, the process would be classified as non-Markovian. In panel **b)**, the process of Example C.2 is shown. Here, three bits are initially prepared as described in the text, and each bit fed out of the process at successive timesteps. The preparation is such that if the second bit is sharply measured to be in the state 0, bits 1 and 3 are perfectly correlated; if the second bit is in state 1, bits 1 and 3 are perfectly anti-correlated; whilst on average, *i. e.,* with respect to the fuzzy measurement coarse-graining over everything in the dashed blue box, bits 1 and 3 are completely uncorrelated. Note that a legend is provided in the rightmost panel **c)**.

other transitions are forbidden. Such a process is clearly Markovian, as knowledge of any current state suffices to deduce the probability of the next. Now suppose that, an experimenter could not distinguish between outcomes $b_k$ and $c_k$, *i.e.*, instead of $x_j$, they observe $y_k \in \{a_k, d_k = b_k \cup c_k\}$. In this case, when the state at some time is $a$, the next state is for sure $d$; while if the state is $d$, with probability $p$ it will transition next to $a$ or with probability $1 - p$ it will remain $d$ (alternating between $b$ and $c$ deterministically, although the experimenter is ignorant of this fact). Conditioned on any consecutive sequence of $j$ observations of outcome $d$ following an observation of $a$, we have

$$\mathbb{P}_k(a_k|d_{k-1}, \ldots, d_{k-j}, a_{k-j-1}) = \begin{cases} 0 & \text{j odd} \\ p & \text{j even,} \end{cases} \tag{C.16}$$

which is different from $\mathbb{P}_k(a_k|d_{k-1}) = p$. Thus, with only the fuzzy measurement apparatus at hand, the experimenter would consider the process to be non-Markovian. Lastly note that given a faulty instrument that alternatively measures $x$ and then $y$ at each consecutive pair of timesteps, the experimenter would determine the Markov order to be $\ell = 2$.

Interestingly, we can also have the opposite scenario occur, *i. e.,* a process can display finite Markov order with respect to a fuzzy measurement sequence, but given access to the system at a finer resolution, the experimenter would attribute a longer memory length to the process, as we now show.

**Example C.2** (Fuzzy measurements can decrease classical Markov order)**.** Consider the classical process depicted in the middle panel of Fig. C.1. Here, three bits $x_j = \{0, 1\}$ are output by some process in succession over three timesteps $\{t_1, t_2, t_3\}$. Suppose that these bits are initially prepared according to the probability distribution $\mathbb{P}_{3:1}(x_3, x_2, x_1)$ which is such that $\mathbb{P}_{3:1}(0,0,0) = \mathbb{P}_{3:1}(1,0,1) = \mathbb{P}_{3:1}(0,1,1) = \mathbb{P}_{3:1}(1,1,0) = \frac{1}{4}$, and the rest of the possibilities vanish. The process thus constructed is such that if the bit output at the second step is measured to be 0, then the first and third bits are perfectly (classically) correlated; whilst if bit at the second step is measured to be 1, then the first and third bits are perfectly (classically) anti-correlated. Thus, the process is perceived to be non-Markovian with respect to sharp measurements of the second bit value. On the other hand, on average, there is no correlation between the first and third bits; thus, with respect to a coarse-grained measurement that sums over outcomes of the second bit value, the process is perceived to be Markovian.

In either of the above cases, the perceived memory length of the process is instrument-dependent: the first example is a process that is Markovian, but exhibits non-Markovian statistics to the experimenter; whilst the second example is a non-Markovian process that looks Markovian on average, *i.e.,* with respect to the coarse-graining instrument.

## C.4 MEMORY LENGTH OF A GENERALISED COLLISION MODEL WITH MEMORY VIA REPEATED SYSTEM-ANCILLA INTERACTIONS

In Section 4.3 we introduced a type of underlying system-environment dynamics that arises from a generalised collision model where the system interacts $\ell$ times with each ancilla in the order depicted in Fig. 4.3. We claimed that the state of the system subject to such dynamics interspersed with the application of $\ell$ trash-and-prepare operations can be expressed as a function of only the last $\ell$ preparations. Here, we explicitly prove this statement.

Consider, without loss of generality, the case for $\ell = 2$ (the extension to larger $\ell$ is straightforward). Writing out the specific form of the collision model dynamics considered here explicitly, the final output state of the system following two trash-and-prepare instruments, with the re-preparations of the system state at time $t_j$ represented by $\sigma_j^S$, is given by

$$\rho_3^S = \text{tr}_{A_4A_3} \left[ \mathcal{U}_{3:2}^{SA_3}\mathcal{U}_{3:2}^{SA_4}\sigma_2^S \text{tr}_{SA_2} \left[ \mathcal{U}_{2:1}^{SA_2}\mathcal{U}_{2:1}^{SA_3}\sigma_1^S \text{tr}_{SA_1} \left[ \mathcal{U}_{1:0}^{SA_1}\mathcal{U}_{1:0}^{SA_2}\rho_0^S \otimes \tau^E \right] \right] \right]$$
$$= \text{tr}_{A_4A_3} \left[ \mathcal{U}_{3:2}^{SA_3}\mathcal{U}_{3:2}^{SA_4}\sigma_2^S \otimes \tau^{A_4} \ldots \text{tr}_{SA_1} \left[ \mathcal{U}_{1:0}^{SA_1}\tau^{A_1} \otimes \mathcal{U}_{1:0}^{SA_2}\rho_0^S \otimes \tau^{A_2} \right] \right], \quad \text{(C.17)}$$

where the initial environment is composed of ancillas, $\tau^E = \tau^{A_1} \otimes \tau^{A_2} \otimes \tau^{A_3} \otimes \tau^{A_4}$, and in the second line we pulled these individual ancillas through the maps that do not act upon them.

Now note that we can write the joint $SA_2$ state after the first interaction, *i.e.*, $\mathcal{U}_{1:0}^{SA_2}\rho_0^S \otimes \tau^{A_2}$, as $\overline{\rho}_0^S(\rho_0^S, \tau^{A_2}) \otimes \overline{\tau}^{A_2}(\rho_0^S, \tau^{A_2})$, where $\overline{\rho}_0^S(\rho_0^S, \tau^{A_2}) := \text{tr}_{A_2} \left[ \mathcal{U}_{1:0}^{SA_2}\rho_0^S \otimes \tau^{A_2} \right]$ and similarly for $\overline{\tau}^{A_2}(\rho_0^S, \tau^{A_2})$. This simply expresses the post-interaction marginal states (marked with the overline) as a linear map acting on the pre-interaction states. Importantly, despite the tensor product, this notation does not imply a product state of $SA_2$, because of the cross-dependency of the input states; to make this clear, we clearly track this dependency of states through the process with respect to arbitrary unitary interactions, as we are interested in understanding how far into the future their influence can persist. Continuing from above and repeatedly applying this method, we yield

$$\rho_3^S = \text{tr}_{A_4A_3} \left[ \mathcal{U}_{3:2}^{SA_3}\mathcal{U}_{3:2}^{SA_4}\sigma_2^S \otimes \tau^{A_4} \text{tr}_{SA_2} \left[ \mathcal{U}_{2:1}^{SA_2}\overline{\tau}^{A_2}(\rho_0^S, \tau^{A_2}) \otimes \mathcal{U}_{2:1}^{SA_3}\sigma_1^S \otimes \tau^{A_3} \right] \right]$$
$$= \text{tr}_{A_4A_3} \left[ \mathcal{U}_{3:2}^{SA_3}\overline{\tau}^{A_3}(\sigma_1^S, \tau^{A_3}) \otimes \mathcal{U}_{3:2}^{SA_4}\sigma_2^S \otimes \tau^{A_4} \right]$$
$$= \text{tr}_{A_4A_3} \left[ \mathcal{U}_{3:2}^{SA_3}\text{tr}_S \left[ \mathcal{U}_{2:1}^{SA_3}\sigma_1^S \otimes \tau^{A_3} \right] \otimes \mathcal{U}_{3:2}^{SA_4}\sigma_2^S \otimes \tau^{A_4} \right]$$
$$= \mathcal{M}(\sigma_1^S, \sigma_2^S). \quad \text{(C.18)}$$

Here, in the penultimate line, we re-expanded $\overline{\tau}^{A_3}(\sigma_1^S, \tau^{A_3})$ to make explicit the fact that $\rho_3^S$ is a function of only the previously 2 prepared states, which can be written as a linear map $\mathcal{M} : \mathsf{BL}(\mathcal{H}_1^S \otimes \mathcal{H}_2^S) \to \mathsf{BL}(\mathcal{H}_3^S)$ as in the final line, with no dependency on prior historic states such as $\rho_0^S$. Through time-translational invariance, the proof method holds for arbitrary timesteps and the extension to longer $\ell$ is immediate. Indeed, the process depicted in Fig. 4.3 has a length-$\ell$ memory with respect to the trash-and-prepare protocol.

If, on the other hand, one were to apply a different instrument, then the output state here, denoted $\rho_3'^S$, would in general show dependence on the historic state $\rho_0^S$. Consider for concreteness that an experimenter were to first apply a trash-and-prepare instrument and then at the second timestep a measurement on the system of some outcome $m$ followed by an independent re-preparation of the system into the state $\sigma_2^S$. Changing the second operation to a measurement and re-preparation amounts to introducing the local system measurement operator, $\Pi_2^{(m)}$, into Eq. (C.18) directly after the joint unitary dynamics $\widetilde{\mathcal{U}}_{2:1}$ as follows

$$\rho_3'^S = \tag{C.19}$$
$$\text{tr}_{A_4 A_3}\left[\mathcal{U}_{3:2}^{SA_3}\mathcal{U}_{3:2}^{SA_4}\sigma_2^S \otimes \tau^{A_4}\text{tr}_{SA_2}\left[\Pi_2^{(m)}\mathcal{U}_{2:1}^{SA_2}\overline{\tau}^{A_2}(\rho_0^S,\tau^{A_2}) \otimes \mathcal{U}_{2:1}^{SA_3}\sigma_1^S \otimes \tau^{A_3}\right]\right].$$

However, since the system and ancillas $A_2$ and $A_3$, in general, build up correlations during the interactions $\mathcal{U}_{2:1}^{SA_2}$ and $\mathcal{U}_{2:1}^{SA_3}$, the ancillary state of $A_3$ that feeds forward into the next step of dynamics will be conditioned upon the measurement outcome $m$, which implicitly depends upon the initial system state $\rho_0^S$; indeed, the future dynamics proceeds differently for distinct histories. Explicitly, the furthest we can proceed is to write

$$\rho_3'^S = \text{tr}_{A_4 A_3}\left[\mathcal{U}_{3:2}^{SA_3}\overline{\tau}^{A_3}(m;\rho_0^S,\sigma_1^S,\tau^{A_2},\tau^{A_3}) \otimes \mathcal{U}_{3:2}^{SA_4}\sigma_2^S \otimes \tau^{A_4}\right], \tag{C.20}$$

where

$$\overline{\tau}^{A_3}(m;\rho_0^S,\sigma_1^S,\tau^{A_2},\tau^{A_3}) \tag{C.21}$$
$$:= \text{tr}_{SA_2}\left[\Pi_2^{(m)}\mathcal{U}_{2:1}^{SA_2}\overline{\tau}^{A_2}(\rho_0^S,\tau^{A_2}) \otimes \mathcal{U}_{2:1}^{SA_3}\sigma_0^S \otimes \tau^{A_3}\right].$$

Without knowledge of $\rho_0^S$, the output state $\rho_3'^S$ when this instrument sequence is applied cannot be specified and hence the process displays memory effects that persist longer than $\ell$ timesteps when a specific measurement, rather than an averaging over such measurements, is recorded.

## C.5 OTHER GENERALISED COLLISION MODELS WITH MEMORY

The example introduced in Section 4.3 presents a generalisation of a collision model to include the possibility of memory effects; however, its construction provides by no means the only way to build memory into collision models, which we now briefly explore for the curious reader. A discrete-time, $n$-step memoryless collision model consists of a system $S$ interacting with an environment $E$ that has a particular structure: it is made up of a number of constituent ancillary subsystems, $A_j$, with the dynamics proceeding through successive unitary collisions between the system and ancillas (see the top panel in Fig. 4.2). A memoryless collision model assumes the following three key points

1. The system only interacts with each ancilla once.
2. There are no ancilla-ancilla interactions.
3. The ancillas are initially uncorrelated.

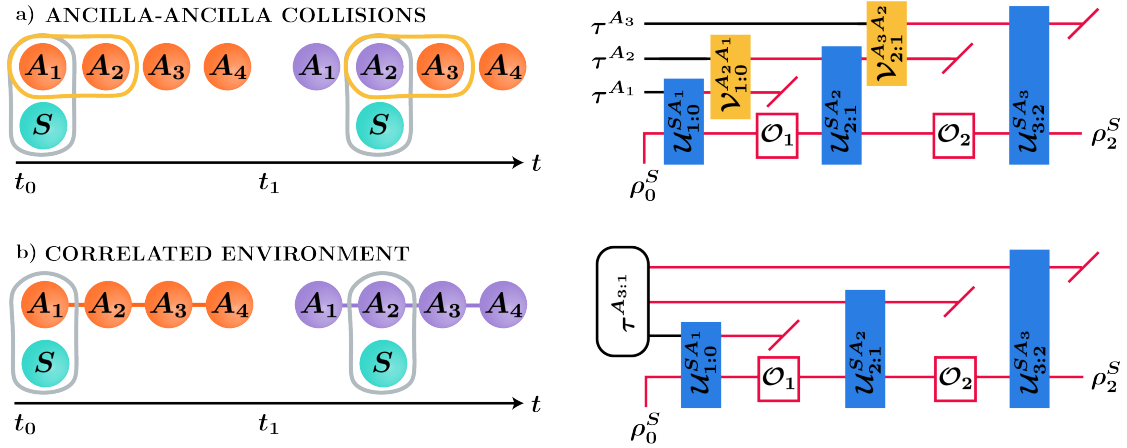

**Figure C.2:** *Generalised collision models with memory.* Memory can also be built into collision models by allowing for: **a)** ancilla-ancilla interactions (top row) and **b)** an initially correlated environment (bottom row). (The legend is as per Figs. 4.2 and 4.3). The top-left panel depicts a schematic of the dynamics where ancilla-ancilla interactions (yellow boundary) are interleaved between the system-ancilla collisions (grey boundary). After $t_1$, the ancilla $A_2$ already has knowledge of the state of the initial system state mediated via the $A_1A_2$ interaction, and so the future dynamics is conditioned on the initial system state. The top-right panel displays the corresponding circuit diagram. Here, for arbitrary operations on the system $\mathcal{O}_j$, it is clear that the ancilla-ancilla interactions provide a possible path of influence from the history to the future state; hence, such a process generically displays infinite Markov order with respect to any instrument sequence (as shown by the red path). The bottom-left panel depicts a schematic of the dynamics where the ancillas constituting the environment begin in a correlated state (represented by the orange line connecting them). As soon as the system interacts with a part of the correlated environment state, all other ancillas can store information about the initial system state, and therefore can influencing the future dynamics. The bottom-right panel displays the corresponding circuit diagram for this case. Again, the initial correlations in the environment provide a mechanism for the history to influence the future over an infinite length of time.

Such a model has surprising power in describing dynamics which, in the continuous-time limit, are governed by a Lindbladian master equation as per Eq. (2.39) [160, 162]. Breaking any one of the above assumptions, whilst maintaining satisfaction of the other two, endows the process with a different type of memory mechanism [174] (see Figs. 4.2, 4.3 and C.2 for illustration). We now examine such memory effects in terms of the structure of the underlying dilation, without any assumptions on the action of the unitaries.

**Case 1: Repeated System-Ancilla Interactions.** As shown in Section 4.3 and Appendix C.4, in the case where one allows for repeated system-ancilla interactions, as in Refs. [170, 171], the memory effect depends on the nature of these repeated collisions. For example, if they occur in the nested order depicted in Fig. 4.3 then the process has Markov order $\ell$ with respect to the trash-and-prepare protocol. If the interactions are simply repeated between the system and a given ancilla multiple times between each timestep, then the process is Markovian on an appropriate coarse-graining of timesteps, *i.e.*, grouping together blocks of $\ell$ timesteps as one. In general, however, repeated system-ancilla interactions give rise to infinite-length memory (even with respect to the trash-and-prepare protocol). This can be seen by considering the dynamics depicted in Fig. 4.3 with the order of any pair of joint unitary operations flipped: now, a continuous path can be drawn from the history to the future across a length-$\ell$ trash-and-prepare protocol, indicating a possible historic influence on the future dynamics.

**Case 2: Ancilla-Ancilla Interactions.** This includes the scenarios considered in Refs. [165–169] and is depicted in the top row of Fig. C.2. In the case where ancilla-ancilla interactions are allowed, the historic influence can, in principle, last forever, since it can permeate continuously through the environment by ancilla-ancilla interactions. Consider specifically the case where at the first step, $S$ is swapped with $A_1$ through the swap map $\mathcal{U}_{1:0}^{SA_1} = \mathcal{G}^{SA_1}$, then during each successive ancilla-ancilla interaction, the initial system state is continually swapped into the next ancilla via $\mathcal{G}^{A_j A_{j-1}}$, before, finally, $A_n$, which now stores the initial system state, is swapped back to the system level through $\mathcal{U}_{n:n-1}^{SA_n} = \mathcal{G}^{SA_n}$. Suppose that all but the first and last system-ancilla interactions are identity transformations and we allow for the application of arbitrary probing operations on the system at each timestep in between. These are represented by $\mathcal{O}_j$, which could, *e.g.*, be trash-and-prepare operations. It is clear that the output system is (trivially) a function of its initial state, regardless of whatever intermediary operations an experimenter applies on the system

$$\begin{aligned}
\rho_n^S &= \mathrm{tr}_{A_n} \left( \mathcal{G}^{SA_n} \mathcal{O}_{n-1}^S \mathrm{tr}_{A_{n-1}} \left( \mathcal{G}^{A_n A_{n-1}} \mathcal{O}_{n-2}^S \ldots \right. \right. \\
&\qquad \ldots \mathrm{tr}_{A_2} \left( \mathcal{G}^{A_3 A_2} \mathcal{O}_1^S \left( \mathcal{G}^{A_2 A_1} \mathcal{G}^{SA_1} \rho_0^S \otimes \tau^{A_1} \otimes \ldots \otimes \tau^{A_n} \right) \right) \Big) \Big) \\
&= \mathrm{tr}_{A_n:A_1} \left[ \mathcal{G}^{SA_n} \mathcal{G}^{A_n A_{n-1}} \ldots \mathcal{G}^{A_2 A_1} \mathcal{O}_{n-1}^S \ldots \mathcal{O}_1^S \mathcal{G}^{SA_1} \rho_0^S \otimes \tau^{A_1} \otimes \ldots \otimes \tau^{A_n} \right] \\
&= \mathrm{tr}_{A_1} \left[ \mathcal{G}^{SA_1} \rho_0^{A_1} \mathcal{O}_{n-1:1}^S \bar{\tau}^S \right] = \rho_0^S.
\end{aligned} \qquad (\text{C.22})$$

Here, we made use of the composition property of the swap map $\mathcal{G}^{AB}\mathcal{G}^{BC} = \mathcal{G}^{AC}$, compressed the description of the operation sequence as $\mathcal{O}_{n-1:1}^S := \mathcal{O}_{n-1}^S \ldots \mathcal{O}_1^S$ and defined $\bar{\tau}^S := \mathrm{tr}_{A_1} \left[ \mathcal{G}^{SA_1} \rho_0^S \otimes \tau^{A_1} \right]$ is the initial state of $A_1$ that is swapped into the system space during the first joint interaction.

Despite the generally infinite-length memory, from the perspective of simulation, this type of memory is not complex: here, given control over part of the environment, one only needs to track one additional ancilla to efficiently simulate such processes, hence the classification of a *memory depth* of 1 [169], even though the memory length here is infinite. Memory depth is the number of additional ancillas required to embed a non-Markovian process as a Markovian one; in other words, a process with a single ancilla-ancilla interaction between timesteps evolves in a Markovian fashion with respect to treating the system and the ancilla it interacts with at each timestep together as a single larger system of interest. In distinction, memory length concerns the number of timesteps back one needs to store information about the state of the system that could influence future dynamics. The notion of memory depth is indeed interesting for further pursuit in regards to understanding the complexity of the underlying memory at hand.

**Case 3: Initially Correlated Environment.** Lastly, consider the case of an initially correlated environment, as is studied in Refs. [163, 164] and is depicted in the bottom row of Fig. C.2. Again, there is no generic way to erase the influence of the state's history on its future evolution by action on the system alone: this is because the ancillary states in the environment begin correlated, and so as soon as the system interacts with the first ancilla, in principle all of the ancillas that *will* interact with the system at some time in the future already store knowledge of the initial system state. Thus, through later interactions, this information can feed-back to dictate the future evolution of the system, giving rise to non-Markovian behaviour.

In the case of an initially correlated environment, one requires control over the entire collection of ancillas to simulate general processes. Again, consider the situation where, $A_1$ and $A_{n-1}$ begin correlated, and at the first interaction $S$ and $A_1$ are swapped. Due to the $A_1$–$A_{n-1}$ correlation, $A_{n-1}$ also stores knowledge of the initial system state, which can be swapped back to the system level at the final interaction to give the final output. At all intermediate timesteps the dynamics looks like the initial state of $A_1$ interacting with each other ancilla pairwise in succession. It is clear that, as in Case 2 above, the final state of the system will be identical to its initial state, regardless of the operations one might perform. However, in contrast, simulation of such processes is generically highly complex, as it requires control over a large number of ancillary subsystems in the environment.

# D

## PROCESSES WITH FINITE MEMORY LENGTH

### D.1 PROCESS WITH FINITE MARKOV ORDER AND NON-ZERO QUANTUM CMI

Here we construct the process tensor for a simple dynamics and show that, whilst it has Markov order 1 with respect to a POVM measurement, the quantum CMI is non-vanishing.

**Example D.1** (Finite Markov order does not imply vanishing quantum conditional mutual information). Consider the case of a three-step process on a qubit, where Alice and Bob have access to the first and second steps respectively, and the final output state is accessible to Charlie (depicted in Fig. D.1). Initially, the following tripartite state is constructed

$$\rho_{ABC} = \sum_b \frac{1}{4} \rho_A^{(b)} \otimes \Delta_B^{(b)} \otimes \rho_C^{(b)}, \tag{D.1}$$

where, for each value of $b = \{1,2,3,4\}$, $\Delta_B^{(b)} := \frac{1}{2}(\mathbb{1} + \sqrt{3} \sum_i \beta_i^{(b)} \sigma^i)$ is defined in terms of Pauli matrices $\{\sigma_i\}$ with tetrahedral coefficient vectors $\{\beta^{(b)}\} = \{(1,1,1), (1,-1,-1), (-1,1,-1), (-1,-1,1)\}$. These objects forms the dual set to the following POVM $\Pi_B$, comprising elements $\Pi^{(b)} := \frac{1}{4}(\mathbb{1} + \frac{1}{\sqrt{3}} \sum_i \beta_i^{(b)} \sigma^i)$. We define the states $\rho_X^{(b)} = \frac{3}{8}\mathbb{1} + \frac{1}{2}\Pi^{(b)}$, with $X = \{A, C\}$ in terms of these POVM elements, before finally normalising the overall tripartite state. The process is such that the $A$ subsystem of the state constructed in Eq. (D.1) is first given to Alice, who can make any operation that she likes. After this, Alice's part is discarded, and the $B$ part of the state above is given to Bob, who can make any operation that he likes. Lastly, Bob's part is discarded, and the $C$ part of the state is given to Charlie. The process tensor is thus given by $\Upsilon_{ABC} = \rho_{ABC}^{\texttt{i}} \otimes \mathbb{1}_{AB}^{\circ}$, where the identity operators on the output spaces of Alice and Bob signify that whatever they feed back into the process is discarded.

Now, suppose Bob chooses to measure the POVM $\Pi_B$ as his instrument (potentially with an arbitrary update to the state after the measurement). Then, Eq. (4.4) holds for each outcome and Alice and Charlie's states are conditionally independent; however, if he chooses any other instrument, Alice and Charlie's states remain correlated (at least for some outcomes). Thus, with respect to the instrument defined by $\Pi_B$, the process has Markov order 1, whereas it has larger Markov order for generic instruments. Importantly, the POVM elements of Bob's measurement are non-orthogonal, so the corresponding instrument has no classical counterpart. Lastly, the quantum CMI of the process tensor does not vanish $I(A:C|B) \approx 0.059$. Nonetheless, knowing Bob's measurement outcome with respect to $\Pi_B$ allows one to reconstruct the entire $ABC$ state and therefore the process.

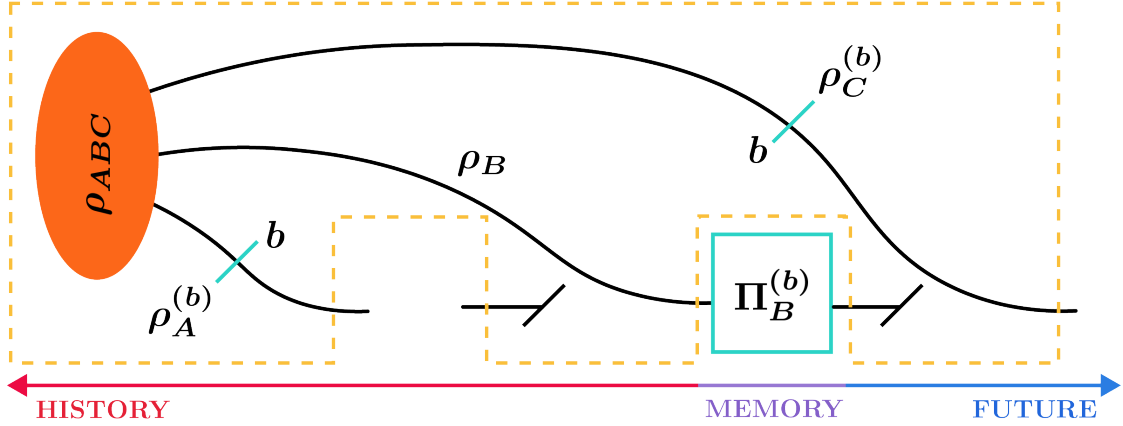

**Figure D.1:** *Process with finite quantum Markov order but non-vanishing quantum CMI.* The process is as described in the text. Temporally, we trivialise the output spaces, so what Alice receives denotes the history (red); Bob's measurement denotes the memory (purple); and what Charlie receives denotes the future (blue). For arbitrary instruments of Bob's choosing, Alice and Charlie's states are, in general correlated; except for when Bob measures with the specific POVM $\Pi_B$. In this case, measurement of any outcome $b$ has the effect of breaking the correlations between Alice and Charlie's subsystems, rendering them in the conditional product state $\rho_{AC}^{(b)} = \rho_A^{(b)} \otimes \rho_C^{(b)}$. Importantly, Bob's instrument is distinct from anything he could implement classically and $I(A:C|B) \neq 0$.

More generally, the example above is a particular case of the following construction: we can consider any $\Upsilon_{FMH} = \sum_x \Upsilon_F^{(x)} \otimes \Delta_M^{(x)} \otimes \widetilde{\Upsilon}_H^{(x)}$ such that $\Upsilon_{FMH} \geq 0$ and $\text{tr}\left[\Delta_M^{(x)\dagger} \mathsf{O}_M^{(y)}\right] = \delta_{xy} \ \forall \ x, y$, where each $\mathsf{O}_M^{(y)}$ is a CP map in a collection that forms an instrument $\mathcal{J}_M = \{\mathsf{O}_M^{(y)}\}$. Such processes can have non-vanishing quantum CMI, $I(F:H|M) > 0$, when the Choi operators of the $\mathsf{O}_M^{(y)}$ do not all commute; indeed, $I(F:H|M)$ is not monotonic with respect to instruments in $M$ (it can increase or decrease, even on average), and is therefore a poor quantifier for memory strength.

Nonetheless, such processes have finite Markov order with respect to the instrument $\mathcal{J}_M$. This observation directly leads to the proposition of Theorem 5.3.

## D.2 FINITE MARKOV ORDER DOES NOT IMPLY VANISHING QUANTUM CMI

*Proof of Theorem 5.3.* From the structure of Eq. (5.9) it is clear that there exists a history-blocking instrument sequence, namely that comprising the projectors onto each of the $m$ orthogonal subspaces. Begin by rewriting Eq. (5.9) as a regular sum by projecting onto the constituent orthogonal subspaces of the decomposition

$$\Upsilon_{FMH}^{\text{CMI}=0} = \bigoplus_m p_m \widetilde{\Upsilon}_{FM^L}^{(m)} \otimes \widetilde{\Upsilon}_{M^RH}^{(m)} \tag{D.2}$$
$$= \sum_m p_m \Pi_{M^L}^{(m)} \widetilde{\Upsilon}_{FM^L}^{(m)} \Pi_{M^L}^{(m)} \otimes \Pi_{M^R}^{(m)} \widetilde{\Upsilon}_{M^RH}^{(m)} \Pi_{M^R}^{(m)}.$$

Consider now the instrument made up of the projectors in the above decomposition, *i.e.*, $\mathcal{J}_M = \{\Pi_{M^L}^{(m)} \otimes \Pi_{M^R}^{(m)}\}$. This constitutes a valid instrument sequence as it sums to an identity operator on $\mathsf{BL}(\mathcal{H}_M)$ which is CPTP. It also constitutes a history-blocking sequence for the process described by $\Upsilon_{FMH}^{\text{CMI}=0}$, as for each realisation of the instrument, the future and history are conditionally independent

$$\text{tr}_M\left[\left(\Pi_{M^L}^{(m')} \otimes \Pi_{M^R}^{(m')}\right)^\text{T} \Upsilon_{FMH}^{\text{CMI}=0}\right] \tag{D.3}$$
$$= \text{tr}_M\left[\sum_m p_m \Pi_{M^L}^{(m)} \widetilde{\Upsilon}_{FM^L}^{(m)} \Pi_{M^L}^{(m)} \otimes \Pi_{M^R}^{(m)} \widetilde{\Upsilon}_{M^RH}^{(m)} \Pi_{M^R}^{(m)} \delta_{mm'}\right]$$
$$= p_{m'} \text{tr}_M\left[\widetilde{\Upsilon}_{FM^L}^{(m')} \Pi_{M^L}^{(m')} \otimes \widetilde{\Upsilon}_{M^RH}^{(m')} \Pi_{M^R}^{(m')}\right]$$
$$= p_{m'} \text{tr}_{M^L}\left[\widetilde{\Upsilon}_{FM^L}^{(m')}\right] \otimes \text{tr}_{M^R}\left[\widetilde{\Upsilon}_{M^RH}^{(m')}\right] = \Upsilon_F^{(m')} \otimes \widetilde{\Upsilon}_H^{(m')},$$

where we use the orthogonal projector identity $\Pi^{(m)}\Pi^{(m')} = \delta_{mm'}\Pi^{(m)}$ and the trace properties of cyclicity and linearity, and in the final line absorbed the probability into $\widetilde{\Upsilon}_H^{(m')}$ to yield the exact form of Eq. (4.4).

We now examine the structure of vanishing quantum CMI processes in further detail: this serves to illuminate the connection between processes with finite Markov order with respect to instruments comprising only orthogonal projectors and those with vanishing quantum CMI, which we explore in Section 5.2. Continuing from Eq. (D.2), note that the projectors in the decomposition are not necessarily rank-1; we can thus expand the conditional process tensor parts in terms of a basis within each $m$ subspace as

$$\widetilde{\Upsilon}_{FM^L}^{(m)} \otimes \widetilde{\Upsilon}_{M^R H}^{(m)} = \Upsilon_F^{(m)} \otimes \Pi_{M^L}^{(m)} \otimes \Pi_{M^R}^{(m)} \otimes \widetilde{\Upsilon}_H^{(m)} \qquad (D.4)$$
$$+ \sum_{ss'} \widetilde{\Upsilon}_F^{(m,s)} \otimes \xi_{M^L}^{(s)} \otimes \xi_{M^R}^{(s')} \otimes \widetilde{\Upsilon}_H^{(m,s')}.$$

The $\xi_{M^{L/R}}$ encode the off-diagonal elements within each $m$ subspace (since the projector $\Pi_{M^L}^{(m)} \otimes \Pi_{M^R}^{(m)}$ encodes all of the diagonal elements), and can therefore be chosen such that $\operatorname{tr}\left[\xi_{M^{L/R}}^{(s)}\right] = 0$ and $\Pi_{M^{L/R}}^{(m)} \xi_{M^{L/R}}^{(s)} = \delta_{ms} \xi_{M^{L/R}}^{(m)} \; \forall \; m, s$. In this expansion, neither $\widetilde{\Upsilon}_F^{(m,s)}$ nor $\widetilde{\Upsilon}_H^{(m,s')}$ are required to be proper process tensors, since the $\xi_{M^{L/R}}$ do not necessarily represent physical operators. We therefore have

$$\Upsilon_{FMH}^{\mathrm{CMI}=0} = \sum_m p_m \Upsilon_F^{(m)} \otimes \Pi_{M^L}^{(m)} \otimes \Pi_{M^R}^{(m)} \otimes \widetilde{\Upsilon}_H^{(m)} \qquad (D.5)$$
$$+ \sum_{m,s,s'} p_m \widetilde{\Upsilon}_F^{(m,s)} \otimes \xi_{M^L}^{(m,s)} \otimes \xi_{M^R}^{(m,s')} \otimes \widetilde{\Upsilon}_H^{(m,s')}.$$

Note that if the $M$ subspaces in the decomposition of Eq. (5.9) (see also Eq. (D.2)) are all 1-dimensional, *i.e.*, the projectors in Eq. (D.2) are all rank-1, we only have the first term in the above equation

$$\Upsilon_{FMH}^{\mathrm{CMI}=0} = \sum_m p_m \Upsilon_F^{(m)} \otimes \Pi_{M^L}^{(m)} \otimes \Pi_{M^R}^{(m)} \otimes \widetilde{\Upsilon}_H^{(m)}. \qquad (D.6)$$

Regarding the converse statement of Theorem 5.3, we have shown examples of processes with finite Markov order with non-vanishing quantum CMI (Examples 5.1, 5.2 and the generalised collision model of Section 4.3 all display this feature, and for an explicit calculation see Ex. D.1). The analysis above shows that the structural constraint required to guarantee vanishing quantum CMI is strict; processes with finite Markov order must only satisfy the more relaxed structure outlined in Theorem 5.1, and it is therefore insufficient to conclude that such processes have vanishing quantum CMI. Furthermore, even if a given process has finite Markov order with respect to an instrument sequence comprising only rank-1, orthogonal projectors, the process can still have non-vanishing quantum CMI. In this case, since any such set of projectors form a self-dual set, we can reconstruct the process via Eq. (5.2) as

$$\Upsilon_{FMH} = \sum_{x=1}^d \Upsilon_F^{(x)} \otimes \Pi_M^{(x)} \otimes \widetilde{\Upsilon}_H^{(x)} + \sum_y \underline{\widetilde{\Upsilon}}_{FMH}^{(y)}, \qquad (D.7)$$

with $\operatorname{tr}\left[\Pi_M^{(x)} \underline{\widetilde{\Upsilon}}_M^{(y)}\right] = 0 \; \forall \; x, y$. Even though the projectors in the history-blocking instrument are not necessarily the same as those that project onto the subspaces defined in the decomposition of Eq. (5.9), this condition does not imply that the process tensor is block-diagonal in some basis of $\mathsf{BL}(\mathcal{H}_M)$; rather, the process can have off-diagonal elements with respect to the subspaces defined by $\{\Pi_M^{(x)}\}$ and satisfy Eq. (D.7). This implies that there are processes with non-vanishing quantum CMI but finite Markov order. □

To summarise, the salient points from this analysis are as follows. Firstly, suppose that a process has finite Markov order with respect to an instrument sequence comprising only orthogonal projectors that are not rank-1: in this case, there is no reason that the future-history correlations within each $m$ subspace must obey the product structure outlined in Eq. (D.6), and hence the process can have non-vanishing quantum CMI. This is shown explicitly in Example D.2 of Appendix D.3. However, similar behaviour also arises in an operational interpretation of classical stochastic processes, as discussed in Section 4.2: if an experimenter cannot measure realisations of the process sharply, *i.e.*, with sequences of rank-1 projectors, then the statistics observed do not necessarily have vanishing classical CMI, even if the true underlying process is one of finite Markov order (see the examples of Appendix C.3).

Secondly, suppose that a process has finite Markov order with respect to an instrument sequence comprising only sharp, orthogonal projectors. The condition $\mathrm{tr}\left[\Pi_M^{(x)}\widetilde{\underline{\Upsilon}}_M^{(y)}\right] = 0 \; \forall \; x,y$ of Eq. (D.7) does not imply that the process must be block-diagonal in some basis of $\mathsf{BL}(\mathcal{H}_M)$, as is required for the quantum CMI to vanish (see Eq. (D.2)), and it follows that there exist such processes with non-vanishing quantum CMI. In contrast to the earlier point regarding instrument sequences comprising higher-rank projectors, the present statement is indeed a fundamentally quantum mechanical phenomenon. In the classical setting, finite Markov order with respect to sharp realisations of the process and the classical CMI vanishing are equivalent statements (see Subsection 2.1.6).

It is lastly interesting to consider why these two notions are equivalent in the classical setting but not for quantum processes. Suppose that a classical process has finite Markov order with respect to the sequence of sharp projectors $\{\Pi_M^{(m)}\}$; then, the process can be written of the form in Eq. (D.7). However, in the classical setting, where there can be no off-diagonal terms, $\mathrm{tr}\left[\Pi_M^{(x)}\widetilde{\underline{\Upsilon}}_M^{(y)}\right] = 0 \; \forall \; x,y$ indeed implies that $\widetilde{\Upsilon}_M^{(y)} = 0$. Alternatively, $d$ orthogonal projectors are informationally-complete in the classical setting; thus, the process must be of the form of Corollary 5.2, with the projectors on the $M$ block. In either case, the process is then of the form of Eq. (D.6) (by choosing either $\mathcal{H}_{M^L}$ or $\mathcal{H}_{M^R}$ to be trivial), meaning the quantum CMI vanishes.

## D.3 FUZZY ORTHOGONAL PROJECTIVE MEASUREMENTS ON A QUANTUM PROCESS

For the sake of completeness, here we provide a quantum mechanical analog of Example C.1. As in the classical case, when fuzzy projective measurements are allowed and such a sequence can block the effect of history on the future, the CMI over the statistics observed does not necessarily vanish.

**Example D.2** (Process with non-vanishing quantum conditional mutual information but finite Markov order for a sequence of fuzzy, orthogonal projectors). Consider the process depicted in Fig. D.2. Begin with the four two-qubit Werner states defined as

$$\rho^{(x)}_{3^i 1^i}(r) := r\beta^{(x)} + (1-r)\frac{\mathbb{1}}{2}, \tag{D.8}$$

where $r \in (0,1)$ and each $\beta^{(x)} \in \mathsf{BL}(\mathcal{H}_{3^i} \otimes \mathcal{H}_{1^i})$ is the projector of one of the four Bell pairs

$$|\psi^\pm\rangle := (|00\rangle \pm |11\rangle)/\sqrt{2} \quad \text{and} \quad |\phi^\pm\rangle := (|01\rangle \pm |10\rangle)/\sqrt{2}. \tag{D.9}$$

Now take some symmetric, IC qubit POVM $\{\Pi^{(x)}_{2^i}\}$, such as the tetrahedral measurement defined in Example D.1. In terms of its dual set $\{\Delta^{(x)}_{2^i}\}$, construct the following state

$$\mu_{3^i 2^i 1^i}(r) := \sum_x \frac{1}{4} \rho^{(x)}_{3^i 1^i}(r) \otimes \Delta^{(x)}_{2^i}. \tag{D.10}$$

This object is positive, and therefore a valid quantum state, only for $r \in (0, 1/3]$, which correspond to the values for which the Werner states defined in Eq. (D.8) are separable. Now, suppose that the system associated to $\mathcal{H}_{2^i}$ represents a qutrit: the first two levels are described by Eq. (D.10), the state of which is mixed with probability $q \in (0,1)$ with an arbitrary tensor product state $\sigma_{3^i} \otimes \sigma_{1^i}$ in product with the third level basis state $|2\rangle$, giving the overall initial system-environment state

$$\rho_{3^i 2^i 1^i}(q, r) = q\mu_{3^i 2^i 1^i}(r) + (1-q)\sigma_{3^i} \otimes |2\rangle\langle 2|_{2^i} \otimes \sigma_{1^i}. \tag{D.11}$$

The process proceeds by initially preparing this state and feeding out the $\rho_{j^i}$ marginal state at each timestep $t_j = \{t_1, t_2, t_3\}$. No matter what operations are implemented on the system at these timesteps, the process acts to discard whatever is fed back into it; therefore, it has trivial output spaces and the corresponding process tensor is

$$\Upsilon_{3^i:1^i}(q,r) = \rho_{3^i 2^i 1^i}(q,r) \otimes \mathbb{1}_{2^\circ 1^\circ}. \tag{D.12}$$

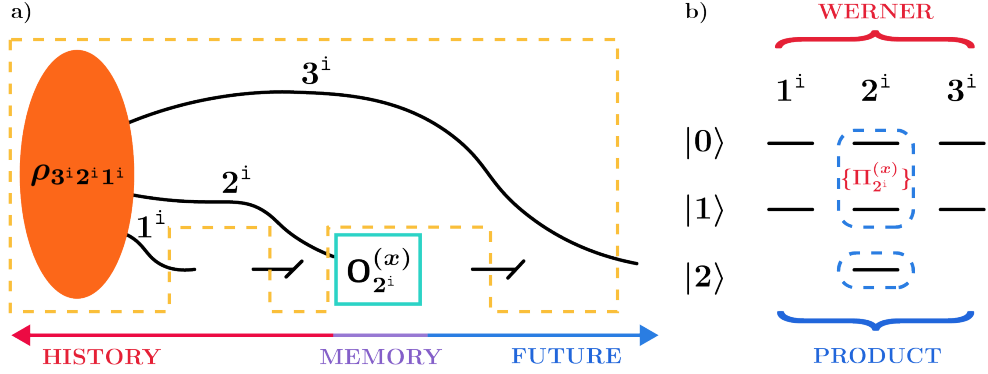

**Figure D.2:** *Process with non-vanishing quantum CMI but finite Markov order with respect to fuzzy, orthogonal projectors.* The tripartite state $\rho_{3^i 2^i 1^i}$ as defined in Eq. (D.11) is depicted on the left. Here, if an experimenter cannot distinguish between measurement outcomes in the $\{|0\rangle, |1\rangle\}$ subspace of $\mathcal{H}_{2^i}$, represented by the dashed, blue boxes on the right, then the conditional state $\rho_{3^i 1^i}^{(x)}$ for each outcome is product. If, on the other hand, the experimenter can resolve sharp measurements in the $\{|0\rangle, |1\rangle\}$ subspace and implement, *e.g.*, the operations $\{\mathsf{O}_{2^i}^{(x)}\} = \{\Pi_{2^i}^{(x)}\}$, then for each outcome realised, the conditional state $\rho_{3^i 1^i}^{(x)}$ is a correlated Werner state, defined in Eq. (D.8). The fuzzy orthogonal measurement at timestep $2^i$ blocks the influence of history on the future, although a sharp measurement resolving all three outcomes does not. Lastly, the quantum CMI for this process does not vanish.

Now, consider the instrument made up of the following two fuzzy, orthogonal operations $\mathsf{O}_{2^i}^{(1)} = (\mathbb{1} - |2\rangle\langle 2|)_{2^i}$ and $\mathsf{O}_{2^i}^{(2)} = |2\rangle\langle 2|_{2^i}$. With respect to this instrument, the conditional process tensors for each outcome are

$$\Upsilon_{3^i 2^o 1^o 1^i}^{(1)} = \frac{\mathbb{1}_{3^i}}{2} \otimes \mathbb{1}_{2^o 1^o} \otimes \frac{\mathbb{1}_{1^i}}{2} \quad \text{and} \quad \Upsilon_{3^i 1^o 1^i}^{(2)} = \sigma_{3^i} \otimes \mathbb{1}_{2^o 1^o} \otimes \sigma_{1^i}. \tag{D.13}$$

Thus, Eq. (4.4) is satisfied and the process has Markov order 1 with respect to this instrument comprising only fuzzy orthogonal projectors. (Note that this process is not Markovian, as an IC instrument of causal breaks does not block the history.) However, had the experimenter been able to resolve measurements in the $\{|0\rangle, |1\rangle\}$ subspace of $\mathcal{H}_{2^i}$, *e.g.*, apply the instrument comprising the operations $\mathsf{O}_{2^i}^{(x)} = \Pi_{2^i}^{(x)}$ for $x \in \{1, 2, 3, 4\}$ and $\mathsf{O}_{2^i}^{(5)} = |2\rangle\langle 2|_{2^i}$, then the conditional process tensors for each outcome are

$$\Upsilon_{3^i 1^o 1^i}^{(x)} = \psi_{3^i 1^i}^{(x)} \otimes \mathbb{1}_{2^o 1^o} \quad \text{and} \quad \Upsilon_{3^i 1^o 1^i}^{(5)} = \sigma_{3^i} \otimes \mathbb{1}_{2^o 1^o} \otimes \sigma_{1^i}. \tag{D.14}$$

For each outcome $x$ observed in the $\{|0\rangle, |1\rangle\}$ subspace, the conditional future and history processes exhibit correlations via one of the four Werner states, which are separable, but not product, and therefore correlated. Similarly, if the experimenter applied the sharp projectors that make up the fuzzy history-blocking instrument, *i.e.*, measure the three outcomes associated to $\{|0\rangle\langle 0|, |1\rangle\langle 1|, |2\rangle\langle 2|\}$ at $t_2$, the conditional states for outcomes 0 and 1 are again correlated. Lastly, note that this process has non-vanishing quantum CMI: a straightforward calculation shows that $I(F:H|M) = q$ for $\Upsilon_{3^i;1^i}(q,r)$ defined in Eq. (D.12).

## DECLARATION

I hereby declare that this thesis contains no material which has been accepted for the award of any other degree or diploma at any university or equivalent institution and that, to the best of my knowledge and belief, this thesis contains no material previously published or written by another person, except where due reference is made in the text of the thesis. This thesis includes ideas presented in two original articles that have been accepted for publication, namely Ref. [1]: "Quantum Markov Order" and Ref. [2]: "The Structure of Quantum Stochastic Processes with Finite Markov Order", and one preprint article Ref. [3]: "Bounding Quantum Process Recoverability with Non-Markovian Memory Strength". The core theme of the thesis is Memory Effects in Quantum Stochastic Processes. The ideas, development and writing up of all the papers in the thesis were the principal responsibility of myself, the student, working within the Monash Quantum Information Science group under the supervision of Dr. Kavan Modi and Dr. Felix A. Pollock. (The inclusion of co-authors reflects the fact that the work came from active collaboration between researchers and acknowledges input into team-based research.) I have renumbered sections of submitted or published papers in order to generate a consistent presentation within the thesis.

*September 10, 2019.*

Philip Taranto